\documentclass[letterpaper,11pt,dvipsnames]{report}

\usepackage{graphicx} 
\usepackage{fullpage}
\usepackage{wrapfig}
\usepackage{xcolor}
\usepackage{tcolorbox}
\usepackage{lipsum}
\usepackage{hyperref}
\usepackage{caption}
\hypersetup{
    colorlinks=true,
    citecolor=blue,
    filecolor=blue,
    linkcolor=blue,
    urlcolor=blue
}
\usepackage{sidecap}
\usepackage{layout}
\usepackage{lineno}
\usepackage{eso-pic,graphicx}
\usepackage[intlimits]{amsmath}
\usepackage{pdfpages}
\usepackage{enumerate}
\usepackage{dcolumn}
\usepackage{makeidx}
\usepackage{ifthen}
\usepackage{adjustbox}
\usepackage{color,soul}
\usepackage{amssymb}
\usepackage{latexsym}
\usepackage{epsfig}
\usepackage{psfig}
\usepackage{tabularx}
\usepackage{longtable}
\usepackage{hhline}
\usepackage{subfigure}
\usepackage{fnpct}

\begin{document}
\normalsize

\thispagestyle{empty}

\AddToShipoutPictureBG*{\includegraphics[width=\paperwidth,height=\paperheight]{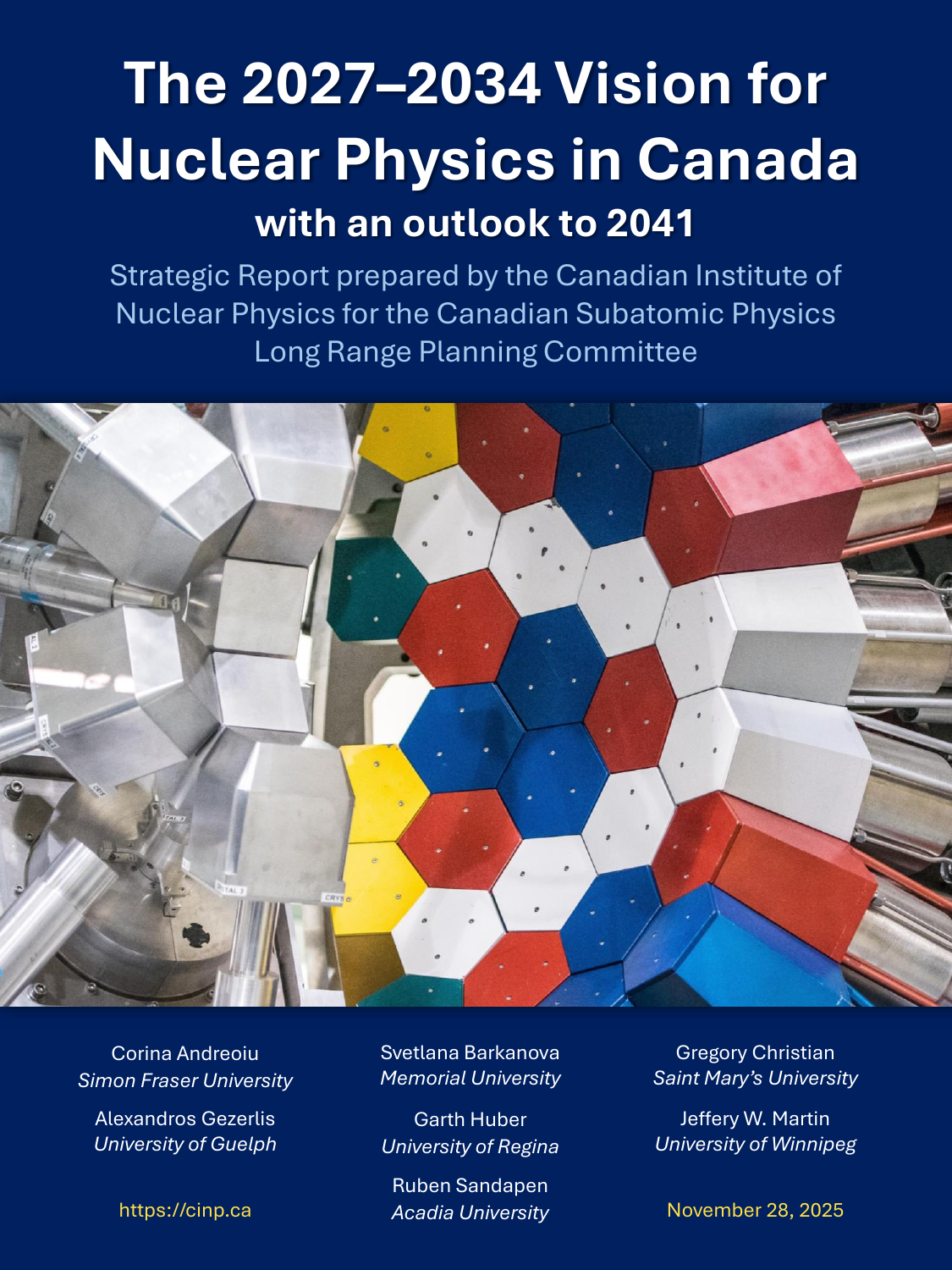}}

%
%
%
%

%
%

\pagenumbering{arabic}

\begin{center}
{\Large
}
\end{center}
\clearpage

\clearpage

\tableofcontents
\newpage

\chapter{Executive summary of recommendations}

The Canadian nuclear physics community engages in a broad and ambitious research program which
address key questions identified by broad international consensus as being of major importance in
understanding the origin, evolution, and structure of visible matter in the universe. 
Researchers take advantage of exciting
and unique opportunities to carry out these endeavours at TRIUMF, SNOLAB, and international
facilities.
The Strategic Report Committee members participated in
the process described in Appendix~\ref{sec:CINP_proc} to
ensure that their information and conclusions 
reflect the consensus of the CINP community. 
In Section~\ref{sec:prio_grid} we provide a grid of
different Canadian nuclear-physics projects and the 
associated timelines.
In Section~\ref{sec:prio_budget} we discuss different budget
scenarios and the impact they are envisioned to have on
different projects.
Our specific recommendations for
maximizing Canadian scientific output in nuclear physics, together with the rationale 
justifying them, are spelled out in Section~\ref{sec:prio_reco}; 
they are listed below in shorter form for convenience.

\begin{enumerate}

\item{Top priority: increasing the NSERC Subatomic Physics envelope:}
The NSERC Subatomic Physics envelope is essential to provide long-term support for national
and international research projects in nuclear physics and should be enhanced. Our analysis
in Sec. \ref{sec:prio_budget} indicated a need of 66\% more funds over the next 7 years. There are several
pressing issues caused by the current funding situation:

\begin{enumerate}

\item
{\em Competitive remuneration needed for students, PDFs and technicians.}  Increased HQP stipends will ensure that we can compensate junior researchers with a living wage, and enhance recruitment and retention of top talent.

\item 
{\em Operating funds for experimental programs.}
 It is essential that research funds permit experiments to operate with the critical number of HQP needed for program success.

\item {\em Opportunities for training in transformative technologies.} Applications of artificial intelligence, quantum science, and superconductors to the pursuit of subatomic physics discovery provide exceptional opportunities for the training of HQP in areas of strategic importance to Canada.

\item 
{\em Increased RTI funds.}  
The RTI-1 ceiling needs to be increased, and more avenues are needed for the funding of medium-size projects (up to \$1M).

\item 
{\em MRS support restored.}  
These cuts are alarming, and we require the envelope to expand to enable this support to be restored.
\end{enumerate}

\item{How Canadian subatomic physics support should be directed:}

\begin{enumerate}

\item 
{\em Support high-profile research addressing the ``big questions'' of nuclear physics.}
It is essential that NSERC and CFI support a broad and diverse research program in nuclear physics that leads to high-profile research in the respective sub-fields in all funding scenarios.

\item 
{\em Increased nuclear theory support.}
Given the close linkage between progress in nuclear theory and experiment, it is essential that nuclear theory research be allocated increased funds, to allow for support of postdocs and graduate students, and to maintain and further develop a leading edge in nuclear physics.

\item 
{\em Support of small-scale impactful research.}
 Resources should be allocated fairly to smaller projects with scientific promise; doing otherwise undermines Canada’s potential to foster innovation in fundamental science.

\item
{\em Leverage the scientific opportunities enabled by the completion of ARIEL.}  
Given the support recently announced by the Government
of Canada, new positions associated with the opportunities associated with ARIEL (both
proton and electron driven beams) are needed, including experiment and theory support.
ARIEL beam delivery will be tripled, and the community’s operating funds to acquire, analyze and interpret the data will need to keep pace.

\item 
{\em Support for Canadians playing integral roles in international research endeavours.}\\
Canada’s excellent international reputation stems not only
from programs based onshore, but also from our involvement in global projects. 
Given the strong budget pressure within the NSERC SAP envelope, it is essential that 
Canadian significant contributions to international nuclear physics research should be supported for both SAP-IN and SAP-PJ.
Regarding two major international projects on the horizon:

\begin{enumerate}

\item 
A substantial involvement in the Electron-Ion Collider (EIC) project will confirm Canada’s leadership role in scientific research and development.

\item 
We encourage the realization of a global nEXO 2.0 experiment utilizing $^{136}$Xe in a timely fashion at SNOLAB.

\end{enumerate}

\item  
{\em Support for an inclusive Canadian nuclear physics community.}
CINP strongly recommends equitable access to available resources and the promotion of equity, diversity and inclusion in the nuclear physics ecosystem in Canada. 

\end{enumerate}

\item{Increased support to enable transformational technologies:}

\begin{enumerate}

\item  
{\em Quantum sensors and high-temperature superconductors.} Canadian nuclear physics researchers lead the development of quantum technologies, superconducting technologies, and associated cryogenic engineering.  New funding mechanisms should be developed to support this important work, which will have benefits to Canadian society.

\item 
{\em High-performance computing, quantum computing, and AI.}  Nuclear physics is driving the development of quantum and AI applications.  It is essential to have ready access to large-scale high-performance parallel computers, including national access to quantum computing infrastructure. The JELF program should continue to support HPC infrastructure upgrades.

\end{enumerate}
\item{The vital roles of TRIUMF:}

\begin{enumerate}

\item  
{\em TRIUMF’s infrastructure role.}
TRIUMF has an important role in supporting Canada’s international physics program, and it is vital that TRIUMF be provided adequate resources for to act
in this capacity, including detector and accelerator development.

\item  
{\em Potential TRIUMF upgrades should be supported in a staged approach:}

\begin{enumerate}

\item 
{\em TRIUMF Storage Ring (TRISR)} would be a unique facility, 
leveraging the connection to the existing ISAC.

\item
{\em Two-step RIB} would be an ambitious but important project, building directly upon
ARIEL after the latter is fully implemented.
\end{enumerate}

\end{enumerate}

\item{Funding mechanism improvements:}

\begin{enumerate}

\item{Enhancements to SAPES operation:}

\begin{enumerate}

\item 
{\em In-person SAPES meetings.}  
Virtual meetings tend to lead participants back to their (pre-determined) entrenched  positions. Thus, 
 we strongly urge SAPES to resume in-person meetings, with the possibility of remote participation when special constraints are at play (paid from the Subatomic Physics envelope, if necessary).

\item 
{\em Enhanced SAPES voting.}
We recommend an increase to seven reviewers per SAPES application.

\item  
{\em Stronger role for academics within NSERC.}
The placement of academics seconded to NSERC could allow these personnel to play a strong 
coordinating role.

\item 
{\em Longer Project Grants.}  
 NSERC should extend the maximum duration of Subatomic Physics Project Grants to 5 years for fully operational and ongoing research programs. 

\item 
{\em SAP envelope management.}
SAPES should have full control of the management of the envelope (duration and amount of funds awarded for both SAP-PJ and SAP-IN grants, and carry forward funds for upcoming projects foreseen in the LRP).  

\end{enumerate}

\item{CFI infrastructure support:}

\begin{enumerate}

\item 
{\em Sufficient CFI funding to sustain the international leadership of Canadian researchers.}
Canadian nuclear physics researchers to play a visible leadership role internationally, both at home and abroad.  To maintain this Canadian leadership, current CFI funding levels should be maintained and enhanced.

\item 
{\em More frequent Innovation Fund (IF) competitions.}
 Predictable and more frequent (annual) CFI IF time lines are essential to allow planning for submissions in the context of large international collaborations.

\item 
{\em Flexibility to handle purchases that are not off-the-shelf.}  
When IF funds are awarded, it is essential that funding rules are sufficiently flexible to allow the necessary R\&D for program success.

\item 
{\em Need for a enchanced mid-scale CFI funding.}
 CFI should provide a nation-wide competition for infrastructure costing below the IF threshold and open to researchers in all stages of their career.

\end{enumerate}

\item  
{\em A better mechanism is needed for large-scale, long-term projects.}
Canada should develop, within ISED, a mechanism to fund large-scale, long-term projects in subatomic physics, potentially in the form of a new national institute with its own dedicated funding and management.  This institute should also have the authority to speak for Canada with our international scientific partners.

\end{enumerate}

\end{enumerate}

\clearpage

\chapter{The big questions of nuclear physics}
\label{sec:big_questions}

\begin{SCfigure}
\centering
     \includegraphics[width=0.4\textwidth]{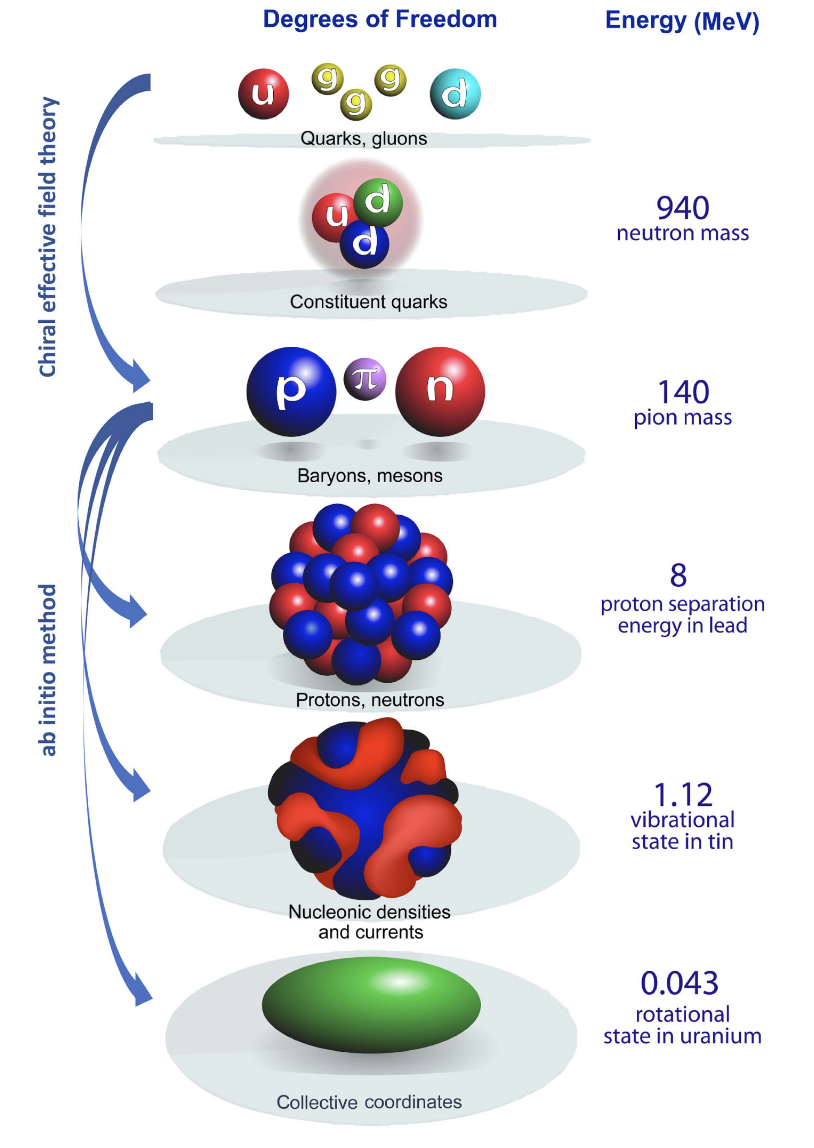}
\caption{The degrees of freedom and energy
scales in nuclear physics, broadly construed.
Source: Z. H. Sun \textit{et al}., Phys. Rev. X 15, 011028 (2025).
} \label{fig:bigquestions_scales}
\end{SCfigure}

Nuclear physics, at the most general level,
tackles problems ranging involving quarks and gluons,
neutrons and protons, entire nuclei, or compact stars.
The big questions introduced in the present chapter
and extensively discussed in the rest of this document
can be viewed along several conceptual axes.
One such way of organizing the discussion
is that shown in Fig.~\ref{fig:bigquestions_scales},
where the relevant degrees of freedom and the corresponding
energy scale are listed (ranging from a thousand
MeV down to less than 0.1 MeV). Schematically, quantum chromodynamics
employs quarks and gluons to build up baryons
and mesons. In chiral
EFT, interactions between nucleons (via the 
exchange of pions) are set up to be consistent
with the symmetries of QCD. \textit{Ab initio}
quantum many-body theories are employed to build
nuclei from scratch (starting from individual
neutrons and protons). Mean-field-like theories
employ nucleonic densities and currents,
while some problems necessitate collective coordinates describing rotations and
vibrations of the nucleus as a whole.
The bulk of the present strategic report focuses on 
the experimental efforts and facilities that 
will allow us to develop a deep understanding
of how nature behaves at all these scales.

\section{How do quarks and gluons give rise to the properties of strongly interacting matter?}

The strong nuclear force governs the interactions
between quarks and gluons (collectively known as
partons) and is described within the Standard Model by
Quantum Chromodynamics (QCD). One remarkable feature of
QCD is that the force between quarks is small at close
distances but grows larger as the quarks separate.
QCD is a complex force with three colour charges
(labeled red, green, and blue).  Gluons are
exchanged between the quarks and carry the
strong nuclear force. Gluons can interact not only
with quarks, but also with each other, leading to interesting consequences. 
Adding to the complexity are the sea quarks, which are quark–antiquark pairs
that are created and destroyed on very short time scales. 
Quarks,
antiquarks, and gluons form particles known as hadrons. The proton and neutron (together known as nucleons) are the most familiar and ubiquitous hadrons in nature. Other shorter-lived
hadrons, including mesons such as the pion, illustrate the varied ways in which QCD manifests in nature.  
One of the great challenges in studying these systems is that 
the colour degrees of freedom carried by quarks and gluons are {\em confined} within the hadron. A full understanding of the strong force at long distances,
where confinement dominates, and how this leads to the observed properties of nuclear matter, remains one of the major unsolved problems of subatomic physics. 

A recent success is the development of {\em nucleon femtography}, the 3D imaging of the internal dynamics of the nucleon.  This is made possible by electron scattering measurements.  The goal is to understand the position and
momentum distributions of the constituent quarks and gluons in the nucleon. 
This will lend clarity to the origin of properties such as the proton's spin, and provide the first picture of the force distribution within the nucleon.
Another approach is the search for hybrid mesons predicted to
exist by lattice QCD calculations as a means to understand how the quark and gluonic degrees
of freedom that are present in the fundamental QCD Lagrangian manifest themselves in the
spectrum of hadrons. 
Measurements of the
electromagnetic form factors of mesons, such as the charged pion and kaon, elucidate the
role of confinement and chiral symmetry breaking in fixing the hadron’s size and mass as well as
the transition from the perturbative-QCD to strongly-coupled domains (short to long distances).
Exotic matter can also be created by colliding nuclei at relativistic energies, creating conditions
similar to those that existed shortly after the Big Bang, which informs the construction of the
phase diagram of nuclear matter.

\emph{Partonic structure of nuclei:} There is a deep connection between the fields of hadronic and nuclear structure physics.  
A notable discovery in the 1980's was the EMC Effect, which found that quarks and gluons (partons) 
have significantly different momentum distributions in a bound nucleon within a nucleus, compared to a free nucleon.  The measured parton distributions in fact are sensitive to nuclear structure details, i.e. surprisingly, the partons know exactly what nucleus they are bound within.
By colliding polarized electrons with a wide range of ions (up to uranium) the EIC will probe how the nuclear environment modifies the internal structure of protons and neutrons, revealing for the first time the 3D structure of the quarks and gluons inside nucleons bound within a nucleus.  This research directly addresses fundamental questions at the intersection of two sub-fields of nuclear physics, such as how the collective phenomena of nuclear physics (such as gluon saturation or the parton distribution modifications of the EMC Effect) emerge from the underlying quark-gluon dynamics of QCD. This presents an opportunity to join Canada's strengths in hadronic structure and nuclear structure in the exploration of this new frontier.

\subsection{Canadian contributions and achievements}


\begin{itemize}
    \item The Canadian theoretical community leverages a range of calculational approaches, including Lattice QCD, Light Front Holographic QCD and Chiral Pertubation Theory to advance the field and to support the Canadian experimental efforts. For example, recent achievements include systematically improvable 
    constraints on the shear and bulk viscosity of 
    the quark-gluon plasma, through a combination of \textit{ab initio} modelling and machine-learning techniques (see Section~\ref{sec:nuctheo_qcd}).
  
\item Canadians have leadership roles in a number of experiments at offshore facilities, including detailed measurements of proton, kaon and pion structure, and investigations of the spectrum of hybrid mesons containing explicit gluonic degrees of freedom. The GlueX project currently taking place at Jefferson Lab aims to measure the properties of hybrid mesons produced through photo-production. Canadians have maintained
responsibility for the gain calibrations of the silicon PMTs for the Barrel calorimeter which
was designed and built in Canada. The pion form factor program at Jefferson Lab has been led by Canadians, garnering over
1000 citations for their work collecting and analyzing the data from various experiments.

\item Canadians are making significant contributions to the Electron-Ion Collider (EIC), a flagship facility under development at Brookhaven National Laboratory in the United States. Canadian physicists and institutions are playing key roles in leadership and the scientific and technical aspects of the project, including detector design, theoretical modeling, and data analysis.  Canadian teams are leading the development of superconducting radio-frequency (RF) crab cavities essential for achieving high luminosity, building on Canada's expertise from the High-Luminosity LHC upgrade. Another major Canadian-led effort is the design and construction of a high-granularity hybrid silicon and lead/scintillating fiber (Pb/ScFi) barrel calorimeter, drawing on Canada's prior experience with the GlueX Barrel Calorimeter at Jefferson Lab. 

\item Canadians continue to play a pivotal role in the investigation of ultra-relativistic heavy ion
collisions in general, and the properties of the quark-gluon plasma in particular, through
the calculations of relevant experimental observables using hydrodynamics techniques and
the development of this formalism.
    
\end{itemize}

\section{How do the structures of atomic nuclei emerge from nuclear forces?}

Atomic nuclei, the core of all visible matter, constitute unique many body systems of strongly
interacting fermions. 
Nuclei are
made of tens, or even hundreds, of neutrons and protons whose interactions, structure, and dynamics are
governed by the interplay of three of nature’s fundamental forces: the strong and weak nuclear interactions and electromagnetism. These forces produce
a tremendous diversity and complexity of nuclear
phenomena, including the organization of neutrons and protons into shells much like electrons in an atom; regular sequences of energy levels caused by rotations
and vibrations that involve many, if not all, nucleons
acting collectively; and clustered states in which protons and neutrons group into substructures.
The core objectives of the nuclear
structure field is to arrive at
a predictive understanding of the properties of atomic nuclei, the limits of their existence, and their behaviour in nuclear decays and reactions.

The properties and structure of nuclei are of paramount importance to many
aspects of physics, at scales from $10^{-15}$ m (the proton radius) to $10^4$ m (neutron star radius), and
to the evolutionary history of the universe. 
Many of the phenomena encountered in atomic nuclei also
share common basic physics ingredients with other mesoscopic systems, thus making nuclear
structure research relevant to other areas of contemporary science, e.g. condensed
matter and atomic physics.
A wide variety of nuclei exist in the universe, but traditional nuclear models are based on
the properties of those that exist on Earth or can be created artificially with relatively long
half-lives. The rare isotopes, with nuclei towards the limit of nuclear binding, provide a new
window into nuclear structure. Their observed properties show unexpected deviations
from current models, thereby challenging our fundamental understanding of nature’s principles in building these many-body quantum systems.

\subsection{Canadian contributions and achievements}

\begin{itemize}
    \item \emph{Theoretical nuclear physics:} Canada is a world-leader in the theoretical descriptions of atomic nuclei from first principles. The ultimate goal of these efforts is to develop a predictive ab-initio theory of nuclear structure and nuclear reactions, to understand nuclei studied at rare isotope facilities.  
    Nuclear-structure studies in Canada also employ mean-field-like techniques for problems outside the
    current reach of \textit{ab initio} techniques, like the confluence of deformation and pairing in heavy nuclei; 
    this involved a proposal of a new state of matter (mixed-spin pairing) to be experimentally
    probed at TRIUMF and elsewhere (Section~\ref{sec:nuctheo_struc}).

    \item \emph{Evolution of single-particle states in nuclei as a function of both neutron and proton numbers:} Experiments with TRIUMF’s advanced facilities have revealed that the filling of specific orbitals can drive nuclei from spherical configurations into prolate, oblate, or triaxial shapes, often resulting in shape coexistence at low excitation energies. In the N = 40 region, high-precision spectroscopy of Zn and Ge isotopes shows rapid transitions and competing shapes linked to neutron excitations across the g$_{9/2}$ orbital, while mass and spectroscopy measurements of heavy Sn isotopes beyond N = 82 reveal weakening of spherical closures and the onset of deformation. Coulomb excitation, transfer reactions, $\beta$-decay studies and mass measurements with radioactive beams—enabled by EMMA, GRIFFIN, TIGRESS, and TITAN—have provided precise observables such as transition strengths, quadrupole moments, and mass surface trends that anchor theoretical models. These benchmarks not only validate state-of-the-art shell-model and \emph{ab-initio} approaches but also establish predictive power for shape evolution across the nuclear chart, directly informing astrophysical nucleosynthesis pathways (Section~\ref{sec:NS-Evolution}).

    \item \emph{Light neutron drip-line nuclei and halos:} First direct evidence of strong quadrupole deformation in the most neutron-rich drip-line nucleus $^8_2$He at the N = 6 comes from a proton inelastic scattering experiment performed at TRIUMF with the IRIS facility that reveals an unbound 2$^+$ state at 3.54(6) MeV with large deformation of $\beta_2$ = 0.40(3), in excellent agreement with ab initio no-core shell predictions. \footnote{M.\ Holl \emph{et al.},\ Phys.\ Lett.\ B 822 136710 (2021).} This finding supports the notion of a closed subshell at N = 6 and open a new paradigm for future investigations (Section~\ref{sec:NS-Evolution}).

    \item \emph{Heavy neutron drip-line nuclei beyond N = 82:} A study on neutron-rich tin (Z =50) isotopes beyond the doubly closed shell of N = 82 through high-precision mass measurements, including the first-ever measurements of the masses of $^{136,137,138}$Sn isotopes was performed with TITAN at TRIUMF.  The new masses of $^{136,137,138}$Sn are used to estimate the final abundances of mass numbers A = 135 and 137 in $r$-process network calculations. \footnote{A. Mollaebrahimi \emph{et al.}, Phys. Rev. Lett. 134, 232701 (2025).} On the nuclear structure side, the measurements reveal changes in pairing gaps and two-neutron separation energies that are in agreement with ab initio predictions in this unexplored region (Section~\ref{sec:NS-Evolution}).
    
    \item \emph{Collective states and shape coexistence in atomic nuclei:} Canadian researchers pursue an influential program exploring shape coexistence in many regions of the nuclear chart both at TRIUMF and at laboratories around the world, using a broad range of techniques. 
    For example, the excited states of $^{74}$Zn$_{44}$ were investigated via $\gamma$-ray spectroscopy following $^{74}$Cu $\beta$ decay with GRIFFIN.\footnote{M. Rocchini \emph{et al.}, Phys. Rev. Lett. 130, 122502 (2023).} The data reveal the first experimental evidence of triaxial deformation and configuration coexistence in this nucleus. Large-scale shell-model calculations reproduce the results and show that the ground state and an excited 0$^{+}$ band have similar average shapes but markedly different softness in the $\gamma$ degree of freedom. These results have broad implications for nuclear-structure theory and for astrophysical models that depend on precise nuclear shape and mass predictions (see Section~\ref{sec:NS-ShapeCoexistence}).

\end{itemize}

\section{How do nuclear processes drive stellar evolution and element formation?}

Humanity has long sought to understand the origin of visible matter and the abundance of the
known stable and long-lived nuclei. Nuclear processes began shaping the universe a few minutes after the Big
Bang, and from the beginning of the cosmos to the
present epoch, they have governed the birth, life, and
death of stars and the physics of some of the most
exotic matter in the universe. 
It has been firmly established that the synthesis of
elements in the universe occurs through a variety of nuclear processes, from quiescent stellar
burning, to dynamic conditions involving the remnants of stellar explosions, to violent neutron star mergers.  

Nuclear astrophysics
is intrinsically an interdisciplinary field, with nuclear
physics at its heart.
A coherent experimental and theoretical effort in
nuclear astrophysics is required for the interpretation of
observational multi-messenger signatures carried
by photons, 
gravitational waves,
neutrinos, and cosmic rays. 
An extremely important event for the nuclear astrophysics community 
was the detection of the gravitational wave signal GW170817 from a binary neutron star merger, followed about 1.7 s later by a short $\gamma$-ray burst (GRB170817A). This triggered a world-wide unique
effort that coined the term ``multi-messenger astronomy'', and about 11 h later the astronomical transient (AT2017gfo), a ``kilonova'', was under close observation by dozens of telescopes in various wavelengths from radio to X-rays. 
The detection of the gravitational wave signal 
along with the electromagnetic transients from the same astrophysical event 
provided the first direct evidence for the production of heavy $r$-process elements in neutron-star mergers, a decades-old proposed site, 
and firmly linked neutron matter to the universe, posing 
new challenges for nuclear physics.
Looking forward, advances in observational capabilities, such as next-generation $\gamma$-ray telescopes, are anticipated to provide a wealth of new observational data which will require significant input from the nuclear physics community to fully understand.

\subsection{Canadian contributions and achievements}

\begin{itemize}
   \item \emph{Theoretical nuclear astrophysics:} Canada has recently grown its efforts
   targeting the theoretical study of nuclear astrophysics.
   This is especially important in today's new multi-messenger era, after the first confirmation of a neutron-star merger via both gravitational-wave and electromagnetic signals.
   A recent example involved using $^{208}$Tl
   as a real-time indicator of astrophysical heavy element production (see Section~\ref{sec:nuctheo_astro}). This is important for the
   $r$-process in neutron-star mergers occurring nearby in Local Group galaxies, as well as other events such as rapidly accreting white dwarfs or special types of AGB stars which may produce ${}^{208}$Tl via other neutron-capture processes.
    

    \item \emph{Domestic direct measurements:} Highlights of the domestic direct-measurement program are: 1) The successful measurement of the ${}^{83}\mathrm{Rb}(p,\gamma){}^{84}\mathrm{Sr}$ reaction at energies relevant to $p$-process nucleosynthesis, using a radioactive ${}^{83}\mathrm{Rb}$ beam and the EMMA recoil separator coupled with TIGRESS $\gamma$-ray detectors. \footnote{G.\ Lotay \emph{et al.}, Phys.\ Rev.\ Lett.\ 127, 112701 (2021).}\textsuperscript{,}\footnote{M.\ Williams \emph{et al.}, Phys.\ Rev.\ C 107, 035803 (2023).} This is the first-ever published direct measurement of a $p$-process radiative-capture reaction using a radioactive beam. 2) A direct measurement of the key astrophysical resonance in the ${}^{26m}\mathrm{Al}(p,\gamma){}^{27}\mathrm{Si}$ reaction using DRAGON, the first-ever direct radiative capture measurement using an isomeric radioactive beam.\footnote{A.~Lennarz \emph{et al}, Phys.\ Rev.\ Lett. 128, 042701 (2022).} (See Section~\ref{sec:AstroDirect} for more details on both experiments.)
    
    \item \emph{Domestic indirect measurements:} A highlight of the TRIUMF indirect measurement program is an experiment using the ${}^{23}\mathrm{Ne}(d,p){}^{24}\mathrm{Ne}$ reaction to measure spectroscopic factors of the mirror states of ${}^{23}\mathrm{Al}(p,\gamma){}^{24}\mathrm{Si}$ reaction. This experiment reduced uncertainties in the strength of the key $E_{cm} = 157$~keV resonance, significantly affecting energy production and the preservation of hydrogen fuel in all but the most energetic $X$-ray bursts (Section~\ref{sec:AstroIndirect}).\footnote{G.\ Lotay \emph{et al.}, Phys.\ Lett.\ B 833, 137361 (2021).}
    
    \item \emph{Offshore measurements:} A highlight of the offshore nuclear astrophysics program is the recent measurement of the bound-state $\beta$~decay of ${}^{205}\mathrm{Tl}^{81+}$ decay rate using the ESR storage ring at GSI. The decay rate for the fully-stripped ions was found to be much longer than the value previously used in astrophysical models, 
    with significant implications for the solar neutrino spectrometry and for the time it took for the solar cloud to collapse in the early Solar System.
    \footnote{G.\ Leckenby \emph{et al.}, Nature 635, 321 (2024).}\textsuperscript{,}\footnote{R.\ S.\ Sidhu \emph{et al.}, Phys.\ Rev.\ Lett.\ 133, 232701 (2024).}

\end{itemize}



\section{What lies beyond the Standard Model?}
\label{sec:bigquestion}

The Standard Model (SM) of particle physics has been spectacularly
successful. Its predictions have been confirmed repeatedly, as
generations of increasingly sophisticated experiments in particle,
nuclear, and atomic physics have pushed down the limits for possible
deviations. At the {\em energy frontier} of this endeavour, collider
physics has not observed new particles beyond the Higgs boson.
Neither have we found any clear signatures of physics beyond the SM at
the {\em precision frontier} with precision measurements that search
for tiny discepancies from expectation that could be hints of new
physics.

Yet there are extraordinarily compelling reasons to believe that the
SM should not be the final answer; it can explain neither dark matter
nor dark energy, it gives no explanation of the predominance of matter
over antimatter in our universe, and it has yet to incorporate a
quantum theory of gravity.  In addition, aspects of the SM, while
reproducing observations correctly, seem contrived, indicating that we
lack deeper understanding.  What determines the masses of the SM
particles?  Why are there three generations of quarks and leptons?
Why does only the weak interaction violate parity, and why is the
violation maximal, while charge-parity (CP) violation seems unnaturally small?  In
addition, we have only started to unravel the details of the neutrino
sector. We press on to find answers to these fundamental questions.

Nuclear physics experiments at the low-energy, {\em precision
  frontier}, have played an important role in trying to answer these
questions, in a complementary way to high-energy
techniques. Advantages of this community are the diversity of efforts,
nimble response to the changing landscape, relatively modest resource
requirements, and diverse HQP training opportunities. Increasingly, a
connection is forming to the emerging field of ``quantum sensing'', or
more broadly ``quantum technology'', promising major gains in
experimental sensitivity.

\subsection{Canadian contributions and achievements}

\begin{itemize}
\item Canadian efforts on beyond-the-standard model physics continue
  to address electroweak radiative corrections of relevance to
  experiments like MOLLER, but have now expanded to also include
  nuclear-structure corrections to fundamental-symmetries questions.
  For example, a recent study was the first-ever \textit{ab initio}
  calculation of such corrections for the $^{10}$C $\rightarrow$
  $^{10}$B superallowed transition, paving the way for a precise
  extraction of the CKM matrix element $V_{ud}$
  (see Section~\ref{sec:nuctheo_funsym}).


\item ALPHA

The ALPHA collaboration demonstrated laser cooling of antihydrogen
atoms \footnote{Nature 592, 35 (2021)} and observed the gravitational
free fall of antihydrogen \footnote{Nature 621, 716 (2023)}. These
accomplishments, realized with significant Canadian leadership, were
once considered only a dream in the field and now open entirely new
opportunities in precision symmetry tests.

\item EDM$^3$

A uniquely Canadian experiment on the electron electric dipole moment
(eEDM) EDM$^3$ has demonstrated the steps required for an eEDM
measurement including detection \footnote{Hessels {\it et al.},
arXiv:2410.04591(2024)}, and optical pumping \footnote{Hessels {\it et
  al.}, arXiv:2410.04605(2024)}.
  
\item TUCAN

  The TUCAN collaboration produced first ultracold neutrons using
  their new source at TRIUMF.  In doing so, they broke their own
  record for ultracold neutron production \footnote{B. Algohi {\it et
    al.}, arXiv:2509.02916}, and are poised to break the world record
  in an upcoming run.  This will open up new possibilities in
  precision for fundamental symmetry tests.  First on the menu will be
  measurements of the neutron electric dipole moment and the neutron
  lifetime.

\item TRINAT

  The TRIUMF neutral atom trap (TRINAT) measured analog-antianalog
  mixing in the $\beta^-$ decay of $^{47}$K \footnote{Kootte {\it et
    al.}, Phys. Rev. C 109, L052501 (2024)}.  This work supports a search
  for violations of time-reversal symmetry in the $\beta$ decay of
  $^{47}$K.
  

\item BeEST     

  Using newly developed quantum technology (superconducting tunnel
  junction detectors), the BeEST experiment provided direct
  experimental constraints on the spatial extent of a neutrino
  wavepacket in the decay of $^7$Be \footnote{J. Smolsky {\it et al.},
  Nature 638, 640–644 (2025)}.  The experiment was conducted at the
  TRIUMF-ISAC facility, the data collection performed in collaboration
  with US labs.
  
\item MOLLER leadership

  In the MOLLER experiment, conducted at Jefferson Lab, the Canadian
  group is currently constructing the 224 main integrating detectors,
  including the supports, PMT bases, and testing the components, and
  is overseeing the construction of the spectrometer magnets which
  they designed.  The Canadian team has also established leadership
  roles in simulation and analysis software.

\item nEXO leadership

  \newcommand{\0}{\ensuremath{0\nu\beta\beta}}

  With the US DOE's recent decision to move forward in the near term
  with an alternate \0 experiment, Canada has been thrust forward into
  the overall leadership of nEXO, a Xenon-based \0 experiment to be
  conducted at SNOLAB.  Canadian scientists have enthusiastically
  accepted responsibility for leading, growing and strengthening the
  collaboration, which benefits from the well-developed nEXO concept.
  
\end{itemize}

\chapter{The Human Element: training HQP for Canada's emerging quantum economy}

The Canadian nuclear physics community positively impacts society through technological innovations, the training of highly qualified personnel (HQP) and science outreach. The technological benefits include, but are not limited to, the diagnosis and treatment of disease through isotope production, radiation therapies, semiconductor detectors, ultra-high vacuum, high-power cryogenic systems, and of course commercial power generation. By participating in a fundamental research project which is technically demanding and contextually fascinating, students have the opportunity to broaden their experience and hone their skills in preparation for the transition to the workforce. In particular, nuclear physics HQP are poised  for a seamless transition into the strategically important fields of AI and emerging quantum technologies. As explorers of the fundamental forces and symmetries that shape our universe, subatomic physicists are quite successful in promoting science to the general public. Through outreach, science becomes a vibrant part of the cultural landscape, sparking curiosity and inspiring young minds to explore STEM (science, technology, engineering and mathematics) fields. As we expand the frontiers of knowledge and share new discoveries, we endeavor not to repeat the mistakes of the past. Recent scapegoating and rollbacks of equity, diversity and inclusion (EDI) initiatives in North America and elsewhere illustrates the all-too-easy recurrence of injustices that historically thrived in academia for persons who are ``different or not of the right kind''. We acknowledge the need to remain proactive in overcoming inequities and to eliminate systemic barriers caused by explicit or unconscious biases. Canadian physicists have the unique responsibility to promote EDI, especially when other collaborators are unable to do so.

\section{Nuclear physics for AI and Quantum Technologies}

The United Nations Educational, Scientific and Cultural Organization (UNESCO) declared 2025 as the International Year of Quantum Science and Technology to celebrate 100 years of Quantum Mechanics and the emergence of new quantum technologies in computing, communication and sensing. These quantum technologies, which harness the fundamental quantum principles of superposition and entanglement, are expected to deliver major economic benefits. According to a study\footnote{Update of the 2017 Report: Quantum Canada: Socio-Economic Impact Assessment, Doyletech, 2020.} commissioned by the National Research Council (NRC) of Canada, the quantum sector will become a $139$ billion industry in Canada with more than 200,000 jobs and $42$ billion in returns by 2045, potentially contributing $3\%$ to Canada's gross domestic product (GDP). AI and quantum computing are expected to propel each other forward, driving innovation. Canada's National Quantum Strategy explicitly calls for efforts to build and retain a \textit{cross-disciplinary} quantum-literate workforce from diverse backgrounds\footnote{Canada’s National Quantum Strategy, Cat. No. Iu4-414/2022E-PDF, ISBN 978-0-660-44884-8}. 

Subatomic physics faculty across Canada play a critical role in addressing the overarching goals of Canada's National Quantum Strategy. Through world-class research, cutting-edge curriculum development, and hands-on student training, they equip the next generation with the skills needed in the forthcoming quantum economy. Their contributions span from foundational science—such as precision measurements and theoretical modeling—to applications of AI and quantum computing in data analysis and simulations. Faculty at institutions, large and small, ensure broad access to expertise, while active participation in national collaborations, such as TRIUMF and SNOLAB, provides students with unique opportunities for interdisciplinary and applied research. By fostering talent, innovation, and inclusive participation, the subatomic physics community is strengthening Canada’s leadership and workforce development in both AI and quantum technologies.

Specifically, a new Scientific Computing department was established in 2020 at TRIUMF and currently includes two primary groups: Big Data $\&$ Distributed Computing  and Machine Learning (ML) $\&$ Quantum Information Systems, under a common umbrella of activities. For example, project CaloQVAE is an application of quantum-assisted generative AI using D-Wave quantum annealers for fast simulation of calorimeters at the high-luminosity Large Hadron Collider (LHC) experiments–a computing load projected to consume millions of CPU years annually in the ‘business-as-usual’ projections. For the EIC, detector layouts are informed by AI studies at URegina,  and event reconstruction algorithms using AI are being developed at UManitoba. The nEXO experiment fosters interdisciplinary skills through hands-on experience including AI-based reconstruction algorithms. TITAN’s ion-trapping and precision measurements are relevant for the quantum computing industry. A former TRINAT postdoctoral fellow now serves as a key NRC scientist in the frequency standards division—an example of true quantum HQP. The RadMol program  uses quantum-enabled technology and pushes experimental sensitivities by exploiting quantum control in molecules potentially all the way to quantum logic spectroscopy. The PIONEER collaboration develops expertise in cryogenic techniques which is highly coveted by the quantum computing industry and provides training in sophisticated data analysis techniques, encompassing ML and simulation methods. The Nab/pNAB experiment requires advanced data analysis skills which are now moving toward implementing AI. A strategic vision (with $15+$ signatories from $8+$ institutions) for leveraging quantum sensing techniques promises to catalyze a new era of fundamental subatomic physics searches in Canada with worldwide impact, in the next decade and beyond, training the next generation of quantum researchers, and establish Canada as a global hub for precision quantum sensing in subatomic physics. 

The McGill Nuclear Theory group offers training in AI and quantum computing for use in relativistic heavy-ion collisions. Randy Lewis at YorkU is training HQP to work on various implementations of lattice gauge theory on available quantum computers. Nicole Vassh’s nuclear astrophysics program at TRIUMF trains HQP on high performance computing with large datasets, statistical methods to link nuclear properties and observables, and has pioneered the first application of ML to classify stellar enrichment\footnote{N. Vassh et al, ApJ 992, 36 (2025)}. At UGuelph, Alex Gezerlis plans to extend non perturbative approaches using Quantum Monte Carlo methods to medium-mass and heavy nuclei, using state-of-the-art AI techniques. Petr Navratil (TRIUMF) also plans to implement and develop quantum algorithms for ab initio nuclear theory. Liliana Caballero (UGuelph) plans to develop new software tools that incorporate ML techniques to solve high-dimensional, nonlinear problems in nuclear astrophysics—such as equation-of-state modeling, reaction network optimization, and parameter inference from observational data. 

The ALPHA program heavily leverages both AI and Quantum Technologies in HQP training. With the help of AI experts in TRIUMF Scientific Computing department, a Deep Learning based particle tracing method has been developed for the ALPHA-g radial Time Projection Chamber, resulting in a publication with undergrad students being the first authors \footnote{J. High Energ. Phys. 2025, 250 (2025)}. Also, techniques required for precision antihydrogen spectroscopy and gravity measurements are directly relevant to quantum sensing and quantum information technologies.

SNOLAB offers a unique facility for studying quantum technologies. For example, superconducting qubits are being operated in the underground laboratory to characterize their performance in a low-radiation environment, in collaboration with the Institute for Quantum Computing at the University of Waterloo. Such efforts are a natural extension of the dark matter research at SNOLAB and strengthen ties between our research community and Canada's quantum sector, leading to development of mutually beneficial technologies and training of in-demand HQP.

Subatomic physicists also contribute to  AI/Quantum awareness and training for a broader audience. For example, Svetlana Barkanova (MemorialU), a subatomic theorist and NSERC Chair for Inclusion in Science and Engineering (CISE), participates in the Quantum Dialogues project led by Institut Quantique (IQ) at the Université de Sherbrooke, in collaboration with the Centre for Responsible Quantum Innovation and Technology (CRQIT), which brings together quantum scientists, social scientists, industry leaders, policymakers, and civil society representatives to collaboratively anticipate and guide the societal impacts of emerging quantum technologies. We offer webinars and YouTube videos on quantum science and technology in English and French, featuring a diverse group of Canadian experts including CINP’s Ruben Sandapen and Svetlana Barkanova, and highlight broader themes of inclusion in STEM. Canadian nuclear physicists also contribute to quantum training abroad. For example, YorkU faculty Randy Lewis and $4^{\mathrm{th}}$ year BSc student, Sarah Powell, were invited in 2023 at JLab in the US to teach at a summer boot camp in quantum computing. Ruben Sandapen, supported by an African Diaspora Fellowship from the Carnegie Corporation of New York, has organized, in 2025, a quantum workshop in Mauritius, with hands-on tutorials on Qiskit (IBM's software for quantum computing).

Subatomic Physics HQP are already landing in Quantum/AI: 
\begin{itemize}
\item Nathan Evetts who obtained a PhD from UBC, working on microwaves and magnetometry with ALPHA, is now a Senior Quantum Engineer at Photonics Inc.
\item Ashley Ferreira, one of the inaugural recipients of the Azuma Fellowship at TRIUMF, played a critical role in the machine learning/AI analysis for the ALPHA-g radial Time Projection chamber.
\item Nawar Ismail (MSc, UGuelph), after publishing
a Phys. Rev. C paper on ML for strongly
interacting fermions, is now a data analyst  at FirstPrinciples, an AI-first non-profit research organization.

 \item  Ali Kavaki (PhD YorkU), who worked on 
the implementation of lattice gauge theory on available quantum computers, is now a research scientist at 1QBit Information Technologies Inc. 

\item YorkU's Sarah Powell is now doing a PhD at InQubator for Quantum Simulation at the University of Washington.

\item Shuze Shi (fomer postdoc in the McGill Nuclear Theory Group) is now an Assistant Professor at Tsinghua University, where he is using ML and quantum computing in subatomic research. 
\item Former McGill visiting PhD student, Jing-An Sun, who contributed to the development of the first end-to-end AI-driven event generator for heavy-ion collisions, has secured an internship at Shanghai AI Laboratory. 

\item Melissa Valdez (MSc, YorkU), a former member of the ALPHA collaboration, is now a Tech-consultant  of AI products at Zafin.
\item	SNOLAB undergraduate co-op students Yusuf Ahmed and Ayesha Iqbal, who  worked on installing superconducting qubits into the CUTE facility underground at SNOLAB, are now pursuing graduate studies at the Institute for Quantum Computing (University of Waterloo).
\end{itemize}

\begin{figure}
\centering
\begin{tcolorbox}[
  colback=white,
  colframe=gray!50!black,
  coltitle=black,
  colbacktitle=blue!20,
  width=0.8\textwidth,
  boxrule=0.8pt,
  arc=4pt,
  auto outer arc,
  fonttitle=\bfseries,
  title=Yusuf Ahmed: from SNOLAB to the IQC]
{\small\baselineskip=14truept 
``I appreciate SNOLAB’s environment of learning. In my experience, SNOLAB encourages its students to explore new skills, ideas and to take initiatives, and scientists are welcoming to questions and willing to devote time to help you understand a concept and/or give you feedback.''
}
\end{tcolorbox}
\end{figure}

\begin{figure}
\centering
\begin{tcolorbox}[
  colback=white,
  colframe=gray!50!black,
  coltitle=black,
  colbacktitle=blue!20,
  width=0.8\textwidth,
  boxrule=0.8pt,
  arc=4pt,
  auto outer arc,
  fonttitle=\bfseries,
  title=Shuze Shi: working at the intersection of Quantum and AI]

\begin{minipage}{0.99\textwidth}

\begin{wrapfigure}{l}{0.2\textwidth}
  \vspace{-10pt}
  \includegraphics[width=0.2\textwidth]{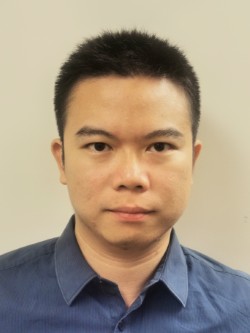} 
  \vspace{-10pt}
\end{wrapfigure}
  
{\small\baselineskip=14truept 
``My postdoctoral training in the Nuclear Theory Group at McGill equipped me with cutting-edge expertise in quantum computation and machine learning. Daily in-depth discussions on challenging nuclear physics problems inspired me to explore novel computational techniques, while my rigorous training in fundamental quantum physics and mathematics provided the foundation to master these advanced tools. This dual expertise enabled a seamless transition into my current role as an Assistant Professor at Tsinghua University, where I developed a machine learning framework to solve inverse problems in nuclear physics and leveraged quantum computing to investigate real-time thermalization in strongly coupled quantum field theories. The interdisciplinary environment at McGill’s Nuclear Theory Group was instrumental in shaping my ability to work at the intersection of quantum science and AI — a testament to its leadership in cultivating talent for these transformative fields."

}
\end{minipage}
\end{tcolorbox}
\end{figure}

\begin{figure}
\centering
\begin{tcolorbox}[
  colback=white,
  colframe=gray!50!black,
  coltitle=black,
  colbacktitle=blue!20,
  width=0.8\textwidth,
  boxrule=0.8pt,
  arc=4pt,
  auto outer arc,
  fonttitle=\bfseries,
  title=Melissa Valdez: from ALPHA to AI and Quantum]
\begin{minipage}{0.99\textwidth}

\begin{wrapfigure}{l}{0.4\textwidth}
  \vspace{-10pt}
  \includegraphics[width=0.4\textwidth]{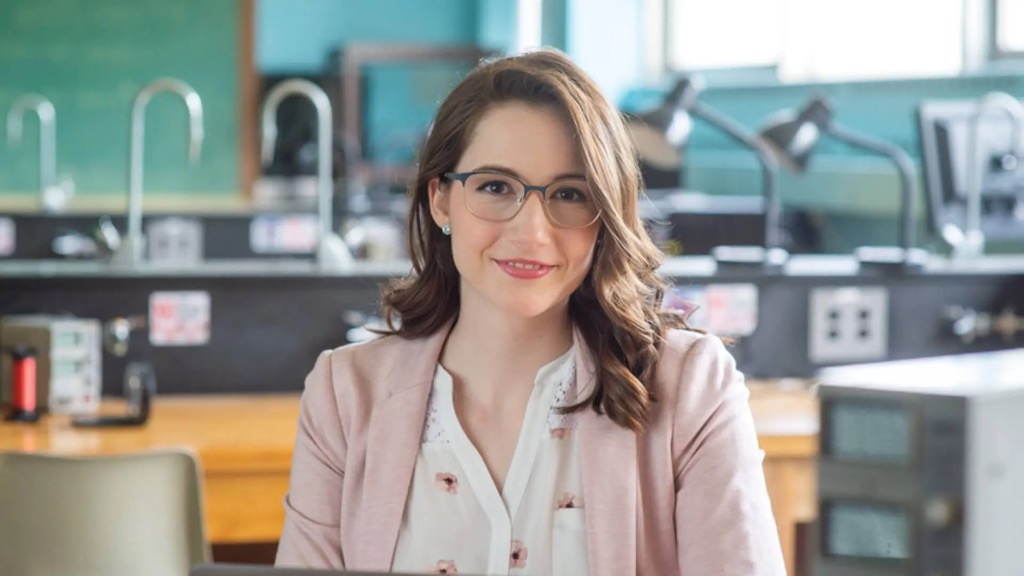} 
  \vspace{-10pt}
\end{wrapfigure}

{\small\baselineskip=14truept 
``Particle physics trained me to solve complex, ill-defined problems, work with massive datasets, and think probabilistically — skills that have become the foundation of my technology career in AI and Quantum Computation. The global, collaborative nature of physics research gave me experience working across cultures and time zones, while also sharpening my ability to communicate technical ideas to diverse audiences. This skillset is one I also use to give back to my community through outreach to children and youth across Canada. 

Most recently I had the pleasure of speaking in the CISE-Atlantic quantum talks series for rural high schools during the International Year of Quantum Science and Technology.”
 }
 \end{minipage}
 
\end{tcolorbox}
\end{figure}

\begin{figure}
\centering
\begin{tcolorbox}[
  colback=white,
  colframe=gray!50!black,
  coltitle=black,
  colbacktitle=blue!20,
  width=0.8\textwidth,
  boxrule=0.8pt,
  arc=4pt,
  auto outer arc,
  fonttitle=\bfseries,
  title=Jing-An Sun: pioneering an AI-driven event generator]
{\small\baselineskip=14truept 
``As a visiting PhD student in the McGill Nuclear Theory Group, I gained invaluable experience, particularly at the cutting edge intersection of AI and nuclear physics. During my time at McGill, I delved into diffusion generative models. The rigorous training in physics was instrumental in deeply grasping the principles behind these models, which themselves have profound physical underpinnings. The collaborative environment, combined with guidance from faculty and peers, enabled us to pioneer the first end-to-end AI-driven event generator for heavy-ion collisions. This unique training at McGill not only honed my research skills but also directly paved the way for my current internship at the Shanghai AI Laboratory and my future aspirations to continue research in ‘AI for Science’. I am deeply grateful for the training and opportunities I received at McGill."
}
\end{tcolorbox}
\end{figure}

\section{Transferable skills}
Canada’s nuclear physics community plays a vital role in training highly skilled personnel equipped with a broad range of transferable skills. Through research in nuclear theory, experiments, detector development, and data analysis, students acquire advanced capabilities in programming, computational modeling, statistics, and precision measurement. Participation in collaborative, often international, projects fosters teamwork, leadership, scientific communication, and project management. These skills are highly valued across sectors, enabling graduates to pursue successful careers not only in academia and national laboratories, but also in medical physics, data science, finance, and other innovation-driven industries.  

Typically, projects are carefully selected, adopted and scaled to match the background and experience of the HQP involved, ranging from introductory, hands-on tasks for high school students, to progressively more complex and independent research responsibilities for undergraduates, graduate students, and postdoctoral fellows. This tiered approach ensures that each trainee is both challenged at an appropriate level and positioned to make meaningful contributions to the broader research program. It also allow students to identify their strengths and acquire diverse transferable skills. Even if they do not pursue graduate studies in subatomic physics, undergraduate HQP acquire skills that make them attractive to the workforce.

HQP also acquire skills in nuclear energy and security. Semiconductor detectors and other radiation detection techniques are needed to scan persons and cargo in a fast and timely manner, to determine the location of any unexpected activity, and to identify the source without false positives for naturally radioactive products. Looking ahead, more time can be spent studying the fission mechanism and mapping its peaks with ARIEL.

\begin{figure}
\centering
\begin{tcolorbox}[
  colback=white,
  colframe=gray!50!black,
  coltitle=black,
  colbacktitle=yellow!20,
  width=0.8\textwidth,
  boxrule=0.8pt,
  arc=4pt,
  auto outer arc,
  fonttitle=\bfseries,
  title=Taraneh Andalib: uncovering insights from large data sets]
\begin{minipage}{0.99\textwidth}

\begin{wrapfigure}{l}{0.28\textwidth}
  \vspace{-10pt}
  \includegraphics[width=0.28\textwidth]{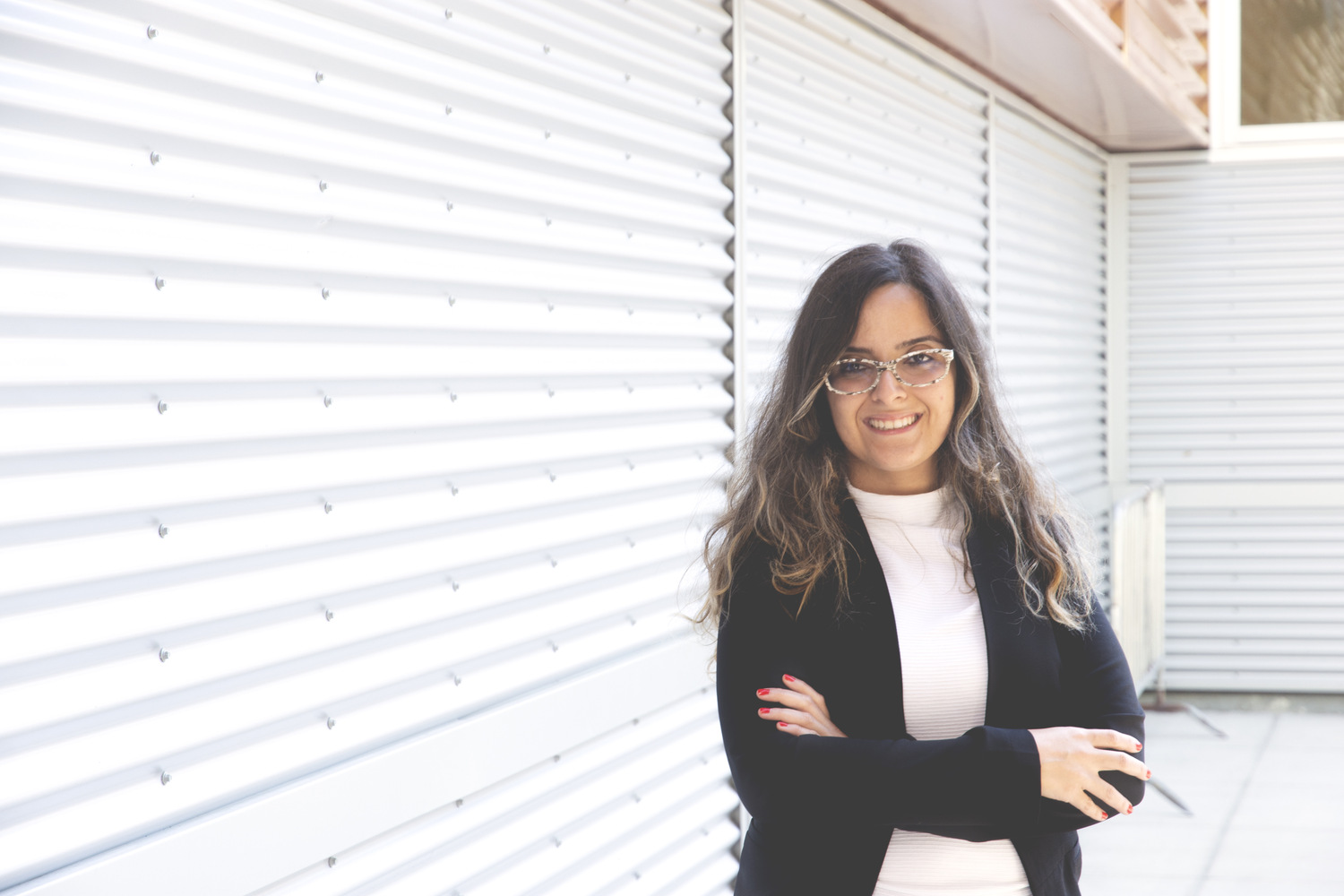} 
  \vspace{-10pt}
\end{wrapfigure}

{\small\baselineskip=14truept 
``My field of study has had a profound impact on my career choice. While at first glance subatomic physics and data science may appear very different, the analytical and methodological skills I developed throughout my training in physics have helped me immensely in my day to day work. The ability to approach and solve complex problems, design and conduct experiments, engage with academic literature, and translate theoretical techniques into practical applications are all direct extensions of my background in subatomic physics. 
Moreover, my experience working with large datasets to uncover fundamental insights aligns perfectly with the needs of data science. This foundation not only informs my daily work but also enables me to maintain a rigorous, research-driven approach that continues to set me apart in the field."

}
\end{minipage}
\end{tcolorbox}
\end{figure}

\begin{figure}
\centering
\begin{tcolorbox}[
  colback=white,
  colframe=gray!50!black,
  coltitle=black,
  colbacktitle=yellow!20,
  width=0.8\textwidth,
  boxrule=0.8pt,
  arc=4pt,
  auto outer arc,
  fonttitle=\bfseries,
  title= Andrew Harrison: the value of summer research]
\begin{minipage}{0.99\textwidth}
\begin{wrapfigure}{l}{0.2\textwidth}
  \vspace{-10pt}
  \includegraphics[width=0.2\textwidth]{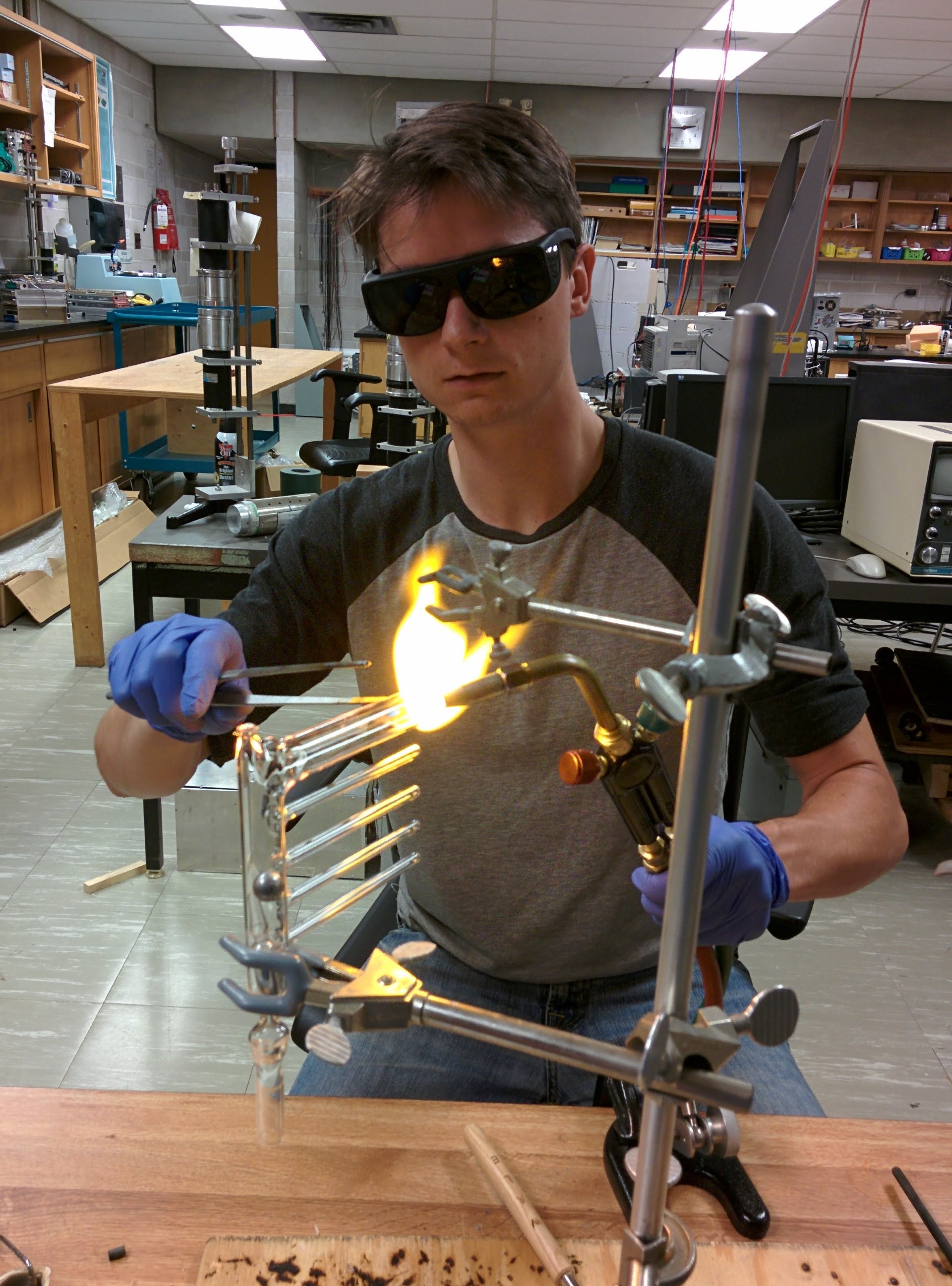} 
  \vspace{-10pt}
\end{wrapfigure}
  {\small\baselineskip=14pt 
``My summer research at the University of Winnipeg was pivotal in securing my internship in smart glass technology. I gained specialized expertise in scientific glass blowing, ultimately crafting a rubidium vapor cell. This work provided experience with materials and optical systems that directly aligned with the innovative demands of the smart glass industry. My demonstrated ability to master complex techniques and apply them to cutting-edge applications made me a compelling candidate for the internship."

}
\end{minipage}
\end{tcolorbox}
\end{figure}

\begin{figure}
\centering
\begin{tcolorbox}[
  colback=white,
  colframe=gray!50!black,
  coltitle=black,
  colbacktitle=yellow!20,
  width=0.85\textwidth,
  boxrule=0.8pt,
  arc=4pt,
  auto outer arc,
  fonttitle=\bfseries,
  title=Kiera Pond Grehan: the value of undergraduate research]

\begin{minipage}{0.99\textwidth}

\begin{wrapfigure}{l}{0.20\textwidth}
  \vspace{-10pt}
  \includegraphics[width=0.20\textwidth]{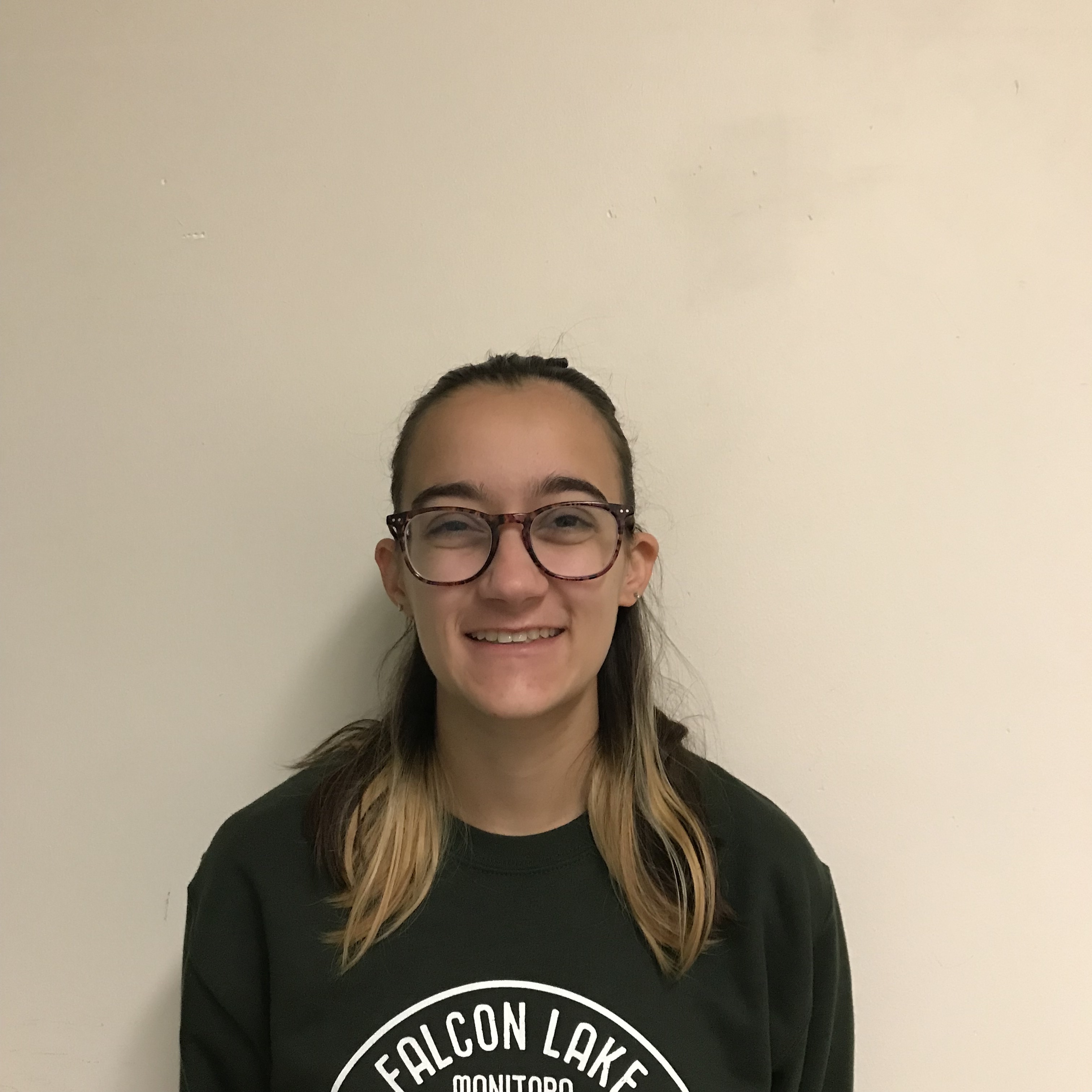} 
  \vspace{-10pt}
\end{wrapfigure}

{\small\baselineskip=14pt
``The subatomic research I conducted during the final two years of my undergraduate education (at UWinnipeg) solidified my fascination with atomic physics. Through this work, I gained experimental experience with coding, optics, and electronics, as well as valuable skills in both small-group collaboration and large international research teams. 

My work with Dr. Martin laid the foundation for my development as a scientist and helped me thrive in my current PhD program (at UToronto). Although my ongoing PhD research is in a different subfield, the skills I acquired during my undergraduate research have proven invaluable early in my physics career.”
}

\end{minipage}

\end{tcolorbox}
\end{figure}

\begin{figure}
\centering
\begin{tcolorbox}[
  colback=white,
  colframe=gray!50!black,
  coltitle=black,
  colbacktitle=yellow!20,
  width=0.85\textwidth,
  boxrule=0.8pt,
  arc=4pt,
  auto outer arc,
  fonttitle=\bfseries,
  title=Frank Wu: a very broad PhD training
]

\begin{minipage}{0.99\textwidth}

\begin{wrapfigure}{l}{0.4\textwidth}
  \vspace{-10pt}
  \includegraphics[width=0.4\textwidth]{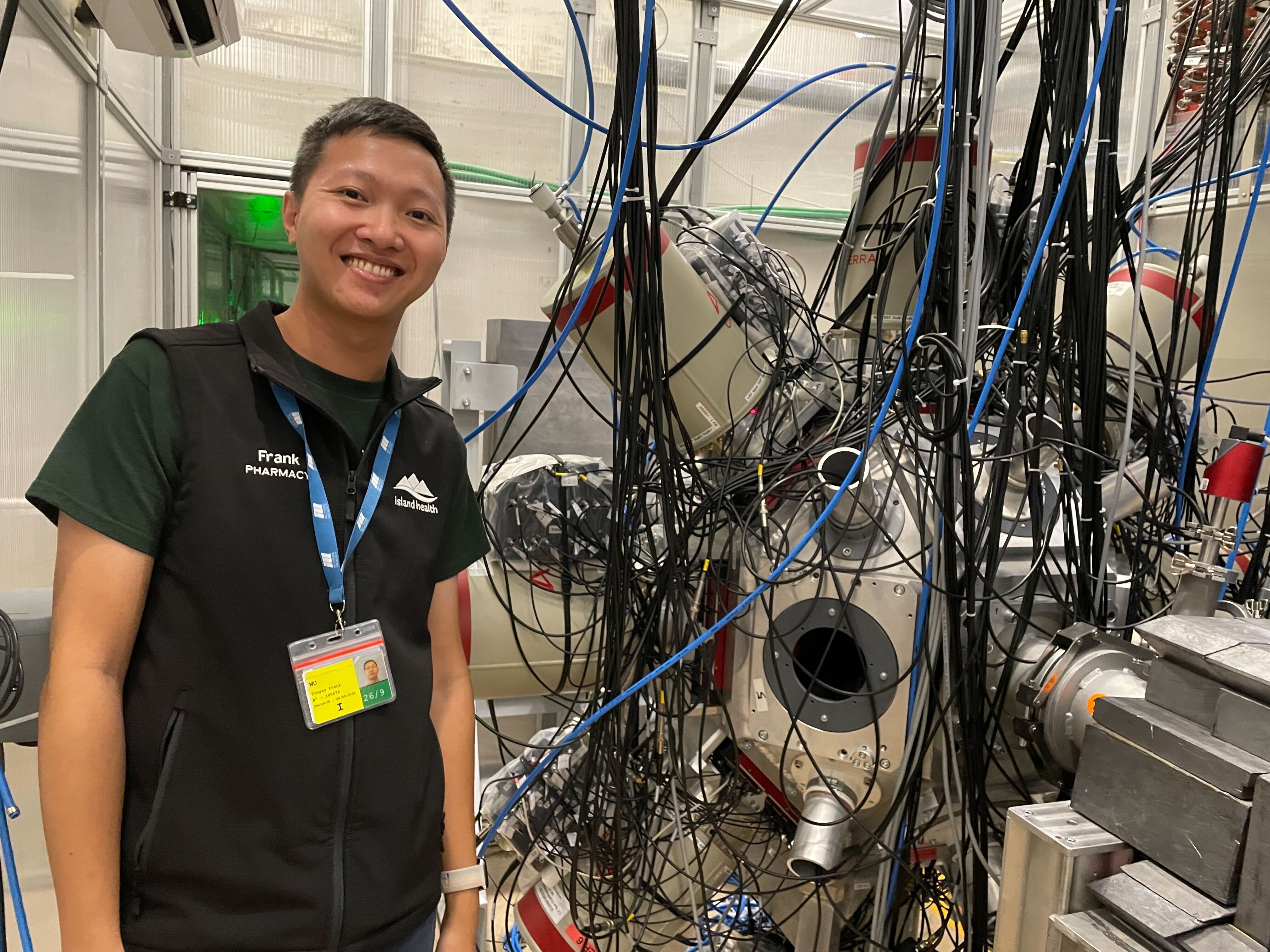} 
  \vspace{-10pt}
\end{wrapfigure}

{\small\baselineskip=14pt
``As a PhD student at SFU, I investigate the shape and structure of tin isotopes by 
measuring their electromagnetic transition rates. While making the first measurements 
of previously unknown observables is always exciting and gratifying, the highlight of 
my PhD experience has been the continued mentorship, friendship, and genuine collegiality 
from not only my local groups, but also the wider international nuclear-physics community. 
I have always felt supported by my peers, whether analyzing experimental data or drafting 
new proposals, and that sense of collaboration is invaluable to me. 

In addition to my thesis work, I have participated in experiments at TRIUMF and other 
leading facilities, and I have deepened my expertise through advanced summer schools: 
GOSIA at HIL in Warsaw; EBSS at FRIB; NUSYS at Fudan University in Shanghai; and TSI 
at TRIUMF. These experiences have provided me with a broader view of nuclear physics 
and enabled me to maximize the impact of my current research and design new high-impact 
experiments.''
}

\end{minipage}

\end{tcolorbox}
\end{figure}

\begin{figure}
\centering
\begin{tcolorbox}[
  colback=white,
  colframe=gray!50!black,
  coltitle=black,
  colbacktitle=yellow!20,
  width=0.8\textwidth,
  boxrule=0.8pt,
  arc=4pt,
  auto outer arc,
  fonttitle=\bfseries,
  title=Rouzbeh Modarresi-Yazdi: nuclear physics to software]
{\small\baselineskip=14truept 
``I use on a daily basis the skills I acquired and refined during my PhD studies (McGill) in nuclear physics. These mostly concern computational and programming skills, where as a part of my studies, I had to familiarize myself with using high-performance computing resources and cloud computing, as well as learning to program in various programming languages. This focus on the computational aspect of nuclear physics has shaped my current career path as a software developer in the private sector. Beyond the computational skills, the training I received in how to conduct independent research, present findings, and collaborate with others has also proven very useful."
}

\end{tcolorbox}
\end{figure}

\begin{figure}
\centering
\begin{tcolorbox}[
  colback=white,
  colframe=gray!50!black,
  coltitle=black,
  colbacktitle=yellow!20,
  width=0.8\textwidth,
  boxrule=0.8pt,
  arc=4pt,
  auto outer arc,
  fonttitle=\bfseries,
  title=Love Preet: from EIC  simulations to small modular reactors]
{\small\baselineskip=14truept 
``While developing a Monte Carlo event generator to perform simulation studies of specific
subatomic physics reactions at the Electron-Ion Collider (EIC)—a cutting-edge, US $2$ billion
next-generation accelerator facility—I acquired a diverse set of skills. These included
proficiency in various programming languages, optimization techniques, statistical analysis,
data integrity assurance, data visualization, and comprehensive data analysis workflows.
Beyond computational work, I also gained valuable hands-on laboratory experience. I
participated in conducting optical fiber tests and in prototype testing (ZDC, calorimeter) for
the EIC, which gave me the opportunity to work directly with advanced detectors and
instrumentation technologies. These experiences not only exposed me to real-world technical
challenges but also enhanced my critical thinking, encouraging me to approach problems from
multiple perspectives—including, at times, thinking outside the box.
These achievements would not have been possible without the generous support of my
supervisor, Dr. Garth Huber, postdoctoral mentor Dr. Stephen Kay, and my colleagues and
collaborators.
Building on this strong foundation of both hardware and software expertise, I was fortunate to
be accepted into a new role—an opportunity for which I am deeply grateful to Dr. Arthur
Situm. In my current position, I conduct simulation studies to analyze the long-term corrosion
behavior of spent nuclear fuel containers in a deep geological repository. While the subject
matter differs from my previous work in subatomic physics, I have been able to successfully
transfer and apply my modeling and research skills. The knowledge and expertise gained during
my earlier work remain immensely valuable in my present role."
}
\end{tcolorbox}
\end{figure}

\begin{figure}
\centering
\begin{tcolorbox}[
  colback=white,
  colframe=gray!50!black,
  coltitle=black,
  colbacktitle=yellow!20,
  width=0.8\textwidth,
  boxrule=0.8pt,
  arc=4pt,
  auto outer arc,
  fonttitle=\bfseries,
  title=Katy Hartling: from subatomic theory to nuclear industry]
\begin{minipage}{0.99\textwidth}
\begin{wrapfigure}{l}{0.3\textwidth}
  \vspace{-10pt}
  \includegraphics[width=0.3\textwidth]{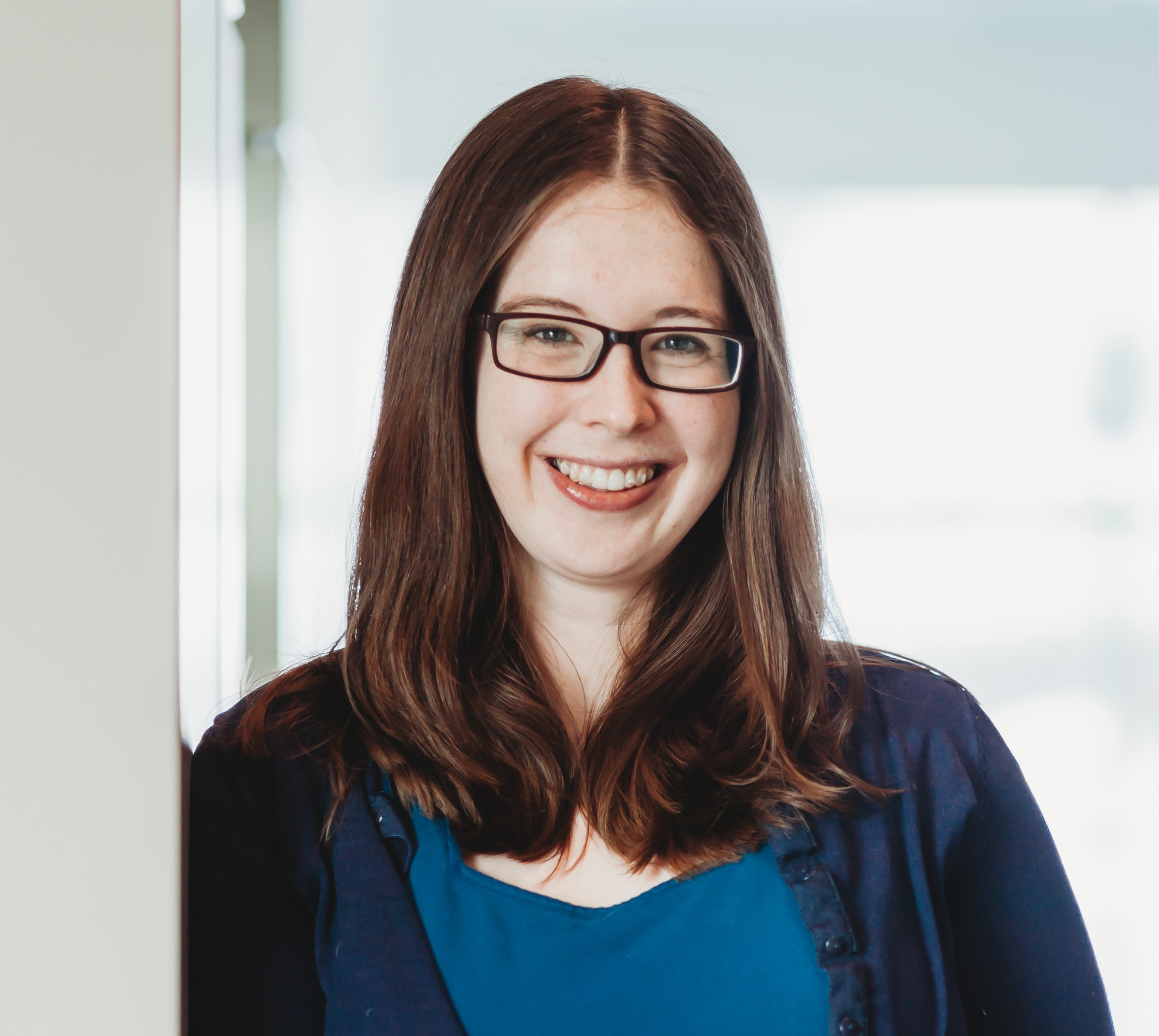} 
  \vspace{-10pt}
\end{wrapfigure}

{\small\baselineskip=14truept 
``My undergraduate (Acadia University 2005-2009, TRIUMF 2008-2009) and graduate (Carleton University 2009-2015) studies and research experience in theoretical nuclear and particle physics provided a solid foundation for my current work in the nuclear industry. Despite a focus on very different subject matter -- e.g. Beyond-the-Standard-Model theories of the Higgs boson and dark matter -- my studies allowed me to develop a varied skillset of data analysis, analytical modeling, simulation, and algorithm development techniques. 

I currently apply that skillset in a wide range of commercial and federal R$\&$D projects at Canadian Nuclear Laboratories, including the development of novel radiation detectors (e.g. gamma, neutron, muon), medical isotope production techniques, shielding design for space exploration, and systems to monitor reactor physics and operations. My work has practical applications in safety and security, medicine and health, and energy generation."
}
\end{minipage}
\end{tcolorbox}
\end{figure}

\begin{figure}
    \centering
    \includegraphics[width=1.0\linewidth]{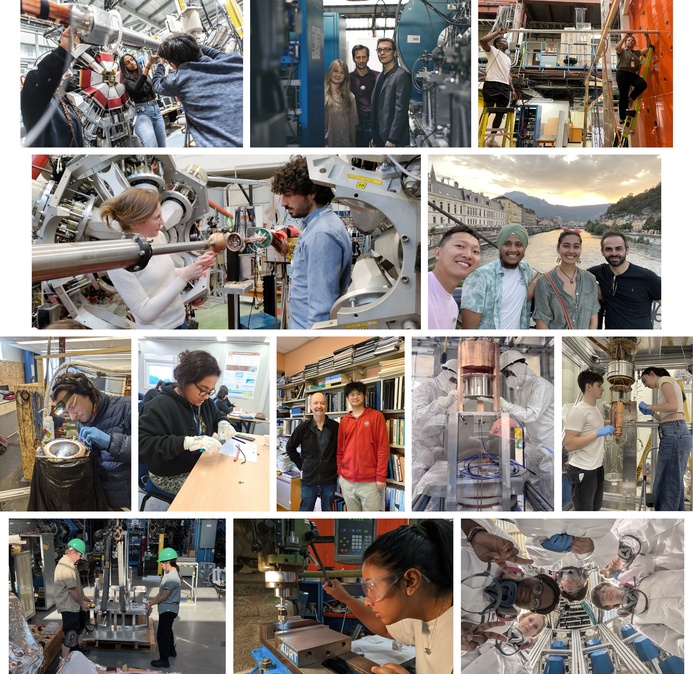}

\caption[HQP ecosystem]{The HQP ecosystem: Knowledge and skills flow not only from faculty and technicians to students, but also from
research assistants (RAs) and graduate students to undergraduate and co\mbox{-}op students,
within a diverse, resourceful environment in Canada and at offshore facilities.
}
 \label{fig:HQP-collage}
\end{figure}

\newpage
The HQP ecosystem, depicted in Fig. \ref{fig:HQP-collage}, encompasses a broad range of knowledge and skills:
\begin{enumerate}
\item{Top row (left to right):
\begin{itemize}
\item TRIUMF and UBC Graduate student, Rashmi Umashankar (left), works with undergraduate co-op students on preparing the GRIFFIN spectrometer for experiments at TRIUMF.   
\item Former PhD students, Eleanor Dunling (left) and
Erich Leistenschneider (middle), and postdoc Moritz Pascal Reiter (right) at the TITAN experiment.  
\item UWinnipeg undergraduate student, Modeste Katotoka (left), and UManitoba PhD student Amala Jaison (right), mapping the TUCAN magnetically shielded room.
\end{itemize}
}
\item Second row (left to right):
\begin{itemize}
\item Dr Kasia Wrzosek-Lipska, Heavy Ion Laboratory (left), and Dr Marco Rocchini, GuelphU (right), making adjustments during a Coulomb excitation experiment in Warsaw. 
\item A stroll through town after a successful neutron capture experiment at ILL Grenoble in France. From left to right: PhD students Frank Wu (SFU) and Sangeet-Pal Pannu (UGuelph), summer student Emily Taddei (SFU), and postdoc Pietro Spagnoletti (SFU).
\end{itemize}
\item Third row (left to right):
\begin{itemize}
    \item Graduate student, Nick Macsai (now a postdoc), performing a continuity test of the Nab Si detector pixel readout pins.        
    \item GuelphU PhD student and CINP graduate fellow, Zarin Ahmed, undergoing training at CERN.
    \item  Nuclear theorist, Charles Gale (left), and visiting PhD student, Jing-An Sun (right), work on AI-driven event generators for heavy-ion collisions at McGill.  
   \item Undergraduate co\mbox{-}op students Ayesha Iqbal and Yusaf Ahmed installing superconducting qubits into the CUTE facility underground at SNOLAB. 
   
   \item Undergraduate student N. Folk and graduate student M. Willett working on Dilution Fridge at TRIUMF.   
   
   \end{itemize}
   \item Bottom row (left to right): 
   \begin{itemize}
          
   \item TRIUMF technician, Shaun Georges, and undergraduate student, Kaylee Directo,  working with the mounts for GRIFFIN detectors at a DRAGON experiment.
    \item Undergraduate co\mbox{-}op student, Serene Rodriguez, manufacturing a holding structure for the acceptance testing of silicon photomultiplier light sensors assemblies as part of construction of ARIES.
    \item Graduate and undergraduate students working on the assembly of the Wall configuration of DESCANT and GRIFFIN at TRIUMF.  
  \end{itemize}  
    \end{enumerate}

\section{HQP: trends and excellence}

\begin{figure}
    \centering
    \includegraphics[width=1.0\linewidth]{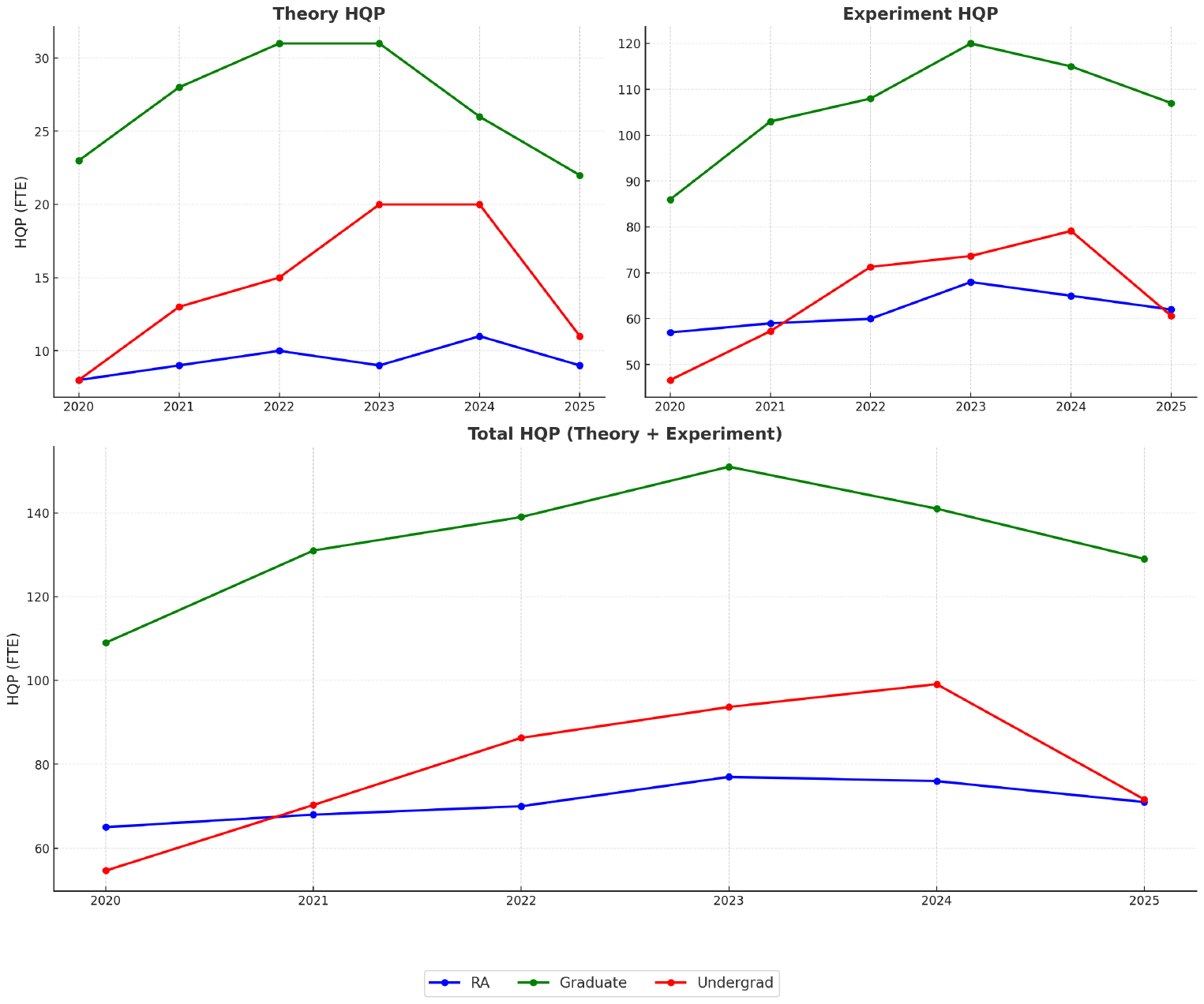}
    \caption{HQP trends for the past 5 years.}
    \label{fig:HQP}
\end{figure}

As a part of our consultations for the Canadian Subatomic Physics Long Range Plan, the CINP requested information from the nuclear physics community to gauge the dynamics and the demographics of the field in respect to the education and training of new personnel in the period of 2020-2025. Fig. \ref{fig:HQP} showing the HQP (full-time equivalent, or FTE) counts per category  (Research Associates (RA), Graduates and Undergraduates), is generated using the data provided with 33 briefs (7 theory, 26 experiment). Note that we did not verify independently these data. Despite new hires in the last years, implying that the community could accommodate a larger HQP workforce, we can see a drop across all $3$ categories in the last year, especially pronounced for undergraduate HQP, both in theory and experiment. 
Such a downward trend was \textit{not} observed in our previous  \href{https://cinp.ca/sites/default/files/2020-12/CINP_2020_LRP_report-600dpi.pdf}{report} (Fig. 5.7) for the period 2015-2019. This highlights the need for an envelope that supports capacity for HQP training.

The excellence of nuclear physics HQP training across Canada is reflected in numerous national and international awards at the PhD, MSc, and undergraduate levels. Among PhD students, standout achievements include: at UBC, Antoine Belley (PhD, 2024), who earned the Carl H. Westcott Fellowship (the first ever awarded to a theorist), the Alexander Graham Bell Canada Graduate Scholarship at both the MSc and PhD levels, UBC’s President’s Academic Excellence Initiative PhD Award, and Best Presentation in Theoretical Physics at WNPPC (Winter Nuclear and Particle Physics Conference) 2021. UBC/TRIUMF PhD student Guy Leckenby won the GSI Exotic Nuclei Community (GENCO) Young Scientists Award and the APPA–SPARC (Atomic, Plasma Physics and Applications; Stored Particles Atomic Physics Research Collaboration) Thesis Prize in 2025, following a Mitacs Globalink Research Award to work at the GSI Darmstadt storage ring. Also at UBC, PhD student Maude Larivière holds a three-year NSERC CGR Scholarship. At the University of Calgary, PhD students Adam Powell (PhD, 2024) and Abby Swadling have each been recognized: Powell received several fellowships, including the CINP Graduate Fellowship, and his thesis was selected for Springer’s Best of the Best Theses Series—the second ALPHA-Canada thesis to earn this distinction—while Swadling, who began as an undergraduate with ALPHA after several summers at CERN, was recently awarded the Izaak Walton Killam Memorial Scholarship. At the University of Regina, PhD student Karthik Suresh won the prestigious 2024 Jefferson Science Associates PhD Thesis Prize. At SFU, PhD student M. Martin received a MITACS Globalink Research Award (2021) and an NSERC-PGS D (Postgraduate Scholarship - Doctoral) (2020), and PhD student Frank Wu won second prize in the Best Student Oral Presentation competition at the 2022 Canadian Association of Physicists (CAP) Congress and the CINP Graduate Fellowship. At the University of Regina, PhD student Alicia Postuma, who participates in Hall C JLab experiments and began her research career with David Hornidge’s group at Mount Allison, was awarded both the 2024–27 Vanier Fellowship and the 2023 CINP Graduate Fellowship.

At the MSc level, notable accomplishments include Ryan Curry, who received the 2024 Governor General’s Gold Medal as the top MSc student across all departments at the University of Guelph, and SFU MSc student Syeda Noor-e-Kainat, who earned the highly competitive 2023–25 IAEA Marie Slodowska-Curie Fellowship. At the undergraduate level, MSc student Andrew Redey (then an undergraduate) earned a distinction for thesis work based on the SCI-CASTER (Segmented Conductor Ionization Chamber for Advanced Spectroscopy
Targeting Emitters of alpha Radiation) project at the NSL (Nuclear Science Laboratory, at SFU) and later received the 2022 Kaiser Foundation Graduate Scholarship in Engineering Science. UBC co-op student Emily Lowe won several awards for a talk on ab-initio calculations at the Canadian Undergraduate Physics Conference.

\section{Equity, Diversity, Inclusion}

Diversity is one of Canada’s greatest strengths, fostering positive work environments and enabling the attraction and retention of top-tier talent. The CINP proudly endorses the Government of Canada’s Dimensions Charter\footnote{Government of Canada 2025, Dimensions: equity, diversity and inclusion Canada, accessed, \href{http://www.nserc-crsng.gc.ca/InterAgency-Interorganismes/EDI-EDI/Dimensions-Charter_Dimensions-Charte_eng.asp}{\today}.} recognizing that a shared national commitment to equity, diversity, and inclusion (EDI) is essential to ensuring that the Highly Qualified Personnel (HQP) pipeline reflects the full diversity of Canada’s population. Such a commitment strengthens the country’s long-term capacity in critical science and technology fields.
Across the country, the nuclear physics community continues to play a significant leadership role in advancing EDI. This includes building inclusive climates within research groups and facilities, providing effective mentoring and coaching to trainees at all stages, and taking proactive steps to recognize and address implicit bias in academic and research environments.

\subsection{EDI initiatives in large collaborations}

Many Canadian large-scale research teams are embedding equity, diversity, and inclusion into decision-making, leadership roles, and trainee support. In 2024, TRIUMF, together with the Laurier Center for Women in Science, completed an arms-length EDI \href{https://triumf.ca/wp-content/uploads/2025/06/TRIUMF-WinS-EDI-Survey-Phase-2-Report.pdf}{survey} sampled over 80$\%$ of their community, showing strengths and deeper structural biases at TRIUMF including representation in leadership and scientific positions, career advancement, and concerns on interpersonal behaviors. Diversity is an important factor in group composition. Therefore, recruitment should be guided by a vision that includes creating dedicated funding opportunities for members of underrepresented groups and widely advertising scholarships. In addition, diversity in terms of nationality must be considered, as the accelerator science community is inherently global. Programs for unconscious bias training in hiring and recruitment, group dynamics, and supervision/mentorship of staff and students have been implemented.  A student-led initiative has matched TRIUMF scientists with early-career HQP for 1:1 mentorship and career guidance. Also at TRIUMF, a new EDI Special Advisor hire at the senior-management level is reformulating policies and best practices on how to engage underrepresented persons and communities in physics from a grassroots level.

The nEXO-Canada team has implemented a Code of Conduct and elected ombudspersons to address EDI concerns impartially. The collaboration is international, and the Canadian team includes a mix of early-career researchers and minority-identifying scientists. Senior team members actively mentor junior colleagues, through a program started by the nEXO EDI Committee. Several team members have received recognition for their mentorship and leadership, reinforcing an inclusive culture. Canadian team members have been elected ombudspersons (Brunner, Caden) and many serve in the EDI Committee. The team also engages in regular EDI training and workshops, further embedding inclusiveness into daily research practices. As “\href{https://arxiv.org/abs/2511.21525}{Striving for Equity in Canadian Physics},” submitted to the International Conference on Women in Physics (ICWIP 2023) proceedings, was proud to report, several Canadian experimental collaborations have now incorporated EDI into their governance frameworks, including nEXO and EIC Canada.

The subatomic community also collaborates with the NSERC Chairs for Inclusion in Science and Engineering to advance the participation and retention of a broad range of equity-deserving groups, including women, Indigenous Peoples, neurodiverse persons and persons with disabilities, members of visible minorities and racialized communities, 2SLGBTQI+ individuals, and students from rural and remote areas. For example, TRIUMF was instrumental in co-developing an innovative program which aims to improve access to engaging physics education in underserved communities, boost student interest and confidence in science, highlight real-world STEM applications and careers, and support teachers in delivering complex content. The program connects students in Grades 7–12 with STEM professionals through live video sessions designed to be engaging and curriculum-relevant and delivered by a diverse group of STEM professionals, many of whom serve as role models for equity-priority students. Presenters may also incorporate Indigenous knowledge and locally relevant examples where appropriate, creating space for culturally responsive science education. 

\subsection{EDI initiatives at universities}

There is a significant number of EDI initiatives led by CINP members at universities across Canada. These actions—typically spearheaded by one or more faculty members—support a wide range of equity-deserving groups, including women and gender minorities, Indigenous Peoples, members of visible minorities/racialized groups, youth in rural and remote communities, and students from non-traditional educational backgrounds. At the University of Winnipeg, Indigenous HQP have served as instructors in the Pathways to Graduate Studies (P2GS) program, an initiative that encourages early-career Indigenous undergraduates to pursue scientific research. One of the first P2GS instructors, Sean Hansen-Romu—who identifies as Indigenous—completed his PhD in 2023 and now works as a data scientist at Scotiabank.
At the University of Regina, Gwen Grinyer is a nationally recognized leader in 2SLGBTQ+ advocacy. Her work has been published in Nature\footnote{G. F. Grinyer, Nature 62, S92 (2022)}, Discourse Magazine, and has been acknowledged through major honours, including the YWCA Regina Woman of Distinction Award for Breaking Barriers (2025), the University of Regina EDI Award (2024), and election as a Fellow of the Canadian Association of Physicists (2024). She integrates this expertise directly into her research program and uses EDI principles as a foundational tool in building her research team.
At TRIUMF, Nicole Vassh focuses on mentoring young women by helping them develop problem-solving abilities, communication skills, and confidence, along with strategies for professional development. In 2023, Svetlana Barkanova (MUN) was selected as one of inaugural NSERC Chairs for Inclusion in Science and Engineering in Atlantic Canada. Her programs include Indigenous youth initiatives, STEM outreach programs in rural and remote communities, and curriculum development for Grades 7–12 with an emphasis on Quantum Science and Technology, Astronomy, Astrophysics, and Subatomic Physics.
At the University of Manitoba, Michael Gericke has proposed a new undergraduate program designed to welcome students with non-traditional skill sets into physics through a more practical, hands-on curriculum. Recognizing that students from underrepresented groups often come with different educational experiences, this program aims to broaden access to physics and has already been partially implemented by the Faculty of Science.
CINP members are also actively engaged in expanding diversity in terms of nationality and global inclusion. Jason Holt (TRIUMF) and, independently, Ruben Sandapen (Acadia University and Carnegie African Diaspora Fellow) have worked to build awareness and create pathways for African HQP. At the University of Guelph, Alex Gezerlis has been instrumental in promoting access for HQP with refugee status in Canada.


\begin{figure}
\centering
\begin{tcolorbox}[
  colback=white,
  colframe=gray!50!black,
  coltitle=black,
  colbacktitle=pink!50,
  width=0.8\textwidth,
  boxrule=0.8pt,
  arc=4pt,
  auto outer arc,
  fonttitle=\bfseries,
  title= Jonathan Barrett: from Indigenous youth initiatives to summer research at CERN]
\begin{minipage}{0.99\textwidth}

\begin{wrapfigure}{l}{0.25\textwidth}
  \vspace{-10pt}
  \includegraphics[width=0.25\textwidth]{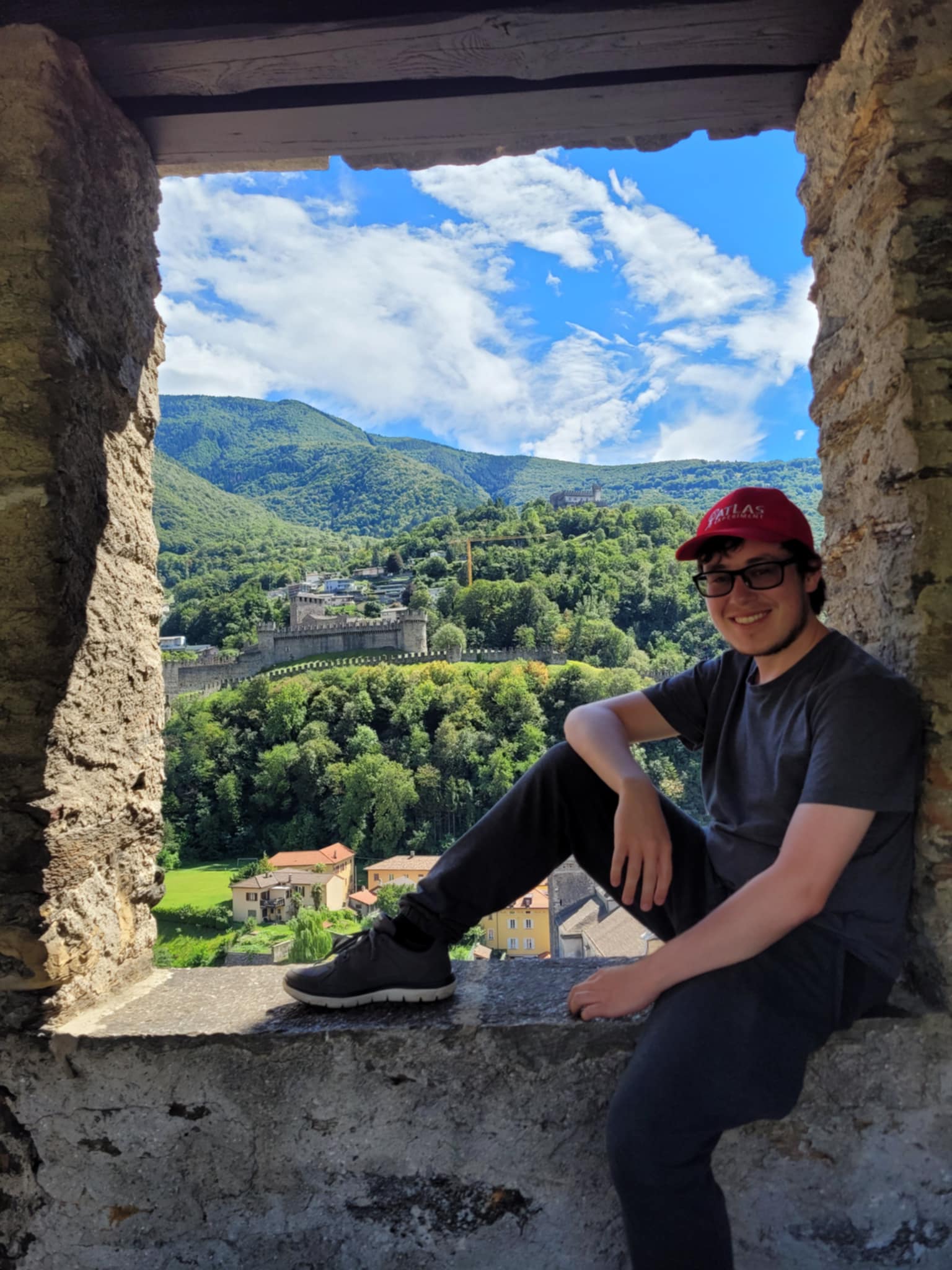} 
  \vspace{-10pt}
\end{wrapfigure}

{\small\baselineskip=14truept 
``As I pursue a master’s degree in physics at Memorial University of Newfoundland and Labrador, I am grateful for my undergraduate experience at MUN’s Grenfell Campus in Corner Brook. Supportive instructors and small class sizes helped me develop strong research skills and the confidence to grow as a scientist and communicator. This foundation led to exceptional opportunities, from multiple research projects and outreach events.   My research experiences ranged from developing a fully immersive virtual reality (VR) simulation to particle physics research at the European Organisation for Nuclear Research (CERN) in Geneva, Switzerland. These projects gave me valuable experience in both physics education and particle physics research. Equally rewarding was sharing my passion for science through outreach. I delivered school presentations and online talks, and assisted with on-campus events as part of Dr. Barkanova’s outreach program. 

In 2022, I served as a panelist and role model at the Indigenous Youth Gathering in St. John’s, NL. This event brought together Indigenous high school students from across the province to explore careers in science and engineering, sparking curiosity and bolstering confidence in the students. These experiences inspired me to continue promoting physics education and engaging in meaningful outreach throughout my career."
}
\end{minipage}
\end{tcolorbox}
\end{figure}

\begin{figure}
\centering
\begin{tcolorbox}[
  colback=white,
  colframe=gray!50!black,
  coltitle=black,
  colbacktitle=pink!50,
  width=0.8\textwidth,
  boxrule=0.8pt,
  arc=4pt,
  auto outer arc,
  fonttitle=\bfseries,
  title= Victoria Vedia: strong and supportive female role models at TRIUMF]
{\small\baselineskip=14truept 

``This experience was fundamental in shaping my identity as an independent scientist revealing my great leadership skills in addition to my technical proficiency and academic knowledge. I led international collaborations, supervised over 15 students across all academic levels- from undergrad to PhD, and spearheaded the ARIES project—designing, building, and implementing GRIFFIN’s flagship ancillary detector, featuring state-of-the-art technology unique to our setup. I contributed to a broad range of experimental campaigns in nuclear structure and decay spectroscopy. Beyond research, I took on a leadership role as Facility Coordinator for six months (covering the parental leave), further strengthening my project management and organizational skills. Collaborators from Simon Fraser University and the University of Guelph fostered a vibrant and collegial research environment. Their expertise and collaborative spirit significantly enriched the scientific output and made my time at TRIUMF particularly rewarding. I am especially grateful for the strong and supportive female role models at TRIUMF, whose guidance and encouragement
were a continual source of inspiration throughout my journey.
Everyone’s journey is different, and mine hasn’t been easy—but I’m proud of how much I’ve
grown and how much I’ve learned along the way.”
}
\end{tcolorbox}
\end{figure}

\section{Outreach}
Outreach in subatomic physics connects cutting-edge research on the building blocks of matter with learners of all ages, inspiring curiosity and showcasing Canada’s leadership in fundamental science. Subatomic physics outreach also serves as a national model for inclusive science education—strengthening equity, accessibility, and scientific literacy across all regions of Canada. Many programs intentionally centre diverse learners historically underrepresented in STEM. By pooling expertise, resources, and partnerships across universities, national laboratories, and outreach networks, we offer high-quality programming that reaches thousands of learners annually coast-to-coast-to-coast.

\subsection{National and cross-regional}
As a well-connected community with extensive experience in collaboration, Canadian nuclear physicists collectively deliver STEM outreach that is cost-effective, scalable, and impactful. 
TRIUMF and SNOLAB  play  key roles in educational programming, offering outreach materials and visitor experiences that showcase Canada’s leadership in underground science.  Members of TRIUMF regularly spearhead, participate in, or collaboratively co-lead numerous scientific outreach events for the wider Canadian community including (a) Science Rendezvous (reaching approximately 3000 people), (b) ``Cosmic Night" events with the H.R. MacMillan Space Centre (approximately 400 people) 2-3 times/year, (c) many local community festivals and celebrations (approximately 2000 people or more), (d) science events for Indigenous and unrepresented youth and communities, (e) smaller local school or community events. Members of TRIUMF, including postdocs and students, participate in multiple types of tours for VIP, scientific visitors or the general public. nEXO team members collaborate with science museums and planetariums to develop educational content focused on neutrinos and cosmic questions. These activities not only inspire future scientists but also highlight the broader societal value of fundamental research. By connecting cutting-edge physics to public curiosity, the nEXO collaboration strengthens public support for Canadian science and fosters a scientifically literate society.

Antimatter research provides an exceptional opportunity to inspire and educate the public through outreach as the topic naturally draws significant public interest. ALPHA-Canada faculty and HQP have been actively engaged in outreach activities, including public lectures at science museums, local libraries, and community groups such as Rotary Clubs, as well as events like Cafe Scientifique and Saturday Lectures. Impacts of ALPHA research have not only been widely recognized by publications in top journals and prestigious prizes and awards for students and faculty but also widely covered in the media. ALPHA-Canada members have had extensive media engagement with both national and international outlets, including CBC (The National, The Current, Quirks $\&$ Quarks), Radio-Canada, The Globe and Mail, Maclean's, The New York Times, and BBC. ALPHA's demonstration of laser cooling of antihydrogen was featured on the cover of Nature and as one of the Top 10 Breakthroughs of the Year in 2021 by Physics World magazine. The gravity experiment was selected as one of Physics World’s Top 10 Breakthroughs of 2023, and as 2 most significant science story of 2023 by CBC.

The Canadian high-resolution $\gamma$-ray spectroscopy group members organize events and activities with local high schools, the Educating Youth in Engineering and Science (EYES) program at URegina, and Nobel Prize talks at SFU in collaboration with the Science World in Vancouver. The group believes that maintaining strong connections with our communities is part of the responsibility that comes with being scientists and that our leadership and visibility serve as an inspiration to young scientists of all backgrounds. PIONEER PI Chloe Malbrunot has a wealth of experience in outreach: interviews to national newspapers and broadcasts in both English and French (Nature news, La recherche, Le Monde, German National Radio, New scientist, Science et Vie Junior, Radio Canada etc),  creating a TED-Ed animation ($>$ 0.5 million views), advising for CERN flagship “science gateway” and speaking for the Esade immersion program for industry CEOs at CERN.

\subsection{Regional}
By integrating culturally grounded approaches and Universal Design for Learning, the community demonstrates how world-class science can be shared in inclusive ways.
At UWinnipeg, Russell Mammei is responsible for outreach activities including student group tours of the university, Science Rendezvous, and visits to schools for demonstrations. Nicole Vassh (TRIUMF) lectured on the astrophysical origin of elements at outreach events for the public via two UBC Saturday Morning Lectures given at both UBC and SFU, along with an upcoming public lecture at the Royal Astronomical Society of Canada Sunshine Coast and an invited talk discussing the possible astrophysical production of the superheavy elements at PacifiChem, a gathering of the chemical societies of 5+ countries. Jason Holt (TRIUMF) has given a TEDx talk, been invited to emcee another TEDx event, and been a panelist on a Theatre and Physics symposium with the National Arts Centre of Canada on the play Copenhagen, which was curated into an episode of Ideas on CBC Radio,  and gave a public talk at the Vancouver Space Centre about new searches for dark matter. Alex Gezerlis (UGuelph) has organized outreach presentations by local undergrads to Ontario high school science classes, prefaced by an introduction by his on neutron stars and subatomic physics. Liliana Caballero has co-organized Nuclear Science Day, STEM Week at the City of Guelph museum and Let’s talk Science at the UGuelph. Ruben Sandapen leads outreach activities at AcadiaU which included a music-physics collaboration for a public event "Acadia in the stars" and a talk on Oppenheimer’s contributions to quantum physics, preceding the screening of Oppenheimer at the Al Whittle Theatre in Wolfville. In collaboration with the SFU Faculty of Science and Science World in Vancouver, Corina Andreoiu organized annual public talks from 2010 to 2024 to explain the science behind the Nobel Prize awards and their implications for society. MUN's group engages with thousands of students annually through online outreach programs, inviting the entire Canadian research community to help meet the growing demand. TRIUMF has been one of the most active partners, with a diverse team of presenters offering talks in English and French. This online delivery model is reaching a broad range of underserved and remote communities and is now attracting national attention. In 2023, Svetlana Barkanova developed a physics course for the Nunavut Teacher Education Program — the first university-level physics course offered in Nunavut — combining virtual laboratories with real-time online instruction for students in multiple northern communities, and is  building on this experience to expand physics curriculum for \href{ Connected North}{Connected North}.

\chapter{Nuclear physics as a driver of future technologies}

A crucial aspect of nuclear physics research is the development of new
technologies.  These developments enable the breakthroughs made in
Canadian scientific research.  In this Chapter, recent developments in
high-performance computing, quantum computing, detector development,
quantum sensor development, and particle accelerators are described.
These technological developments also have the potential for
additional benefits to Canadian society through spin-off applications.

\section{High-performance \& quantum Computing}
\label{sec:HPC}


Computing is an integral part of nuclear-physics research: our use ranges from medium to large scale computing; storage
of petabyte-scale data samples; and cloud and high-performance computing. High-speed networks are an essential
part of the infrastructure as they enable us to transfer large volumes of data around the world. Today our
community uses many thousands of compute cores and many tens of petabytes of data storage. Our software
manages complex distributed heterogeneous systems of compute and storage resources. Further, the computing
resources are a critical element for the training of highly qualified personnel (HQP). Our graduate students and
research associates become skilled in a wide range of areas from data analysis, writing complex software, or using
large high-performance computers or compute clouds. They find employment in research and in a wide range of
industries from medical physics to financial modeling.

The subatomic physics community has been a large user of computing for many decades. We use the high-
throughput (general purpose) and high-performance computing facilities, and the storage resources of the
Digital Research Alliance of Canada (the Alliance). Our projects span years, and more often, decades. 
Much of this infrastructure has been funded
in a somewhat \textit{ad hoc} manner by CFI (and provincial funding organizations). This makes it difficult to make long-
term international commitments to our projects. A secure and continuing source of funding is critical to the community's future success.

The use of artificial intelligence (AI) technologies
has been growing dramatically in subatomic physics,
in both theoretical and experimental work. 
An important recent development in this connection is the Government of Canada's 
\$705 million AI Sovereign Compute Infrastructure Program
(AI SCIP), for which the Alliance has submitted 
a Statement of Interest. The AI SCIP will lead
to the building of a state-of-the-art national high-performance
AI-specific supercomputing system; 
nuclear physicicists are well placed to both guide
the development and benefit from the deployment
of such cutting-edge Canadian-based computing infrastructure.

Nuclear-physics researchers have faced challenges using existing computing resources. Of utmost priority, is that the Alliance
operate its resources on a 24x7 basis with well-established operating procedures. No resource or centre should
be offline for any length of time. This requires a technical design that ensures redundancy and the ability to take
selected hardware components offline without impacting operations. A 24x7 service requires that system
administration teams be available at any time. Currently, the Alliance operates on a 9x5 basis and issues are
addressed during working hours. Frequently sites are taken down with little warning or are down for extended
periods; to make matters worse, sometimes multiple sites are down at the same time, making the availability of resources at the national level scarce. Our research environment is very competitive and other computing facilities around the world operate
on a 24x7 basis. The Alliance needs to ensure that the facilities are designed and managed to be fully operational
around the clock, and to ensure that scheduled downtimes do not result in loss of access to resources for any user.

Our vision is that the Alliance turn into a national entity of compute, storage, software, networks and personnel resources
providing 24x7 services to the research community.
Currently, Compute Canada sites operate as independent organizations with staff being employees of the local
site. Each site has different rules and operating conditions. There is minimal interoperability and
interchangeability between infrastructure elements of the sites. The ideal future setup would involve unified infrastructure
and personnel, something which may require a significant reorganization.

It is important that the Alliance, at management and operational levels, consult with international organizations that run
large computer systems such as National Energy Research Scientific Computing Center (NERSC) in the USA and
CERN in Europe. These facilities have successfully deployed large systems with well-defined operational
procedures. A review of the best practices at other sites will help the Alliance manage its resources and personnel. It
has been our observation that the management, operations and organizational aspects have historically been a weak point of
Compute Canada, so this is a natural point on which 
to improve in the coming decade.

The Alliance should also explore the use of commercial cloud computing resources that could, for example, meet peak
demands that exceed existing resources. The commercial cloud systems are extremely reliable and their technical
teams are highly responsive to their customers' requirements. Members of our community have demonstrated that
commercial clouds can be effectively utilized. The Canadian research community would benefit from improved
ties with the commercial cloud computing industry.

The Canadian subatomic physics community should support the development of new research software and data management tools. No organization
can predict the software and services that will be adopted by the researchers, and the Alliance should invest in many
pilot projects rather than trying to select a system or service for everyone. Funds in these areas should be focused
on our highest priority projects that have significant international impact based on peer review.

The above summarizes somewhat ``traditional''
aspects of computing in nuclear physics. Two 
emerging directions are also noteworthy.

First, many important scenarios in the context of 
lattice gauge theory cannot be handled by the standard lattice
approach. Examples include particle collisions as a function of time, and nuclear matter at high densities (such as in neutron stars or the early universe). In this connection, a forward-thinking recommendation for Canada's subatomic physics community is to create a plan for providing national access to quantum computing infrastructure. This research area is growing rapidly in other countries (Europe, US, and elsewhere). Canada has the advantage of a longstanding strength in quantum information science, so our subatomic physicists are poised to be leaders in the use of quantum computing for nuclear and particle physics (both experiment and theory, and not just lattice). A national plan for remote access to quantum computers would allow Canadians to be leaders in this new approach to subatomic physics.

Second, it may be worthwhile to explore 
the development of vendor-agnostic languages in order 
to leverage heterogeneous computing.
Specifically, an intriguing prospect is to explore how to pack calculations, at mixed and reduced precision, for hardware accelerators using the Single Instruction Multiple Data (SIMD) paradigm to be deployed for timely completion of computations. Such hardware is already commercially available (within CPUs, GPUs, or AI accelerators), so the goal would be to devise how to repurpose this hardware to accelerate physics simulations. This research also includes exploring the potential for a new computer language to be devised — in collaboration with mathematicians, physicists, computer scientists, and industry partners — that alleviate the burden of maintaining and upgrading physics-based simulation or data analysis software utilizing accelerated computing. Such a language would also benefit all who wish to use heterogeneous computing environments, spanning from hand-held devices to supercomputers, in a vendor-agnostic way.

Finally, next-generation facilities like the Electron-Ion Collider (EIC) will generate unprecedented data volumes for nuclear physics, at the scale of the Hi-Lumi LHC program, demanding innovation in high-throughput computing and AI-driven analysis to fully realize its discovery potential. Already, Canadian facilities are leading contributors of both human and compute power to the development of EIC software and computing. Canadian investment in computing infrastructure is therefore essential to capitalize on our detector contributions to this project.

\section{Detector development \& quantum sensors}

The core principle of experimental nuclear physics is the use of radiation detectors to measure the products of nuclear reactions and decays. As a result, developments in radiation detector technology go hand-in-hand with developments in experimental nuclear physics. This is a two-way street. In many cases, advances in detector technology are exploited to develop new experimental methods or to enhance sensitivity. In other cases, nuclear physicists are the drivers of new technologies that can be exploited in a  variety of related fields.
The development and exploitation of new detector technologies has significant implications beyond nuclear physics, in fields such as medical physics, fast neutron imaging, security, and energy exploration. The Canadian nuclear physics program includes a significant number of efforts that make use of novel detector technologies with broad societal applications.

\begin{itemize}
    \item \textbf{Quantum sensors (superconducting devices)} - The
      Canadian nuclear physics community is exploiting superconducting
      tunnel-junction (STJ) quantum sensors to measure low-energy
      radiation, such as the products of ${}^{7}$Be decay, with
      unprecedented sensitivity. While these and related quantum
      sensors have broad application in nuclear physics, there are
      also potential benefits in other fields, such as medical
      physics, particle astrophysics, and commercial applications. For
      example, the high energy resolution of these sensors is expected
      to enable the characterization of Auger electron decays in
      medical isotopes developed in-house at TRIUMF. With a dedicated
      R\&D program to boost the overall device active volume, the
      low-energy threshold of these detectors could allow searches for
      ultra-low mass dark matter.  Lastly, operating these devices at
      TRIUMF will enable full studies of the impact of particle
      interactions in these types of superconducting quantum sensors,
      which will strongly benefit commercial applications such as
      quantum computing.

    \item \textbf{Quantum sensors (atomic, molecular, and optical
      techniques)} - Another class of quantum sensors are based on
      atomic, molecular and optical systems. Canadian projects such
      as, EDM$^3$, TUCAN, RadMol, ALPHA/HAICU are exploiting high
      sensitivities enabled by these quantum techniques to probe
      fundamental symmetries. The relevant techniques include
      phase-sensitive Ramsey techniques, quantum state manipulation,
      spin-based magnetometry, particle traps, and atomic
      fountains. These techniques are workhorses of quantum sensing
      and information technologies, which have potential applications
      delivering new quantum technologies to Canadian society at
      large.
      
    \item \textbf{Quantum sensors at SNOLAB} - SNOLAB offers a unique facility for studying quantum technologies. For example, superconducting qubits are being operated in the underground laboratory to characterize their performance in a low-radiation environment, in collaboration with the Institute for Quantum Computing at the University of Waterloo. Such efforts are a natural extension of the dark matter research at SNOLAB and strengthen ties between our research community and Canada's quantum sector, leading to development of mutually beneficial technologies and training of in-demand HQP.
    
    \item \textbf{Time-projection chambers (TPCs)} - Canada, and
      TRIUMF in particular, has a long tradition in the development of
      novel gaseous detectors. This includes the first adaptation of
      time projection chamber in a physics experiment. Recent example
      includes the design, construction, and successful use of the
      radial Time Projection Chamber for the ALPHA-g experiment. A
      novel radial drift scheme was developed to address the severe
      experiment requirements for antihydrogen gravity experiment.
      %
      %
      
      One of the broad applications of the TPC is the development of a
      generic set of electronics required to read out the necessarily
      large number of channels. The General Electronics for TPC’s
      (GET) system was developed specifically for the ACTAR device
      (operated in an experimental campaign at TRIUMF in 2025) but was
      sufficiently versatile that nearly 120k channels have been
      deployed at other TPCs and silicon detector projects
      worldwide.\footnote{E.\ C.\ Pollacco \emph{et al.},
      Nucl.\ Instrum.\ and Meth.\ in Phys.\ Res.\ A 887, 81 (2018).}
      Future upgrades to this system are anticipated to take advantage
      of intelligent triggering, potentially with the help of AI. The
      requirement of high channel density electronics has broad
      applications in other fields such as medical imaging.

\item \textbf{High-granularity detectors for the AI era} -
Modern nuclear physics experiments are increasingly designed in and for the era of artificial intelligence (AI). This design philosophy involves building high-granularity detectors with a large number of readout channels that are optimized for AI-driven algorithms. An example of this approach is the Barrel Imaging Calorimeter (BIC), a major Canadian contribution to the Electron-Ion Collider (EIC) portrayed in Fig. 
\ref{fig:imagingcal}.
The BIC is a hybrid detector with layers of high-granularity silicon pixel sensors interleaved with lead and scintillating fiber (PbSciFi). This design provides detailed 3D tracking information of particle showers, creating rich datasets ideal for machine learning analysis. By integrating detector technology with advanced data science from the ground up, this approach represents a significant leap forward, providing powerful new capabilities for discovery while training HQP in the strategically important fields of detector hardware and AI.

    \item \textbf{Organic glass scintillators (OGS)} - Canadian researchers are making use of recently-developed OGS detectors to measure neutrons produced by $(\alpha,n)$ reactions in coincidence with heavy-ion recoils detected using the DRAGON and EMMA recoil separators at TRIUMF. Detector development work on position-sensitive OGS neutron-detector bars (Fig.~\ref{fig:OGSBar}) is also taking place at Saint Mary's University.    
    OGS is a new material, developed at Sandia National Laboratory (USA), which offers sub-ns timing resolution, excellent $n/\gamma$ pulse shape discrimination, and high detection efficiency for MeV-scale neutrons.\footnote{J.\ S.\ Carlson and P.\ L.\ Feng, Nucl.\ Instrum.\ and Meth.\ in Phys.\ Res.\ A 832, 152 (2016).} The structure of the material as an organic solid alleviates many of the safety challenges encountered with traditional liquid scintillator neutron detectors, while also allowing for fabrication into a variety of shapes that are difficult or impossible with other organic solids. These properties are an excellent match to the needs of the TRIUMF $(\alpha,n)$ experiments. At the same time, the advantages of OGS are increasingly being recognized outside of fundamental nuclear physics, with potential applications in fast-neutron imaging, nuclear security, and $\gamma$-ray astronomy using Compton imaging.

\begin{SCfigure}
{\includegraphics[width=0.6\textwidth]{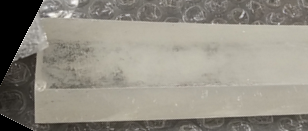}
\caption{Photograph of an OGS scintillator bar being developed into a position-sensitive neutron detector at Saint Mary's University.}
\label{fig:OGSBar}
}
\end{SCfigure}

  \item \textbf{Fast timing $\gamma$-ray detectors} -  The DRAGON collaboration has plans to upgrade its existing BGO array to a new scintillator array with fast timing capabilities and improved energy resolution. This will enable new techniques such as the resonance timing method to be used to study astrophysical radiative capture reactions. DRAGON's requirements for fast-timing and highly efficient $\gamma$-ray detectors dovetail with those of medical imaging. In particular, TOF-PET imaging techniques have similar requirements for fast timing resolution and high detection efficiency. Many of the same materials being considered for the DRAGON upgrade are materials also exploited for TOF PET. The prime candidate for the DRAGON upgrade, LaBr$_3$ is currently used for TOF-PET studies, and other materials being considered such as LYSO or GAGG:Ce were originally developed for medical physics applications.

    \item \textbf{Magnet technologies} - Magnet technologies lie at
      the heart of modern fundamental physics. In experiments such as
      those exploring the properties of antimatter, or symmetries in
      neutrons, atoms or molecules, powerful superconducting magnets
      enable scientists to trap, cool, and precisely control charged
      and neutral particles. These same technologies---born from the
      desire to understand the basic laws of nature---have also driven
      applications across society. Advances in superconducting
      materials, cryogenics, and magnetic-field control developed for
      research accelerators and precision measurement devices now are
      the basis for technologies such as magnetic resonance imaging
      (MRI), material characterization, and emerging quantum-sensing
      systems. Continued innovation in magnets, such as
      high-temperature superconductors, not only deepens our
      understanding of the fundamental physics but also advance the
      next generation of practical technologies that benefit health,
      energy, and information sciences.
\end{itemize}





\begin{figure}[h]
\centering
\begin{tcolorbox}[
before upper={\parindent 15pt},
  colback=teal!30,
  colframe=gray!50!black,
  coltitle=black,
  colbacktitle=gray!20,
  width=\textwidth,
  boxrule=0.8pt,
  arc=4pt,
  auto outer arc,
  fonttitle=\bfseries,
  title=Water quality in Indigenous communities
]
{\small\baselineskip=14truept 


Subatomic physics detectors have broader applications that benefit society. An example is a drinking water monitoring project, which combines technologies adapted from water monitoring systems originally developed for large water Cherenkov neutrino detectors.

Problems with drinking water disproportionately affect remote Indigenous communities. Subatomic physics researchers from HyperK-Canada
in collaboration with TRIUMF Science-Technology department, developed a water monitoring system for the HyperK detector.
They
began a collaboration with Prof. Arzu Sardarli of the First Nations University of Canada, who introduced them to the band council, elders, and the water facility operator of the Cowessess First Nation in Saskatchewan, and are planning the deployment of a water monitoring system at their facility. Prof. Sardarli is also enthusiastic about STEM education programs of Indigenous communities, and an memorandum of understanding (MOU) was recently signed between TRIUMF and the First Nations University based on this collaboration. The group is hoping to develop an internship program on water monitoring for Indigenous education students, who could return to their communities and teach science. The group is also communicating with the Kirkness program regarding participation in hands-on training for Indigenous high-school students using the water monitoring equipment.

The group has added water quality specialists to their team from U. Victoria and U. Regina, which led to the deployment of sensors at water facilities in Victoria, BC and Weyburn, SK, and they are now working on a deployment at the UBC water distribution facility.

The system to be deployed is based on two complementary detection technologies. The first is a transmission detector that measures light attenuation through a water sample to quantify dissolved impurities. By tuning its optical sensitivity, it can detect harmful substances such as cyanotoxins from algal blooms in lakes, organic mercury released from thawing permafrost, and selenium contamination from coal mining effluent. The second system is a scattering detector that uses pulsed-laser Mie scattering to detect and count individual particles in the water. With a ring-imaging Cherenkov method, it measures the angular distribution of scattered light, enabling the precise determination of particle size one by one. This capability makes it well-suited for monitoring particulate pollutants such as E. coli bacteria or microplastics, providing real-time particle counts and size profiles. Together, the attenuation and Mie scattering systems offer a powerful, multi-modal approach to environmental water quality assessment, demonstrating how innovations from fundamental physics research can be applied to address critical environmental challenges.}
\end{tcolorbox}
\end{figure}

\section{Particle accelerators}

\begin{SCfigure}
{\includegraphics[width=12.0cm]{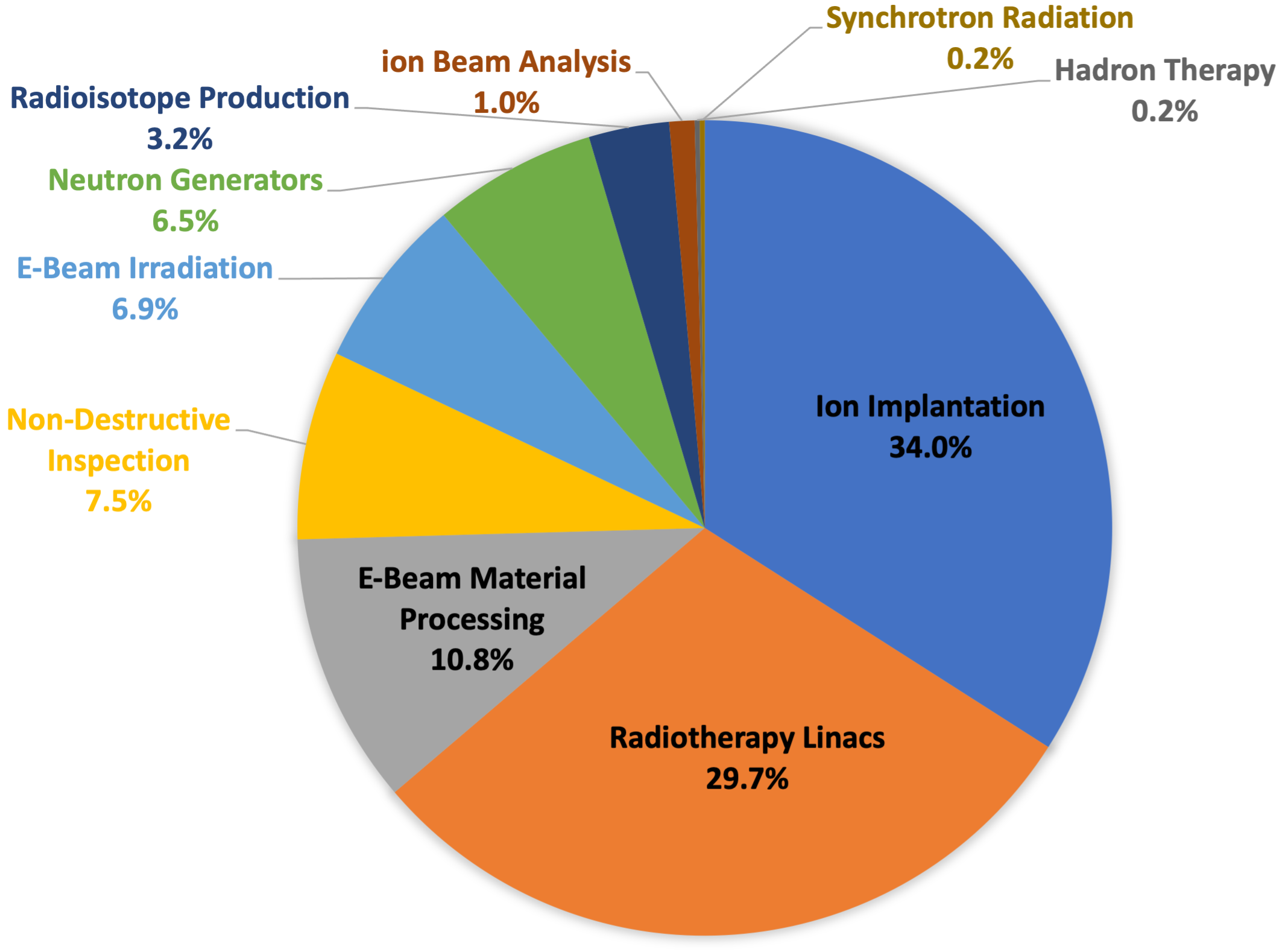}
\caption{Distribution of accelerators worldwide by common applications in 2019. Data sourced from: B.L. Doyle, F.D. McDaniel, R. Hamm, {\em The Future of Industrial Accelerators and Applications}, World Scientific 2019.}
\label{fig:accelerators}
}
\end{SCfigure}

The use of accelerators is a common feature in many spin-off
technologies from subatomic physics research.  It is estimated that
more than 50,000 accelerators are used today in a variety of
applications in physics, chemistry, medicine and structure analysis.
Fig.\ \ref{fig:accelerators} shows the breakdown of accelerator usage
by application as of 2019.  Those used for subatomic physics research
account for $<$1\% of the total and are not shown; 64\% of the total
are used as ion implanters or radiotherapy.  Accelerators for medical
and industrial represent an industry estimated to be worth \$5
billion/year in 2019, and the value of the treated products exceeds
\$50 billion/year.

Below are some of the most common applications including medical and industrial uses, ion beam analysis, synchrotron light sources and neutron sources.
\begin{itemize}
\item {\bf Cancer Therapy} - Radiation therapy using particle beams
  (external beam therapy) was developed from subatomic physics
  research.  There are about 10,000 accelerators worldwide presently
  devoted to radiation therapy, with millions of successful
  treatments.  Due to Canada's strong background in experimental
  nuclear physics, many of the radiation physics techniques used
  worldwide in this therapy were also developed here.
\item \textbf{Medical isotopes} - Particle accelerator technology
  originally developed for subatomic physics research currently
  supplies about 10\% of the world supply of medical isotopes.  In
  Vancouver, Nordion Inc.\ operates three medical cyclotrons for the
  production of medical isotopes, which are exported
  worldwide. Looking forward, the Institute for Advanced Medical
  Isotopes (IAMI) at TRIUMF is expected to significantly increase
  production and distribution of critical medical isotopes such as
  ${}^{99m}$Tc and ${}^{18}$F.
 %
%
Additionally, advances in high-power electron accelerators (driven by nuclear physics), combined with $\gamma$-ray activation methods, open a new pathway for industrial-scale production of medical isotopes.

\item {\bf Ion implantation for semiconductors} -
The most common use in industry of ions is in the precise implantation of dopants in semiconductors, which results in the precise modification of electrical, optical and other properties, enhancing device performance and creating new functionalities
\footnote{S. Felch et al., {\em Ion implantation for semiconductor devices:\ the largest use of
industrial accelerators} (2013) pp. 740–744.}

\item {\bf Wastewater irradiation and other sterilization techniques} - Electron beams can be used as a sterilisation technique, and this is used extensively in industries such
as medical equipment sterilisation, food processing and pharmaceuticals. Manufacturers of medical
disposables have to ensure that syringes, bandages, surgical tools and other gear do not carry pathogens,
without altering the material itself.

\item {\bf Non-destructive testing} - Non-destructive testing frequently involves electron linear accelerators up to a few MeV in beam energy.
These electrons are used to generate X-rays for inspection of high-density items or objects which are
large and difficult to inspect in other ways.
In addition, radioactive
  threat detection technology involves the detection of neutrons (nuclear waste materials) and
  $\gamma$~rays (dirty bombs) during routine X-ray inspections of cargo and at
  border crossings. Subatomic physics techniques are also used to detect land
  mines and improvised explosive devices.

\end{itemize}

Particle accelerators have an astonishing number of uses that benefit
society.  Most of these uses were never foreseen when the basic
accelerator technologies were originally developed for subatomic
physics research, highlighting the importance and potential of
pursuing research in fundamental science.

\clearpage
\chapter{Science case}

Canada possesses an exiciting nuclear physics program, and a vibrant
nuclear physics community.  Canadians lead major experimental efforts
both at home and worldwide, and generate exciting theoretical physics
results that drive the science forward.  In this Chapter, exciting new
results from the nuclear physics community are described, along with
plans for the next seven years (2027-2034), and beyond.

\section{Hadronic physics and QCD}

\subsection{The Canadian program (current and next 7 years)}
\subsubsection{Introduction}
The building blocks of the atomic nucleus—the protons, neutrons, and other particles (such as mesons) that bind them—are collectively known as hadrons. These hadrons are composed of even more fundamental constituents: quarks and gluons. However, due to the exceptionally strong nature of the quark-gluon interaction at long distances, these constituents are permanently confined within hadrons. Understanding this quark confinement is one of the greatest unsolved problems in modern physics. On the theoretical front, Quantum Chromodynamics (QCD) provides a rigorous framework that describes the interactions of quarks and gluons with great accuracy at short distances or high energies—where perturbative methods apply. However, QCD becomes extremely challenging in the low-energy, long-distance regime where confinement dominates, and perturbative approaches break down. In this regime, theoretical progress relies on either phenomenological models or numerical techniques such as Lattice QCD, both of which require careful comparison with precision experimental data. An indispensable tool in this effort is Chiral Perturbation Theory (ChPT), a low-energy effective theory of QCD that allows systematic expansions in quark masses and momenta. ChPT is essential for connecting theoretical predictions to observables in processes such as pion-nucleon scattering, meson decays, and Compton scattering, and it provides a crucial interface between experiment and Lattice QCD (see Section~\ref{sec:nuctheo_qcd}).

Key properties of hadrons—such as their masses, spins, and excitation spectra—are best studied through electron scattering and photoproduction experiments, such as those conducted at Jefferson Lab (USA) and Mainz Microtron (Germany), which employ high-duty-factor continuous-wave electron beams. Experiments involving inclusive electron scattering at high momentum and energy transfer have yielded important insights into the internal dynamics of quarks and gluons, but our understanding remains incomplete. To address these gaps, the upcoming Electron-Ion Collider (EIC)—currently under construction at Brookhaven National Laboratory (USA)—is expected to transform our knowledge of the inner structure of hadrons and nuclei. The EIC will allow precision studies of the 3D structure of hadrons, gluon saturation, and the dynamics of quark-gluon interactions across a wide kinematic range. Canada is contributing significantly to the EIC through both theoretical advances in QCD and leadership in detector development, including the Barrel Imaging Calorimeter (BIC) and superconducting RF crab cavities. These contributions will help ensure the EIC delivers the critical experimental data needed to confront QCD in its most challenging, non-perturbative regime.
Canadian researchers have long played a leading role in hadron structure studies and continue to advance our understanding of the strong force through innovative theory, modeling, and precision experimentation.

\subsubsection{Thomas Jefferson National Accelerator Facility (Jefferson Lab, JLab)}

\paragraph{Major Scientific Goals}
\begin{itemize}

	\item Due to the charged pion's relatively simple quark-antiquark
	($q\bar{q}$) valence structure and its experimental accessibility, the pion
	elastic form factor ($F_{\pi}$) offers our best hope of directly observing
	QCD's transition from colour-confinement at long-distance scales to
	asymptotic freedom at short distances.  The $K^+$ form factor experiment at
	JLab, where the $\bar{d}$-quark is replaced with the $\bar{s}$-quark,
	provides a vital second study case.  Data were acquired in 2018-22 and are under analysis. 
        Due to the enormous quantity of data collected, this work will continue through 2030.

	\item The search for hybrid mesons, conceived as $q\bar{q} g$ states, via
	the GlueX Experiment at JLab in the light-quark sector, is a pivotal
	component of the worldwide experimental hadron spectroscopy effort in
	mapping out and understanding the hadron spectrum in terms of its quantum
	numbers. Hadron spectroscopy addresses the question of how quarks and
	gluons give rise to the hadronic properties and the phases of hadronic
	matter. An extension of GlueX will probe the role of scalar meson dynamics
	in chiral perturbation theory, tighten the uncertainty in the light quark
	mass ratio, search for sub-GeV dark gauge bosons and provide direct probes
	for C-violating, P-conserving new forces.  These efforts will also continue
	through 2027.

	\item Generalized Parton Distributions (GPDs) unify the concepts of parton
	distributions and hadronic form factors, and are ``universal objects,''
	which will allow a tomographic (3D) view of the nucleon to be built up.
	The Solenoidal Large Intensity Detector (SoLID) at JLab will enable crucial
	new GPD data to be acquired.  Other core components of the SoLID scientific program are investigations of gluonic contributions to the origin of hadronic mass through threshold $J/\psi$ electroproduction, and investigations of the nucleon's transverse momentum and spin structure through semi-inclusive deep inelastic scattering (SIDIS) from polarized targets.

\end{itemize}

\paragraph{Gluonic Excitations Experiment, GlueX, and Jefferson Lab Eta Factory, JEF
\label{sec:jef}}
\noindent{(JLab) University of Regina, Mount Allison University; {\em{Armenia, Chile, China, Germany, Greece,
Russia, UK, USA}}}

A key question of QCD is to understand how quark and gluonic degrees of freedom 
manifest themselves in the spectrum of hadrons.
Mesons are grouped in {\em multiplets}, each characterized by a given $J^{PC}$
combination. Hybrid mesons result from the addition of angular momentum
quantum numbers from a gluonic component 
and are
visualized as $q\bar{q}g$ states. Among the hybrids, a subset is predicted to
have exotic $J^{PC}$ combinations, not allowed in the simple quark
model. 
GlueX is one of the flagship experiments of 12-GeV
JLab, and is searching for exotic particles where the {\em glue} is in an energetically excited state.  Hadron spectroscopy ties in with the
2017--2021 Canadian Subatomic Physics LRP under section 5, ``How do Quarks and
Gluons Give Rise to the Hadronic Properties and the Phases of Hadronic
Matter?''

The Regina-GlueX team monitors and calibrates the Barrel Calorimeter (BCAL), a 24-ton detector designed and built in Regina, and has carried out machine learning studies to improve shower reconstruction for BCAL and the Forward Calorimeter (FCAL, upgraded to FCAL2 for JEF).  The team led analysis and publication efforts on photon-beam asymmetry for $\eta^{(\prime)}$, which indicated the dominance of natural parity exchange.
In addition, the extraction of $\eta$ cross sections 
has been completed with the publication nearing internal review.
The most challenging analysis is to extract moments and partial waves of the $b_1$(1235) meson in its
dominant $\omega \pi$ decay, to look at hybrid-meson decays into the $b_1 \pi$ state and to extract cross sections.
 Three Ph.D. students, a M.Sc. student and a postdoc are working on these projects at Regina.
David Hornidge at Mount Allison University has also joined GlueX's neutral/charged pion polarizabilities efforts, which dovetails with the MAMI polarizability program.

JEF involves a significant upgrade of the GlueX base instrumentation 
and will allow experiments to explore fundamental topics with sensitivities not previously achievable. 
Specifically, these features will: (1) enhance the GlueX exotic hybrid meson search with better particle
identification;  
(2) enable the measurement of $\eta/\eta^{\prime}$ decays such as $\eta \rightarrow \pi^0 \gamma \gamma$ reaction to access ${\cal O}(p^6)$ chiral perturbation theory 
and $\eta \rightarrow \pi^+\pi^-e^+e^-$ reaction to probe doubly-virtual transition form factors, $(g-2)_{\mu}$, $P$/$CP$ violation, and axion-like particles; (3) tighten the uncertainty in the light-quark mass ratio
extracted from $\eta \rightarrow 3\pi$; and (4) search for sub-GeV dark gauge bosons 
by improving the existing bounds more than two orders of magnitude.   The JEF program will ramp up and continue into the next five years.

\paragraph{Pion and Kaon Form Factors as probe of emergent mass generation in
hadrons}
\noindent (JLab) University of Regina; {\em{Armenia, Croatia, France, Korea, UK, USA}
\label{sec:hallc}}

The elastic electromagnetic form factors of the charged pion and kaon,
$F_{\pi}(Q^2)$ and $F_K(Q^2)$, are a rich source of insights into basic
features of hadron structure, such as the roles played by confinement and
Dynamical Chiral Symmetry Breaking (DCSB) in fixing the hadron’s size,
determining its mass, and defining the transition from the strong- to
perturbative-QCD domains.  
The pion's properties in particular are intimately
connected with DCSB, which explains the origin of more than 98\% of the mass
of the visible matter in the universe.  Furthermore, owning to the intimate
connection between DCSB and the pion, the properties of nature's lightest
hadron provide the most direct access to QCD's momentum-dependent effective quark mass
\footnote{T. Horn, C.D. Roberts, {\em The pion: an enigma within the Standard Model}, 
J.Phys.G. 43 073001 (2016)}.  
The measurement of the pion form factor
presents an extraordinary opportunity to observe QCD's transition from
confinement-dominated physics at large length-scales to short-distance scales
where the aspects of perturbative QCD become apparent.
The $K^+$ form factor, where the $d$ anti-quark is replaced with the $s$
anti-quark, provides a vital second study case. The $K^+$ structure also has
significant influence from DCSB, but the greater mass of
the $\bar{s}$ also has influence from the Higgs mechanism.  
The understanding
of the structure of both mesons, within a unified theoretical framework, will
greatly assist with our understanding of the origin of hadronic mass. 
The Regina group leads the meson form factor program at Jefferson Lab.
Their work has had significant impact, gathering over 1400 citations to date,
and appearing in various review articles.  The work has confirmed that at a
photon virtuality of $Q^2=2.45$~GeV$^2$, one is still far from the resolution
region where the pion behaves like a simple $q\bar{q}$ pair, i.e. far from the
``asymptotic'' limit
\footnote{F. Gross et al., {\em 50 Years of QCD}, Eur.Phys.J. C 83 1125 (2023); G.M. Huber et al., Phys.Rev.C 78 045203 (2008)}.  
The extension of these studies to shorter distance scales is
underway.  Data taking took place 2018--22.  Data analysis is underway, but due to the quantity of data taken, and the fact that the new Hall ~C infrastructure had to be understood in great detail, the current group of PhD students and PDF will only analyze a fraction of these data.  Another generation of PhD students is being recruited to continue this effort, with the goal to have the analysis substantially complete by 2030. 
 This work will yield high quality $F_{\pi}$ data up to
$Q^2=6.0$ GeV$^2$, and lower quality data up to $Q^2=8.5$ GeV$^2$, and will
provide the world's best data set on this fundamental quantity for many years.  
If the $K^+$ data
confirm that the scattering from the virtual $K^+$ in the nucleon dominates at
low four-momentum transfer to the target $|t|\ll m_p^2$, similar to as was
done for the pion, the Regina group will also publish the world's first quality data
for $F_K$ above $Q^2>0.2$~GeV$^2$.  The group hopes to release at least two $K^+$ form factor publications in 2026--27.  
Not all of the approved $K^+$ data were acquired in 2019, and the Canadian team plans to submit a request to complete the data taking once the first  form factor results are released.

\paragraph{Solenoidal Large Intensity Detector, SoLID
\label{sec:solid}}
\noindent{(JLab) Regina; {\em{Australia, China, France, Germany, Israel, Italy,
Korea, Mexico, Russia, Slovenia, Ukraine, UK, USA}}}

The development of the Generalized Parton Distribution (GPD) formalism in the
last 20 years is a notable advance in our understanding of the structure of
the nucleon
\footnote{D. Mueller, {\em Generalized Parton Distributions: Visions, Basics,
and Realities}, Few-Body Syst. 55 317 (2014)}.  Unifying the concepts of parton distributions and hadronic form
factors, GPDs are ``universal objects'' which provide a comprehensive
framework for describing the quark and gluon structure of the nucleon.
Knowledge of GPDs would allow a tomographic 3D understanding of the nucleon to
be built up.  They are probed experimentally through a variety of Deep
Exclusive reactions.
The four lowest-order GPDs are parameterized in terms of quark chirality, with
$\tilde{E}$ involving the difference between left and right handed quarks.
The Regina-led experiment will use SoLID, in conjunction with a polarized
$^3$He target, to probe the poorest-known GPD, $\tilde{E}$.  
$\tilde{E}$
cannot be related to already known parton distributions, so experimental
information about $\tilde{E}$ via Deep Exclusive Meson Production (DEMP)
can provide new information on nucleon
structure unlikely to be available from any other source.  The
experimental access to $\tilde{E}$ is through the azimuthal variation of the
emitted pions from a transversely polarized 
$^3$He target.

\begin{figure}[h!]
\begin{center}
\includegraphics[width=4.5in]{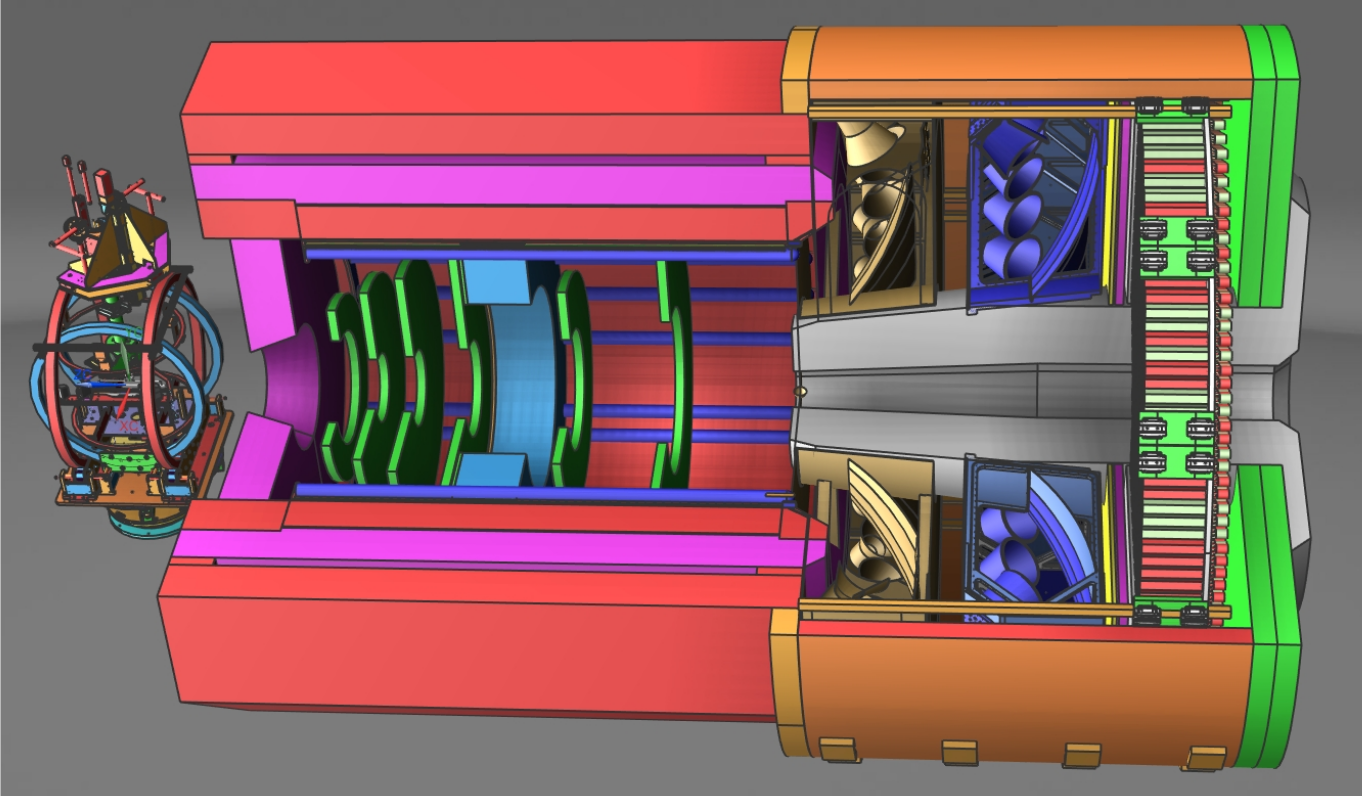}
\vspace{-0.2cm}
\caption{\label{fig:solid}
The Solenoidal Large Intensity Device (SoLID), planned for JLab Hall A.  The
polarized $^3$He target is at far left, and
the Heavy Gas Cherenkov detector is second from right (violet).}
\end{center}
\end{figure}

The essential idea of SoLID is to use the latest technology detector and
readout to enable an increase in luminosity by a factor of ten compared to
existing detectors (such as CLAS-12)
\footnote{J. Arrington \emph{et al.}, {\em The solenoidal large intensity device (SoLID) for JLab 12 GeV}, J.Phys.G. 50 110501 (2023)}. The construction of the SoLID detector,
an international project estimated at ~US\$80M, has received very favorable
feedback in scientific reviews conducted by the U.S. Department of Energy
(US-DOE).  The SoLID Collaboration has $\sim$300 members from 70+ institutions in 13
countries, and is in an advanced stage of project planning.  Jefferson Lab has identified SoLID as its next flagship project, and has proposed a \$30M redirect of lab funds to SoLID reduce the total ask to US-DOE.  In addition, the new US-DOE secretary has announced a simplified approval scheme for projects with value $<$\$200M, which means that SoLID would no longer have to go through the Critical Decision path.  It is expected that both developments will allow SoLID construction and installation to begin in the near future, with the physics program planned to begin in FY32.
The Regina group,
in partnership with Duke University, is responsible for the Heavy Gas
Cherenkov (HGC) detector, which will be used for forward angle pion identification.
The full HGC detector is projected to cost $\sim$US\$5M.  The newly-awarded CFI-IF funds (2023 competition) will pay for the entirety of the gas pressure vessel for this detector, with the mirrors and photo sensors expected to come from US-DOE funds.  The HGC pressure vessel has been awarded to be constructed by Dyna Industries in Regina, providing local machine shop jobs and improving their high quality fabrication expertise.

\subsubsection{Polarizability experiments at Mainz and Duke}

A central problem of modern physics research is the solution to QCD in the
non-perturbative regime.  One method of testing QCD in this low-energy region
is by measuring certain structure constants of hadrons --- called
polarizabilities --- that show particular promise of allowing a direct
connection to the underlying quark/gluon dynamics through comparison to modern
QCD-inspired model calculations, and to solutions of QCD done computationally
on the lattice.

The Mainz laboratory provides a high-quality, high-flux, continuous-wave,
1.5-GeV electron beam providing a beam of polarized photons, a near-$4\pi$
CB-TAPS detector system, and a frozen-spin polarized proton target\footnote{Schlimme \emph{et al.}, arXiv:2402.01027 [nucl-ex] (2024)}.  The HIGS
facility at Duke provides similar targets but with lower-energy, higher-flux
beams, complementary to Mainz but with higher linear and circular
polarization\footnote{C.R. Howell \emph{et al.}, International Workshop on Next Generation Gamma-Ray Source, J. Phys. G: Nucl. Part. Phys. 49, 010502 (2022)}.  Both allow for experiments with unique access to high-precision
measurements of nucleon structure.  Obtaining new, precise nucleon-structure
data is the aim of each of these experiments.

\paragraph{Research Goals}

Due to the absence of free-neutron targets, neutron polarizabilities—especially spin polarizabilities—are poorly known. A small group led by David Hornidge at Mount Allison University in New Brunswick is developing a high-pressure helium target for elastic Compton scattering experiments on both $^3$He and $^4$He, using the CB-TAPS system, complemented by low-energy measurements at HIGS. Supported by Chiral Perturbation Theory (ChPT), this work aims to enable the first accurate extraction of neutron scalar polarizabilities and will continue through the decade. The same group is also conducting measurements of both charged- and neutral-pion polarizabilities using the modified GlueX detector at JLab—key tests of low-energy QCD. 

The new MAGIX detector at the MESA accelerator facility in Mainz will be used to measure elastic electron-proton scattering at low momentum transfer and varied kinematics, enabling a model-independent determination of the proton's magnetic form factor and the extraction of the Zemach radius. This program is expected to continue into the next decade. In parallel, the high-flux beam from MESA will support high-precision quasi-real Compton scattering on proton and nuclear targets, aiming to significantly reduce uncertainties in both proton and neutron polarizabilities.

\subsubsection{Canadian participation in the Electron-Ion Collider
\label{sec:eic}}

The Electron-Ion Collider (EIC) will be a new US\$2.8B particle collider facility to be built at the start of the next decade at Brookhaven National Laboratory (BNL), Long Island, New York, by the US Department of Energy (US-DOE). The EIC is the only new collider worldwide to be built in the foreseeable future and the first in this century, which naturally makes it the next subatomic physics {\em discovery} machine. At the EIC, polarized electrons will collide with polarized protons, light ions, and heavy nuclei at luminosities far beyond what is presently available (see Fig.~\ref{fig:eic}). The facility will answer fundamental questions about the origin of mass and spin, and about gluon dynamics. During the construction phase (2027--2034), EIC Canada aims to construct (with support from CFI-IF) the end-of-sector readout electronics and the interaction region crab cavities for the proton storage ring. During the initial operations phase (2034-2041), EIC Canada will lead a physics program studying the origin of mass through meson form factors, the strong dynamics giving rise to exotic hadrons, and the search for new physics in the electroweak sector; all topics enabled by the high luminosity, high polarization, and excellent resolution of the EIC and ePIC detector. 

\paragraph{Research Goals}

The EIC will uniquely address profound questions about nucleons and how they are assembled to form nuclei.  In addition, the EIC presents significant opportunities that connect to neutrino, high-energy and particle physics, as well as astrophysics \footnote{{\em EIC Yellow Report: Science Requirements and Detector Concepts
for the EIC}, Nucl.Phys.A 1026 122447 (2021)}. 

\subparagraph{How does the mass of the nucleon arise?}
While gluons have no mass and $u$, $d$ quarks are nearly massless, the total mass of a nucleon (proton or neutron) is 100 times greater than the mass of its three valence quarks. The largest contribution to the nucleon mass originates from the gluon field energy. 
The EIC will 
determine, for the first time, the relative spatial size of the distributions of valence quarks, sea quarks, and gluons, and the spatial structure of the different contributions to the energy density and pressure forces in the nucleon. Studies of $\pi$ and $K$ structure over a broad $Q^2$ range will probe the contributions of quark and gluon energy to the hadronic mass, and exclusive production of $J/\psi$ and $\Upsilon$ will probe contributions of the trace anomaly.

\subparagraph{How does the spin of the nucleon arise?}
Understanding nucleon spin in terms of quark and gluon spin and angular momentum contributions has been an essential goal since the discovery by the EMC Experiment that quark polarization contributions comprise only $\sim$30\% of nucleon spin.
The EIC will dramatically improve our understanding by studying the resolution dependence of polarized Deep Inelastic Scattering (DIS), and by measuring exclusive reactions, where precise knowledge of the spin of gluons combined with Generalized Parton Distribution (GPD) sum rules allows isolating the contribution of the orbital angular momentum of gluons.

\subparagraph{What are the emergent properties of dense systems of gluons?}
The nature of gluons in matter, i.e.~their arrangements or states, and the details of how they hold matter together, is not well known. Experiments at the EIC will be able to explore modifications of the quark distributions in nuclei in the limit of low Bjorken-x, $x_B$, where the number of gluons in the target is very large.

\subparagraph{What is the composition of exotic hadrons?}
The EIC has the potential to produce hadronic resonances, which are exotic in nature, in photoproduction reactions in the heavy-quark sector. Such studies could confirm observations from previous experiments and provide complementary insight into their composition (tetraquarks, pentaquarks, molecules, etc.).

\subparagraph{What lies beyond the Standard Model?}
The study of electroweak and beyond the Standard Model physics program relies on the high luminosities enabled by Canadian crab cavities. It includes the precision study of neutral current inclusive scattering to test the Standard Model to extract neutral current inference structure functions and the search for physics beyond the Standard Model in the form of charged-lepton flavor violation.

\paragraph{Methodology}

\begin{figure}[h!]
\centering
 \includegraphics[width=0.8\textwidth]{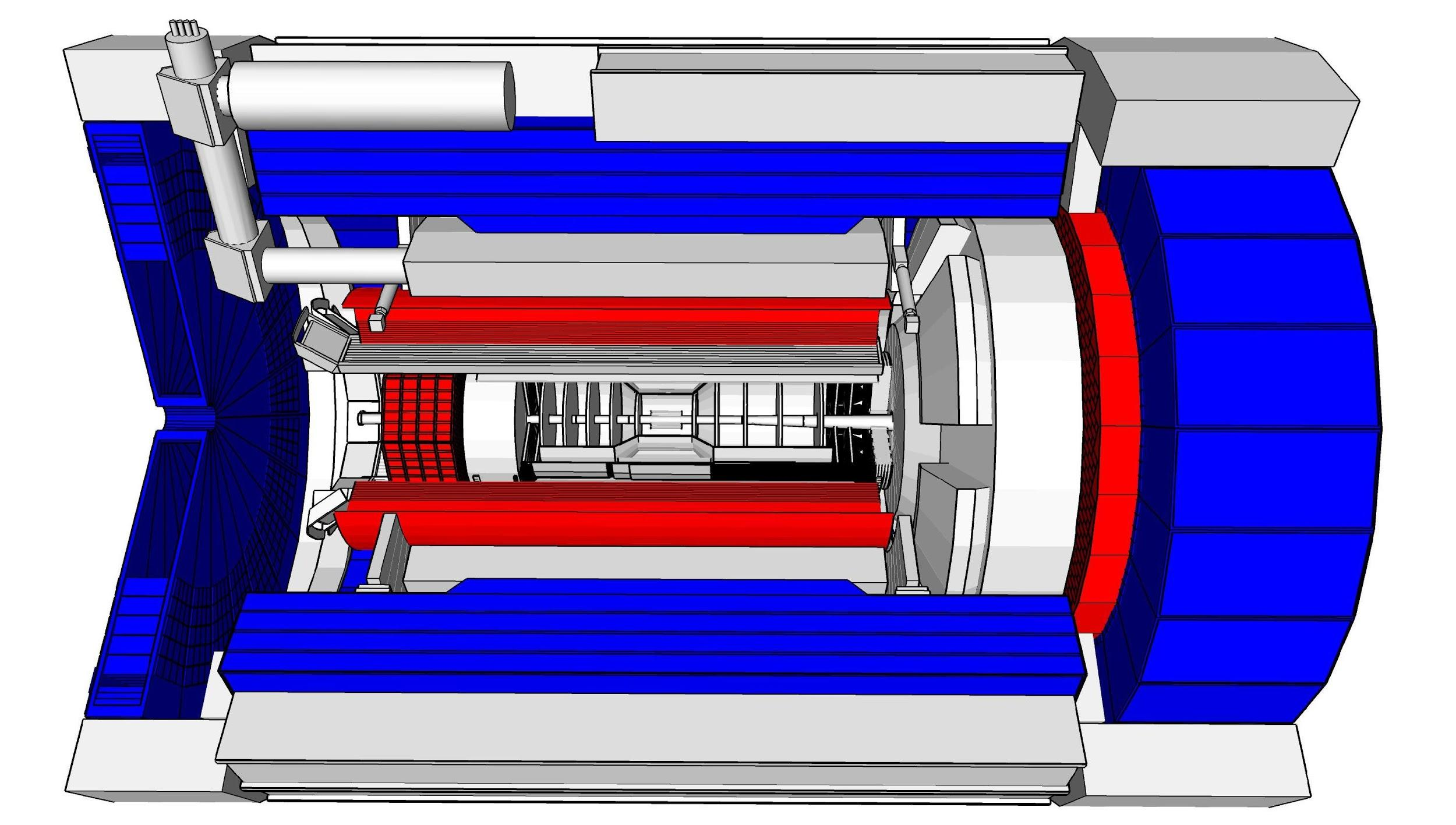}
 \caption{Schematic drawing of the 9\,m-long ePIC detector. The central red calorimeter detector is the Barrel Imaging Calorimeter (BIC).}
 \label{fig:epic}
\end{figure}

Answering the above questions relies crucially on the \emph{1000-fold increase of collision intensity} compared to previous electron-proton collider facilities (enabled in part by the accelerator infrastructure in this project) and on the \emph{polarization of both the electron and ion beams} which has not been realized at any previous collider. The EIC is unique in several other aspects as well:

\begin{itemize}
    \item The \emph{collision center-of-mass energy will be variable} by almost a factor 10, from 29 to 140 GeV. This capability is necessary to study the proton at multiple energy and distance scales and has not been available at previous colliders.
    \item The collision of an electron (a well-understood point particle) with a proton or ion (the less-understood object under study) \emph{differs fundamentally} from the collision of two protons or ions with each other, or of two electrons with uncontrolled creation of secondary particles, which are the only modalities available at colliders currently in operation or planned for the next decade.
    \item The ePIC detector (see Fig.~\ref{fig:epic}) will be the first subatomic physics detector \emph{designed in and for the era of artificial intelligence (AI)}. Design parameters were optimized with AI, and high-granularity technologies are chosen in the anticipation of data-driven AI algorithms for electron/pion identification.
\end{itemize}
The development of tomographic images of the structure of nucleons requires the separation of reactions along multiple kinematic dimensions. Instead of only the longitudinal momentum of the struck quark, or only the transverse impact parameter of the collision, or only the momentum transfer from the electron to the nucleon, a full tomography of the proton requires separating along all dimensions simultaneously. 

\paragraph{Seven-year year outlook}

The EIC project is moving rapidly and steadily.
Over the next seven years, the EIC community plans to achieve the next critical decisions (CD): CD-3C (third long lead procurement, mid-2026), CD-2 (performance baseline, June 2026), and CD-3 (start of construction, late 2027). 

\begin{figure}[h!]
\centering
\includegraphics[width=3.0in]{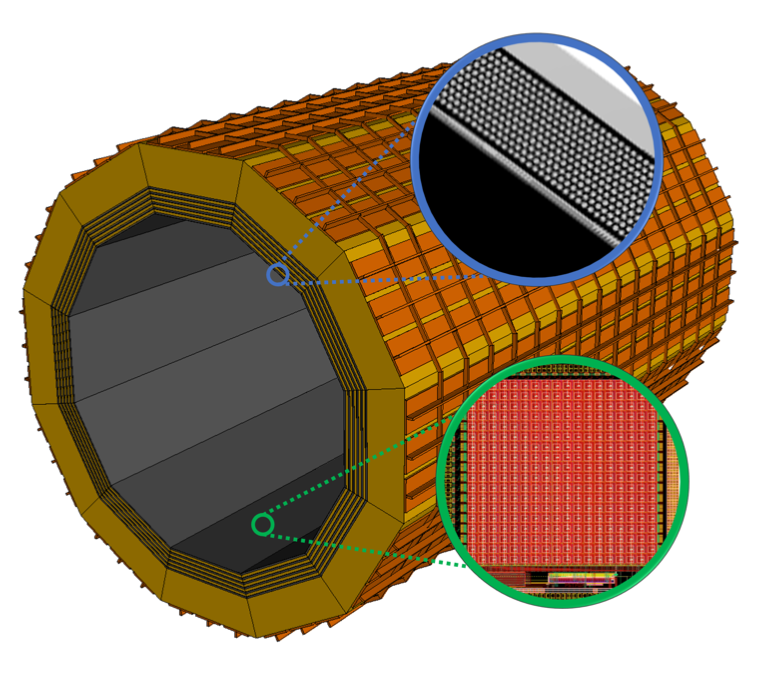}
\caption{\label{fig:imagingcal}
  A sketch of the Barrel Imaging Calorimeter geometry. It is comprised of an outer ``bulk'' PbSciFi section, with five 2-cm-thick ``imaging layers'' of PbSciFi interleaved with six AstroPix tracking layers, shown in the zoomed circles, respectively. The BIC will be 435~cm long, with inner and outer radii of 82~cm and 122~cm, respectively, and will weigh $\sim$40 metric tons. The bulk PbSciFi section is modelled after the GlueX BCAL\@, built by U.~Regina. Two-sided silicon photomultiplier readout will be implemented for spatial resolution along the z-coordinate (or pseudorapidity $\eta$).
  }
\end{figure}

EIC Canada is concentrating its efforts on several high priority areas:
\paragraph{Barrel Imaging Calorimeter (BIC)}
The EIC physics topics lead to unique requirements for the electromagnetic calorimeter design, as nearly all channels require detection of the scattered electron for momentum or energy reconstruction and particle identification. 
The Barrel (electromagnetic) Imaging Calorimeter (BIC) is a hybrid imaging calorimeter (see Fig.~\ref{fig:imagingcal}) 
that is cost-effective in relation to its excellent performance in energy and spatial reconstruction and particle identification, and its design is driven by Argonne National Lab (ANL), U Regina, and U Manitoba
Evaluation of the Response to Electrons and Pions in the Scintillating Fiber and Lead Calorimeter for the Future Electron-Ion Collider
\footnote{
H, Klest et al., {\em Evaluation of the Response to Electrons and Pions in the Scintillating Fiber and Lead Calorimeter for the Future EIC}, J.Instr. 20 P07028 (2025); S. Oresic, Univ. Regina PhD July 2025.}
. The Canadian EIC team has submitted a CFI-IF application to fund the Canadian contributions to the BIC 
and expects to complete this construction by 2029.

\subparagraph{Superconducting RF Crab Cavities}
The accelerator work is part of a separate submission.  
TRIUMF’s accelerator physicists and engineers have the expertise to support the construction of the advanced EIC accelerator complex as a Canadian in-kind contribution. The main contribution described will be the unique 394 MHz crab cavity cryomodule system for the hadron storage ring (HSR) and the electron storage ring (ESR).  The submitted CFI-IF application also includes the Canadian contributions to this crab cavity cryomodcule system.

\subparagraph{Simulations in support of the EIC physics program}

The Regina group is leading feasibility studies of charged pion and kaon form factor measurements at the EIC, in support of the ``mass of the nucleon'' scientific question \footnote{J Arrington et al., {\em Revealing the structure of light pseudoscalar mesons at the EIC}, J.Phys.G 48 075106 (2021)}.
In addition to their intrinsic physics value, our $\pi^+$ and $K^+$ form factor feasibility
studies are important for detector design validation, as they provide well-defined but challenging final
states that test the far-forward event reconstruction, and help the
Collaboration understand the far-forward detector requirements.  
This work will be ongoing for the next 2-4 years.

The Manitoba group is performing studies of projected electroweak (EW) and beyond the Standard Model (BSM) physics at the EIC. It relies on the high luminosities enabled by the Canadian crab cavities described above.  
The group is developing machine learning-based event selection of exotic processes, and evaluating the systematic uncertainties in measurements of electroweak structure functions.
Because of their topology (balanced jets, large missing $p_T$), Charged Lepton Flavor Violating (CLFV) events mimic neutral current DIS, charged current DIS, and photo-production backgrounds. The Manitoba group is therefore approaching event selection as a multi-variate event selection problem with machine learning solutions.

\subsection{Beyond the next 7 years}

The JLab work continues in Hall~C (electron scattering and spectrometer)
and GlueX in Hall~D (linearly polarized tagged photons and a hermetic
detector) into the next decade.  The team is significantly involved in
data-taking and lead experimental efforts in the analysis and publication of
data, including Canadian-led publications.  The JLab Eta Factory (JEF)
experiment (an upgrade of the base GlueX apparatus) was completed with
Regina participation and took its first data in 2025.  SoLID's design is mature and has been listed as Jefferson Lab's next major scientific priority.  Construction and installation in Hall~A is expected for FY27--32, after the completion of the
MOLLER experiment (Section \ref{sec:moller}).  The Regina group was recently awarded CFI-Innovation Fund support for SoLID Heavy Gas Cherenkov detector construction, with delivery to JLab planned in 2028.

A portion of the experimental efforts of the team will transition to EIC as
several JLab projects are completed.  The exception is the SoLID experiment at
JLab that will use the latest detector and readout technology to enable an
increase in luminosity by a factor of ten compared to existing detectors.
 The SoLID experiment will see
continuing activity through the end of the new Canadian Subatomic Physics Long Range Plan (2041).

\begin{figure}
\centerline{\includegraphics[width=\textwidth]{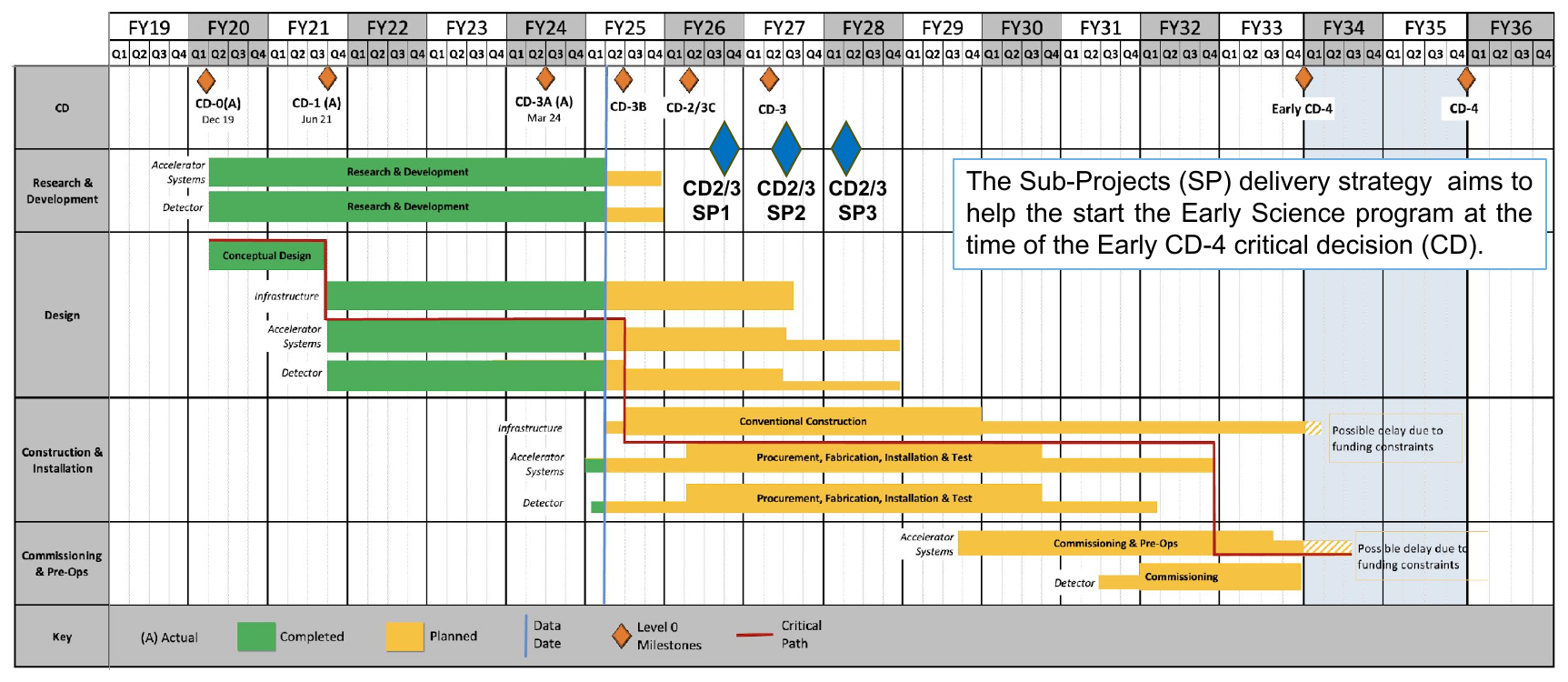}}
\caption{Top-level project schedule of the EIC with indication of Critical Decisions (CD) and Sub-Projects (SP), as of the January 2025 CD-3B review. Early physics studies will be possible during the commissioning and pre-operations phases, with increasing luminosity as time goes on.  The US fiscal year (Oct 1-Sept 30) is used.}
\label{fig:eic:schedule}
\end{figure}

A recent EIC project schedule is shown in Fig. \ref{fig:eic:schedule}.  Currently, the EIC physics program is projected to begin in FY34, and is expected to continue for $\sim$20 years.
The Canadian EIC group projects that the Canadian detector construction for the EIC and commissioning efforts will result in an increase to 15 HQP by 2029, and the start of physics data taking will result in an increase to 21 HQP supervised by an integrated 5.6 FTE investigators and funded researchers. The start of the first North American collider of this century will be associated with extraordinary scientific interest, marking a transformative milestone for particle and nuclear physics. In the first years of the 2030s, significant new results will be published by the detector collaboration(s), both in the form of instrumentation performance publications and initial physics results with the ePIC detector. In the years after 2034, when the EIC will have reached full design parameters including hadron cooling and double polarization, numerous other publications will result.

During the 2034--41 period, JLab intends two major upgrades of its capabilities.  The first is a positron beam upgrade motivated by the fact that interferences between many reaction amplitudes change sign between $e^-$ and $e^+$ beams, allowing reaction phase structures to be uncovered.  Such ``charge asymmetry'' data from HERMES proved to be vital in pioneering SIDIS and GPD investigations, and 12 GeV positron beam from JLab would enable a vastly improved data set to be acquired.  The positron beam upgrade would occur FY32--35, with the physics program planned for FY36--38.  The second major JLab capability upgrade is the 22 GeV upgrade project intended to complement the EIC program by filling in the gap between the JLab 12 GeV physics reach and the EIC. While the positron program is running, advanced R\&D for the 22 GeV upgrade would be carried out, with the intent for construction to occur in FY35--40 and 22 GeV physics to begin in FY41.  One of the Canadian team members has helped to organize two workshops on the proposed 22 GeV physics program, to help set the stage for the next generation of physicists.  SoLID, Hall~C and GlueX would all make use of the positron and 22 GeV beams.
Future theory efforts will rely on tight integration of theory with experimental design, precision calculations, advanced simulations, and global data analyses. Canadian experts will make critical contributions to hadronic/QCD physics in both theory and experiment while training the next generation of HQP, foster inclusive research environments, and strengthening Canada’s international leadership.

\subsection{Summary}

The research community engaged in hadronic and QCD physics in Canada continues to pursue a rich and diverse program aimed at probing the strong interaction across all its facets—from understanding the many-body dynamics of QCD at zero and finite temperatures to exploring the transition between hadronic and partonic degrees of freedom. Close collaboration between theory and experiment is essential to fully realize the scientific impact of next-generation facilities, and Canadian researchers are exceptionally well positioned to play a leadership role in these integrated theory–experiment efforts. Canadian researchers have taken on key responsibilities in major experimental collaborations and have made fundamental contributions to theoretical developments, including precision QCD calculations and effective field theory approaches. A major recent milestone is the successful operation of the upgraded Jefferson Lab 12 GeV facility, which is now delivering unique insights into nucleon and nuclear structure. Looking ahead, planned upgrades at Jefferson Lab will significantly extend the hadron structure program while providing a critical physics and methodology bridge to the Electron–Ion Collider. The EIC has now moved beyond the planning stage and is entering construction at Brookhaven National Laboratory. As construction proceeds, both experimental and theoretical programs are ramping up to ensure that the EIC will deliver its full scientific potential from the beginning of operations. The Canadian hadronic/QCD community is already playing an active role in detector design and R\&D—most notably in projects such as the Barrel Imaging Calorimeter and superconducting RF crab cavities—as well as in experiment planning and theoretical modeling. To fully realize the scientific potential of these new-generation facilities and to sustain a vigorous theoretical program, continued strong support for Canadian researchers and strategic investment in the training and retention of HQP will be essential.

\newpage
\section{Nuclear Structure}

\subsection{The Canadian program (current and next 7 years)}

\subsubsection{Introduction}
At the heart of every atom lies the atomic nucleus, a dense core composed of protons (Z) and neutrons (N), collectively known as nucleons, bound together by the strong nuclear force — one of the four fundamental forces of nature. Despite its minuscule size, the nucleus is one of the most complex quantum systems in the universe and is central to answering some of the most profound questions in science. From the nature of neutrinos and dark matter, to the origin of matter-antimatter asymmetry, to the inner workings of neutron stars, and the astrophysical origins of the chemical elements, nuclear physics plays a crucial role.

The field of nuclear structure investigates how nucleons are arranged and interact within the nucleus, and how they decay. Nuclear Structure also explores how simple interactions between particles give rise to the complex and collective behaviors of the nuclear many-body system (as seen in Fig.~\ref{fig:bigquestions_scales}), and seeks to understand the limits of nuclear existence — how many protons and neutrons can be held together before a nucleus becomes unstable.

While it is well established that nearly all elements heavier than helium are synthesized in astrophysical environments — such as novae, supernovae, and neutron star mergers, the precise pathways of nucleosynthesis seen in Fig.~\ref{fig:Nucleosynthesis} intrinsically linked with specific nuclear properties, still pose major open questions in astrophysics. To fully understand the cosmic origin of the elements, we must know the detailed properties of nuclei from masses and binding energies to half-lives and decay modes, and even the excitation energies and structure of nuclear states. These properties serve as critical inputs for astrophysical models, helping us unravel the conditions in which the elements were formed and understand the elemental and isotopic abundances observed in the Universe today.


To interpret experimental data and make predictions at the limits of the nuclear stability, physicists use a range of complex theoretical models. The nuclear shell model treats nucleons as moving in discrete quantum levels within a potential well, successfully explaining magic numbers and nuclear stability. Modern \emph{ab-initio} models aim to describe nuclei from first principles, starting from the basic interactions between nucleons — often derived from quantum chromodynamics through effective field theories. These calculations, powered by high-performance computing, seek a predictive framework applicable across the nuclear landscape. Together, these models — from phenomenological to fundamental — guide and complement experimental investigations, advancing our understanding of the building blocks of matter and the forces that shape them.

\begin{figure}
\centering
\begin{tcolorbox}[
  colback=teal!30,
  colframe=gray!50!black,
  coltitle=black,
  colbacktitle=gray!20,
  width=0.8\textwidth,
  boxrule=0.8pt,
  arc=4pt,
  auto outer arc,
  fonttitle=\bfseries,
  title=Nuclear structure in Canada and its position in the world
]
{\small\baselineskip=14truept Canadian researchers are internationally recognized leaders in nuclear structure science. Through TRIUMF — Canada’s particle accelerator centre — and its suite of world-class instruments (DRAGON, EMMA, GRIFFIN, IRIS, TIGRESS, TITAN, and now the ARIEL facility), Canadians have mastered measurements of exotic isotopes with beams as weak as a few hundred particles per second, developed novel detector technologies, and pushed the boundaries of theoretical modeling in collaboration with leading groups worldwide. These investments have positioned Canada at the forefront of rare-isotope science, the global frontier of nuclear physics.}
\end{tcolorbox}
\end{figure}

\subsubsection{Evolution of the single-particle states in nuclei as a function of both neutron- (N) and proton-number (Z)}
\label{sec:NS-Evolution}
Our understanding of the shell structure has been largely based on nuclei on or near the valley of stability, where the Nuclear Shell Model has successfully predicted the magic numbers that confer extra stability to nucleons consisting of combinations of Z and N corresponding to 2, 8, 20, 28, 50, 82, and N = 126 nucleons (marked by grey horizontal and vertical lines in Fig.~\ref{fig:Nucleosynthesis}). However, theoretical models have predicted that the relative energies of the single-particle orbitals can shift significantly, in some cases causing the disappearance of the traditional magic numbers and the appearance of new ones, especially in the neutron-rich regions of the nuclear chart.\footnote{T.\ Otsuka \emph{et al.}, \ Rev.\ Mod.\ Phys.\ 92, 015002 (2020).}

Halo nuclei represent a particularly interesting class of nuclei in the nuclidic chart. Found at the driplines, these systems consist of a compact nuclear core surrounded by orbiting protons or neutrons, resulting in radii that are significantly larger than predicted by the classical $A^{1/3}$ dependence valid near stability. Among them, the most neutron-rich dripline nucleus, $^8$He—traditionally considered a 'doubly magic' nucleus due to its shell closures—was investigated at the TRIUMF Isotope Separator and ACcelerator (ISAC) experimental facility using IRIS, short for ISAC Charged Particle Reaction Spectroscopy Station facility (Fig.~\ref{fig:IRIS-facility}). The experiment revealed a large deformation in its lowest 2$^{+}$ resonance state. Despite its doubly magic character, $^8$He shows clear evidence of neutron deformation, in agreement with \emph{ab initio} no-core shell model calculations that successfully reproduce the resonance energy, width, and transition density. This study establishes a novel paradigm in which even doubly magic nuclei can exhibit deformation in excited states.\footnote{M. Holl et al., Phys. Lett. B 822, 136710 (2021).} The results received international attention, with media coverage in Physics World, Phys.org, and Apple News.

\begin{figure}
    \centering
    \includegraphics[width=0.45\linewidth]{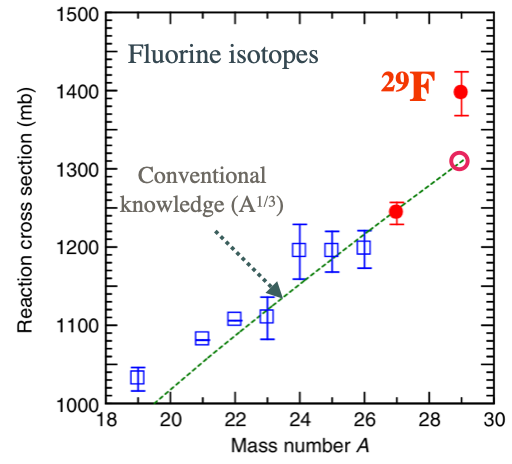}
    \includegraphics[width=0.45\linewidth]{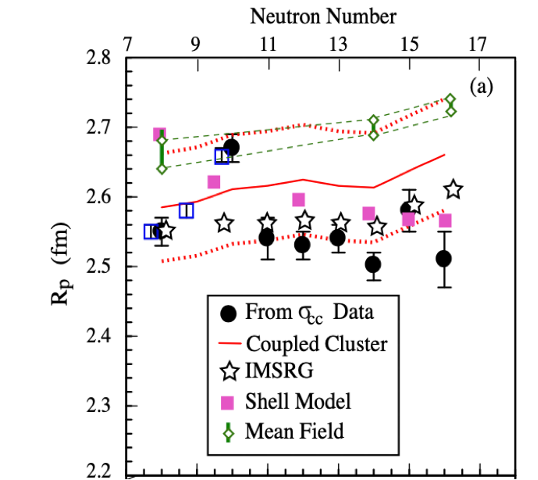}
    \caption{(Left) Measured reaction cross sections of $^{27,29}$F isotopes in red compared to others in blue. The dashed line shows the trend of A$^{2/3}$ relative evolution normalized for best fit to $^{19}$$^{-}$$^{27}$F [S. Bagchi \emph{et al.}, Phys. Rev. Lett. 124 222504 (2020)]. (Right) Experimental root mean square proton radii (R$_{p}$) (black circles) compared to various models.}
    \label{fig:F-29_RIKEN}
\end{figure}

In parallel with the local program at TRIUMF, an experiment at RIKEN, RIBF in Japan has unveiled the heaviest two-neutron Borromean halo to date, in the last bound N = 20 nucleus, $^{29}$F (see Fig.~\ref{fig:F-29_RIKEN}). This finding also demonstrated the breakdown of the N = 20 and 28 traditional shell closures. It showed that for both N = 8 and 20 the breakdown of traditional shell closure with the appearance of a Borromean neutron halo holds. This suggests that such a phenomenon might exist in the higher shell closures too. \footnote{S. Bagchi \emph{et al.}, Phys. Rev. Lett. 124 222504 (2020).}

Complementary to RIKEN, a research program to extract the root mean square proton distribution radii of neutron-rich nuclei at the FRagment Separator (FRS) facility at GSI in Germany using a new technique measured the charge changing cross section of neutron-rich oxygen isotopes from the stable $^{16}$O up to the dripline nucleus $^{24}$O revealing a local minima, which signals the existence of unconventional shell closures at N = 14 and 16 due to the attractive proton-neutron monopole tensor interaction. \footnote{Phys. Rev. Lett. 129 142502 (2022).}

In the light regions, in an attempt to connect the islands of inversion at N = 20 and 28, $^{31-35}$Na and $^{31-35}$Mg were measured in the TITAN's MR-TOF. The mass of $^{33}$Na was improved by two orders of magnitude, and a millisecond isomer was discovered in $^{32}$Na. These results show that the N = 20 shell reaches a minimum strength at $^{32}$Mg and rises at N = 31 hinting an enhanced shell strength at unbound $^{28}$O and were compared to state-of-the-art \emph{ab initio} nuclear predictions.\footnote{M. Lykiardopoulou \emph{et al.}, Phys. Rev. Lett. 134 052503 (2025).}

The power of the GRIFFIN spectrometer (shown in Fig.~\ref{fig:GRIFFIN-all}) enables Canadian researchers to pursue detailed studies of nuclei into the very neutron-rich region, in some cases approaching the $r$-process path. For example, studies of nuclei near $^{132}$Sn have provided not only crucial information on nuclear structure, but also precise measurements of $\beta$-decay lifetimes, where the high $\gamma$-ray energy resolution of GRIFFIN allows accurate distinction between different $\beta$-decaying parent states.\footnote{R.\ Dunlop \emph{et al.},\ Phys.\ Rev.\ C 99, 045805 (2019), K. \ Ortner \emph{et al.},\ Phys.\ Rev.\ C 102, 024327 (2020), K. Whitmore \emph{et al.},\ Phys.\ Rev.\ C 103, 024310 (2021).} With the enhancements of the CANREB and ARIEL facilities, these measurements will push further into the neutron-rich regions while obtaining higher-quality data.

Other experiments proved the persistence of the N = 82 shell with almost unmodified shell gap energies established up to the proton dripline by using high-accuracy, high-precision mass measurements of neutron-deficient Yb isotopes with TITAN. This allowed first-time measurements of the ground state masses of $^{150,153}$Yb and the excitation energy of $^{151m}$Yb.\footnote{S.\ Beck \emph{et al.},\ Phys.\ Rev.\ Lett.\ 127, 112501 (2021).} As a result, mean field calculations were performed to unravel the puzzling systematics of the h$_{11/2}$ excited states of the N = 81 isotones. Remarkably, the measurements of the two most neutron-deficient isotopes of Yb were only possible by the demonstration of a new in-situ beam purification technique at TITAN, which allows for dynamic ranges of 1.0$\times 10^4$ to 1.0$\times 10^8$.

\begin{figure}[ht]
\centering
  \includegraphics[width=3.0in]{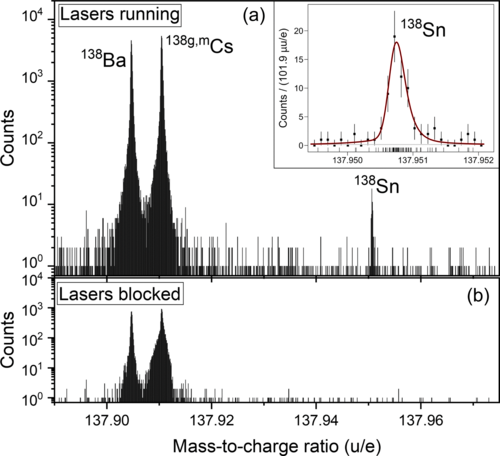}
	\caption{First-time mass measurement of $^{138}$Sn in the TITAN MR-TOF. [A. Mollaebrahimi \emph{et al.}, Phys. Rev. Lett. 134, 232701 (2025).]} 
    \label{fig:Sn-138-MRTOF}
\end{figure}

TITAN also pushed the limit of their experiments beyond the doubly-magic $^{132}$Sn nucleus and performed the first-time mass measurement of $^{136,137,138}$Sn using its MR-TOF showcasing the MR-TOF’s in-situ beam purification (spectrum shown in Fig.~\ref{fig:Sn-138-MRTOF}), disentangling the Sn from overwhelming isobaric contamination intrinsic to RIB production. These measurements enhance our understanding of the nuclear structure and provide input for astrophysical nucleosynthesis in this previously unexplored region. \footnote{A. Mollaebrahimi \emph{et al.}, Phys. Rev. Lett. 134, 232701 (2025).} Three methodologies to measure nuclear half-lives ($>$ a few ms) in the TITAN MR-TOF have been developed and tested for $^{100}$Rb and $^{54}$K, and neutron-rich isotopes of Yb. \footnote{I. Mukul \emph{et al.}, J. Phys.: Conf. Ser. 1643 012057 (2020), C. Walls, Univ. Manitoba, PhD August 2025, C. Walls, \emph{et al.}, to be submitted to PRC (2025).}

\begin{figure}[ht]
    \centering
    \includegraphics[width=0.40\linewidth]{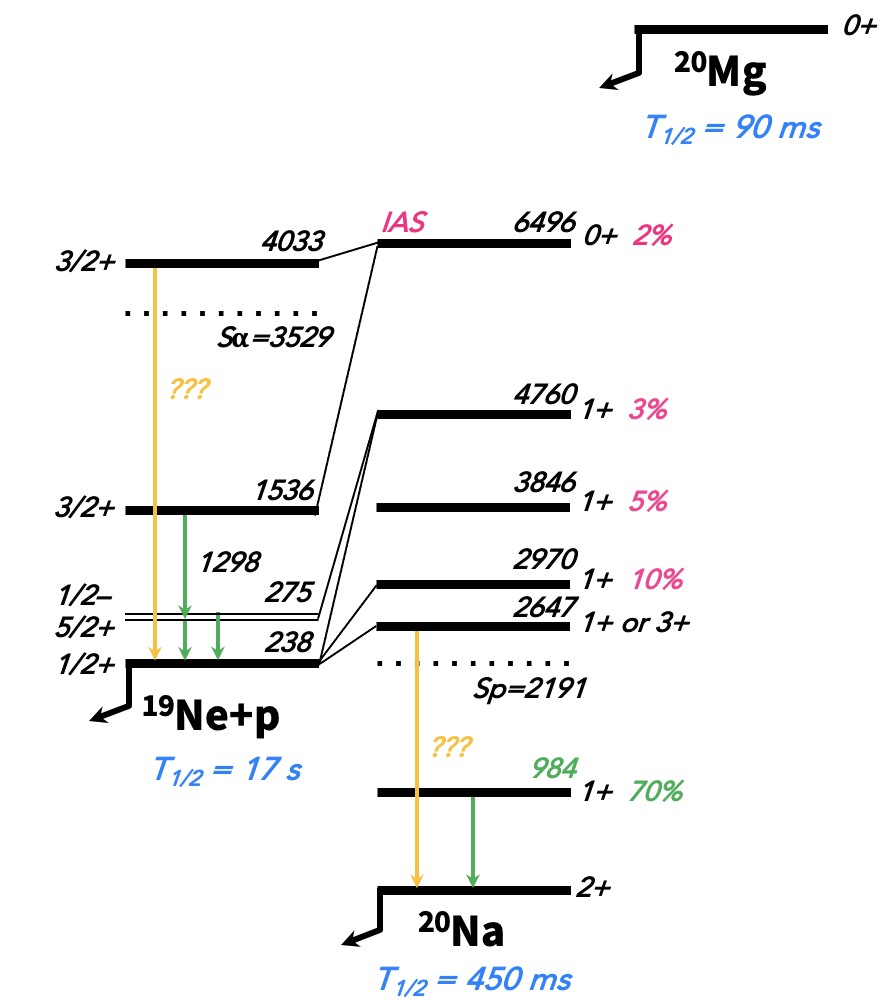}
    \includegraphics[width=0.40\linewidth]{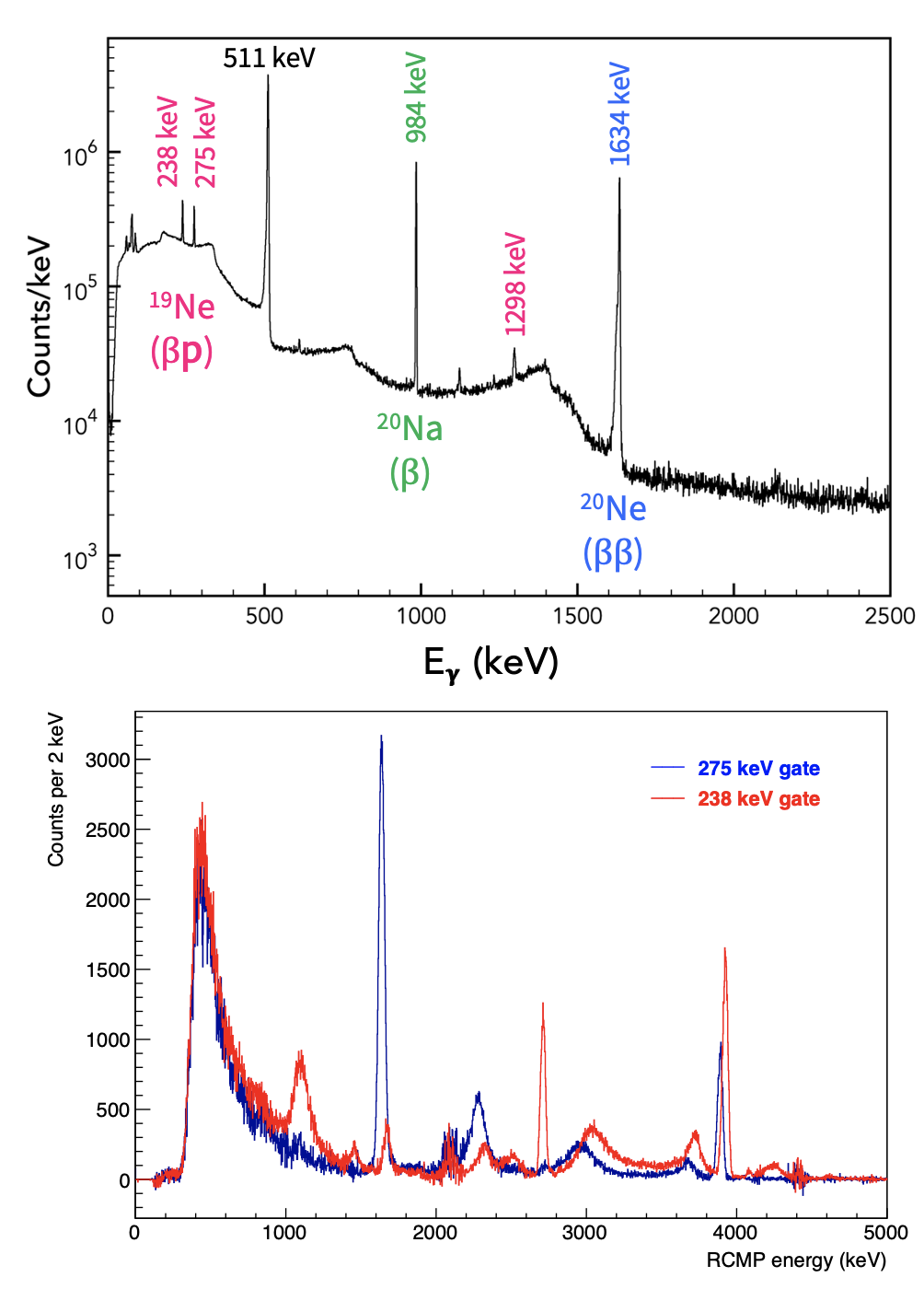}
    \caption{(Left) The decay level scheme of $^{20}$Mg investigated using GRIFFIN and the new RCMP detector in 2024. (Right) Top: The $\gamma$-ray spectrum obtained with GRIFFIN showing the 238 and 275 keV peaks marked in red from the $^{20}$Mg $\beta$ delayed proton decay daughter $^{19}$Ne. Bottom: Gates on these two transitions produces the overlap proton spectra. The biggest peak in the  the 275 keV gated spectrum in blue around 1.6 MeV was observed for the first time in this experiment.}
    \label{fig:Mg20}
\end{figure}

New experiments at TRIUMF to study proton-drip line nuclei were performed with the newly commissioned RCMP detector build by URegina shown in Fig.~\ref{fig:RCMP-ARIES} (left), and GANIL ACTAR TPC detector shown in Fig.~\ref{fig:ACTAR collaboration}. Figure~\ref{fig:Mg20} shows spectra from the successful RCMP commissioning experiment to study the $\beta$-delayed charged-particle emission from $^{20}$Mg; it highlights the power of coupling the RCMP with GRIFFIN and exploiting the proton-gamma coincidences which reveals new transitions unseen in previous experiments.

The ACTAR TPC detector and all of its associated electronics and equipment were shipped from GANIL to TRIUMF to perform three experiments in 2025. This campaign took combined advantage of the fact that such a detector does not yet exist in Canada and operated it with rare-isotope beams that are only available at the ISAC-II facility at intensities required for the experiments. Part of the 2025 ACTAR Collaboration is seen in Fig.~\ref{fig:ACTAR collaboration} (right) after the installation of the array.


Proton unbound single-particle states in the proton dripline nucleus $^{21}$Al was studied in the $^{20}$Mg($p,p^{'}$) reaction using the GANIL ACTAR TPC to constrain nuclear theory. A second experiment proposed the $^{17}$F($p,p^{'}$) reaction to populate and search for $\alpha$ and 2p decay from the the 6.15 MeV
resonance in $^{18}$Ne to constrain reaction modeling of the breakout of the Hot Carbon-Nitrogen-Oxygen (HCNO) cycle in explosive astrophysics. A deuterium gas target was also used to study the $^{11}$Li(d,p)$^{12}$Li reaction to test continuum coupling effects in the limits of extreme neutron-rich matter. These experiments relied exclusively on the combination of TRIUMF’s high-quality beams, delivered at the energies and intensities required to make measurements with ACTAR feasible, and the exceptional sensitivity of the ACTAR active target detector. Thus, this synergy resulted in a distinctive and highly successful collaboration. The data analysis of these experiments is in progress.

To firmly establish single-particle energies, systematic studies of single-nucleon transfer reactions are essential. These challenging experiments require the combination of TIGRESS with charged-particle detectors such as SHARC, allowing the measurement of cross sections to specific excited states and the extraction of particle angular distributions needed to determine spectroscopic strengths. Measurements in the neutron-rich Ca region are particularly valuable. Monte Carlo Shell Model calculations\footnote{T. Togashi et al., Phys. Rev. Lett. 117, 172502 (2016).} predict a so-called Type-II shell evolution, where nuclei self-organize into states with different shapes and altered single-particle energies. This effect is expected to be most pronounced in the N = 60 Sr–Zr isotopes, where a sharp ground-state shape transition is observed. Transfer reactions on Sr isotopes near N = 60 will provide a sensitive probe of this phenomenon, requiring long-duration experiments that will be enabled by the future ARIEL facility.

In future, IRIS will engage in new experiments with medium heavy nuclei using beams from ARIEL and CANREB. A particular focus will be to investigate neutron capture cross sections via the indirect (d,p) nucleon transfer reaction in
neutron-rich nuclei relevant for astrophysical processes. A new active target TPC
(called EXACT-TPC) to add new capability of studying nuclear reactions and exotic decays which
will be complementary to currently existing infrastructure at TRIUMF will be built and added to this facility. Development of an Active Target TPC will
bring complementary advantages to further expand the scientific program, especially enabling
reactions with $^{3,4}$He and detecting rare particle emitting decay modes. In this context, the 2025 ACTAR campaign at TRIUMF served as a crucial first step towards the realization of EXACT-TPC at TRIUMF.


\subsubsection{Shape coexistence in atomic nuclei}
\label{sec:NS-ShapeCoexistence}

Beyond the single-particle picture, nuclei often exhibit collective behaviors that emerge from the correlated motion of many nucleons. Among the most fascinating manifestations is shape coexistence, where the same nucleus can support multiple configurations with distinctly different shapes — spherical, prolate (rugby-ball), or oblate (disk-like) — at nearly the same excitation energy. They provide perhaps the most demanding tests of nuclear structure models, and permit the extraction of correlation energy as a function of deformation on within the same nucleus.~\footnote{K. Heyde and J.L. Wood, Rev.\ Mod.\ Phys.\ 83, 1467 (2011), P.E. Garrett, M. Zieli\'nska and Cl\'ement, Prog.\ Part.\ Nucl.\ Phys.\ 124, 103931 (2022).}  At TRIUMF-ISAC, Canadian researchers lead an ongoing program exploring shape coexistence in many regions of the nuclear chart \footnote{M. Rocchini \emph{et al.}, Phys. Rev. Lett. 130, 122502 (2023), K. Mashtakov \emph{et al.}, Acta Phys. Pol. B 18, 2-A21 (2025), D. Kalaydjieva \emph{et al.}, Acta Phys. Pol. B 16, 4-A15 (2023), P.E. Garrett \emph{et al.}, Phys. Rev. Lett. 123, 142502 (2019), K. Ortner \emph{et al.}, Phys. Rev. C 102, 024323 (2020), S.A. Gillespie \emph{et al.}, Phys. Rev. C 105, 044313 (2021), J. Smallcombe \emph{et al.}, Phys. Rev. C 110 024318 (2024).} in which the emergence of multiple coexisting shapes and the nature of quantum phase transitions in the nuclear shape are studied. Data obtained from this program have been key in a paradigm shift away from interpretations involving vibrations of the nuclear surface dominating the low-energy excitation spectrum towards deformation and rotations. Coulomb excitation, lifetime measurements, and detailed $\gamma$-ray and conversion electron spectroscopic studies are all required to build evidence for the coexisting shapes.

At the same time, the experimental studies go hand in hand with the contemporary calculations of nuclear structure \footnote{S.R. Stroberg \emph{et al.}, Phys. Rev. C 105, 034333 (2022).} to perform experimental tests of their predictions both in the context of the traditional phenomenological shell model \footnote{S. Cruz \emph{et al.}, Phys. Rev. C 102 024335 (2020).} but also \emph{ab initio} treatments of nuclear forces well suited to modern many-body techniques with special emphasis on the {\it sd} shell \footnote{J. Henderson \emph{et al.}, Phys. Rev. C 105 034332 (2022), J. Williams \emph{et al.}, Phys. Rev. C 102 064302 (2020).} and cross-shell excitations in the {\it sdpf} valence space approaching the $A\sim 30$ island
of inversion \footnote{J. Williams \emph{et al.}, Phys. Rev. C 108 L051305 (2023), M.S. Martin \emph{et al.}, Phys. Rev. C 110 034314 (2024).} and the extension of the northern border of the N = 40 island of inversion. \footnote{M. Rocchini \emph{et al.}, Phys. Rev. Lett. 130, 122502 (2023).}

The neutron-rich nuclei in the region of N = 50 to 60 also exhibit interesting evolution of nuclear shapes and shape coexistence. A one-neutron transfer reaction study using the $^{93}$Kr(d,p)$^{94}$Kr reaction led to the first observation of a low-energy 0$^{+}$ excited state in $^{94}$Kr, providing a clear signature of shape coexistence in this neutron-rich nucleus. The measurement was made possible by the use of the IRIS solid deuteron target, enabling transfer reaction studies with extremely low-intensity ISAC-II beams ($\sim$200 pps), demonstrating IRIS’s unique capabilities for rare-isotope research. A picture of IRIS can be seen in Fig.\ref{fig:IRIS-facility}. The newly observed 0$^{+}$ state exhibits a large s-wave spectroscopic factor ($\sim$2.14(82)), suggesting a more spherical configuration, while the non-observation of the ground state points toward a more deformed structure. These findings are consistent with \emph{ab initio} valence-space in-medium similarity renormalization group (VS-IMSRG) calculations and mean-field model predictions, and help constrain different nuclear force prescriptions. \footnote{D. Walter \emph{et al.}, Phys. Lett. B 862 139352 (2025).}

\begin{figure}
    \centering
    \includegraphics[width=0.30\linewidth]{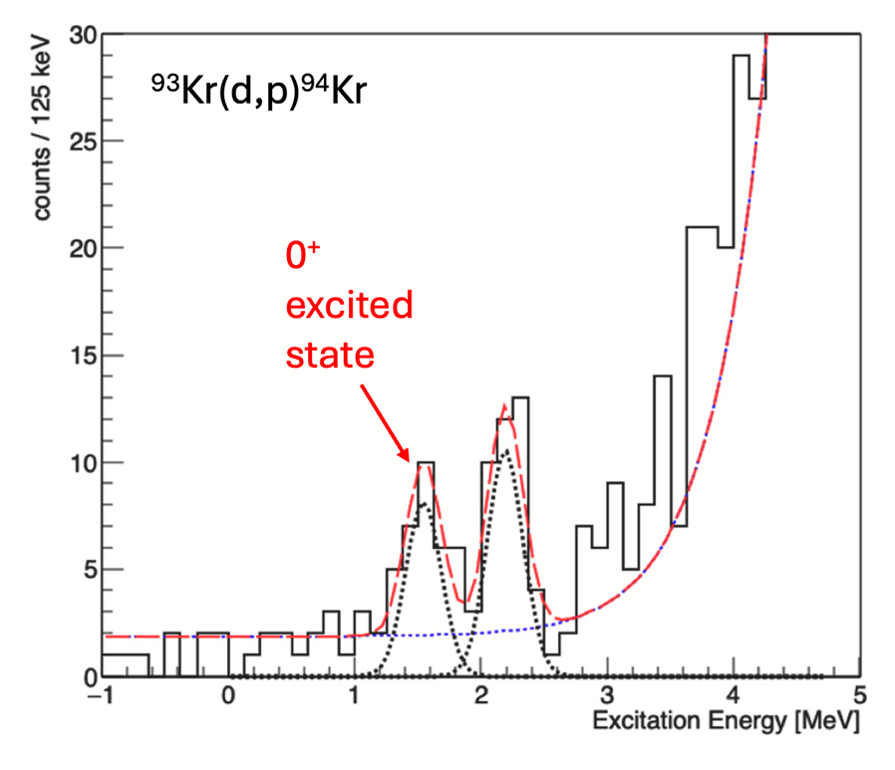}
    \includegraphics[width=0.90\linewidth]{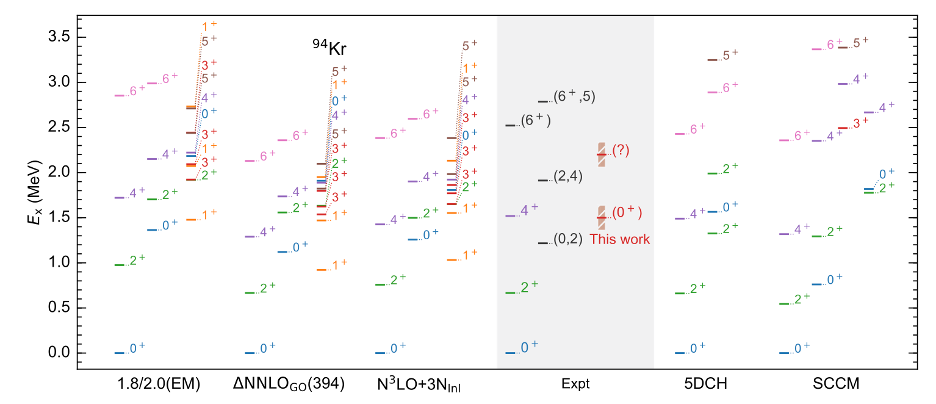}
    \caption{Excitation spectra for $^{94}$Kr showing the scaled background fit (blue), Gaussian fit to the two excited states
    (black dotted curve), and the total fit to the data (red curve). A comparison of the experimental data with \emph{ab initio} valence space in-medium similarity renormalization group (VS-IMSRG) calculations based on three different chiral two- (NN) and three nucleon (3N) interactions [D. Walter \emph{et al.}, Phys. Lett. B 862 139352, (2025)].}
    \label{fig:IRIS-94Kr}
\end{figure}



A complementary program to TRIUMF to study collective states and shape coexistence in nuclei close and near stability at leading laboratories throughout the world concentrating mainly on stable, or near-stable nuclei is pursued by Canadian researchers at UGuelph and SFU using a variety of techniques such as Coulomb excitation, neutron capture reactions, fusion evaporation reactions, $\beta$ decay, and transfer reactions, to enumerate a few.

\begin{figure}
    \centering
        \includegraphics[width=0.30\linewidth]{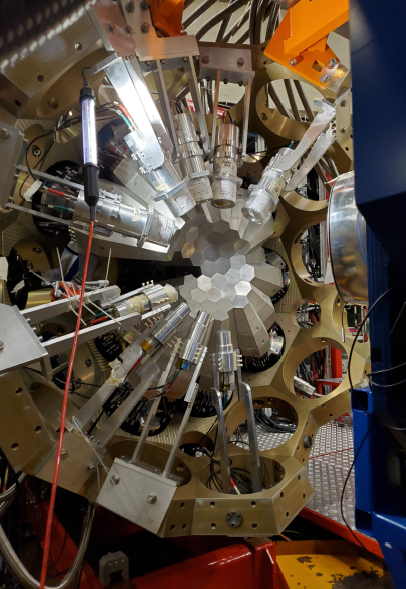}
\includegraphics[width=0.65\linewidth]{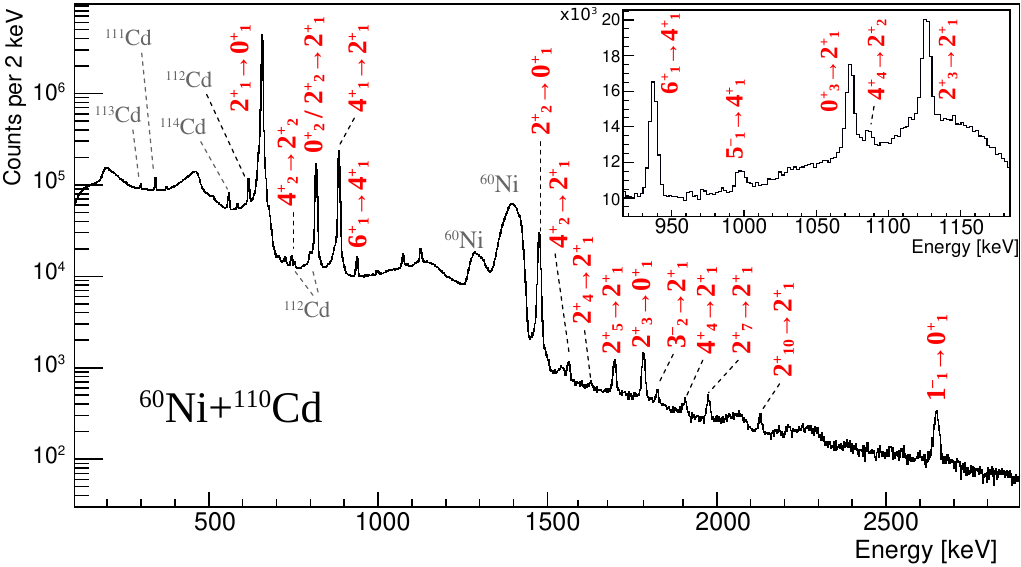}
    \caption{(Left) The $\gamma$-ray tracking array AGATA in its configuration used for Coulomb excitation experiments at the Legnaro National Laboratories, Legnaro, Italy. (Right) Doppler-corrected $\gamma$-ray spectrum recorded with the AGATA spectrometer from the Coulomb-excitation experiment of $^{110}$Cd performed using a 187-MeV $^{60}$Ni beam. The $\gamma$ rays originating from $^{110}$Cd are labelled in red. Transitions resulting from Coulomb excitation of target isotopic impurities ($^{111-}$$^{114}$Cd) and of the $^{60}$Ni projectile are also marked. Inset: Part of the same spectrum expanded around the 1073-keV 0$^{+}_{3}$ $\rightarrow$ 2$^{+}_{1}$ $\gamma$-ray transition. Figure from [I.Z. Pietka \emph{et al.}, Acta Phys. Polon. Proc. Supp. 18, 2-A26 (2025).}
    \label{fig:Cd-110-AGATA}
\end{figure}

Coulomb excitation probes the collective character of nuclear levels, and can be used to determine in
a model independent way the shapes of the states if a sufficient number of transition and diagonal
matrix elements can be determined. \footnote{M. Zieli\'nska, Lect. Notes in Phys. 1005, (2022).} One of the 
world’s most advanced $\gamma$-ray spectrometers used to perform Coulomb
excitation experiments that enable us to probe the shapes of nuclear states in a model independent
way is the Advanced GAmma Tracking Array (AGATA), seen in Fig.~\ref{fig:Cd-110-AGATA}(left). Using AGATA, an unprecedented level of precision and detail can be obtained. A successful programme of study pursuing measurements on the Cd isotopes – nuclei
once believed to be paradigms of spherical vibrational motion \footnote{J. Kern et al., Nucl. Phys. A593,
21 (1995).} have been re-interpreted \footnote{P.E. Garrett \emph{et al.}, Phys. Rev. Lett. 123, 142502 (2019).} as
possessing multiple shape coexistence. The AGATA array is hosted by various laboratories in Europe (GANIL In France,
Legnaro National Laboratory (LNL) in Italy, and GSI in Germany) for campaigns that take advantage of each of the laboratories specialities. Canadian researchers have also taken advantage of the variety of beams available at ANL to perform measurements with
the GRETINA spectrometer, a first-generation, HPGe-based, gamma-ray tracking detector, and will pursue measurements at Argonne National Laboratory (ANL) that cannot be performed at
other facilities. This study is part of a program to determine shapes of states in $^{110}$Cd motivated by experiments performed at the TRIUMF-ISAC facility.  If the conjecture of multiple shapes in the Cd isotopes is proven, it would result in a paradigm shift in our understanding of how collectivity in nuclei evolves and necessitate a rewriting of textbooks on nuclear structure. \footnote{I.Z. Pietka \emph{et al.}, Acta Phys. Polon. Proc. Supp. 18, 2-A26 (2025).}
A Guelph-led Coulomb excitation experiment
performed at ANL using the GRETINA combined with a silicon particle detector called BAMBINO reveals an unprecedented level of sensitivity to excited-state populations being achieved
in the Coulomb excitation of a “vibrational” nucleus. Such studies can continue 
in the medium and longer terms on species such as the Ru, Pd, Sn, and Te isotopes to study shape
coexistence. A key aspect to these experiments is to combine data using multiple binary-reaction
partners to build confidence in the extracted transition matrix elements. To achieve this, a strong collaboration with researchers at the Heavy-Ion Laboratory, University of Warsaw (HIL), using reaction partners such as $^{14}$N, $^{16}$O, and
$^{32}$S to perform the complementary Coulomb excitation experiments. In the long term, odd-A nuclei in the region will be investigated to study the influence and interaction of the single-particle on the collective excitations. These studies are much more challenging, requiring a high degree of spectroscopic knowledge due to the increased level density and the nature of the
decays.

Another component of the offshore program to study shape coexistence is by taking advantage of neutron capture reactions to perform $\gamma$-ray spectroscopy at the world’s highest-flux nuclear reactor at the ILL Grenoble in France. This 
experiments are populating states that are not constrained by the $\beta$ selection rules or the available Q-value from $\beta$ decay experiments performed with GRIFFIN. Using the FIssion Product Prompt gamma–ray Spectrometer (FIPPS), experiments probing
low-spin, highly-excited states in nuclei can be performed. This gives access to states difficult to
observe by other means, permitting us to conduct studies impossible to pursue at other facilities. A
broad program of study is envisioned, including investigations of shape-coexisting states \footnote{C.M. Petrache \emph{et al.}, Phys. Rev. C 99, 024303 (2019), K. Ortner \emph{et al.}, Phys. Rev. C 109, 054317 (2024), F. Wu \emph{et al.}, Phys. Rev. C 111, L051307 (2025), F. Wu \emph{et al.}, Nucl. Phys. A1060, 123105 (2025).} studies of the so-called pygmy 
resonances \footnote{K. Ortner \emph{et al.}, Phys. Rev. C 109, 054317 (2024).}, level densities and strength functions. 
Experiments using radioactive sources with fusion reactions for atomic species that cannot be extracted at
TRIUMF-ISAC or ARIEL, such as the refractory elements, can be produced taking advantage of accelerators and
their $\gamma$-ray spectroscopy facilities at university-based laboratories.

Direct reactions such as single- and two-nucleon transfer provide information on the single-particle
and pairing aspects, respectively, of nuclei. Multiparticle-multihole ($mp \textendash mh$) correlations
or $\alpha$-particle clustering aspects can be probed with $\alpha$-particle transfer reactions. These kinds of
experiments are best performed using magnetic spectrographs coupled to accelerators that can
provide high-intensity with high-energy-resolution beams with excellent spatial focusing on very
thin targets. Such facilities do not exist in Canada, but are available in the US at Florida State University (FSU), the Triangle Universities Nuclear Laboratory (TUNL), and the Australian National University (ANU).
Canadian researchers will utilize transfer reactions investigating the single-particle components of collective
wave functions in the Ru-Cd region of nuclei and mapping the evolution of single-particle states
across isotopic or isotonic chains. An important, and under-explored, reaction process involves
two-proton transfer reactions to understand the proton pairing correlations, and $\alpha$-particle transfer
reactions such as ($^{6}$Li,d) and (d,$^{6}$Li) probing $mp \textendash mh$ structures and $\alpha$-particle correlations. As
part of this program, a new heavy-ion detector for use in the magnetic spectrometer’s focal plane
will need to be constructed.

\begin{figure}
    \centering
    \includegraphics[width=0.65\linewidth]{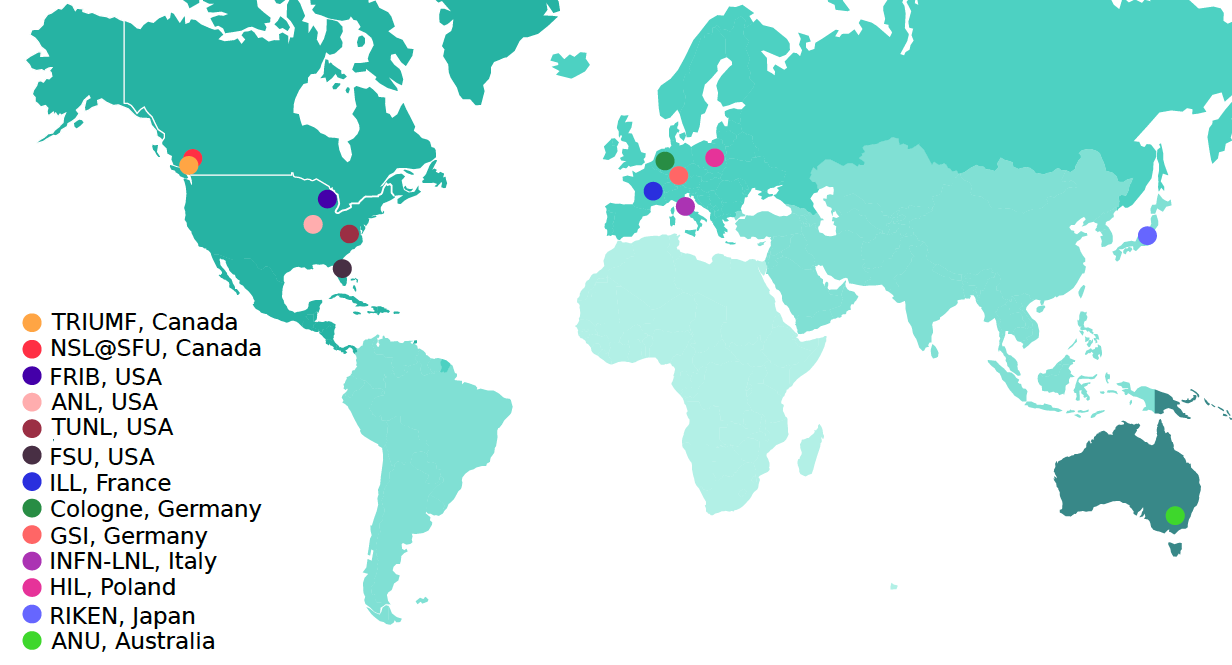}
    \caption{Names and location of the major laboratories around the world where Canadians are leading experiments in nuclear structure.}
    \label{fig:Map}
\end{figure}

\subsubsection{Studies of mirror nuclei and isospin breaking symmetry} 

The charge independence of the nuclear force underpins modern theories of nuclear structure and reactions. Under this assumption, protons and neutrons are regarded as two states of the same particle — the nucleon — in the strong interaction. Extending this principle to nuclei leads to the concept of mirror symmetry, whereby light nuclei exhibit analogous structures when their proton and neutron numbers are interchanged.

Mirror symmetry is, however, broken by the Coulomb interaction and the isospin non-conserving components of the strong force. The extent of mirror symmetry breaking can be investigated through measurements of mirror energy differences (MED) and electromagnetic (EM) transition rates. Violations of charge symmetry of the nuclear force manifest in various observables, including reduced transition strengths, excitation energies, and nuclear masses. Mirror nuclei in several regions along the N = Z line have been investigated with TIGRESS, both as a stand-alone array and in conjunction with EMMA (see Fig.~\ref{fig:EMMA-TIGRESS}), charged-particle detectors, and the TIGRESS Integrated Plunger (TIP), using fusion-evaporation reactions.

In the region near the doubly magic nucleus $^{56}$Ni, mirror nuclei such as $^{55}$Ni and $^{55}$Co have been studied to probe their excited states and their decay patterns. These fusion-evaporation experiments, performed with both stable and radioactive beams at TRIUMF, utilized TIGRESS for $\gamma$-ray detection, SFU’s TIP CsI Ball for charged-particle identification, specialized Ca targetry, and Doppler-shift lifetime techniques \footnote{H. Asch \emph{et al.}, Nucl. Phys. A 1062, 123151 (2025).}. Additional experiments employed EMMA to enhance selectivity in studies of $^{55}$Ni. From these measurements, the energies, spins, and parities of excited states were determined, establishing MED between $^{55}$Ni and $^{55}$Co. Future work will focus on measuring the temperature of the compound nuclei $^{60}$Zn and $^{61}$Ga, determining electromagnetic transition rates in $^{55}$Ni and $^{55}$Co, and providing accurate data to benchmark shell-model calculations of hole states in the $f_{7/2}$ orbital near $^{56}$Ni.

Complementary experiments with TIGRESS and TIP on $^{34}$Si aim to extend investigations around the A = 32 island of inversion, focusing on the identification of negative-parity cross-shell excitations and lifetime measurements to quantify nuclear deformation and test \textit{ab initio} calculations. Similarly, an experiment on $^{20}$Mg will complete the series of studies of neutron-deficient Mg isotopes with TIGRESS and TIP.

Transfer reaction studies have also been performed to probe isospin symmetry breaking in high-lying, weakly bound states within the $A = 22$ isobaric triplet ($^{22}$Ne, $^{22}$Na, $^{22}$Mg) using the $^{21}$Na(d,p/n) and $^{21}$Ne(d,p/n) reactions. These experiments employed TIGRESS for $\gamma$-ray tagging and EMMA for recoil identification, allowing neutron and proton reaction channels to be distinguished. The results have implications for using isobaric analogue states to infer astrophysically relevant reaction rates that cannot be measured directly.

Upcoming proposals include lifetime measurements and E1 transition searches in the $^{29}$P and $^{29}$Si mirror pair using TIGRESS, EMMA, and TIP in fusion-evaporation experiments. E1 transitions, being purely isovector, are forbidden in N = Z nuclei and should exhibit equal strength in mirror systems. Measuring and comparing the excitation energies and E1 transition strengths between these nuclei will provide stringent tests of charge symmetry in the nuclear force and supply valuable benchmarks for theoretical models across the sd–fp shell region.

\subsubsection{Resonant elastic scattering for ab-initio studies}
\label{sec:NS-DRAGON-STRUCTURE}
In addition to astrophysics goals described in the Nuclear Astrophysics section \ref{sec:AstroDirect}, DRAGON (shown in Fig.~\ref{fig:DRAGON_SONIK}) has a program of measurements of scattering observables on short-lived light nuclei, which exists to benchmark the most modern in-house \emph{ab-initio} nuclear theories such as the No-core Shell Model with Continuum (NCSMC) using previously inaccessible but highly fundamental data for the problem.

Much progress has been made recently in \emph{ab-initio} nuclear theory, especially for light nuclei using NCSMC, a TRIUMF-Lawrence Livermore National Laboratory (LLNL) framework for coupling scattering and reaction channels in the p-shell. As seen from recent theory comparisons with data, light-ion scattering data on unstable nuclei is critical to improving these theories. Using resonant elastic scattering at TUDA at sub-Coulomb barrier energies one can achieve high sensitivity to low-energy resonances in the ($p, p$) channel, able to extract resonance energies, widths, and spin-parities through angular distributions. These are simple but powerful experiments that TUDA excels
at, and can be done with modest RIB intensities due to the amplifying nature of the Coulomb-resonant interference effects leading to excellent precision. Data can be fit with multichannel R-Matrix formalism and compared directly with NCSMC calculations which predict $\it{a~priori}$ the resonance spins, widths and energies. TUDA currently has approved experiments to perform measurements of $^6$He($p,p$)$^6$He and $^{11}$C($p,p$)$^{11}$C resonant elastic scattering at below 1 MeV centre-of-mass energies, relative to \emph{ab-initio} resonant state predictions in the $^7$Li and $^{12}$N systems. Data extracted here can also be combined with global R-Matrix fits to all
the available open reaction channel data as has been done with e.g. (Azuma, deBoer \emph{et al.}). The simplicity of these experiments allows this to be done for any light beam with sufficient purity and intensity, and further proposals are being considered for species available at ARIEL.

\begin{figure}
    \centering
    \includegraphics[width=0.45\linewidth]{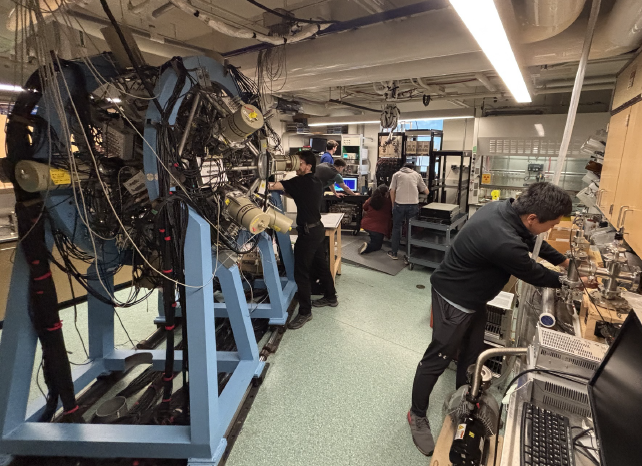}
    \includegraphics[width=0.45\linewidth]{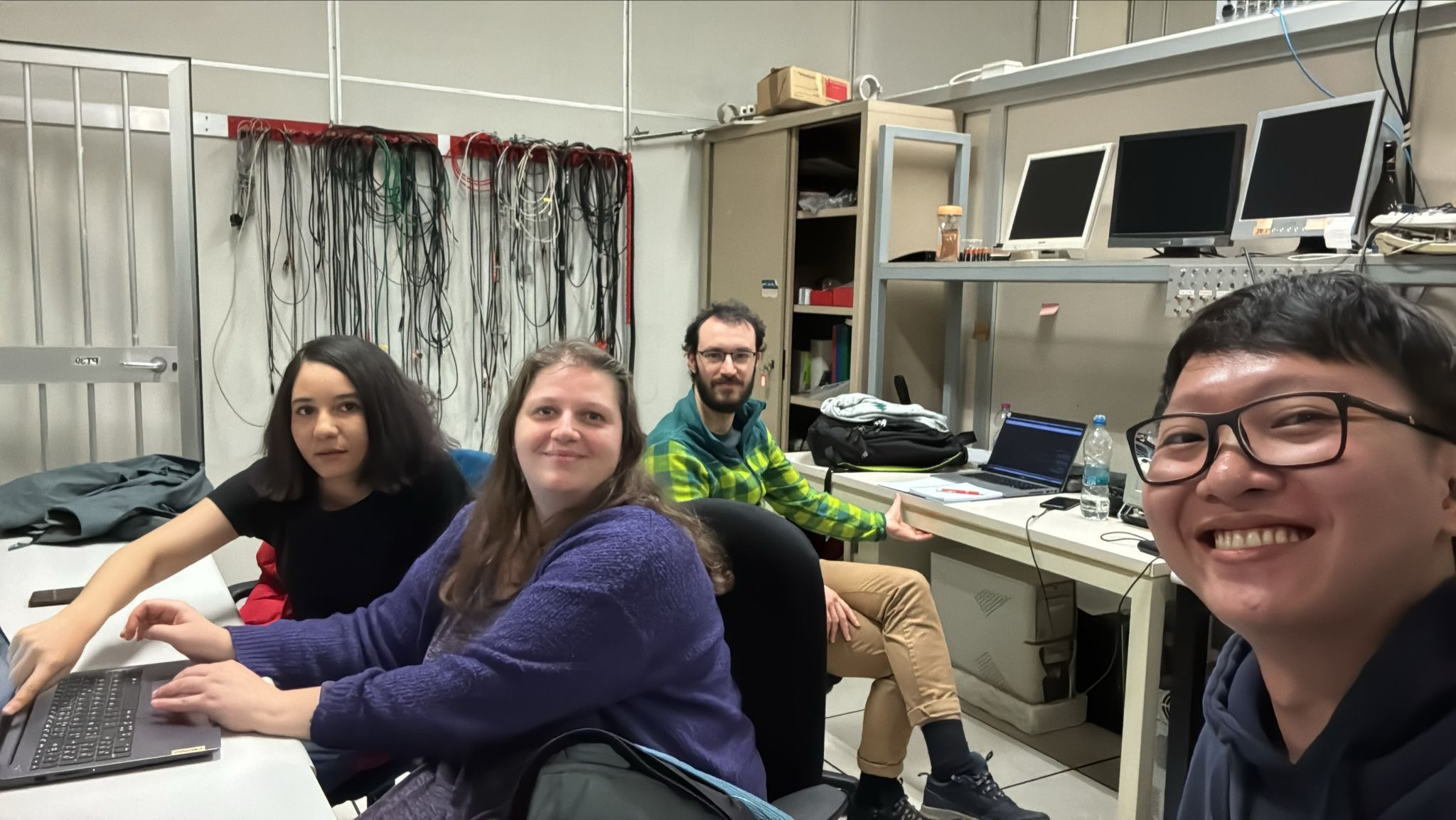}
    \caption{(Left) Students working in the Nuclear Structure Laboratory at SFU, which hosts the 8$\pi$ spectrometer, a 14-MeV neutron generator and a variety of ancillary detectors and electronics. (Right) A group of graduate students and postdoctoral researchers from Guelph and SFU conducting an AGATA experiment at INFN Legnaro in Italy.}
    \label{fig:NSL}
    \end{figure}
    
\subsubsection{Nuclear Structure Laboratory (NSL) at SFU} 
\label{sec:NS-NSL}

The national nuclear structure program is not confined to TRIUMF. The Nuclear Structure Laboratory (NSL) at SFU plays a vital role in both research and training. The NSL provides local experimental facilities for undergraduate, graduate, and postdoctoral researchers, offering hands-on experience before they transition to large-scale projects at TRIUMF or international laboratories.

Scientific goals to be realized at the site of the NSL at SFU in the medium time range between years 2027–2034 are: i.) structure characterization of double-$\beta$ decay candidates through $\beta-\gamma$ decay spectroscopy, ii.)
structure characterization of neutron-rich isotopes accessible through nuclear fission and other modes of nuclear decay, iii.) identification and implementation of novel methods of production and separation of radioisotopes from fast-neutron induced nuclear reactions for fundamental research and applications in medicine and industry.

In the long term time perspective of years 2034-2041 the NSL is planning to continue to be involved in the Canadian and the world-wide effort investigating nuclear interactions with the aim of describing nuclear structure and reactions from first principles. The NSL is planning to capitalize on its proven record of developing highly sensitive apparatus for experimental research in house and in user facilities. As the NSL is set up for irradiations with fast neutrons, significant impact
can be achieved through development of collimated tagged Am/Be neutron sources, acquisition of a high-yield neutron generator, developments of schemes for neutron flux amplification through secondary nuclear reactions, development of schemes for moderation of fast neutrons, and eventually, development of a neutron-generator driven subcritical nuclear reactor.

Crucially, NSL’s role extends beyond research output. By serving as a training hub, it ensures a steady pipeline of highly qualified personnel equipped with the technical skills and experimental mindset required for Canada’s ambitious nuclear structure program. This integration of education and frontier research is a cornerstone of Canada’s leadership in the field.

\subsubsection{The Advanced Rare Isotope Facility (ARIEL) at TRIUMF}
\label{sec:NS-ARIEL-Science}

The ARIEL facility at TRIUMF holds the promise of world-class, transformative research in three major, interlinked research areas related to nuclear structure and nuclear astrophysics:
\begin{enumerate}
\item[]
(a) Elucidating our fundamental understanding of nuclei. A central goal of nuclear physics is
to develop a predictive theoretical framework for all nuclei – a standard model for nuclear
physics. ARIEL experiments will provide decisive input to this quest (inside NSERC SAP).
\item[] 
(b) Searching for new forces in nature. ARIEL will provide both the beam intensities required for
high-precision measurements of the weak interaction, and the multi-user capability to allow
experiments to run for hundreds of days per year to allow for precision experiments (inside NSERC SAP).
\item[] 
(c) Determining how the heavy elements were produced in the universe. ARIEL will enable decisive measurements of the
nuclear properties of the most exotic neutron-rich nuclei that will help, in conjunction with
astronomical observations and astrophysical simulations, to elucidate the so-called $r$ process,
responsible for the production of elements from iron to uranium. (inside NSERC SAP).
\end{enumerate}

\begin{figure}
\centering
\begin{tcolorbox}[
  colback=teal!30,
  colframe=gray!50!black,
  coltitle=black,
  colbacktitle=gray!20,
  width=0.8\textwidth,
  boxrule=0.8pt,
  arc=4pt,
  auto outer arc,
  fonttitle=\bfseries,
  title=ARIEL
]
{\small\baselineskip=14truept As the ARIEL facility begins full operations, Canada is uniquely poised to capitalize on new opportunities. With three simultaneous rare-isotope beams of unprecedented intensity and purity, Canadian researchers will be able to pursue bold experiments at the frontiers of nuclear existence, probing questions of elemental origins, shell evolution, and nuclear collectivity in ways previously impossible. The integrated national program outlined here provides the roadmap for how Canadian leadership in nuclear structure will deliver transformative science over the next 15 years.}
\end{tcolorbox}
\end{figure}

\subsubsection{Future explorations the exotic nuclear landscape with reactions using in-flight radioactive beams}
\label{sec:NS-inflight}
An impactful program of unveiling exotic nuclear forms, neutron halos, neutron
skin and new arrangements of nuclear shells in nuclei approaching the edges of the bound nuclear landscape utilizing reactions with relativistic rare isotope beams takes place at the inflight
rare isotope facilities at GSI-FAIR in Germany, RIBF-RIKEN in Japan and FRIB in
USA. These facilities have been the discovery frontiers unearthing new isotopes with their
novel characteristics and expanding our knowledge of the nuclear landscape. Rare isotope
science is at the international frontline of nuclear physics research.
The thick neutron dominated surface emerging in these nuclei provides
laboratory access to gain knowledge on the equation of state of asymmetric nuclear matter
that describes characteristics of neutron-rich cosmic environments such as neutron stars and
supernovae. The rare isotopes produced at the in-flight facilities with relativistic energies are
crucial to enable these discoveries through direct reactions. A sketch of a typical experiment at RIKEN is given in Fig.~\ref{fig:RIKEN}.

Near term plans will continue ongoing programs at RIBF, GSI/FAIR and at FRIB.
One target of the program will be around the traditional shell closures N = 28, 50 and 82 with
two-fold aim of constraining the symmetry energy from neutron skin thickness measurements
and find signatures of shell structure changes. The most neutron-rich light nuclei will be explored at FRIB and the heavy isotopes will be measured at GSI. The intermediate mass nuclei, as well as nuclei around the proton drip-line will be measured at RIKEN-RIBF.

The measurements will continue for the long-term specifically with added new capacity as both FRIB and FAIR ramps up to full power. As the highest energy beams become available at FAIR our program will extend to the heaviest region of closed shell around N = 126 in the $r$-process path.

\subsubsection{antiProton Unstable Matter Annihilation (PUMA) at CERN}

The antiProton Unstable Matter Annihilation (PUMA) project at CERN aims to investigate the neutron-to-proton density ratios at the surfaces of exotic, short-lived atomic nuclei. By uniquely employing low-energy antiprotons as probes, PUMA seeks to identify and characterize phenomena such as neutron skins and halos—regions where neutrons extend beyond the typical nuclear boundary.
Understanding these structures is crucial for insights into nuclear structure as well as the nuclear equation of state and the internal composition of neutron stars. PUMA represents a pioneering approach in nuclear physics, bringing together two accelerator produced exotic species: antiprotons and short-lived radionuclides. CERN is the only laboratory where this program can take place and provides unique opportunities to Canadian researchers and HQP.

\begin{figure}
\centering
\begin{tcolorbox}[
  colback=teal!30,
  colframe=gray!50!black,
  coltitle=black,
  colbacktitle=gray!20,
  width=0.8\textwidth,
  boxrule=0.8pt,
  arc=4pt,
  auto outer arc,
  fonttitle=\bfseries,
  title=Nuclear Structure in the ARIEL Era
]
{\small\baselineskip=14truept During the 2027–2041 period covered by this Long-Range Plan, all Canadian nuclear structure experiments will benefit from the flagship ARIEL facility at TRIUMF that transitions from
construction and commissioning to the delivery of rare isotope beams for science. This will include
both high-purity charge-bred beams of high-mass (A $>$ 30) isotopes that will have a major impact
on the nuclear structure and nuclear astrophysics research programs with accelerated radioactive
beams from the ISAC-II superconducting linear accelerator, and a wide range of new neutron-rich
isotopes produced through photofission of actinide targets driven by bremsstrahlung from the high power
ARIEL electron linac. Ultimately, the parallel operation of the current ISAC 500 MeV proton
beamline, the ARIEL e-linac, and a second 500 MeV proton beam driving the ARIEL production
targets will provide three simultaneous radioactive beams, tripling the beamtime available to the
suite of state-of-the-art research infrastructure at ISAC and ensuring Canada’s continued leadership
role in ISOL-based rare isotope science for the foreseeable future. }
\end{tcolorbox}
\end{figure}

\subsection{Beyond the next 7 years}
\label{sec:Beyond7Facil}

\subsubsection{Two-step RIB facility with ISOL and in-flight at TRIUMF}
\label{sec:TwoStepRIB}

The access to nuclei with large neutron excess over protons has opened the window to a new view of nuclear structure and understanding the role of these exotic isotopes in our universe. The standard model of nuclear physics, i.e. our traditional concepts laid down with extensive
studies of stable nuclei, start to breakdown in such regions. One of the major goals in science is gaining a complete understanding on the origin of the heavy elements via the $r$ process (see Section~\ref{sec:astro:HeavyElement}).
The investigation of properties of nuclei on the $r$-process path and those that can shed light
on neutron star properties will be at the frontline of international research in nuclear physics. 
The neutron-rich isotones at N = 50 and N = 82 for proton numbers lower than Z = 50 and       
Z = 36, respectively, down in the neutron-rich region of the nuclear chart are two main target
areas of importance to both $r$-process nucleosynthesis and nuclear structure.
It is still unknown if and where these conventional shell closures vanish,
as was found in the light neutron-rich nuclei near the neutron drip-line. The lighter isotones
get progressively more neutron-rich and in such regions neutron skin thickness would rapidly
increase. This allows access to study the symmetry energy of the equation of state of asymmetric nuclear matter with high sensitivity. Are there unknown giant neutron halos and what
drives their existence?
Production of neutron-rich isotopes with N $\geq$ 126 using high intensity beams of isotopes
from ARIEL with Z $\geq$ 82 might also be explored with a two-step RIB facility.
A wide range of measurements of high importance will be enabled by the access to these exotic
isotopes such as masses, beta decay half-lives, beta delayed neutron emission probabilities,
laser spectroscopy, observation of unknown excited states, near-threshold resonances, determination of nuclear radii, investigations of nuclear orbitals, investigations on neutron capture
reactions and potentially tests of fundamental symmetries. These will lead to pioneering
results on nuclear structure and nuclear astrophysics.
The production of these neutron-rich isotopes will build on intense beams from ARIEL and
will be the best suited facility in the world to produce these. This will involve the addition of a
heavy ion accelerator with energy E/A $\geq$ 100 MeV. This capacity for accelerating the ARIEL
isotopes to higher energies also opens the possibility of studying nuclear reactions with large
negative Q values (i.e. threshold energies) such as two-neutron transfer in near the proton
drip-line nuclei and proton transfer in neutron-rich nuclei and some indirect reactions for
proton capture studies.

\begin{figure}[h]
\centering
\begin{tcolorbox}[
  colback=teal!30,
  colframe=gray!50!black,
  coltitle=black,
  colbacktitle=gray!20,
  width=0.8\textwidth,
  boxrule=0.8pt,
  arc=4pt,
  auto outer arc,
  fonttitle=\bfseries,
  title=The 15-year vision for nuclear structure
]
{\small\baselineskip=14truept Over the next 15 years, Canadian nuclear structure research will remain a driver of discovery and innovation. With ARIEL as its flagship, and through continued international collaboration, Canada will lead the exploration of rare isotopes and train a new generation of researchers who will carry forward Canada’s global leadership in nuclear science.}
\end{tcolorbox}
\end{figure}

\subsection{Summary}
\label{sec:NS-Summary}

Nuclear structure research in Canada is a vibrant and impactful field, encompassing a wide range of phenomena studied through both experimental and theoretical approaches that address the most exciting questions in contemporary nuclear physics with an impact in nuclear structure, astrophysics, and fundamental symmetries. While much of the experimental activity is centered at the TRIUMF's ISAC facility, it is also complemented by collaborative experiments at leading laboratories around the world (see the map in Fig.~\ref{fig:Map}). 

The experimental activities are guided and enhanced by world-leading experts working to develop \emph{ab initio} and many body nuclear theories that will replace the phenomenological approaches across the nuclear chart and with nuclear astrophysics theory modeling the connections between nuclei and astrophysical observables. Canada has made substantial capital investments to support and expand this research, including the continued development and exploitation of the ISAC facility, the completion of the flagship ARIEL facility, and the individual experimental setups described in this section.

Looking ahead, the parallel operation of the current ISAC 500 MeV proton beamline, the ARIEL electron linac, and a second 500 MeV proton beam for ARIEL production targets will enable the delivery of three simultaneous radioactive ion beams. This will triple the available beamtime for the cutting-edge experimental infrastructure at ISAC and will further elevate scientific output ensuring Canada’s leadership in ISOL-based rare isotope science well into the future.

\section{Nuclear Astrophysics}

\subsection{The Canadian program (current and next 7 years)}

\subsubsection{Overview}

\begin{SCfigure}
    \centering
    \includegraphics[width=0.8\textwidth]{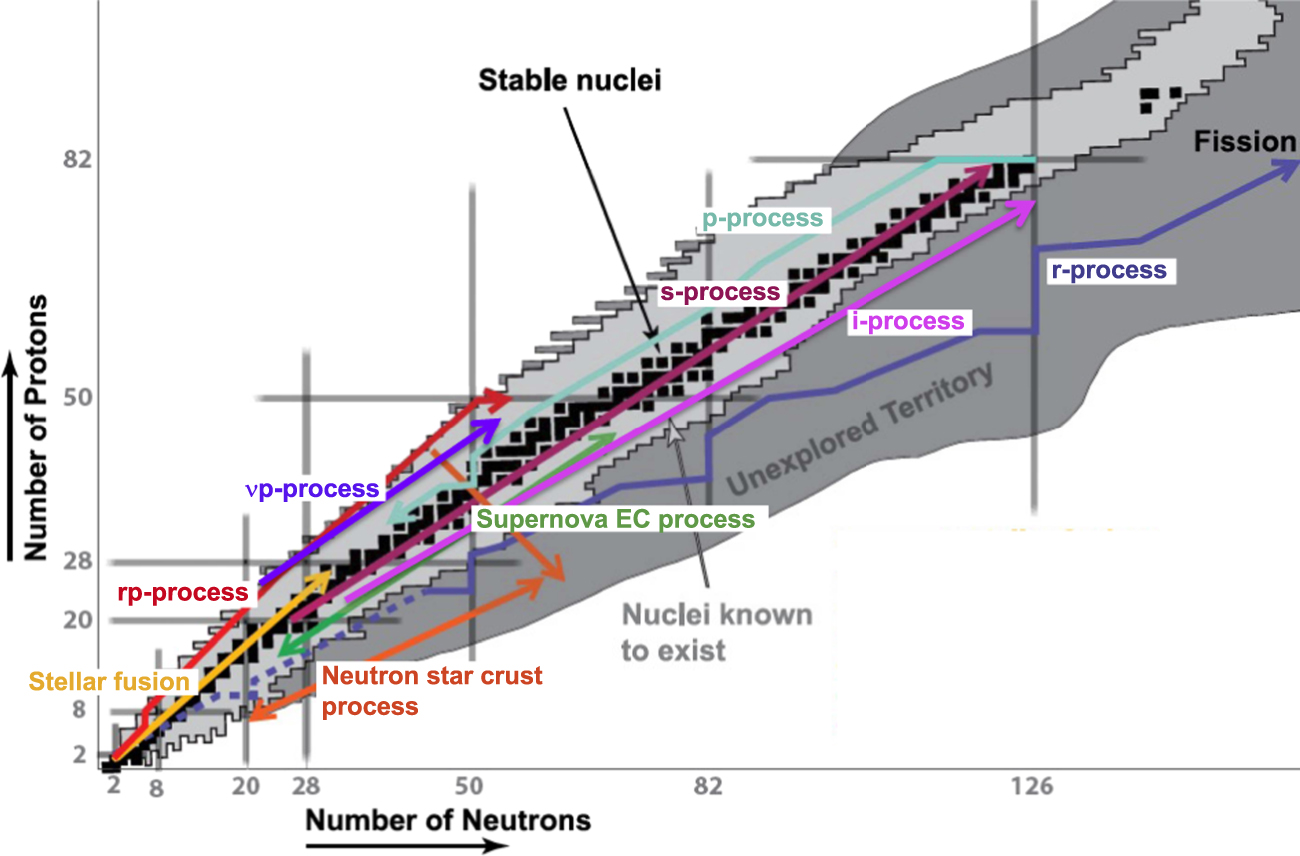}
    \caption{Global view of the various nucleosynthesis pathways contributing to the formation of the chemical elements. Figure credit: H Schatz \emph{et al.}, ``Horizons: nuclear astrophysics in the 2020s and beyond,'' \href{https://doi.org/10.1088/1361-6471/ac8890}{J.\ Phys.\ G: Nucl.\ Part.\ Phys.\ 49 110502 (2022).}}
    \label{fig:Nucleosynthesis}
\end{SCfigure}

The ultimate goal of nuclear astrophysics is to understand the role of nuclei in shaping the universe. This includes understanding the formation of the chemical elements, through the Big Bang, stellar processes, and explosive astrophysical events, with the goal of reliably modeling observed abundance patterns. It also includes understanding the role of nuclei in determining astronomical observables other than elemental abundances, such as $X$-ray burst light curves or stellar $\gamma$~rays.
%
World-wide research efforts aim to place
experimental or theoretical constraints on the nuclear reaction and decay rates along the various nucleosynthesis pathways responsible for element formation (see Fig.~\ref{fig:Nucleosynthesis}). These pathways span nearly the entire range of the nuclear chart, from close to the proton dripline in the $rp$- and $\nu p$-processes to extremely neutron rich species in the $r$ process. The Canadian nuclear astrophysics program has elements involving all of the major nucleosynthesis processes, and can be broadly partitioned into investigations into ``light-element'' ($\leq\mathrm{Fe}$) and ``heavy-element'' ($>\mathrm{Fe}$) nucleosynthesis. The program involves a wide range of experimental methods, ranging from direct measurements of low-yield capture reactions at astrophysical energies, to indirect methods using nuclear structure data to infer reaction rates or extrapolate to astrophysical energies. There is a significant degree of cross-pollination with nuclear structure studies, as many of the same tools are used to study nuclear structure and nuclear astrophysics---sometimes during the same experiment. The Canadian theory program also includes a broad range of nuclear astrophysics efforts (see Section~\ref{sec:nuctheo_astro}).

In addition to studying nuclear properties relevant for nucleosynthesis processes, the Canadian program has an element investigating dense nuclear matter, relevant for neutron stars. These include theoretical studies of nuclear pasta phases, \emph{ab initio} nucleonic matter studies, and efforts at developing a QCD-based nuclear equation of state. Offshore experimental efforts aimed at understanding neutron skins also provide knowledge on the equation of state of asymmetric nuclear matter, improving understanding of neutron stars and supernovae.

\subsubsection{Light element nucleosynthesis}

Aside from a few exceptions, namely the light elements H, He, and Li formed in the Big Bang and isolated nuclides formed through cosmic-ray spallation, the elements lighter than iron are 
formed through a series of nuclear reactions and decays in stars.
The sites for stellar nucleosynthesis are varied and include quiescent stellar burning as well as stellar explosions such as novae and supernovae. A complete understanding of light-element synthesis requires information on the nuclear reaction rates making up these stellar reaction pathways.
Typically, this means nuclear data that constrain $p$ or $\alpha$ capture reactions, often involving radioactive nuclei. In addition to constraining nucleosynthesis predictions, nuclear reaction data are necessary to accurately model astronomical observables such as $X$-ray burst light curves, $\gamma$~rays from astrophysical sources, or solar neutrinos. The Canadian nuclear astrophysics program includes a broad range of experiments studying reactions that have an observable impact on either nucleosynthesis or astronomical observables. A variety of experimental techniques are used at a variety of facilities. Broadly speaking, they can be partitioned into direct measurements---those that measure the reaction cross section of interest at the relevant astrophysical energy, and indirect measurements---those that measure nuclear properties that influence the needed rates.

\paragraph{Direct measurements:}
\label{sec:AstroDirect}

The Canadian direct-measurement program for light-element synthesis is mostly focused at the DRAGON facility, located in the ISAC-I hall at TRIUMF (Fig.~\ref{fig:DRAGON_SONIK}).  DRAGON was built specifically to measure the cross sections of $(p,\gamma)$ and $(\alpha,\gamma)$ reactions in inverse kinematics, with beam energies ranging from $0.15$--$1.5$ MeV/nucleon. DRAGON has been a world leading device in measuring radiative capture reactions for over $20$ years, having made $80\%$ of the world's radiative capture reactions using radioactive beams. DRAGON is also a world leader in measurements of stable-beam reactions, with the inverse kinematics technique pushing into difficult-to-measure regimes only otherwise accessible at underground facilities. These experiments provide an important benchmarking and cross-checking tool against systematics in the normal kinematics measurements, and they also provide additional sensitivity since DRAGON primarily detects recoils.

\begin{SCfigure}
\centering
    \includegraphics[width=0.5\textwidth]{
        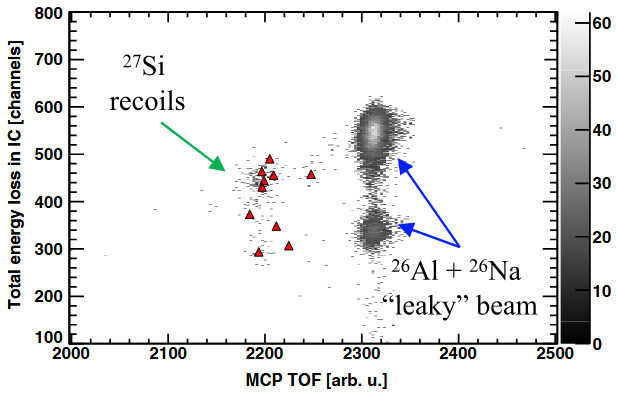
    }
    \caption{Scatter plot of total energy loss in the ionization chamber at the focal plane of DRAGON vs.\ time-of-flight between two micro-channel plates separated by 59~cm. The cluster of 10 clearly identified ${}^{27}$Si recoils was used to determine the strength of the key $E_{cm} = 447$~keV resonance in the ${}^{26m}\mathrm{Al}(p,\gamma){}^{27}\mathrm{Al}$ reaction. Figure credit: A.~Lennarz \emph{et al}, Phys.\ Rev.\ Lett. 128, 042701 (2022).}
    \label{fig:Al26m}
\end{SCfigure}

A highlight of the DRAGON program over the past five years was the measurement of the ${}^{26m}\mathrm{Al}(p,\gamma){}^{27}\mathrm{Al}$ reaction---the first ever direct measurement of radiative capture using an isomeric radioactive beam.\footnote{G.~Lotay \emph{et al}, Phys.\ Rev.\ Lett. 128, 042701 (2022).} In this experiment, a mixed isomeric and ground-state ${}^{26}\mathrm{Al}$ beam was sent to DRAGON, which cleanly separated the ${}^{27}\mathrm{Si}$ recoils resulting from $(p,\gamma)$ reactions from background (see Figure~\ref{fig:Al26m}). After accounting for all possible contributions from ${}^{26g}\mathrm{Al}$-induced reactions, a strength of $432^{+146}_{-226}$~meV for the key $E_{cm} = 447$~keV resonance in ${}^{26m}\mathrm{Al}(p,\gamma){}^{27}\mathrm{Al}$ was determined. This landmark experiment paves the way for future isomeric-beam studies, both at TRIUMF/DRAGON and other recoil separators around the world.

Looking forward to the next seven years, a highlight of the planned DRAGON program is a measurement of the ${}^{7}\mathrm{Be}(p,\gamma){}^{8}\mathrm{B}$ reaction, well-known as a critical component to understanding the solar neutrino spectrum, but also relevant for other scenarios
such as ${}^{7}$Be destruction in novae. This experiment benefits from the intense (${\sim}5\times10^9$~pps) $^{7}$Be beam available at TRIUMF, and it is likely that this reaction will become the subject of more comprehensive
studies with the availability of long beam times and simultaneous RIBs made possible ARIEL. 
A large number of other approved proposals exist, the majority of which are likely to be completed in the next seven years---namely,
$^{15}\mathrm{O}(\alpha, \gamma){}^{19}\mathrm{Ne}$, $^{37}\mathrm{Ar}(p, \gamma){}^{38}\mathrm{Ca}$, $^{10}\mathrm{Be}(p, \gamma){}^{11}\mathrm{B}$, $^{38}\mathrm{Ar}(p, \gamma){}^{39}\mathrm{Ca}$, $^{20}\mathrm{Ne}(\alpha, \gamma){}^{24}\mathrm{Mg}$, and $^{16}\mathrm{O}(\alpha, \gamma){}^{20}\mathrm{Ne}$. Development proposals are also in place for $^{51}\mathrm{Mn}(p, \gamma){}^{52}\mathrm{Cr}$, important for next-generation $\gamma$-ray astronomy, and $^{35}\mathrm{Ar}(p, \gamma){}^{36}\mathrm{Ca}$, which lies along the $rp$-process nucleosynthesis path.
Additionally, it is foreseen that with the availability of longer beam times
at ARIEL, some of the most important reactions already studied at DRAGON will be revisited
with comprehensive and precise measurements---measuring all accessible resonances
(and direct capture where possible) to a precision that completely eliminates the uncertainty
pertaining to the reaction's role in stellar burning or nucleosynthesis. In addition, as beam
intensities are improved and new beams become available at ARIEL, this opens up new reactions that were previously impossible.
%
%
New prospects also exist for measuring reactions where the dominant uncertainty is the energy of key resonances, benefiting from the recently-developed resonance timing technique to determine resonance energies with only handful of events.\footnote{G.\ Christian \emph{et al.}, Nucl.\ Inst.\ and Meth.\ in Phys.\ Res.\ A 1072, 170199 (2025).}

Direct measurements of reactions with a charged-particle exit channels, primarily $(p,\alpha)$ or $(\alpha,p)$, are also part of the Canadian nuclear astrophysics program. These are focused at the TUDA, TACTIC, and IRIS facilities at TRIUMF, each of which has unique strengths. The TUDA facility, a UK-funded device with strong collaborative efforts from Canadian researchers, is able to run in the ISAC-I hall and thus can measure $(p,\alpha)$ or $(\alpha,p)$ reactions at energies below $1.5$ MeV/nucleon. TUDA has four approved $(p,\alpha)$ or $(\alpha,p)$ experiments that are anticipated to run in the next seven years. The TUDA program is also expected to benefit from the recent development of novel implanted He targets, allowing $(\alpha,p)$ measurements without the background induced by gas-cell windows. TACTIC, a recently-commissioned and UK-funded device, is an active target optimized for measuring $(\alpha,p)$ reactions at low energies. Approved TACTIC  proposals currently exist for $^{17}\mathrm{F}(\alpha, p){}^{21}\mathrm{Ne}$ and $^{27}\mathrm{Al}(\alpha, p){}^{30}\mathrm{Si}$.
The IRIS program focuses on direct measurements of $(p,\alpha)$ reactions using radioactive beams, capitalizing on the unique solid hydrogen target to achieve a high target density without low-$Z$ contaminants.  IRIS mostly focuses on measurements at higher beam energies, which are complementary to those carried out at TUDA, and a recent highlight of the IRIS astrophysics program is the direct measurement of the key $rp$-process reaction ${}^{59}\mathrm{Cu}(p,\alpha){}^{56}\mathrm{Ni}$.\footnote{J.\ S.\ Randhawa \emph{et al.}, Phys.\ Rev.\ C 104, L042801 (2021)}

\paragraph{Indirect measurements:}
\label{sec:AstroIndirect}


\begin{SCfigure}
\centering
    \includegraphics[width=0.6\textwidth]{
        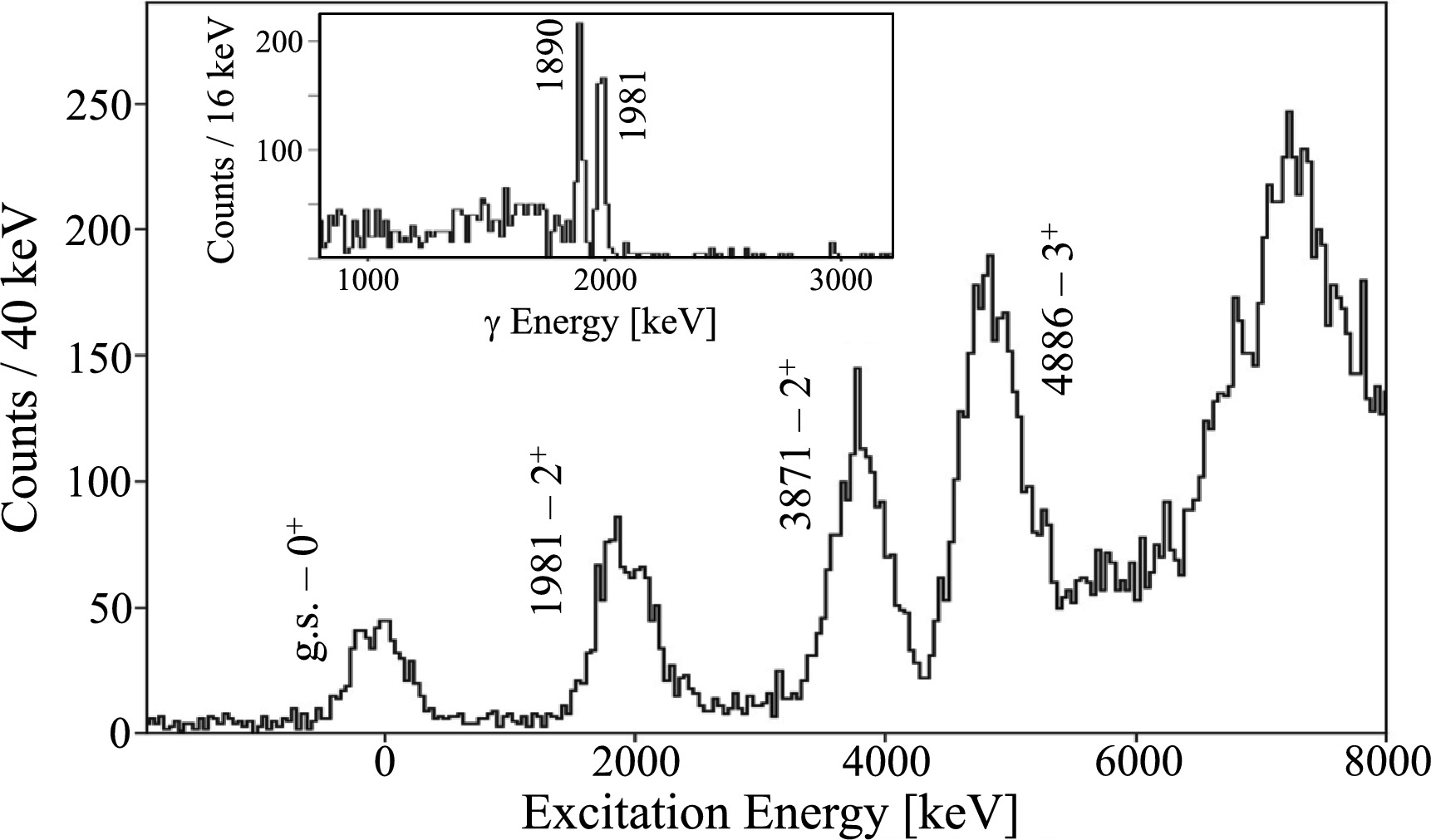
    }
    \caption{\emph{Main figure:} Excitation energy spectrum from the ${}^{23}\mathrm{Na}(d,p){}^{24}\mathrm{Na}$ reaction measured with TIGRESS/SHARC. \emph{Inset:} The $\gamma$-ray spectrum observed in coincidence with the 3871 keV excitation energy peak. Figure credit: G.\ Lotay \emph{et al.}, Phys.\ Lett.\ B 833, 137361 (2021).}
    \label{fig:Na23dp}
\end{SCfigure}

Direct measurements of astrophysical reactions are often not possible---for example, due to limited beam intensities or low cross sections at astrophysical energies. The Canadian nuclear astrophysics program includes a variety of indirect techniques, measuring nuclear properties that determine the reaction rate of interest. One such technique is the use of transfer reactions to determine spectroscopic factors, which in turn give the corresponding particle width. This method is used at both the TUDA and TIGRESS/SHARC facilities, where $(d,p)$ reactions are used to determine the neutron spectroscopic factors of the mirror states of astrophysical $(p,\gamma)$ resonances. Using mirror symmetry, these spectroscopic factors give the requisite proton widths necessary to determine the low-energy $(p,\gamma)$ resonance strengths. Experiments performed at TUDA and TIGRESS are complementary---TUDA provides maximal resolution for charged-particle spectroscopy, while TIGRESS/SHARC can select specific $\gamma$-ray transitions to isolate the astrophysical states.
A recent highlight of this program is the measurement of the  $^{23}\mathrm{Ne}(d, p){}^{24}\mathrm{Ne}$ reaction, for extraction of information on the resonance strength of $^{23}\mathrm{Al}(p, \gamma){}^{24}\mathrm{Al}$ (Fig.~\ref{fig:Na23dp}).\footnote{G.\ Lotay \emph{et al.}, Phys.\ Lett.\ B 833, 137361 (2021).}
A related technique is the use of resonant elastic scattering to determine low-energy phase-shifts, allowing extrapolation of radiative
capture cross sections to astrophysical energies. Such experiments are performed with TUDA, as well as the SONIK target that can be installed at DRAGON (Fig.~\ref{fig:DRAGON_SONIK} inset). Measurements of both $p$ and $\alpha$ elastic scattering on $^{7}\mathrm{Be}$ are planned during the next seven years. The resonant scattering technique is also used to benchmark \emph{ab initio} theory calculations motivated by nuclear structure (see Section~\ref{sec:NS-DRAGON-STRUCTURE}).

Another indirect technique is the use of decay spectroscopy to determine branching ratios and partial widths that determine capture-reaction strengths. These studies are carried out using the new RCMP detector together with the GRIFFIN spectrometer at TRIUMF. 
A recent highlight is the measurement of $\beta$-delayed $p$ and $p\alpha$ emission of $^{20}\mathrm{Mg}$, which constrains both the $^{15}\mathrm{O}(\alpha, \gamma){}^{19}\mathrm{Ne}$ and $^{19}\mathrm{Ne}(p, \gamma){}^{20}\mathrm{Na}$ reactions, important for breakout of the hot CNO cycle in $X$-ray bursts. Looking forward, plans are in place to measure $\beta p\alpha$ emitters like $^{17}\mathrm{Ne}$ and $^{13}\mathrm{O}$, as well as angular correlations in $\beta 2p$ decays, and $\beta$-delayed particle emission from states that do not emit $\gamma$-rays.

\subsubsection{Heavy element nucleosynthesis}
\label{sec:astro:HeavyElement}

The origin of the elements heavier than iron is a long-standing and key question in nuclear astrophysics. Beyond iron, nuclear fusion reactions become energy consuming, meaning that alternative pathways to traditional stellar nucleosynthesis are needed to form the heaviest elements. The vast majority (${\sim}99\%$) of these elements are formed through various neutron capture processes, consisting of sequences of neutron capture followed by $\beta$ decay. These are dominated by the ``rapid'' ($r$) and ``slow'' ($s$) neutron capture processes, which have contributed in roughly equal amounts to our solar system.
The $r$ process occurs in environments with high neutron density, such as neutron star mergers, and involves reactions far from stability. In contrast, the $s$ process occurs in low neutron density environments such as massive and AGB stars, and involves reactions close to stability. In addition, it is now recognized that there is likely an ``intermediate'' ($i$) process which occurs in regions of moderate neutron flux, involving radioactive nuclei closer to stability than the $r$ process.

In addition to neutron capture processes, other pathways are required to fully explain the observed heavy-element abundance patterns. These include the ``proton'' ($p$) process, responsible for the origin of the $30$ or so ``$p$ nuclei''---neutron-deficient stable nuclides that are blocked from formation in neutron-capture processes. It is also now recognized that the abundance patterns of certain heavy elements, specifically the lighter species from Sr--Ag, cannot be fully explained by the $s$, $i$, and $r$ processes.
The abundances of these elements are likely to be enhanced by processes producing so-called ``weak $r$-process'' elements from $A\sim 80$ to $A\sim 130$, such as the so-called ``$\alpha$ process'' operating via a series of $(\alpha,n)$ reactions in core-collapse supernova, or a proper weak $r$-process from a subset of neutron star merger ejecta or other events with accretion disk ejecta.

The Canadian nuclear astrophysics program includes components targeting all of the heavy-element formation process, with experiments carried out at a variety of facilities, both domestically and abroad. The theory program also investigates heavy element synthesis, for example through modeling studies investigating the impacts of nuclear properties on nucleosynthesis networks and structure calculations including nuclei on the nucleosynthesis pathways.

\paragraph{$r$ and $i$ processes:}

The $r$ and $i$ processes both involve sequences of neutron capture reactions and sequences of neutron capture reactions and $\beta$ decays that take place away from stable nuclei. Accurate modeling of the nucleosynthesis occurring in both processes requires knowledge of decay rates as well as neutron capture cross sections. Because the nucleosynthesis pathways are far from stability, direct measurements of the relevant cross sections and decay rates are often not possible. A variety of indirect techniques are needed, together with theory, to calculate the relevant inputs for these processes.  The Canadian program utilizes a number of different experimental techniques, both at TRIUMF and offshore, including decay spectroscopy, mass measurements, and transfer reactions. Efforts are also underway to make direct measurements of $(n,\gamma)$ reactions in a storage ring a reality (see Section~\ref{sec:AstroNext7}), with the installation of a demonstrator at GSI-CRYRING and a conceptual design study for a dedicated future storage ring that could be hosted at an ISOL facility like TRIUMF-ISAC.

%
The Canadian theory program (see Section~\ref{sec:nuctheo_astro}) also has a strong $r$ process element.
These include network calculations by N.\ Vassh to explore neutron capture across the range of $i$ to $r$ process neutron densities, as well as work modeling the $\gamma$-ray emission from ${}^{208}$Tl, a possible signature of real-time synthesis of Pb.
Proposed future projects include studies of the role of isomers in $r$-process nucleosynthesis as well as global \emph{ab initio} models that allow for calculations of nuclear properties with uncertainty estimates.

The decay spectroscopy work includes a domestic component, focused on the investigation of decays of very neutron-rich nuclei that are important for the $r$ process. These studies are performed using the GRIFFIN spectrometer at TRIUMF and benefit from its high $\gamma$-ray detection efficiency, allowing studies with incident beam rates as low as $1$~pps. There is also a significant offshore program focused on $\beta$-delayed neutron emission using the BELEN neutron detectors from the DESPEC collaboration. This program focuses on measurements of the most neutron-rich nuclei possible, approaching, and in some cases reaching, the pathways of the $r$~process. In 2021, this program completed a long campaign of measurements at the RIKEN facility in Japan. 
 The campaign at RIKEN ran from 2016 until 2021 and covered almost 200 isotopes between ${}^{75}$Ni and ${}^{222}$Bi. So far, 19 new half-lives, 64 new $1n$ branching ratios ($P_{1n}$ values), and 31 new $P_{2n}$ values have been published, and data for about 76 $\beta n$ emitters is still being finalized.
In the next seven years, a new campaign is anticipated, focusing on the neutron-rich $A\simeq 190$--$230$ mass region using beams from the new FAIR facility at GSI in Germany. These measurements are expected to reach the $r$-process abundance peak maximum at $N=126$ down to $Z=69$ (${}^{195}$Tm).


The mass measurement program at the TITAN facility at TRIUMF includes high-precision and high-accuracy mass measurements of nuclei away from stability. These mass data are used both as direct inputs into $r$-process nucleosynthesis studies, as well as exacting benchmarks for cutting-edge nuclear models that provide nucleosynthesis inputs not yet available to experiment. The TITAN mass measurements 
are now approaching regions of the nuclear chart that can help to differentiate $r$-process abundance patterns from those of the $i$~process, and recent work has found significant deviations from current model predictions.
These results will help refine the theoretical calculations necessary for extrapolation to $r$-process nuclides.\footnote{A.\ Jacobs, Phys.\ Rev.\ Lett.\ 134, 062701 (2025).}

At present, direct measurements of neutron capture cross sections on short-lived nuclei, necessary to understand the $i$ and $r$ processes, are not possible. Instead, indirect techniques must be used to experimentally constrain these reactions. One such technique is the use of the $(d,p)$ reaction to study the single-particle structure of neutron-rich nuclei, measuring nuclear properties that influence the neutron capture rates. These types of experiments are a focus of the IRIS facility at TRIUMF, which uses a world-unique solid hydrogen target to study $(d,p)$ and other reactions on nuclei far from stability. Plans for the next seven years include studies of $(d,p)$ reactions on  ${}^{81}\mathrm{Ga}$ as well as the neutron-rich tin isotopes, ${}^{135-137}$Sn. Similar methods, adding $\gamma$-ray detection, are employed at TIGRESS/SHARC---highlighted by a successful measurement of the $^{93}\mathrm{Sr}(d, p\gamma){}^{94}\mathrm{Sr}$ reaction, whose analysis is nearly complete. 
Looking forward, approved proposals are in place to measure $^{75}\mathrm{Ga}(d, p){}^{76}\mathrm{Ga}$ and $^{139,142}\mathrm{Cs}(d, p){}^{140,143}\mathrm{Cs}$. The more intense and  pure beams from ARIEL will broaden the scope of these experiments and enable more sensitive measurements with more exotic beams, especially those with $A > 130$.

\paragraph{$s$ processes:}
\label{sec:AstroSprocess}


The $s$ process consists of sequences of neutron capture reactions and $\beta$ decays taking place on stable and near-stable nuclei in low-neutron flux environments, with timescales between successive neutron captures on the order of years. 
A comprehensive understanding of the resulting nucleosynthesis patterns requires knowledge of the constituent neutron-capture and $\beta$ decay rates (in the stellar plasma), as well as the origin of the neutrons driving the process. The Canadian $s$ process program focuses on two distinct aspects of the $s$ process: $\beta$ decay under stellar conditions, and nuclear reactions that influence the $s$ process neutron flux.


\begin{SCfigure}
    \centering
    \includegraphics[width=0.6\linewidth]{
        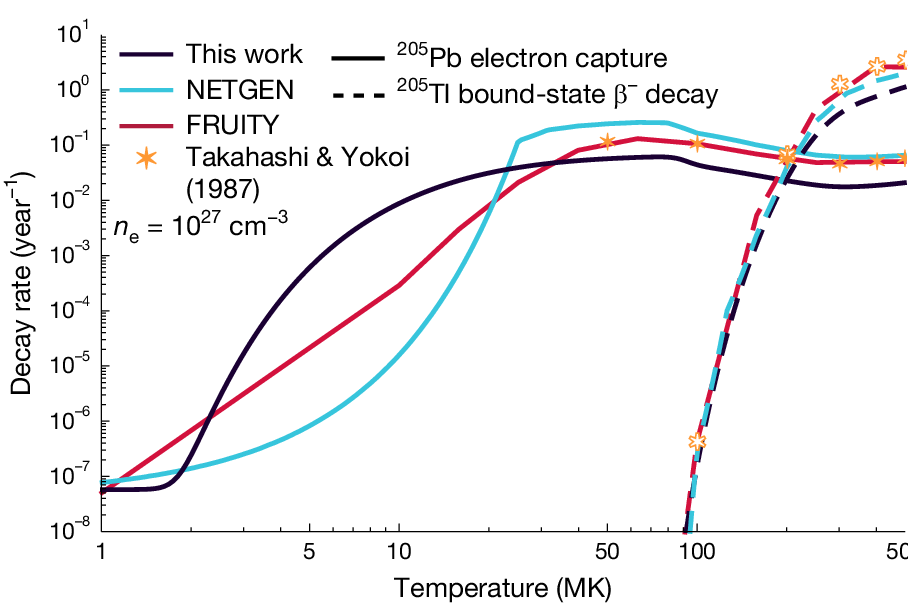
    }
    \caption{Temperature-dependent and density-dependent decay rates for ${}^{205}$Pb and ${}^{205}$Tl. The purple curves show the new rates determined by the Canadian-led measurements of highly-charged ion decay at GSI. Adapted from G.\ Leckenby \emph{et al.}, \href{https://doi.org/10.1016/j.physletb.2022.137361}{Nature 635, 321 (2024).}}
    \label{fig:DecayRates}
\end{SCfigure}

The decay program takes place mostly at the ESR at GSI, within the ILIMA@ESR collaboration, where the Canadian group has constructed two multi-purpose particle detectors.
The focus of the program is the measurement of $\beta$ decay of highly charged ions that circulate in the storage ring. The high charge states of these ions mimic stellar conditions. When certain nuclei are fully stripped, their decay rates can be drastically altered compared to neutral atoms.  Some decay modes, such as electron capture or internal conversion, may be fully or partially blocked, while others, for example bound state $\beta$ decay, may be accelerated by orders of magnitude. A highlight of this program is the recent measurement of the ${}^{205}\mathrm{Tl}^{81+}$ decay rate, which was found to be much longer than the value previously used in astrophysical models.\footnote{G.\ Leckenby \emph{et al.}, Nature 635, 321 (2024).}\textsuperscript{,}\footnote{R.\ S.\ Sidhu \emph{et al.}, Phys.\ Rev.\ Lett.\ 133, 232701 (2024).} The new experimental half-life was used to extract the nuclear matrix element of this transition, which allows for the calculation of accurate astrophysical decay rates of ${}^{205}$Tl and ${}^{205}$Pb in the stellar plasma (Fig.~\ref{fig:DecayRates}). This enables models of the $s$ process in AGB stars to provide accurate ${}^{205}$Pb yields, which are essential for using ${}^{205}$Pb as a cosmochronometer to date processes in the early Solar System.

The neutron flux program focuses on the ${}^{22}\mathrm{Ne}(\alpha,n){}^{25}\mathrm{Mg}$ reaction, which serves as the main $s$ process neutron source in massive stars. The stellar ${}^{22}\mathrm{Ne}(\alpha,n){}^{25}\mathrm{Mg}$ rate is dominated by a single resonance at $E_{cm} = 703$~keV. Recent experiments have shown a discrepancy between direct and indirect determinations of the strength of this resonance.\footnote{A.\ Best \emph{et al.}, Eur.\ Phys.\ J.\ A 61, 99 (2025).} To shed light on this discrepancy, a measurement of this resonance strength using DRAGON coupled with OGS neutron detectors, funded by UK collaborators, in planned for late 2025. This new technique of measuring $(\alpha,n)$ reactions with a recoil separator offers an alternative to normal-kinematics measurements with different systematic uncertainties. A proof of principle experiment, measuring the strong resonance at $E_{cm} = 1213$~keV, has been completed and demonstrates the ability to detect neutrons in the OGS detectors in coincidence with recoils at the DRAGON focal plane. DRAGON has also been used to study the ${}^{22}\mathrm{Ne}(\alpha,\gamma){}^{26}\mathrm{Mg}$ reaction, which competes with the $(\alpha,n)$, as well as  ${}^{18}\mathrm{O}(\alpha,\gamma){}^{22}\mathrm{Ne}$, which influences the amount of ${}^{22}\mathrm{Ne}$ available to initiate ${}^{22}\mathrm{Ne}(\alpha,n){}^{25}\mathrm{Mg}$.

The seven-year plans for the $s$ process include continuation of the highly-charged ion decay program at ESR, with minor detector and facility upgrades planned. The broad capabilities of the Canadian particle detectors allow detection of products of electron capture, $\beta$, $\beta n$, and $\alpha$ decay products, as well as charge-changing products (electron loss or pickup) that would leave the acceptance of the ring.  The neutron-flux program plans to complete the measurement of the $E_{cm} = 703$~keV resonance in late 2025, with plans to measure other relevant $s$ process reactions in the next seven years. These include ${}^{18}\mathrm{O}(\alpha,n){}^{21}\mathrm{Ne}$, which decreases ${}^{22}$Ne production, shuffling flux away from ${}^{22}\mathrm{Ne}(\alpha,n){}^{25}\mathrm{Mg}$, as well as the other main $s$ process neutron source, ${}^{13}\mathrm{C}(\alpha,n){}^{16}\mathrm{O}$. The latter reaction also serves as a background source for deep-underground neutrino oscillation experiments, whose background uncertainties can be reduced through future DRAGON + OGS measurements.

\paragraph{$p$ process:}
\begin{SCfigure}
    \centering
    \includegraphics[width=0.6\textwidth]{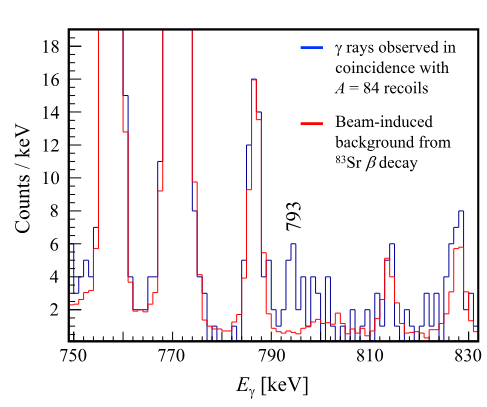}
    \caption{Coincidence and singles $\gamma$-ray spectra observed during the EMMA/TIGRESS ${}^{83}\mathrm{Rb}(p,\gamma){}^{84}\mathrm{Sr}$ experiment. The coincidence $793$~keV line, used to infer the $(p,\gamma)$ cross section, is indicated in the figure. Figure credit: Phys.\ Rev.\ Lett.\ 127, 112701 (2021).
    }
    \label{fig:Rb83}
\end{SCfigure}

The various $p$ nuclides are thought to be formed in core-collapse or thermonuclear supernovae. Initially, $(\gamma,n)$ reactions on the seed nuclei drive the nucleosynthesis pathway towards the neutron-deficient side of stability. 
Once the neutron separation becomes large enough, so-called deflection point nuclei are reached and $(\gamma,\alpha)$ and $(\gamma,p)$ reactions dominate
the reaction flux, necessitating experimental constraints on select $(\gamma,\alpha)$ and $(\gamma,p)$ reactions, typically involving radioactive nuclei. Direct measurements of the inverse reactions, e.g.\ $(p,\gamma)$ or $(\alpha,\gamma)$, which have positive $Q$ values, are recognized as the optimal method for constraining the key $p$-process reactions as this minimizes the impact of thermal excitations within the stellar environment.

The EMMA recoil separator, recently commissioned at TRIUMF, has opened the door for new direct measurements of $(p,\gamma)$ and $(\alpha,\gamma)$ reactions in the mass and energy regions important for $p$-process nucleosynthesis. A recent highlight of this program was the successful direct measurement of the ${}^{83}\mathrm{Rb}(p,\gamma){}^{84}\mathrm{Sr}$ reaction using a radioactive ${}^{83}\mathrm{Rb}$ beam at energies relevant to the $p$-process---the first such measurement ever reported.\footnote{G.\ Lotay \emph{et al.}, Phys.\ Rev.\ Lett.\ 127, 112701 (2021).}\textsuperscript{,}\footnote{M.\ Williams \emph{et al.}, Phys.\ Rev.\ C 107, 035803 (2023).} This experiment coupled EMMA with TIGRESS and observed time-correlated $\gamma$ rays and ${}^{84}$Sr recoils in coincidence. The yield of  $793$~keV $\gamma$~rays from the $2^+_0$ state was used to infer a total cross section (Fig.~\ref{fig:Rb83}), which was smaller than predicted by statistical model calculations. Nucleosynthesis calculations incorporating these results show that the amount of ${}^{84}$Sr synthesized and its uncertainty in both core collapse and Type Ia supernovae are reduced as a consequence.

Looking forward to the next seven years, the Canadian community plans to exploit the existing EMMA infrastructure to continue to make direct cross section measurements of importance for the astrophysical $p$ process, including experiments using the novel thin-film Si-He targets produced by collaborators in Sevilla, Spain. This program will also benefit from the additional radioactive beam hours that will become available when ARIEL comes online, through increased time for beam development as well as longer experiments.

\paragraph{$\alpha$ process:}
While the origin of non-$p$ heavy nuclei can mostly be attributed to the $r$, $s$, and $i$ processes, the observed abundance patterns observed for the lighter heavy elements (Sr--Ag) is not fully explained by these processes alone. For example, observations of metal-poor halo stars indicate over-production of the Sr--Ag elements relative to the solar $r$-process abundance pattern.\footnote{A.\ Frebel, Ann.\ Rev.\ Nucl.\ Part.\ Sci.\ 68, 237 (2018).} 
Possible production mechanisms for these so-called ``weak $r$-process nuclei'' include the $\alpha$ process---a series of $(\alpha,n)$ reactions in the $\nu$-driven winds of core-collapse supernovae, or a more traditional weak $r$~process that only reaches lighter elements in other environments like accretion disk ejecta. Models of the $\alpha$ process in supernovae are sensitive to a variety of $(\alpha,n)$ reaction rates, mostly involving radioactive nuclei, necessitating experimental constraints on these rates for a complete understanding of the production of the  weak $r$-process nuclei.


The Canadian program surrounding the production of weak $r$-process nuclides has a current focus on direct measurements of key $(\alpha,n)$ reactions using the EMMA recoil separator coupled with TIGRESS $\gamma$-ray detectors. This technique relies on the coincident detection of secondary $\gamma$-rays following the primary neutron decay of the compound nucleus, in coincidence with the recoils. These measurements take advantage of the intense radioactive beams in the Sr--Rb region available at TRIUMF, as well as the novel thin-film Si-He targets, developed by collaborators in Spain, which provide a well-characterized ${}^4$He target for the measurements. A recent highlight of this program is the successful measurement of the ${}^{94}\mathrm{Sr}(\alpha,n){}^{97}\mathrm{Kr}$.\footnote{M.\ Williams \emph{et al.}, Phys.\ Rev.\ Lett.\ 134, 112701 (2025).} This experiment successfully measured $\gamma$~rays originating from the ${}^{94}\mathrm{Sr}(\alpha,n){}^{97}\mathrm{Kr}$ reaction in coincidence with the ${}^{97}$Kr recoils, at an energy inside the Gamow window for the $\alpha$~process. The resulting cross section was a factor ${\sim}0.6$ smaller than statistical-model predictions, indicating that Ru yields from $\alpha$-process models should decrease by a factor ${\sim}5$ using the new, experimentally-driven rate.

During the next seven years, it is expected that the $\alpha$-process program will extend to other reactions, and a successful experiment measuring ${}^{93}\mathrm{Sr}(\alpha,n){}^{96}\mathrm{Zr}$ has already been completed. Another area of development is the use of OGS neutron detectors to measure the primary neutrons resulting from the $(\alpha,n)$ reactions. This would enhance the ability to identify $\mathrm{recoil} + \gamma$ coincidences and reduce the model dependence on $\gamma$-ray decay schemes. A first test of this method was carried out in the recent ${}^{93}\mathrm{Sr}(\alpha,n){}^{96}\mathrm{Zr}$ experiment, with four OGS detectors included in the setup. Future upgrades to a dedicated large-area OGS array hold promise for triple-coincidence ($\mathrm{recoil} + n + \gamma$) measurements, which could significantly reduce the model dependence of the extracted cross sections. The $\alpha$-process program will also benefit from the onset of ARIEL, resulting in increased time for measurements and beam development.

\subsection{Beyond the next 7 years}
\label{sec:AstroNext7}

The backbone of the Canadian experimental nuclear astrophysics program is its suite of specialized detector systems, both at TRIUMF and abroad, which are world leaders in performing direct and indirect measurements of astrophysical reactions. Each of these facilities is poised to continue to deliver world-leading science well beyond the next seven years and will remain competitive with, or complementary to, newly-developed devices outside of Canada such as the SECAR recoil separator at FRIB. This continued delivery of world-leading science beyond the next seven years will require development of new beams, or more intense beams of existing species. At TRIUMF, the ARIEL facility, expected to come online in 2027, will ensure the necessary advances in beam delivery are brought to fruition. ARIEL will benefit the Canadian nuclear astrophysics program in two ways, both directly through development of new neutron-rich beams, and indirectly through increased availability of radioactive beam time (see pullout box ``Nuclear Astrophysics in the ARIEL Era''). 
Offshore, the operation of the SuperFRS facility at FAIR will allow a factor 100-1000 higher beam intensities compared to the existing GSI facility. Canadians are leading several efforts at GSI and FAIR, and once the new Collector Ring (CR) at FAIR is constructed, this will ensure a continued leadership of the Canadian program.



\begin{figure}
\centering
\begin{tcolorbox}[
  colback=pink,
  colframe=gray!50!black,
  coltitle=black,
  colbacktitle=gray!20,
  width=0.8\textwidth,
  boxrule=0.8pt,
  arc=4pt,
  auto outer arc,
  fonttitle=\bfseries,
  title=Nuclear Astrophysics in the ARIEL Era
]
{\small\baselineskip=14truept The advent of ARIEL promises to offer intense neutron-rich beams from photofission, as well as parallel-beam capabilities using the existing proton-driven methods. Some nuclear astrophysics programs, such as those studying the $i$ and $r$ processes, will benefit directly from the new neutron-rich ARIEL beams. Others, such as the light-element program, will not use ARIEL beams directly but will benefit indirectly from increased beam time. This will allow longer experiments to study low-yield reactions, or to perform comprehensive studies on higher-yield reactions. ARIEL will also allow significant increases in beam development time. The resulting increases in beam intensity, as well as newly deliverable beam species, will open up a wide range of new experiments that are not currently possible.}
\end{tcolorbox}
\end{figure}

In order to fully capitalize on the beam-delivery advances of ARIEL, a number of modest-scale upgrades and new technique developments are expected for the existing nuclear astrophysics facilities. At DRAGON, plans are in place to upgrade the existing BGO $\gamma$-ray detectors to a material with better timing and energy resolution. LaBr$_3$ is a primary candidate for the new array, although materials with higher detection efficiency such as LYSO or GAGG:Ce are also being considered. DRAGON is also planning an upgrade to a digital DAQ system, which will facilitate future integration with ancillary detectors and alleviate the dead-time issues encountered with certain radioactive beams. It is also expected that the existing DEMAND array of OGS neutron detectors will be expanded, with a potential UK/Canadian joint funding model. The new array would open up a plethora of new experimental possibilities, including studies of $(\alpha,n)$ reactions for the $\alpha$~process and $(d,n)$ reactions to constrain astrophysical proton-capture rates, or to study proton-rich nuclear structure.
Upgrades to the TITAN facility will open the possibility for new in-trap decay studies of longer-lived astrophysical species, such as ${}^{7}\mathrm{Be}$, which complement the offshore in-flight decay studies at GSI.
%

While the existing major facilities are already expected to deliver world leading science in the next seven years and beyond, the Canadian nuclear astrophysics community has plans to develop new major facilities that will ensure it remains a global leader into the second half of the $21^\mathrm{st}$ century.  A major project being pursued by the Canadian nuclear astrophysics community is the TRISR storage ring (Section~\ref{sec:TRISR}). The unique combination of a storage ring and a neutron target would open a completely new program of directly measuring astrophysical neutron capture cross sections on short-lived nuclei for the $s$, $i$, and $r$ processes. At the same time, the new facility will open the door for a wide range of new experiments taking advantage of the six order of magnitude increase in luminosity compared to single-pass experiments to study low-yield reactions.
Together, the new scientific opportunities afforded by TRISR will lead to development of new experimental programs lasting well beyond the next seven years.
Similarly, the construction of a new two-step fragmentation facility at TRIUMF would open new opportunities nuclear astrophysics research in Canada well beyond the next seven years (Section~\ref{sec:TwoStepRIB}). The  neutron-rich beams produced by this facility would be used to study nuclei on or near the $r$-process pathway. The ``intermediate energy" beams (few 100s of MeV per nucleon) delivered by the new facility would also open up a range of new experimental techniques that complement those currently used at TRIUMF.

\subsection{Summary}

The Canadian nuclear astrophysics program targets all of the major cornerstones of nuclear astrophysics, ranging from light-element formation through networks of $p$ and $\alpha$ capture reactions, to solar neutrino production, heavy-element synthesis, and dense nuclear matter. The program makes use of a wide variety of direct and indirect measurement techniques, many of which have close connection to nuclear structure studies, and and there is a strong astrophysics component to the Canadian theory program.

The experimental program capitalizes on past investments in domestic and offshore experimental facilities, many of which are world leading and promise to remain so in the coming seven years and beyond. These facilities are expected to amplify their scientific output in the ARIEL era, taking advantage of the newly-available beams and increased experimental run times. At the same time, new facility developments are planned, such as TRISR and the two-step RIB facility, which will ensure Canadian leadership in this field well beyond the next seven years.
\section{Fundamental symmetries}
\label{sec:FundSym} 

Precise measurements of symmetry violations in nuclear processes could
reveal new physics beyond the Standard Model.  This field of
exploration is known as fundamental symmetries, and is sometimes
referred to as the precision frontier, or intensity frontier.  It
addresses the big question of what lies beyond the Standard Model
(Section \ref{sec:bigquestion}).

\subsection{The Canadian program (current and next 7 years)}

\subsubsection{Time-reversal and CP violation}

Time-reversal symmetry is linked to the symmetry between particles and
their antiparticle counterparts (CP symmetry) through the CPT theorem.
A violation of time-reversal symmetry would therefore indicate a
violation of CP symmetry~\footnote{M.~Pospelov and A.~Ritz,
arXiv:2509.23531 [hep-ph]}.  Experiments search for violations of
time-reversal and CP symmetry in the hope to explain why the universe
is made of matter rather than antimatter.  In nuclear physics in
Canada, these are principally searches for electric dipole moments
(EDMs) of fundamental particles, via neutron and molecular EDM
measurements, which are discussed in this section.  Correlation
measurements in $\beta$-decay can also probe time-reversal violation
(see Section~\ref{sec:PrecisBetaDecayEtc}).

\paragraph{TUCAN}

The new physics scenarios giving rise to a neutron electric dipole
moment (nEDM) usually involve new physics beyond the Standard Model,
and are based on scenarios akin to electroweak
baryogenesis~\footnote{D.E.~Morrissey and M.J. Ramsey-Musolf, New
J.Phys. 14 (2012) 125003}.  The nEDM is also sensitive to the strong
CP problem, the unexplained smallness of CP violation in the QCD
sector, which led to the proposal of Peccei-Quinn symmetry and axions,
a dark-matter candidate.  Measurements of the time-dependence of the
nEDM measurements are now used to constrain the axion parameter space.

The goal of the TUCAN (TRIUMF UltraCold Advanced Neutron) EDM
experiment is to measure the neutron electric dipole moment (nEDM)
with world-leading precision.  The nEDM experiment is projected to be
capable of an uncertainty of $1\times 10^{-27}~e$cm, competitive with
other planned projects worldwide, and a factor of ten more precise
than the present world's best.  The TUCAN Collaboration was originally
formed of physicists from Canada and Japan, and has now grown to
include members from the United States and Mexico.

The technological breakthrough enabling the improvement in precision
is a new spallation-driven superfluid helium ultracold neutron (UCN)
source, which is unique to the TUCAN project.  The production rate in
the source is expected to be $1.4\times 10^7$~UCN/s, and since UCN
losses can be small in superfluid helium, a large number of UCNs can
be accumulated over a period of about 30 seconds.  Ultimately, the
project aims to achieve UCN counts in excess of a factor of 100 over
the previous best nEDM experiment.  The UCN source upgrade and EDM
experimental apparatus were funded by the CFI IF 2017 competition, and
included partner contributions from Japan and TRIUMF.

\begin{figure}
\begin{center}
\includegraphics[width=0.49\textwidth]{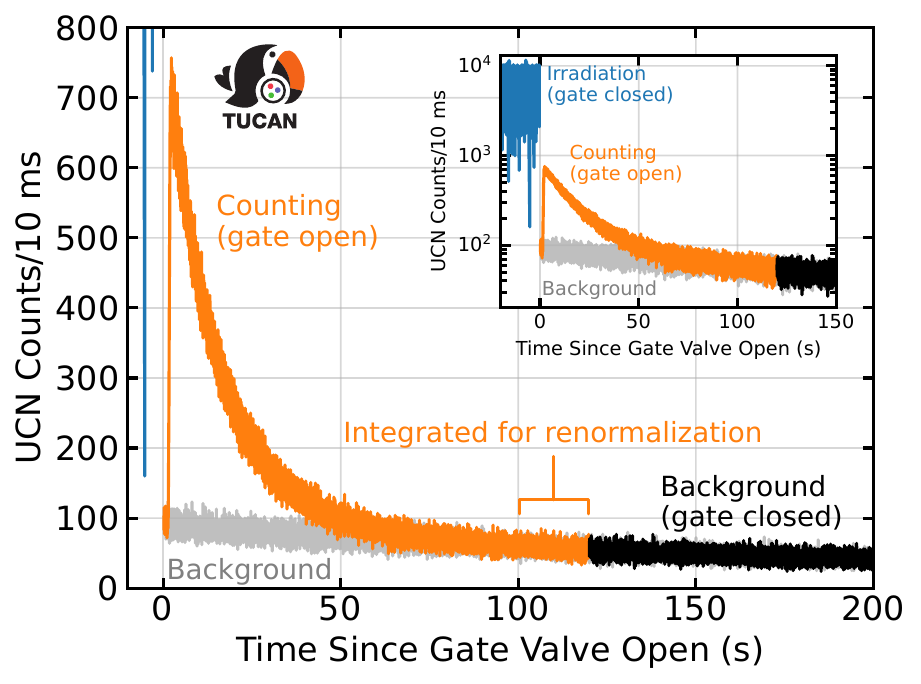}
\includegraphics[width=0.49\textwidth]{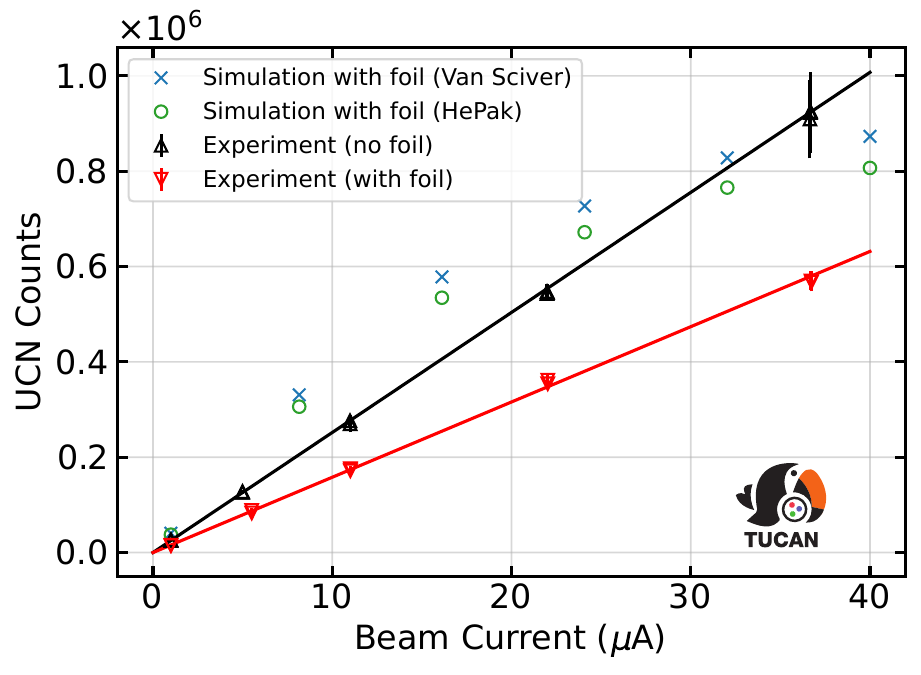}
\caption{Left: UCN counts as a function of time during a measurement
  cycle at TUCAN, showing the new record number of UCNs.  Right:
  Integrated UCN counts measured as a function of the beam current.
  Data are compared with simulations, showing fair agreement with the
  UCN counts observed at high current.}
\label{fig:ucn-vs-current}
\end{center}
\end{figure}

The UCN source upgrade is now in operation throughout 2025.  First UCN
production measurements were conducted in June 2025.  The project
produced a new record number of UCN, compared to the previous UCN
source~\footnote{B.~Algohi {\it et al.} (TUCAN Collaboration),
arXiv:2509.02916 [nucl-ex]}.  This was consistent with expectations
based on simulations (see Fig.~\ref{fig:ucn-vs-current}).  The UCN
source also went through a huge battery of neutron and cryogenic tests
to demonstrate that it meets the requirements to produce a
world-leading number of UCN.  All that remains now is to install the
liquid deuterium moderator, which will make the source the most
intense in the world.

A major milestone for the nEDM experimental apparatus was the
completion of a large magnetically shielded room (MSR) at TRIUMF
(January 2025).  The MSR is now in routine use, in experiments
characterizing its magnetic shielding and stability, testing new
high-precision low-field magnetometers, and testing new coil
assemblies that will produce the highly uniform and stable magnetic
field for the nEDM experiment.

The imminent plans of the project (2027-2034) involve running the nEDM
experiment.  The project has now reached the limit of the helium
liquefier facility in Meson Hall at TRIUMF and have made a request for
a new liquefier based on a turbine system in the CFI IF 2025
competition.  Also requested are modest upgrades to both the UCN
source and EDM experiment, needed to fully exploit the capabilities of
the UCN source.  The PENeLOPE (neutron lifetime) experiment will also
be completed using this facility (see below).

\paragraph{RadMol}

Molecules in which one of its constituting atoms contain a
short-lived, radioactive nuclide were recently introduced as
intriguing objects of research \footnote{R. F. Garcia Ruiz et
al. Nature 581, 396 (2020)}\footnote{G. Arrowsmith-Kron et al.,
Rep. Prog. Phys. 87 084301 (2024)}. RadMol’s initial science program
is focused on the search for permanent electric dipole moments (EDMs)
of atomic nuclei in the form of nuclear Schiff moments, i.e. a
low-order term in the expansion of the nuclear charge distribution
which violates parity (P) and time reversal (T) symmetry. The
existence of such an EDM would imply the violation of time-reversal
symmetry (T) and the discovery of new physics beyond the Standard
Model of particle physics. Identifying new sources of T-violating
phenomena is also critical to explain the matter-antimatter asymmetry
in the universe.  Hence, researchers worldwide are racing to be the
first to detect EDMs in a variety of different systems. RadMol’s
approach relies on three key multipliers:
\begin{enumerate}
\item employ polar molecules which exhibit large amplification factors
  for EDM searches given their enormous internal electric fields,
  1,000 to 10,000 times larger than in free atoms.
\item incorporate rarely found pear-shaped (`octupole-deformed')
  atomic nuclei into these polar molecules. This further enhances a
  molecular signature caused by a T-breaking nuclear EDM by a factor
  of 100-1,000; all currently-known nuclei with (static) octupole-
  deformation are found in radioactive isotopes.
\item trap molecules to boost the observation or coherence time by
  more than a factor of 1,000 relative to conventional beam
  experiments.
\end{enumerate}

The current gold standard in experimental limits on Schiff moments is
found in spin-polarized, neutral (and stable) $^{199}$Hg mercury atoms
\footnote{B. Graner et al. Phys. Rev. Lett. 116, 161601 (2016)}.  The goal of
the next seven years for RadMol is to complete a laboratory
installation at TRIUMF, and improve upon the $^{199}$Hg result by a
factor of 1000, and making developments to go even further using the
techniques mentioned above.

\paragraph{EDM$^3$}

The EDM$^3$ collaboration has the goal of making an ultraprecise
measurement of the electron electric dipole moment (eEDM). Such a
measurement will be a direct search for the additional CP-violating
physics that is required to understand the matter/antimatter asymmetry
in the universe. Modern eEDM measurements use polar molecules and are
limited by the number of molecules that can be observed and the
observation time for each molecule. The EDM3 collaboration is using
polar molecules (BaF and SrF) embedded into cryogenic inert-gas solids
(neon and argon) to dramatically increase both the number of
observations and the duration of each observation, and thus to
dramatically increase the accuracy of eEDM measurements. Equipment
funding at a level of \$8.4 million has been provided through a CFI-IF
grant. An apparatus has been constructed and many of the steps needed
for an eEDM measurement have already been demonstrated. An initial
eEDM measurement is expected within five years, with measurements of
increasing accuracy during the following ten years. At the current
accuracy for eEDM measurements, tests of physics at energy scales of
10 TeV are already possible. By increasing this accuracy by two to
four orders of magnitude, physics on the 100- to 1000-TeV scale will
be tested.

Measurements of the electron electric dipole moment have increased in
precision over the decades and have put strong limits on
beyond-the-Standard-Model theories that have attempted to explain the
matter/antimatter asymmetry in the universe \footnote{Cornell et al., Science
  381,46(2023)}. The EDM$^3$ collaboration is attempting to increase
the accuracy of these measurements (by between two and four orders of
magnitude) by using polar molecules that are embedded into a cryogenic
inert-gas solid \footnote{Hessels et al., Phys. Rev. A
  98,032513(2018)}. Experiments on these molecules employ the same
techniques used for eEDM measurements that use a molecular beam \footnote{ACME
  collaboration, Nature 562, 355 (2018); Hinds et al., 473, 493
  (2011)}, with the advantage that a much larger sample of molecules
can be addressed, since the integrated flux from approximately an hour
of beam is captured in the solid \footnote{Hessels et al., arXiv:2410.04591
  (2024)}. These molecules are fixed in position by the solid
(allowing for long observation times) and also have their orientation
fixed (without the need for an externally applied electric field)
\footnote{Hessels et al., Mol. Phys. 121, e2198044(2023); 121, e2232051(2023)}.

An apparatus has been assembled for performing the EDM$^3$
measurements, and this apparatus continues to be refined as
experimental tests are performed. Funding for the equipment has been
provided from a CFI-IF grant, and funds to initiate the project have
been provided by the Sloan, Moore, and Templeton Foundations. Large
samples of embedded molecules have been assembled in inert-gas
matrices that can be cooled to a temperature of 3 kelvin. Most of the
individual steps required for an eEDM measurement have been
demonstrated, including detection (using laser-induced fluorescence
\footnote{Hessels et al, arXiv:2410.04591(2024)}), optical
pumping \footnote{Hessels et al, arXiv:2410.04605(2024)}, and
radio-frequency transitions. The main remaining challenge for
performing an initial measurement is the production of an ultrapure
sample of embedded molecules.  The inert gas is already purified using
a getter system, and the purity of an undoped solid is tested by
xray-induced fluorescence. Purity of the doping process will require a
deflection of the molecular beam
\footnote{Hessels et al, Phys Rev A 107,032811(2023); 108,012811(2023);
  arXiv:2410.04598(2024)} to separate the polar molecules from other
product produced in the molecular source. With improved purity, an
initial eEDM measurement is planned to take place by 2030.

\paragraph{Nuclear studies for EDMs}

GRIFFIN, TIGRESS, and their auxiliary detection systems will also be
used to explore octupole deformation and collectivity in select
isotopes. The objective is to identify optimal candidates for the
electric dipole moment (EDM) searches proposed by RadMol and evaluate
their sensitivity to CP violation.  TITAN plans to use an Extreme
UltraViolet (EUV) spectrometer \footnote{Y. Wang, MSc thesis, UBC
2022} to measure the absolute nuclear charge radii, which is a crucial
missing ingredient in searches for a electric dipole moment with heavy
radioactive species.

\subsubsection{Tests of the neutral current weak interaction}

By searching for the violation of parity in processes normally
involving the electromagnetic interaction (photon exchange), the tiny
effects of the neutral current weak interaction ($Z^0$ exchange) at
low-energy can be sensed.  The experiments discussed in this section
seek more precise measurements of the weak mixing angle $\theta_W$ by
measuring parity violation in low-energy observables.

\paragraph{MOLLER
\label{sec:moller}}

\begin{SCfigure}
  \includegraphics[width=12.3cm,trim=6mm 6mm 6mm 6mm]{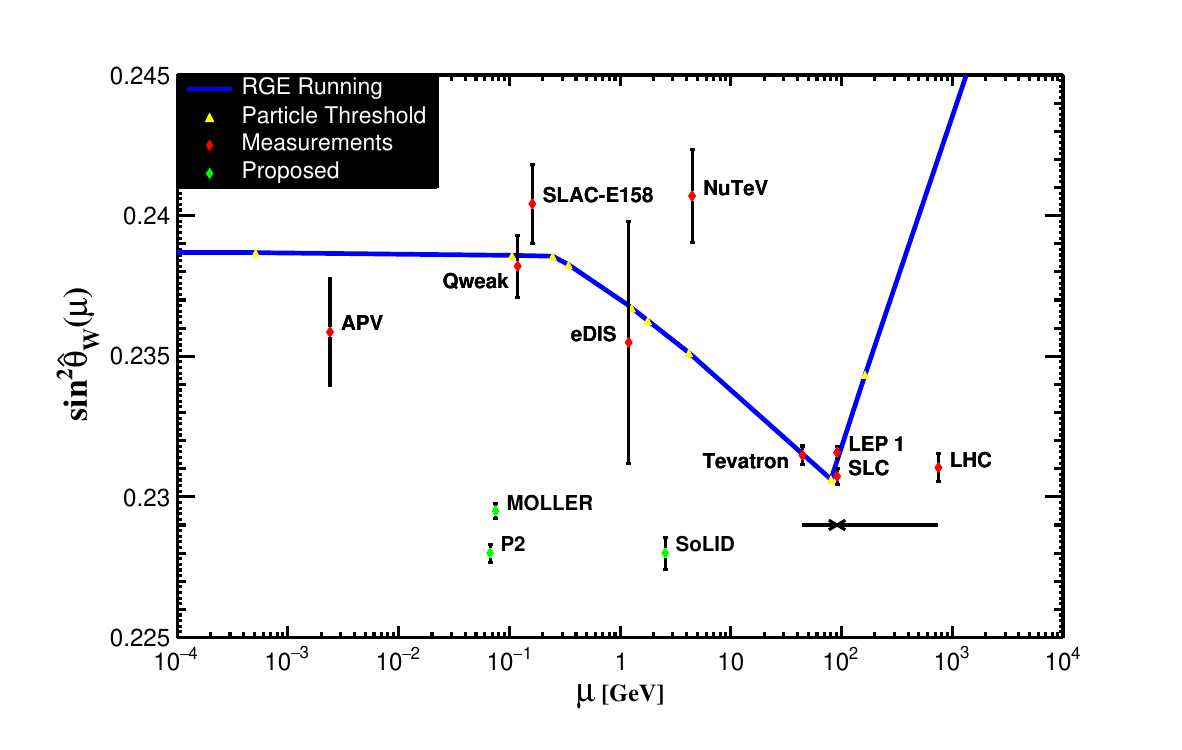}
  \caption{{\it Completed measurements of the weak mixing-angle vs.\ the energy scale
  $\mu$ are shown with red data points. Proposed future measurements are shown with green data points, at the
  appropriate $\mu$ value and with the proposed error bar at arbitrary vertical location. Figure by J.~Erler, with permission.}}
\label{fig:running}
\end{SCfigure}

The MOLLER Experiment is a flagship experiment of the Jefferson Lab
(JLab) 12 GeV program. It will determine the weak
mixing angle via a measurement of the weak charge of the electron,
using parity violating (PV) electron-electron scattering at the 12 GeV
continuous electron beam facility (CEBAF) at JLab. The experiment will
determine the value of the weak mixing angle at low momentum transfer
to very high precision as a sensitive test for physics beyond the
Standard Model (BSM). The current status of the experimental effort to
map out the running of the weak mixing angle is illustrated in
Fig.~\ref{fig:running}, which shows both completed and planned
measurements.

The goal of the MOLLER experiment is to make the world’s most precise
off-resonance measurement of the weak mixing angle, using polarized
electron-electron scattering at JLab in the
USA, as a sensitive test for physics beyond the Standard Model. The
experiment is under development and scheduled to begin data taking in
2027. Since the last LRP, the Canadian team has made significant
progress in leading the design of the magnetic spectrometer and the
integrating detectors and associated electronics. The Canadian team
has also established leadership roles in simulation and analysis
software.

The two most precise measurements at the Z-pole correspond to
$\sin^2\theta_W=0.23070(26)$ \footnote{K. Abe et al., Phys. Rev. Lett. 84 5945
  (2000)} and $\sin^2\theta_W=0.23193(29)$ \footnote{ALEPH Collaboration,
  CDF Collaboration, D0 Collaboration, DELPHI Collaboration, L3
  Collaboration, OPAL Collaboration, SLD Collaboration, LEP
  Electroweak Working Group, Tevatron Electroweak Working Group, SLD
  Electroweak and Heavy Flavour Groups, arXiv:1012.2367 [hep-ex] 2010,
  Phys. Rept. 427, 257 (2006)}, which disagree by more than 3
standard deviations from each other and with the predicted value of
the mixing angle at the $Z$-pole. The highest precision data point
away from the $Z$-pole corresponds to the recently published QWeak
result $\sin^2\theta_W=0.2383(11)$ \footnote{D. Androic et al., Nature 557,
  207-211 (2018)}. Also shown is the previous measurement of the
electron weak charge (SLAC E158 \footnote{P. L. Anthony et al. (SLAC E158
  Collaboration), Phys. Rev. Lett. 95, 081601 (2005)}), among
others. As seen in the figure, MOLLER would provide a significantly
improved error (almost $5\times$ better than the previous measurements
SLAC E158 and QWeak, respectively) for the low energy measurements, on
par with the precise measurements at the $Z$-pole.

MOLLER will measure the parity-violating asymmetry, $A_{PV}$ , arising
due to the interference between photon and $Z^0$ boson exchange.  The
MOLLER $A_{PV}$ is $\sim$ 35 parts per billion (ppb) which will be
measured with a statistical precision of 0.73 ppb. This corresponds to
a 2.3\% measurement of $Q^e_W$ , leading to an accuracy in the
determination of the weak mixing angle of $\sim$ 0.1\%, comparable to
the two best determinations at the $Z$-pole.

MOLLER will run in Hall A of Jefferson Lab between 2026 and 2030 in 3
major run phases.  Nearly 20 HQP have been trained so far on aspects
of the MOLLER experiment. The Canadian group is currently constructing
the 224 main integrating detectors, including the supports, PMT bases,
and testing the components, as well as communicating with engineers
concerning the construction of the spectrometer magnets (all the coils
for the downstream magnet have now been delivered to JLab). Analysis
will likely continue into 2032.

MOLLER is an excellent example of a direct collaboration between
experimental and theoretical physicists, where the theory contribution
is the calculation of electroweak radiative corrections to hone the
new physics interpretation of the measurement (see
Section~\ref{sec:nuctheo_funsym}).

\paragraph{Francium atomic parity violation}

In atoms, extremely weak electric dipole transitions between states of
the same parity are induced by the parity-violating exchange of
Z-bosons between the electrons and the quarks in the nucleus, an
effect known as atomic parity violation (APV). By measuring this
amplitude, one can study neutral-current weak interactions with atomic
physics methods and search for new physics such as extra gauge bosons
and leptoquarks. APV is strongly enhanced in heavy atoms, but the
atomic structure calculations necessary to extract the weak physics
are presently only feasible in alkali atoms. In francium, the APV
effect is 18 times larger than in Cs.  However, Fr has no stable
isotopes, must be produced at a radioactive beam facility such as
ISAC, and needs to be accumulated in a laser trap. The FrPNC (Francium Parity Non-Conservation) team has
established the francium trapping facility at ISAC. A magneto-optical
capture trap receives Fr isotopes from ISAC, and confinement of up to
$\sim 10^6$ atoms of a single Fr isotope in a volume of $\sim
1$~mm$^3$ at micro-Kelvin temperatures has been demonstrated. A
program of precision laser spectroscopy using the $7s-8s$ highly
forbidden transition is underway.  The first goal is to complete the
study of the conventional, parity conserving amplitudes, the Stark
(electric field) induced electric dipole, and the magnetic dipole
(both relativistically and hyperfine induced). The goal is to continue
with a program to observe and eventually measure the parity violating
interference between the Stark amplitude and the Z-boson induced weak
amplitude. This will be accomplished by detection variations (at the
$\sim 10^{-4}$ level) in the $8s$ decay fluorescence under parity
flips such as the reversal of external electric and magnetic fields.

\begin{figure}[h!]
  \centering
  \includegraphics[height=2.2in]{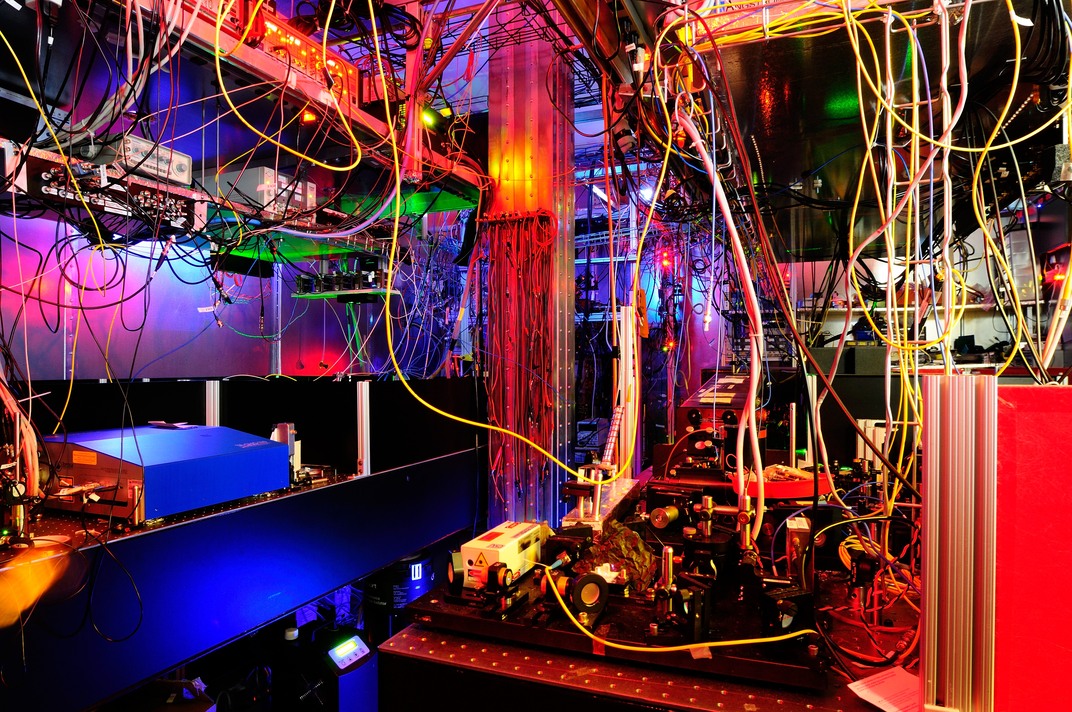}
  \includegraphics[height=2.2in]{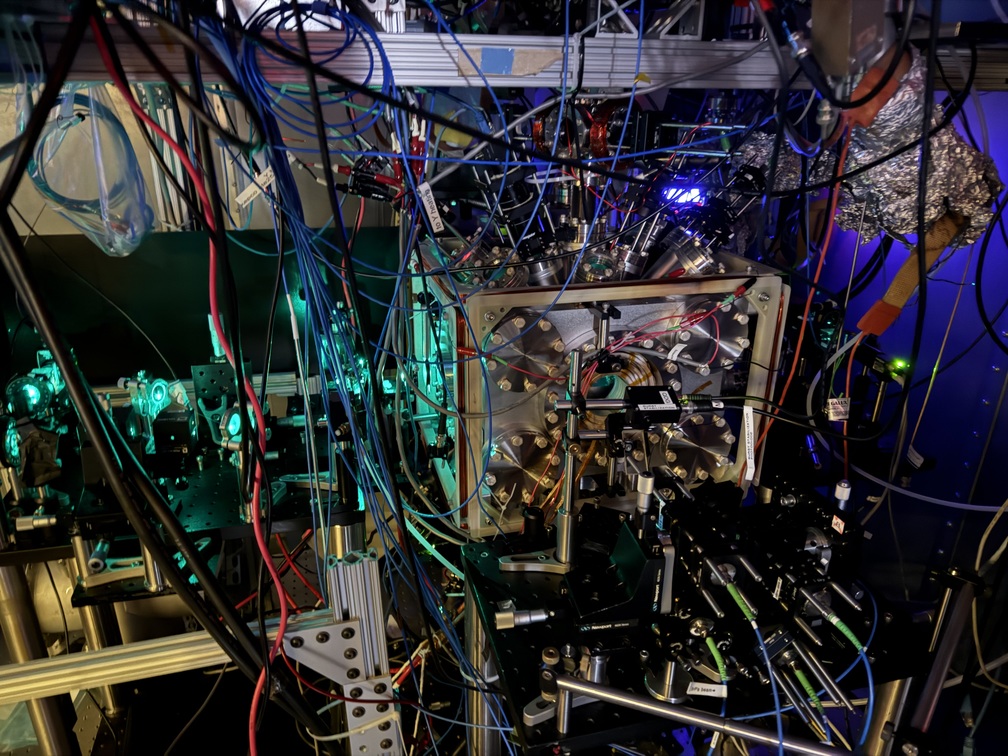}
  \caption{Left: Francium laser trapping facility at ISAC.  Right: Francium spectroscopy chamber online to the ISAC radioactive beam facility.}
  \label{fig:figb}
\end{figure}

The project aims to complete precision measurements of the $7s-8s$
Stark-induced electric dipole and magnetic dipole amplitudes (percent
or sub-percent level) by 2028. In addition to the physics results, an
outcome will be an apparatus that permits statistics-limited
measurements and the ability to optically pump the francium atoms into
unique ground-state hyperfine levels $|F,m_F=\pm F\rangle$.  By 2029
the project should be ready to observe the parity-violating electric
dipole amplitude via its interference with the Stark-induced one. This
will be followed by an upgrade of the apparatus for high-statistics
runs. This involves for example an increase of trapped atoms by up to
an order of magnitude by deploying more trap laser power. Once
multiple targets become operational at ISAC/ARIEL, sufficient beamtime
can be obtained to start precision (percent level) APV measurements.

\paragraph{EIC \label{sec:eic2}}

Searches for new physics beyond the standard model at the EIC will be
conducted using precision measurements in the electroweak sector. It
includes the precision study of neutral current inclusive electron
scattering to test the Standard Model, via measurements of structure
functions using the weak neutral current, and the search for physics
beyond the Standard Model in the form of charged-lepton flavor
violation.  It relies on the high luminosities enabled by the crab
cavities, which will be built by the Canadian group.  Further details
on the EIC plans and physics program are provided in
Section~\ref{sec:eic}.

\subsubsection{CPT, Lorentz symmetry, and the weak equivalence principle}

\paragraph{ALPHA
\label{sec:alpha}}

The ultimate goal of antihydrogen studies is to test fundamental
symmetries between matter and antimatter with the highest possible
precision. Precision tests of CPT and the Weak Equivalence Principle
(WEP) will confront the foundations of modern physics---Quantum Field
Theory and General Relativity. For the past 20 years, ALPHA and
ALPHA-Canada have been leading the field of antihydrogen physics. An
increased antiproton availability with the ELENA facility at CERN, as
well as emerging new technologies such as quantum sensing, vacuum
ultraviolet lasers, high Tc superconducting magnets and AI/Machine
Learning, are creating exciting opportunities in the next decade to
push the precision of antihydrogen measurements to a level comparable
to, or possibly beyond, that of atomic hydrogen---one of the most
precisely studied systems in all of physics.

\begin{SCfigure}
  \includegraphics[width=8cm]{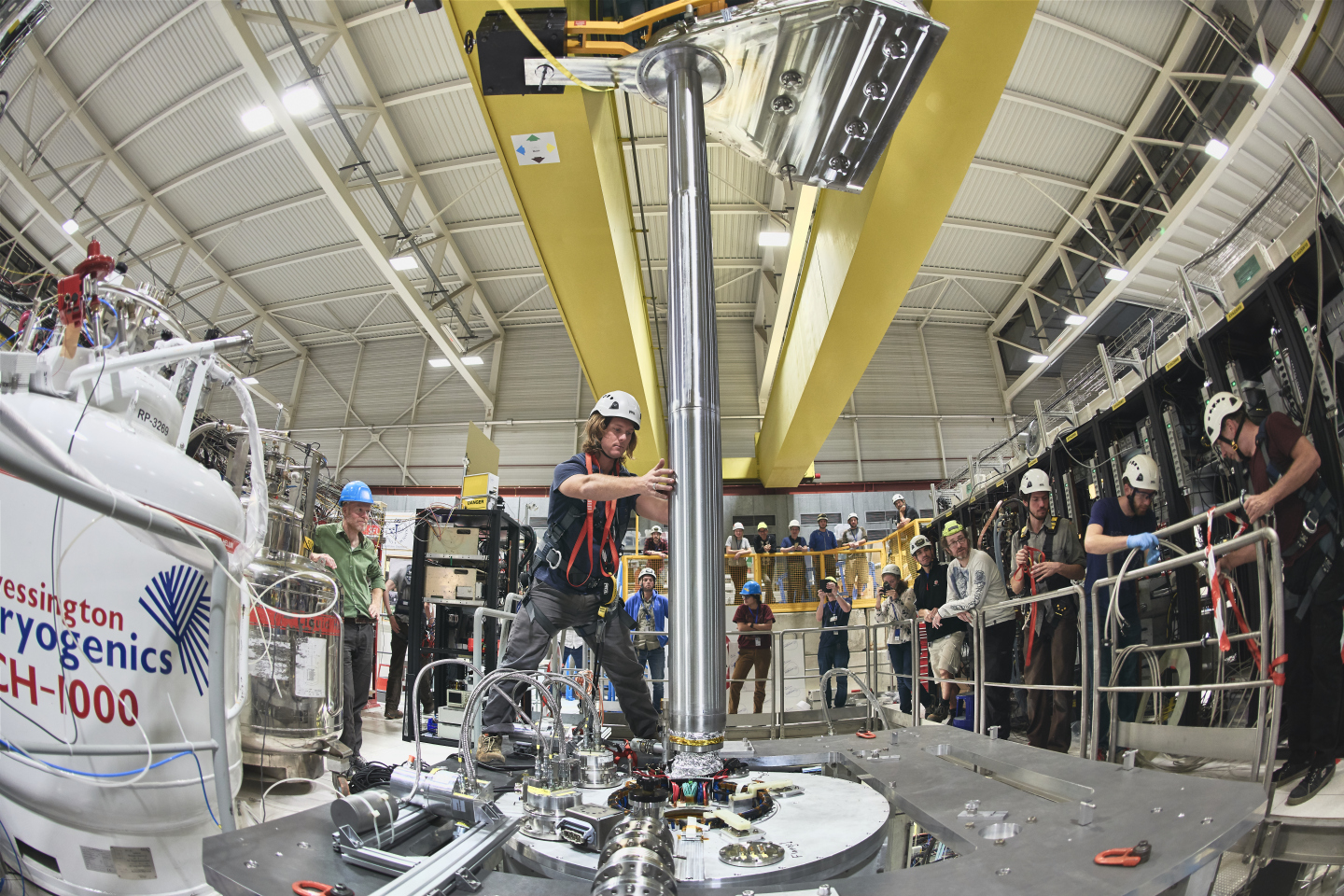}
  \caption{{\it The insertion of ALPHA-g antihydrogen trap into the ALPHA-g detector at CERN, shown with then UBC student Nathan Evetts.}}
\label{fig:alpha-g}
\end{SCfigure}

After over 20 years of development, antihydrogen physics has entered
its precision measurement era. In the past few years, ALPHA has
observed the hyperfine spectrum of antihydrogen \footnote{Nature 548, 66
  (2017)}; measured its splitting to the $10^{-5}$ level \footnote{in
  prep.\ for Nature}; demonstrated laser spectroscopy of antihydrogen
\footnote{Nature 541, 506 (2017)}, measured the 1S-2S interval \footnote{Nature 557, 71
  (2018)} and its hyperfine components \footnote{Nature Physics 21, 201 (2025)}
with $10^{-12}$ level precision; observed 1S-2P transitions in
antihydrogen \footnote{Nature 561, 211 (2018)} and probed the fine structure
splitting and Lamb shift \footnote{Nature 578, 375 (2020)}. Most recently, two
major breakthroughs have been achieved: the demonstration of
antihydrogen laser cooling \footnote{Nature 592, 35 (2021), featured on the
  cover} and the observation of the antihydrogen's gravitational free
fall \footnote{Nature 621, 716 (2023)}.

ALPHA represents arguably one of the most successful examples of
Canadian leadership in an off-shore subatomic physics experiment. The
award-winning Canadian contingent---ALPHA-Canada---constitutes more
than one-third of the full ALPHA collaboration as the large group
within ALPHA, and has led efforts in a number of areas including
particle detection techniques, microwave spectroscopy, laser cooling
and gravity physics; the group has made important contributions to
essentially all of the ALPHA physics program. They have also been
leading major CFI-funded infrastructure initiatives, ALPHA-g and
HAICU.

Building on a long track record of success, ALPHA Canada proposes a
program of fundamental physics with antihydrogen throughout the next
decade.

The short-term objective is to exploit the existing ALPHA-2 and
ALPHA-g apparatus at CERN to produce continuous output of high impact
physics results, while completing the construction of HAICU at
TRIUMF. In the medium-term (2026--2030), taking advantage of CERN's
Long Shutdown, a significant upgrade to the ALPHA apparatus is
planned---the ``ALPHA Next Generation'' project, which is currently
under review at CFI. This will enable simultaneous confinement of and
measurements on antihydrogen and hydrogen, allowing direct comparison
experiments, and alleviating key systematic uncertainties. At the same
time, novel quantum sensing techniques will be applied to both matter
and antimatter, such as atomic fountains and interferometry
techniques, with the HAICU infrastructure in Canada. In the later part
of this planning period (2030--2034), the atomic fountain technique
developed by HAICU will be deployed at CERN, and initiate an
antihydrogen quantum sensing program---breaking new ground in
antimatter physics.

\subsubsection{Precision $\beta$-decay, $V_{ud}$, and searches for violations of symmetry}
\label{sec:PrecisBetaDecayEtc}

Precision measurements of $\beta$-decay can be used to extract the
elements of the Cabibbo-Kobayashi-Maskawa (CKM) matrix, and to search
new physics via discrepancies from the standard electroweak theory.
In nuclear, neutron, and pion decay, the CKM matrix element $V_{ud}$
can be extracted, which is useful for tests of the unitarity of the
CKM matrix.

\paragraph{Superallowed $0^+\rightarrow 0^+$ nuclear decays}

Precision measurements of the $ft$-values for superallowed
$0^+\rightarrow 0^+$ Fermi $\beta$ decays between nuclear isobaric
analogue states provide demanding tests of the electroweak Standard
Model. These measurements have, for example, confirmed the conserved
vector current (CVC) hypothesis at the level of $9\times 10^{-5}$, set
the most stringent limit on a fundamental weak scalar current coupling
to left-handed neutrinos at $(0.00\pm 0.10)$\% of the vector coupling,
and, together with the Fermi coupling constant GF from muon decay,
provide the most precise determination of the $V_{ud}=G_V/G_F$ element
of the Cabibbo-Kobayashi-Maskawa (CKM) quark-mixing
matrix \footnote{J.C. Hardy and I.S. Towner, Phys. Rev. C 102, 045501
(2020)}.  Recent {\it ab initio} nuclear theory results by Canadians
have contributed to controlling errors in the interpretation arising
from nuclear effects (see Section~\ref{sec:theory}).

The Canadian subatomic physics community is uniquely positioned for
scientific impact in this field as the TRIUMF-ISAC/ARIEL facility
produces high-quality beams of many of the superallowed $\beta$
emitters with world-record intensities and hosts a suite of
state-of-the-art spectrometers capable of precision measurements of
all of the experimental quantities of interest in superallowed
decays. These include high-precision half-life measurements through
both $\beta$ counting with the $4\pi$ gas proportional counter at the
ISAC GPS facility and $\gamma$-ray counting with GRIFFIN,
high-precision branching-ratio measurements with GRIFFIN and its
auxiliary detectors, high-precision $Q$-value measurements with the
TITAN mass spectrometer, and charge-radii measurements through
collinear laser spectroscopy required as input to the
isospin-symmetry-breaking calculations.

The precision and performance of TITAN's mass spectrometry will be
boosted by charge breeding in the Electron Beam Ion Trap (EBIT). This
feature will now be fully leveraged with the newly cryogenic Penning
trap \footnote{E.M. Lykiardopoulou, PhD thesis, UBC 2023} and further enhance
by a new measurement methodology \footnote{E.M. Lykiardopoulou, HI 241, 37
  (2020)}. The technique can improve precision by a factor of five for
singly charged ions. The two upgrades will be leveraged for studies in
fundamental symmetries, in particular $Q$-values to test the unitarity
of the quark-mixing matrix, and for neutrino mass and
$0\nu\beta\beta$-decay experiments.

The Canadian gamma-ray group has repeatedly set new precision
benchmarks for half-life and branching-ratio measurements across a
wide range of superallowed Fermi $\beta$ decays from the lightest
($^{10}$C) to the heaviest ($^{74}$Rb) cases currently measured with
high precision. This will continue to be developed at ISAC/ARIEL
throughout the 2027–2034 period as new beams become
available. High-precision half-life and branching-ratio data recorded
with GRIFFIN for the $N=Z-2$ superallowed emitters $^{22}$Mg and
$^{34}$Ar are currently under analysis with the goal of
differentiating between leading models of the nuclear-structure
dependent corrections. The very high $\gamma-\gamma$ coincidence
detection efficiency of GRIFFIN will continue to revolutionize the
precise measurement of superallowed branching ratios for the $A\geq
62$ decays pioneered by the group. As new molecular beams of the heavy
superallowed emitters $^{66}$Ar and $^{70}$Br become available at
ISAC/ARIEL, experiments with GRIFFIN will provide the first precise
branching-ratio measurements for these decays, including measurements
of the weak non-analogue Fermi $\beta$ decay branches to excited $0^+$
states of the daughter nuclei that provide direct experimental tests
of the isospin-symmetry-breaking calculations. GRIFFIN will also
pursue high-precision half-life and branching-ratio measurements for
isospin $T = 1/2$ mirror $\beta$-decays such as $^{15}$O and $^{35}$
Ar, which provide an independent test of the theoretical models used
to calculate the nuclear-structure dependent corrections. A long-term
program of high-precision $\beta$-decay measurements at ISAC/ARIEL
will continue to make major contributions to constraining the
theoretical calculations needed for precision electroweak tests such
as CVC, the determination of $V_{ud}$, and the unitarity of the CKM
quark-mixing matrix.

\paragraph{TRINAT}

TRIUMF's neutral atom trap for $\beta$ decay (TRINAT) plans to
contribute to the determination of V$_{ud}$, the fundamental constant
that nuclear physics determines for the physics community.  The
experiment will also directly measure the helicity of the $\nu$ to
sub-1\% accuracy, the best direct measurements
since \footnote{Goldhaber et al. Phys Rev 109 1015 (1958)}.
Furthermore, TRINAT will search for time reversal-odd but parity-even
(and isospin-breaking) interactions between nucleons, with sensitivity
comparable to NOPTREX and hence complementary to EDM experiments.

TRINAT uses laser trapping, cooling, and spin-polarization techniques
to confine atoms of beta-decaying $^{37}$K, $^{38{\rm m}}$K, $^{38{\rm
    g}}$K, and $^{47}$K to the center of an electric field that
collects the progeny ions. From the momentum of the progeny nucleus
and the $\beta$ momentum, the $\nu$ momentum is determined.  Angular
distributions and correlations are sensitive to the relative and
absolute helicities of the leptons. Time-reversal sensitivity comes
through the scalar triple product of three vectors, a quantity which
is odd under time reversal.

Recently, TRINAT has produced results on the $\beta^-$ decay of
$^{47}$K to $^{47}$Ca, which is dominated by a Gamow-Teller decay with
an isospin-hindered Fermi contribution, measuring the $\beta-\nu$
correlation \footnote{Kootte et al. Phys. Rev. C 109 L052501 (2024)}.  TRINAT
has also measured the energy spectrum of atomic shakeoff electrons
emitted in $\beta^+$ decay \footnote{J. Funk-Frose et al., in prep.}. The
number and energy (5-10\% above 25 eV) is small enough to be ignored,
as expected, compared to Auger electrons from electron capture decay.
TRINAT has recently measured the Fierz interference term in $^{37}$K
decay, achieving complementary sensitivity to the neutron for Lorentz
scalar currents \footnote{M. Anholm thesis 2022 and in preparation}.

In the medium term (2027--2034), TRINAT's plan would be to both
upgrade their electrostatic spectrometer and complete improved
$\beta-\nu$ correlation measurements in $^{37}$K, $^{\rm 38m}$K, $^{\rm
  38g}$K, and $^{47}$K.  A larger collaboration will measure the
charge radius of $^{38m}$K.  This effort would benchmark improved
atomic theory \footnote{Katyal et al. Phys Rev A 111 042813 (2025)} to test
with these isobaric charge radii the Coulomb component of isospin
breaking needed for corrections to V$_{ud}$ \footnote{Seng and Gorchtein, Phys
  Lett B 838 137654 (2023)}.  The statistics of the results are
expected to reach the limit of systematic uncertainties at $\sim$
0.1\% in most observables.

\paragraph{BeEST and SALER}

\begin{figure}   
\centering
\includegraphics[width=0.49\textwidth]{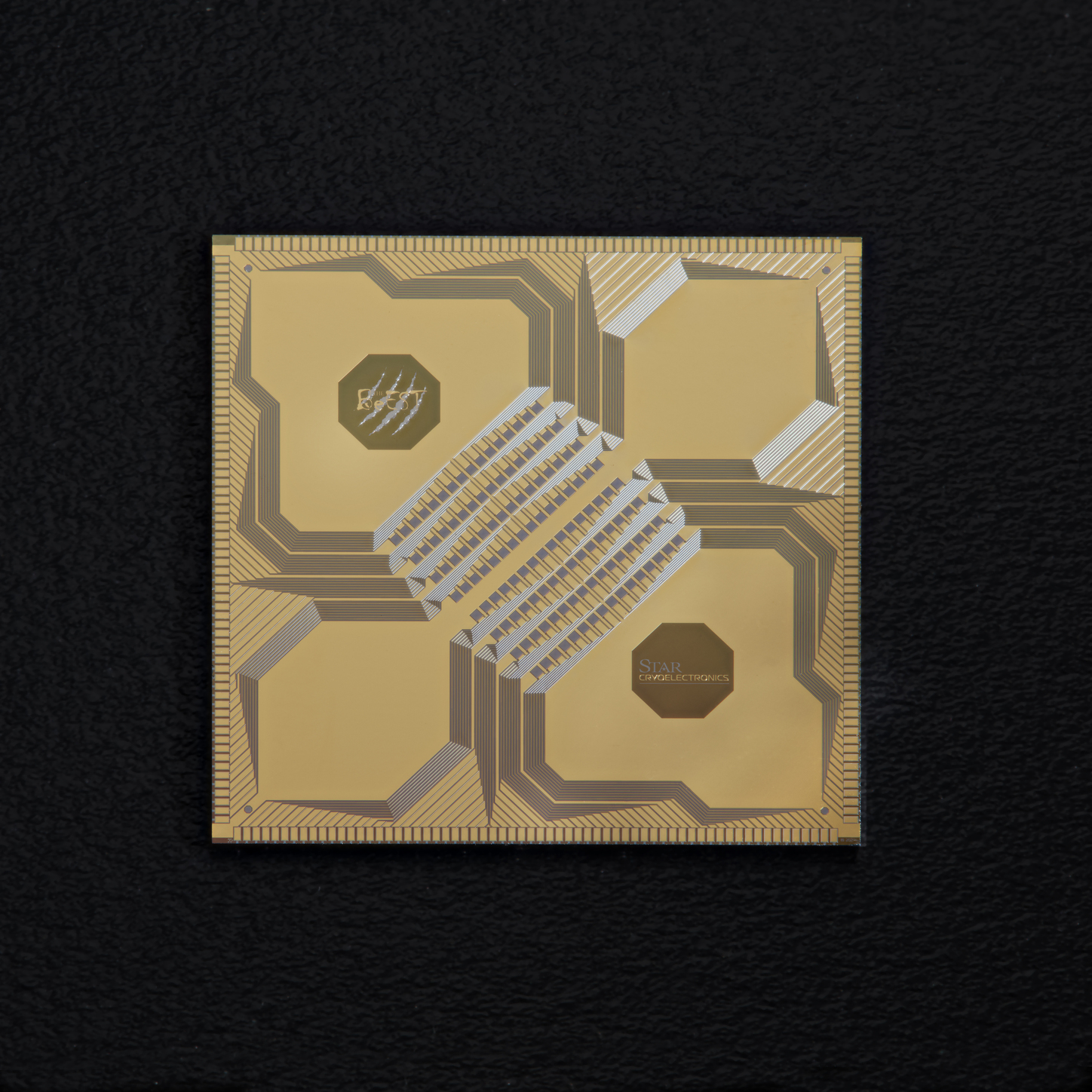}
\includegraphics[width=0.49\textwidth]{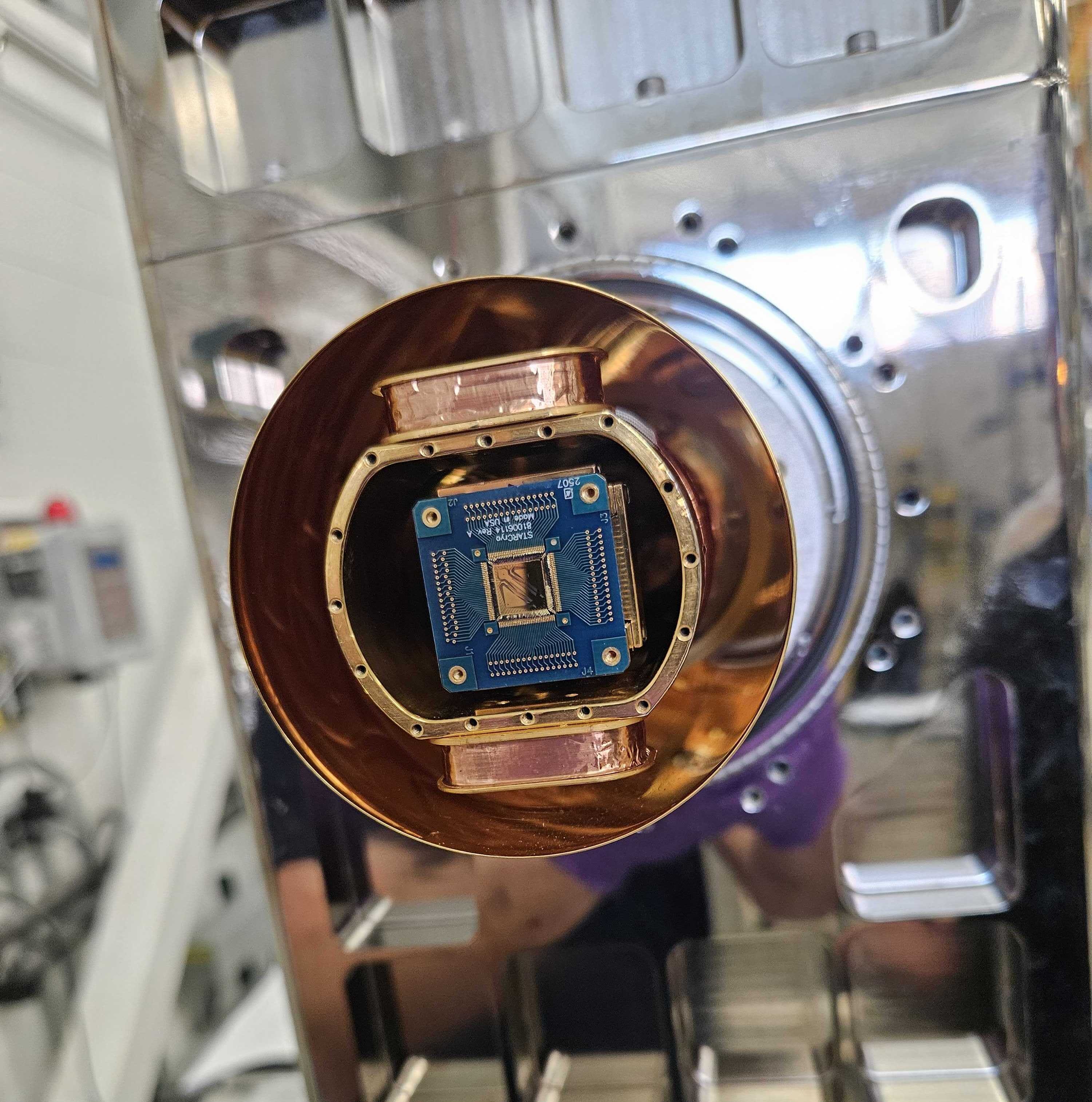}
\caption{Left: Image of 128 pixel STJ detector array used for the
  BeEST experiment.  Right: STJ array installed in SALER
  ADR. \label{fig:beest}}
\end{figure}

The BeEST (Beryllium Electron-capture in Superconducting Tunnel
junctions) and SALER (Superconducting Array for Low-Energy Radiation)
are novel precision nuclear-recoil spectroscopy experiments that use
rare-isotope-doped superconducting tunnel junction (STJ) quantum
sensors (see Fig.~\ref{fig:beest}) to precisely search for beyond the Standard Model (BSM) physics
at the eV–TeV scale. These experiments are the leading edge of a new
revolution in precision subatomic physics and have a strong Canadian
contribution (led at TRIUMF), with key components and leadership in
the US and Europe. The experiments are financially supported by
TRIUMF, NSERC, FRIB (US), the Gordon and Betty Moore Foundation (US),
the EMPIR program (European Metrology Programme for Innovation and Research), and Department of Energy (US).

The primary objective of the BeEST and SALER experiments is to search
for BSM physics in weak nuclear decay of implanted rare isotopes into
STJs. These include world-leading laboratory search for heavy neutral
leptons in the keV mass range, the only direct measurements of the
neutrino wavepacket size, exotic particle searches, applied nuclear
science, and weak nuclear structure observables.

The experiments implant radioisotopes into the active absorber of STJ
sensor arrays (Fig.~\ref{fig:beest}) and measure the eV-scale nuclear
recoil that gets a kick from the emission of a neutrino or other
exotic particles. These are model-independent searches for BSM
physics, as they only rely on energy and momentum conservation in weak
nuclear decay. Precision spectral fits are used to search for excess
or shifted peaks corresponding to beyond-Standard-Model physics.

The program's next phase will complete and publish the blinded
Phase-III BeEST dataset, which contains over 10$^9$ electron-capture
events. This analysis will set new world-leading bounds on sterile
neutrinos, wavepacket localization, and inner bremsstrahlung in weak
decay. BeEST Phase-IV will then deploy a new 128-pixel STJ array with
higher per-pixel activity and reduced background systematics. Isotopes
such as $^7$Be and $^{37}$Ar will be used to extend sensitivity to new
physics in EC decays. The array will be operated in dilution
refrigerators at PNNL, CURIE and TRIUMF, enabling high-statistics,
low-background datasets relevant for axion and dark-matter absorption
searches in the $0.1-1$~keV mass range.

In parallel, the SALER detector will be fully commissioned at
TRIUMF. After offline tests with $^{124}$Sb, its first on-line physics
run will study $^{11}$C $\beta^+$ decay to extract the $\beta-\nu$
angular correlation to 0.1\% precision. Future measurements of
$^{19}$Ne, $^{25}$Al, and other short-lived isotopes will extend this
to a global test of weak current structure.

The two experiments share software, electronics, and HQP across
institutions. This seven-year period will also see advances in
detector microfabrication, readout integration, and the development of
a national user program focused on rare-isotope recoil physics.

\paragraph{Nab}

The goal of the Nab and its polarized version pNAB are to make
precision measurements of the neutron beta decay correlations
parameters to improve the precision of $\lambda=g_a/g_v$ in neutron
$\beta$-decay. When combined with a precision measurement of the
neutron lifetime, this leads to a free neutron measurement of the CKM
matrix element $V_{ud}$, which dominates the first row unitary sum of
the CKM matrix.  Deviations from unitarity would signatures of physics
beyond the Standard Model. Addressing the apparent non-unitarity of
the quark-mixing matrix was emphasized as one of the highest
priorities for the US Fundamental Symmetries, Neutrons, and Neutrino
community from 2023 \footnote{Whitepaper for the 2023 NSAC Long Range
Plan, (2023) arXiv:2304.03451} and is prominently featured in the 2023
USA Nuclear Physics Long Range
Plan\footnote{doi:10.1080/10619127.2024.2303306}.

\begin{SCfigure}
  \includegraphics[width=4in]{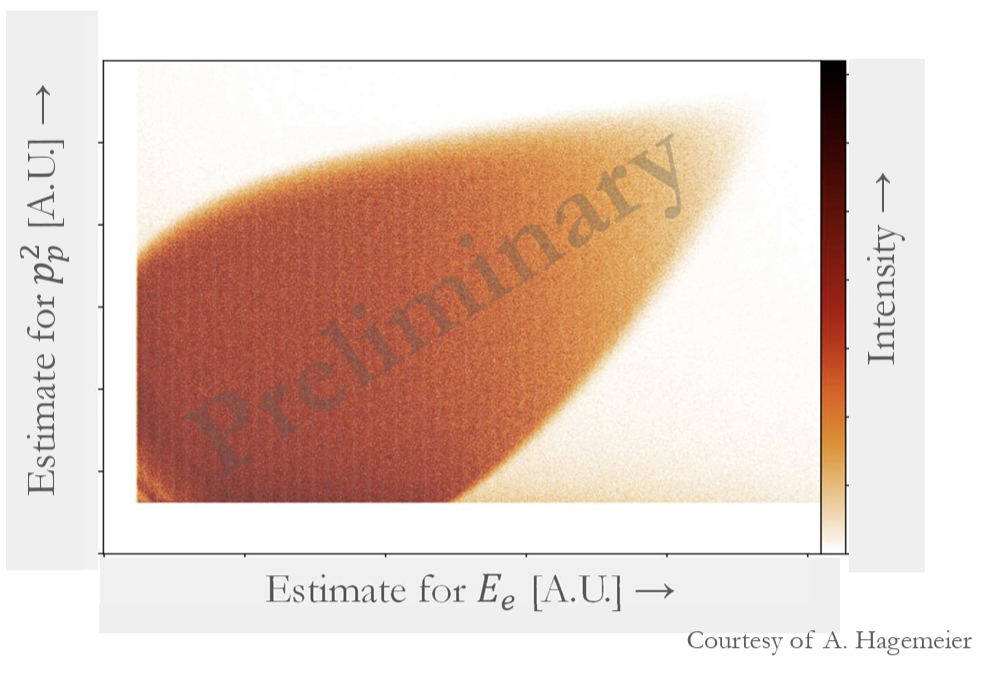}
  \caption{Proton time-of-flight vs. electron energy measured in early
    2025 using the Nab spectrometer.}
    \label{fig:teardrop}
\end{SCfigure}

The Nab apparatus measures neutron beta decay using an asymmetric 5~m
tall time-of-flight (TOF) spectrometer at the Fundamental Physics
Beamline at the Spallation Neutron Source at Oak Ridge National Lab
(US). Each end of the spectrometer employs large area, highly
segmented Si detectors to measure the proton's TOF and electron
energy. The Canadian group supports the Si detector aspects of the
experiment. The group built a 30 keV proton accelerator at the
University of Manitoba and used the apparatus to characterize Nab Si
detectors in 2021-2022. Since then the Canadian group has been
supporting the detector operations during Nab commissioning and
physics data production. The $\beta-\nu$ correlation parameter $a$ is
extracted from a 2D plot of proton TOF vs.~electron energy (see
Fig.~\ref{fig:teardrop})

By 2027, the final Nab ($a$ and $b$) precision data set will be
acquired.  In 2028, neutron polarization studies will begin in support
of introducing polarized neutrons into the Nab apparatus (pNAB) which
will enable additional correlations ($A$ and $B$) to be measured.
By 2031, pNAB data taking will be completed.

\paragraph{PENeLOPE}

The discrepancy in the measured value of neutron lifetime has
significance in big-bang nucleosynthesis and testing the accuracy of
the standard model. A discrepancy of approximately 9 s still persists
between the average value of the neutron lifetime obtained by storing
ultracold neutrons (UCN) in traps and the neutron beam method. This
has sparked discussions about a possible dark neutron decay or excited
neutron states to solve this puzzle.  Furthermore, a new precise
measurement of the neutron lifetime can help to resolve discrepancies
in the extraction of $V_{ud}$ from neutron $\beta$-decay.

\begin{SCfigure}
\includegraphics[width=3.5in, angle=0, page=1,trim=14.5cm 0cm 0cm 0cm,clip]{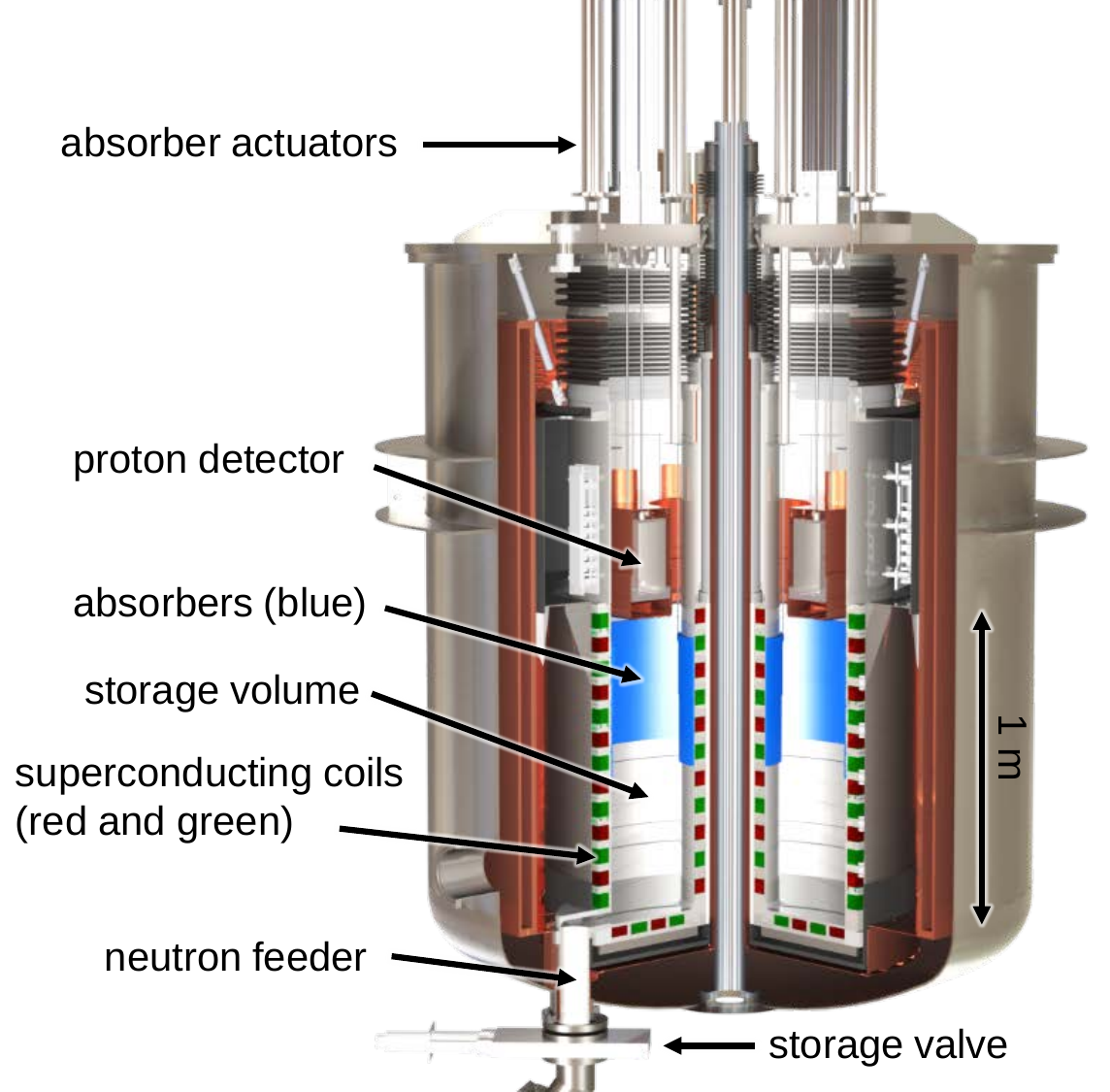}
    \caption{Cutaway view of the PENeLOPE cryostat. The solenoids are shown in red and green indicating alternating current directions. The neutron absorbers are colored blue.}
    \label{fig:rendering}
\end{SCfigure}

The goal of PENeLOPE, the Precision Experiment on the Neutron Lifetime
Operating with Proton Extraction, is to improve the precision of the
trap method to $\Delta \tau < 0.1$~s.  To this end, a large
magneto-gravitational UCN trap was built (see
Fig.~\ref{fig:rendering}).  It consists of a superconducting magnet
made of 24 short solenoids with alternating current direction creating
a region of low magnetic flux density surrounded by a region of large
magnetic flux density.  Through interactions of the neutron magnetic
moment, UCN can be stored without material interactions.  This removes
one of the main systematic effects of previous material storage
experiments: losses during wall interactions.  The top of the trap can
be left open since the gravitational potential confines the UCN's
vertically.  This allows for the extraction and online counting of
decay protons.

The design and construction of the experiment was spearheaded by the
Technical University of Munich.  The magnetic trap was shipped to
TRIUMF in 2024.  Cryogenic commissioning is planned for 2025 and 2026.
After the long TRIUMF shutdown, the experiment can be moved to the
second port of the TUCAN (TRIUMF UltraCold Advanced Neutron) source in
2027.  A statistical precision of 0.1~s (1-$\sigma$) can be reached
within a few months at the TUCAN source.  Systematic studies and
statistical data taking runs can happen towards the end of the decade
approaching 0.1~s statistically.  In the years 2030 to 2032, a proton
detector will be implemented.  This would provide a measurement with
different systematic effects compared to the neutron counting
experiment, allowing a firm 0.1~s measurement towards the end of the
LRP period.

\paragraph{PIONEER}

This is an international effort that has been approved in 2022 with
highest priority at the Paul Scherrer institute in Switzerland (PSI),
host of the world’s most intense beams of pions. Because it addresses
key puzzles of modern subatomic physics through measurements in both
the lepton and quark sectors with unprecedented
precision\footnote{W. Altmannshofer et al., arXiv:2203.01981, (2022)},
PIONEER’s scientific reach and discovery potential is unique and
highly complementary to other leading subatomic physics experiments
worldwide. PIONEER detector construction and the beginning of physics
data taking are anticipated in the period 2028 to 2034.

The primary goal of the experiment is to achieve an order of magnitude
more precise measurement of the branching ratio of pion decays to
electrons versus muons, $R^\pi_{e/\mu}$.  According to the Standard
Model (SM), the principle of lepton universality -- asserting that all
leptons interact identically under the weak force -- leads to an
exceptionally precise prediction for this ratio, at the level of
0.01\%. Currently, the most accurate measurement (0.2\%) comes from
the PIENU experiment at TRIUMF\footnote{A. Aguilar-Arevalo et al.,
  Phys. Rev. Lett. 115, 071801 (2015)}. The theoretical prediction
is however a factor of 15 more precise, leaving a substantial window
in which new physics could appear. Potential new physics contributions to
$R^\pi_{e/\mu}$ span a broad range, from light (MeV-scale) heavy
neutral leptons\footnote{D. A. Bryman and R. Shrock, Phys. Rev. D 100, 073011 (2019)} or dark sector particles, to very heavy (up to $O$(1000 TeV))
scalar or pseudoscalar mediators\footnote{D. Bryman {\it et al}.,
  Ann. Rev. Nucl. Part. Sci. 61, 331–354 (2011)}.

While PIENU has provided the most stringent test of lepton
universality further improvements are strongly motivated. This is
particularly true in light of recent anomalies observed in B-meson and
$\tau$-lepton decays, as well as in the muon $g-2$ measurement. In a
subsequent phase, PIONEER will target a precise measurement of the
ultra-rare pion beta decay process, $\pi^+\rightarrow\pi^0e^+\nu_e$,
which occurs in only one out of every 100 million pion decays, and is
also precisely predicted by the SM. This suppressed process allows
probing recently observed but unconfirmed anomalies in the fundamental
parameters which describe the mixing among quarks and allows a
theoretically pristine and independent measurement of $|V_{ud}|$ free
from the nuclear structure uncertainties that affect superallowed
nuclear beta decay measurements.

\subsubsection{Neutrinoless double beta-decay}

Neutrinoless double beta-decay is a lepton-number violating process
which if observed would establish the Dirac vs.~Majorana nature of
neutrinos, and the absolute mass scale of neutrinos.

\newcommand{\0}{\ensuremath{0\nu\beta\beta}}
\newcommand{\2}{\ensuremath{2\nu\beta\beta}}
\newcommand{\bb}{\beta\beta}
\newcommand{\Q}{\ensuremath{Q_{\beta\beta}}\xspace}
\newcommand{\mbb}{\ensuremath{\langle m_{\bb} \rangle}}
\newcommand{\T}{\ensuremath{T^{0\nu}_{1/2}}}
\newcommand{\isotope}[2]{$^{#2}$#1}

\paragraph{nEXO}

The nEXO project aims to search for neutrinoless double beta decay
(\0) in \isotope{Xe}{136}, probing the quantum nature of neutrinos and
testing the half-life sensitivity beyond $10^{28}$ years.  The US DOE
has chosen to move forward in the near term with an alternate \0
experiment, giving Canada an opportunity to lead a Xenon-based \0
project at SNOLAB.  The nEXO concept is well-developed, representing a
next-generation effort in rare-event physics and complementing
international efforts in the search for lepton number violation.  The
Canadian group has taken a leadership role in growing the
collaboration, which might involve a renaming process. In addition to
\0, nEXO will be able to study solar neutrinos, supernova neutrinos,
and models of fermionic dark matter interactions on xenon.

\begin{figure}[h]
  \centering
  \includegraphics[width=\linewidth]{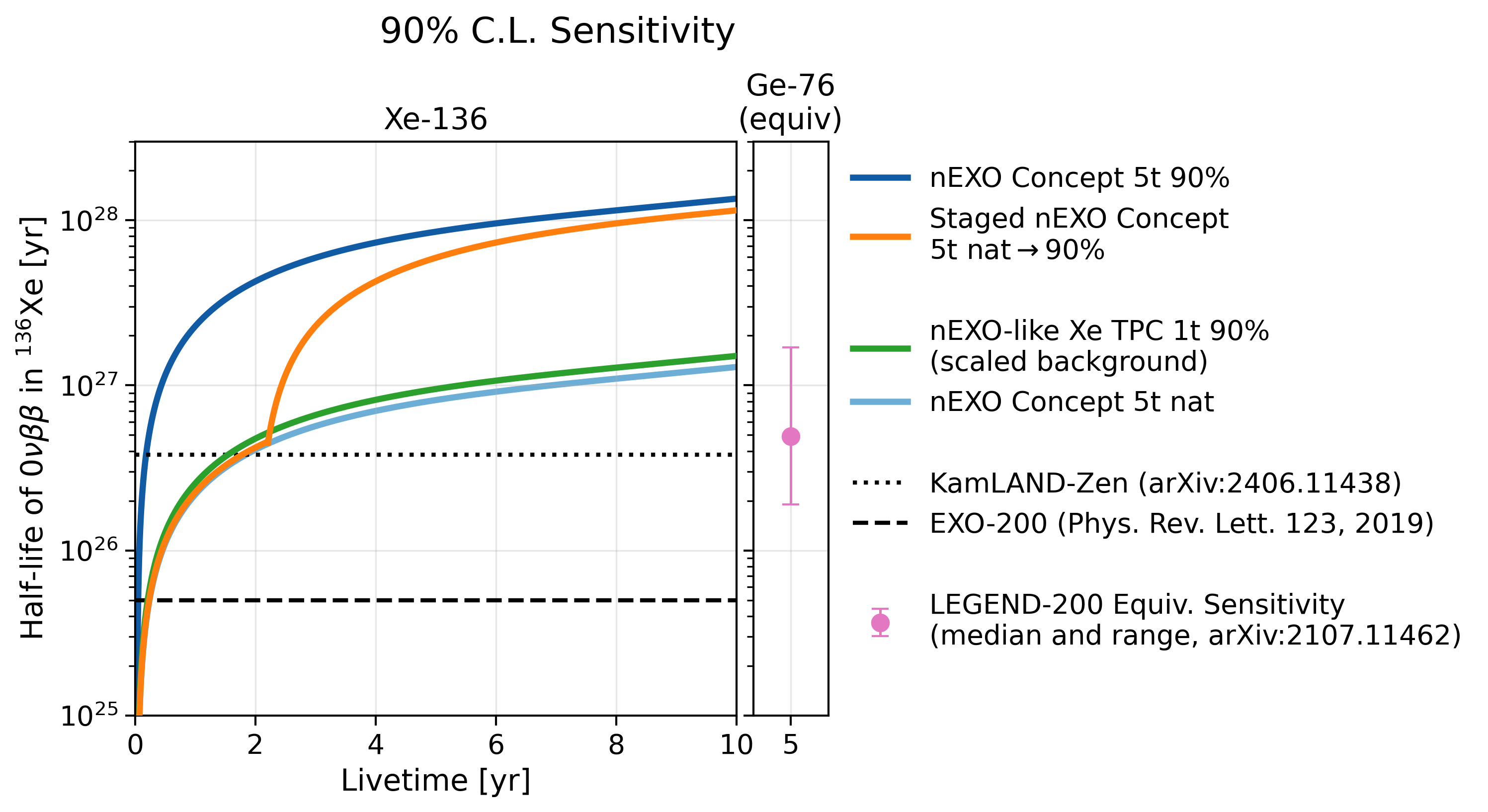}
  \caption{Projected half-life sensitivity of nEXO, with other
    detector concepts and experiments for comparison.  Livetime refers
    to the time of operation of the experiment, in years.}
  \label{fig:Sensitivity}
\end{figure}

The nEXO concept will use a 5-tonne, homogeneous, single-phase
Time-Projection Chamber (TPC) filled with Xenon to detect the
signature of \0 with excellent energy resolution and background
rejection.  The experiment utilizes low-background materials,
topological event reconstruction, and efficient
scintillation/ionization signal collection to reach its analysis
goals.  The projected sensitivity of nEXO is displayed in
Fig.~\ref{fig:Sensitivity}.  An engineering drawing of the proposed
layout of nEXO in the cryopit at SNOLAB is displayed in
Fig~\ref{fig:nexo} in Section~\ref{sec:snolab}.

nEXO is currently building a more global collaboration with additional
stakeholders in the experimental project.  Construction and
integration are expected to take place between 2027 and 2033, with
commissioning beginning shortly thereafter. Initial physics data could
be acquired by 2034, placing nEXO at the forefront of global \0
sensitivity. Concurrently, a phased approach will be evaluated,
allowing for staged enrichment of xenon and progressive sensitivity
increases. The flexibility to start with lower enrichment and upgrade
to 90\% over time provides both scientific impact and risk mitigation.

\paragraph {Other neutrinoless double beta-decay efforts} In addition to nEXO, the landscape of neutrinoless double beta-decay efforts in Canada includes SNO+, XLZD, and Theia. SNO+ is the successor to the Nobel Prize-winning Sudbury Neutrino Observatory (SNO) experiment, currently operating at SNOLAB with 780 tonnes of liquid scintillator. The SNO+ physics program is focused primarily on neutrino measurements, including solar neutrinos, geoneutrinos, reactor antineutrinos, and supernova neutrinos. In 2026, the detector will be loaded with 1.3 tonnes of Te-130, and the experiment will focus on the search for neutrinoless double beta decay. They plan to increase the amount of Te-130 to 4 tonnes by the end of the decade and, subsequently, operate for at least more five years. The collaboration aims to achieve a half-life sensitivity of better than $10^{27}$ years for the \0 decay of Te-130. 

XLZD is the large-scale successor to the XENON and LZ programs. Using a dual-phase time projection chamber (TPC) filled with 60–80 tonnes of liquid xenon, XLZD has a broad physics scope including the search for neutrinoless double beta decay of Xe-136, dark matter, solar axions, and supernova neutrinos. Canada is not yet formally part of the XLZD collaboration, but SNOLAB is one of four shortlisted sites to host the observatory. The observatory will operate with ultra-low backgrounds and high sensitivity, achieving discovery potential for \0 with half-life sensitivity approaching $10^{28}$ years. The project aims to begin construction by 2028, with commissioning and science operations expected to start in the mid-2030s. 

Theia is a proposed hybrid neutrino detector targeting a new, large cavern at SNOLAB, included as part of SNOLAB’s 15-year Facility and Operations plan. Theia combines the directional sensitivity of Cherenkov light with the calorimetric precision of scintillation light. This dual capability enables a uniquely broad and interdisciplinary physics program, including solar neutrinos, geoneutrinos, supernova neutrinos, and neutrinoless double-beta decay. Over the next few years, Theia aims to progress from conceptual to technical design, culminating in a detector on the scale of 25–100 kilotons, depending on cavern availability at SNOLAB. The full breadth of the Theia physics program, including a comparison of the results achieved in 25-kton and 100-kton caverns, is explored in the Theia white paper \footnote{Eur. Phys. Jour. C 80, 416 (2020)}.  Theia permits loading at 10+ ton scale for sensitivity into the range of the normal ordering, with half-life sensitivities well beyond $10^{28}$ years.

\paragraph{Nuclear structure for $0\nu\beta\beta$}

It has also been shown that the nuclear shape can have a direct impact
on the nuclear matrix elements (NMEs) for \0
emitters \footnote{Vaquero, Rodr\'iguez, and Egido, Phys.
Rev. Lett. 111, 142501 (2013)}. If the \0 decay mode were
experimentally observed, it would provide positive proof of the
Majorana nature of the neutrino, and if the NMEs are accurately known,
the rate of the process can used to extract the neutrino effective
mass. In order to build confidence in the values of the NMEs, the
underlying nuclear structure calculations are tested in a variety of
experiments, including Coulomb excitation. A future program will focus
on performing such measurements on the $^{136}$Xe parent and
$^{136}$Ba daughter, beginning with experiments at the HIL (Heavy Ion Laboratory, Warsaw) and
expanding to other facilities (Legnaro National Laboratories and Argonne National Laboratory) using heavier binary
partners.

As a part of a suite of offshore nuclear physics experiments,
P.~Garrett and C.~Andreoiu and a host of international
collaborators are pursuing studies of the intermediate odd-odd nuclei
involved in the \0 decay that test the nuclear structure models and
hence impact calculations of the nuclear matrix elements \footnote{Rebeiro et
  al., Phys. Rev. Lett. 131, 052501 (2023)}. These measurements are
extremely challenging involving long measurement times, and in some
cases the fabrication of radioactive targets.  They have used the
$(d,\alpha)$ transfer reaction to study $^{136}$Cs \footnote{Rebeiro et al., Phys. Rev.  Lett. 131, 052501
  (2023)}, and have established a collaboration to study $^{136}$Cs
via the $^{135}$Cs$(n,\gamma)$ reaction where the first step is the
production of a radioactive $^{135}$Cs target. The nucleus $^{136}$Cs
is particularly interesting since it is the product of charged-current
$\nu_e$ capture reactions on $^{136}$Xe, and thus xenon-based
detectors offer promising avenues to study the low-energy solar
neutrino flux \footnote{Raghavan, Phys. Rev. Lett. 78, 3618 (1997)}, and was
also proposed as a method for dark-matter detection \footnote{Haselschwardt {\it et
  al}., Phys. Rev. Lett. 131, 052502 (2023)}. Complementary experiments
will be required using $\gamma$-ray and conversion electron
spectroscopy following light-ion fusion evaporation reactions, and
also nucleon transfer reactions performed at, for example, the
University of Cologne, FSU or ANU.

Canadian theorists in {\it ab initio} nuclear theory are also
contributing to nuclear matrix element determination for the \0 decay
process (see Section~\ref{sec:theory}.

\subsection{Beyond the next 7 years}

\begin{itemize}
\item TUCAN

For TUCAN, the long-term plans (2034-2041) involve continued
exploitation of the UCN source for other physics experiments
(e.g. neutron gravitational levels experiments, or more precise
neutron lifetime and nEDM experiments).  The medium term will be used
to determine whether further upgrades to the UCN source at TRIUMF
could be useful, or whether the technology could be scaled up for an
even more intense UCN source elsewhere.

\item RadMol

In the long term and once its research infrastructure has been built,
the RadMol collaboration envisions to expand its horizon beyond EDM
searches. For example, radioactive molecules represent an opportunity
to access nuclear anapole moments of short-lived radionuclides. As
nuclear anapole moments violate parity, they provide a unique view on
the interplay between nuclear structure and fundamental symmetries at
play inside the atomic nucleus. Molecules are unique environments
where the nuclear anapole moments can interact with the surrounding
electron cloud, leading to amplified and thus detectable
parity-violating effects at the molecular level \footnote{J. Karthein et al.,
  Phys. Rev. Lett. 133, 033003 (2024)}.

Molecular spectroscopy of cold radioactive molecules will become
another cornerstone of the research program. This is motivated by
various scientific goals, including the importance of their
understanding for quantum chemistry or nuclear medicine. Moreover,
radioactive molecules are becoming increasingly important for
astrophysics \footnote{T. Kaminski et al. Nature Astronomy 2, 778
(2018)} where they provide crucial insights into stellar evolution. An
ambitious spectroscopy program based on (ultra-)cold radioactive
molecules is envisioned.

\item EDM$^3$

For EDM$^3$, continued improvement in purity and lowering of the
temperature of the solid, along with extensive tests for systematic
effects will lead to planned measurements of increasing accuracy over
the period from 2030 to 2040, with a final projected accuracy that
improves on the current accuracy of eEDM measurements by two to four
orders of magnitude.

\item ALPHA

The ALPHA/HAICU program will continue well into the period beyond
2034. The full exploitation of the anti-atomic fountain technique at
CERN will likely require on the order of 10 years of development,
followed by the continual improvements. In parallel, with the HAICU
facility at TRIUMF, ambitious R\&D for new techniques will be pursued,
including optical trapping of antihydrogen, and the production of
first antimatter molecules. These advances could provide extremely
sensitive tests of CPT and Weak Equivalence Principle.

Antihydrogen research is a technology-driven field; it is the
development of new technology which allows progress on the
science. The ALPHA-Canada group has achieved multiple innovations in
antimatter handling techniques. The development of emerging
technologies is anticipated in quantum sensing with novel laser and
microwave techniques, high temperature superconducting magnets for
novel traps, and AI/Machine learning.

%
%

\item BeEST/SALER

In the long term, BeEST and SALER will evolve into a general-purpose
Canadian facility for precision nuclear recoil spectroscopy using
quantum sensors. The long-term vision includes scaling to kilopixel
STJ arrays with high implantation doses and sub-eV resolution, capable
of endpoint studies in ultra-low-$Q$ decays such as $^{163}$Ho and
$^{159}$Dy. These measurements offer a new route to direct neutrino
mass determination and provide complementary data to large-scale
endpoint experiments like KATRIN (Karlsruhe Tritium Neutrino Experiment) and Project 8.

Rare-isotope-doped superconducting sensors will also be used to study
axion-like particle emission in de-excitations, bosonic dark matter
absorption, and nuclear structure effects such as shake-off,
recoil–ion coincidence, and neutron-emission tails. These diverse
observables will be supported through a flexible user program at the
CURIE (Colorado Underground Research Institute) underground lab and through TRIUMF beam access.

The SALER platform will become a mobile precision detector for
deployment at RIB facilities worldwide. With 1,000+ pixel arrays and
high beam compatibility, it will enable high-statistics $\beta$-recoil
correlation measurements across the nuclear chart, and provide new
input for CKM unitarity tests, nuclear astrophysics, and exotic weak
current searches.

Together, the BeEST–SALER program will establish Canada as a world
leader in quantum-enabled nuclear physics. By integrating detector
R\&D, rare-isotope science, and quantum technologies, the program
supports the LRP’s goals of training HQP, developing next-generation
measurement capabilities, and delivering transformative science on a
compact, accessible platform.

Taken together, these goals for BeEST and SALER position Canada at the
forefront of precision weak-interaction physics in one of the newest
experimental areas and represents a flagship table-top particle
physics experimental program. This field of research demonstrates a
compelling path for sustained SAP investment into the 2040s.
  

\item PENeLOPE

The final reach of PENeLOPE will depend on the performance of the
TUCAN source and the magnitude of systematic effects.  If
statistically viable, different methods of feeding neutrons into
PENeLOPE could be explored.  1) A neutron elevator, as implemented in
UCNtau, would allow filling neutrons in while the main magnet is on.
2) A spin flipper inside the trap as in tauSPECT would turn high-field
seeking neutrons into low-field seeking, storable neutrons.  This
could be implemented in the long-term to improve upon the precision of
0.1~s.

\item PIONEER

Between 2035 and 2040, the final data-taking for measurements of the
branching ratio $R^\pi_{e/\mu}$ and searches for sterile neutrinos and
other exotic particles will be completed, concluding Phase I of the
PIONEER experiment.

In parallel, preparations for Phase II, which will operate at higher
pion beam intensities, will begin. This next phase will likely require
detector upgrades to handle increased event rates. Relevant R\&D --
such as the development of optical segmentation in liquid xenon to
mitigate pile-up effects -- is anticipated to develop over the next
five years.

This effort will be pursued in part through the DRD2 collaboration (Detector R\&D on liquid detectors, organized by CERN), in
which Canadian PIONEER PIs are involved. The R\&D effort is expected to
continue in close synergy with other leading noble liquid experiments,
leveraging shared technologies and expertise. The pion beta decay
experiment is expected to run 2040-2045.
  
\item nEXO

The nEXO collaboration is investigating a phased approach to start the
experiment using a 5-tonne TPC filled with natural xenon, and
subsequently increasing the level of the enrichment over time.  nEXO
will perform a decade-scale science run aimed at reaching its
sensitivity goal of \T$>1.35\times10^{28}$ years, which probes the
entire inverted neutrino mass hierarchy and reaches to the normal mass
hierarchy phase space.  While taking data, a detector concept with
increased target isotope will be developed, with the goal to explore
the normal hierarchy phase space towards $m_{\beta\beta}=1$~meV.
  
\end{itemize}

\subsection{Summary}

The Canadian program in fundamental symmetries covers a variety of
different research pathways all aiming to discover new physics beyond
the Standard Model.  Experiments on EDMs are searching for
time-reversal and CP violation and an explanation of the
matter/antimatter imbalance in the universe.  Measurements of
antihydrogen are searching for CPT violation and violation of the weak
equivalence principle.  Precision measurements of $\beta$-decay aim to
extract a precise value of $V_{ud}$, and seek violations of symmetry,
with the goal of finding a violation of the electroweak Standard
Model.  Measurements of parity violation at low energies aim to
extract a precise value of $\theta_W$, with the goal of finding a
deviation from the expected running with energy scale, which would
signify new physics.  A measurement of the lepton number violating
process \0-decay aims to establish the Dirac vs.~Majorana nature of
neutrinos and constrain the absolute scale of neutrino masses.

\section{Nuclear Theory
\label{sec:theory}}

Atomic nuclei are endlessly fascinating, 
being intimately connected to some of the most profound questions in science, such as the nature of neutrinos and dark
matter, the role of fundamental symmetries, the inner workings of neutron stars, and the nucleosynthesis pathways that lead to the observed abundances of elements. Striving to answer the most
fundamental questions in our Universe, theorists work on projects ranging from developing a predictive \textit{ab initio} theory of nuclear structure and nuclear reactions to phenomenological approaches
guided by empirical data in close collaboration with experiment, and on everything in between.
Progress in nuclear theory, both from a fundamental point of view, and in its connection with
experimental measurements, has therefore to proceed on several fronts at once, and it is thus
imperative to maintain a vibrant and diverse theoretical program.

At
high energies most QCD calculations are amenable to perturbative techniques. However, a major challenge in nuclear theory results
from the fact that at the
energy scale where the details of nuclear structure are relevant (recall Fig.~\ref{fig:bigquestions_scales}) QCD is non-perturbative. Lattice
techniques, although greatly improved in recent years, still cannot solve all problems in hadronic
physics. However, there has been remarkable recent progress in effective theories which preserve
the important symmetries of the underlying fundamental theory, and yet are applicable in a given
energy interval. However, for a large class of problems, phenomenological approaches to methods
and modeling are still required.

\subsection{The Canadian program (current and next 7 years)}

Here we briefly summarize recent and near-term theoretical work
done in Canada touching upon the big questions
introduced in Chapter~\ref{sec:big_questions}.

\subsubsection{Hadronic physics and QCD}
\label{sec:nuctheo_qcd}

Collisions of nuclei at relativistic energies are exploring the strong force, described by QCD, in a hot and highly excited state: the quark-gluon plasma (QGP). The latter has previously existed only during the first few microseconds after the Big Bang. Several experimental facilities around the world have a dedicated program studying this state of nuclear matter: RHIC at BNL, FAIR at GSI, and the LHC at CERN, are a few notable examples. To better understand the many-body dynamics involved in the evolution of nuclear matter created in relativistic heavy-ion collisions, both theoretical and phenomenological research is being carried out at URegina (G. Vujanovic) and McGillU (C. Gale and S. Jeon).

\begin{SCfigure}
\centering
     \includegraphics[width=0.7\textwidth]{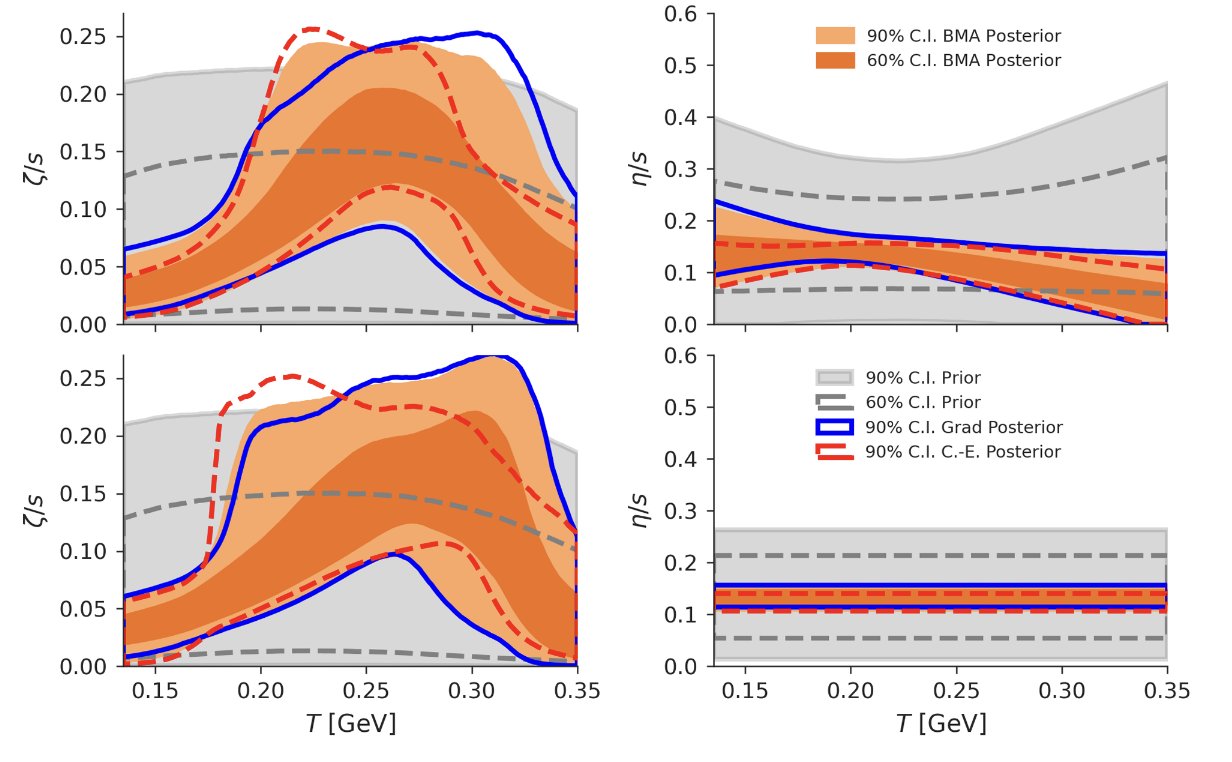}
\caption{Bayes model averaged viscous posterior with shear viscosity allowed to vary with temperature (top panels) and the Bayes
model averaged viscous posterior when the shear viscosity is constant with temperature (bottom panels). Individual 90\% credible
intervals for the Grad and Chapman-Enskog viscous corrections that underlie the averaging are shown in blue and red, respectively. Source: M. Heffernan \textit{et al}., Phys. Rev. Lett. 132, 252301 (2024).
} \label{fig:nuctheo_qcd}
\end{SCfigure}

When describing the dynamical evolution of the QGP at vanishing net chemical potential, sophisticated multistage simulations are now used, wherein relativistic dissipative hydrodynamics plays a crucial role. The formalism that underlies hydrodynamics, as well as its extension to magnetohydrodynamics,\footnote{D. Ye, S. Jeon, and C. Gale, Phys. Rev. C 110, 024907 (2024); L. Tinti \textit{et al}., Phys. Rev. D 99, 016009 (2019).} has received theoretical attention to better elucidate the physical mechanisms that underpin this description. Some of the most notable transport coefficients of the QGP, i.e., its viscosities, have been constrained phenomenologically in a more reliable, and systematically improvable, manner than before; see also Fig.~\ref{fig:nuctheo_qcd}.\footnote{D. Everett \textit{et al}., Phys. Rev. Lett. 126, 242301 (2021); M. Heffernan \textit{et al}., Phys. Rev. Lett. 132, 252301 (2024).} 

Tomographic probes, such as collimated sprays of high-energy particles known as QCD jets together with electromagnetic radiation, both real and virtual, have also been used to gain further insight into the dynamics of the QCD medium. For longer-lived (lower virtuality) partons, jet-medium interactions have been considered at leading and next-to-leading order using the Hard Thermal Loop (HTL) and Landau-Pomeranchuck-Migdal (LPM) resummation, including non-perturbative effects\footnote{R. Modarresi-Yazdi \textit{et al}., Phys. Rev. C 111, 054904 (2025).}. Furthermore, studying jets at the upcoming Electron-Ion Collider (EIC), will probe the physics of gluon saturation in cold nuclear matter. Electromagnetic probes produced from heavy-ion collisions provide a unique window into jet-medium interactions and QGP dynamics, owing to the weaker electromagnetic coupling compared to QCD coupling. The HT formalism has been used to compute photon production from highly virtual quarks,\footnote{A. Kumar and G. Vujanovic, arXiv:2502.02667.} which makes photons an additional probe in a comprehensive study of jet quenching. The next generation
analysis of heavy-ion data from RHIC and LHC will yield a more reliable understanding of the hot QGP.

The next phase of exploration will focus on better understanding the dynamical evolution of flavors in nuclear matter. From a gluon-dominated initial state, how do quark flavors reach a hydrodynamical description? How does this evolution change if the initial state already has non-zero flavor? Studying the hydrodynamical regime itself will also explore the phase diagram of the QCD Equation Of State (EOS), its critical exponents near phase transition(s), along with its transport coefficients, at non-zero net chemical potential and temperature. Understanding the EOS and its transport coefficients also requires further calculations using QCD-based effective Lagrangians of in-medium hadron spectral functions\footnote{G. Vujanovic \textit{et al}., Phys. Rev. C 101, 044904 (2020).} and hadron properties at finite chemical potential, with direct applications to nuclear matter in neutron stars.
Tomographic probes will play a central role in highlighting how quarks and gluons approach hydrodynamical behavior and how they evolve once the hydrodynamic regime is reached. To thoroughly study these dynamics within the coming years, theoretical approaches used to describe the high- and low-virtuality parton interaction with the QGP (i.e. higher-twist and HTL-LPM formalisms, respectively) are to be generalized to interpolate between the high- and low-virtuality regimes and more reliably explain medium interactions of intermediate-virtuality partons.

YorkU's R. Lewis has been studying
exotic hadrons (such as tetraquarks) that each contain a pair of heavy quarks surrounded by lighter quarks and gluons,
dozens of which have been in several experiments
recently. A complete understanding of these exotics 
will require creative new applications of lattice gauge theory.\footnote{R. J. Hudspith \textit{et al}., Phys. Rev. D 102, 114506 (2020).} 
Although lattice gauge theory has become indispensable for modern subatomic physics, important scenarios 
(such as particle collisions as a function of time, and nuclear matter at high densities) 
cannot be handled by the standard lattice
approach. Use of the Hamiltonian formalism on quantum computers could make these scenarios accessible to lattice methods.\footnote{A. T. Than \textit{et al}., arXiv:2501.00579.}

The Atlantic theory effort (R. Sandapen
and M. Ahmady) focuses on confinement in hadrons, guided by the holographic principle, thereby
providing a link between formal theory (like Perimeter) and experimental groups (like the Canadian-JLab group). 
Projects carried out include
an analysis of
the low energy pion data, revealing hints of holographic duality.\footnote{J. R. Forshaw and R. Sandapen, Phys. Rev. D 111 (2025) 3, 034024.}
While the confinement dynamics in light mesons is constrained by chiral symmetry, that
in heavy mesons is constrained by heavy quark symmetry; the kaon is somewhat on the
boundary between light and heavy mesons, so a similar
study promises to be enlightening. A further
avenue is to 
indirectly search for Vector-like quark via their contributions to rare B meson decays, where there
exist tensions between the SM predictions and experimental measurements.

\subsubsection{Nuclear structure}
\label{sec:nuctheo_struc}

Previous-generation theory calculations used nuclear interactions that were purely phenomenological:
they reproduced nucleon-nucleon scattering experimental data
very accurately, but contained several \textit{ad hoc} terms. 
More recently, the nuclear-physics community has embraced chiral Effective Field Theory (EFT)
interactions, in an attempt to connect with the symmetries of the underlying
fundamental theory (QCD).
 Chiral EFT interactions employ
 a separation of scales, that between pion and vector meson masses, and
 attempt to systematically expand in a small parameter that is the ratio
 between the two. This provides a hierarchy of forces controlled by the power of the expansion.
Deriving and fitting nuclear forces to experiment is only the first step in 
describing nuclei or nucleonic matter: such forces need to then be used in 
the framework of a quantum many-body approach. Theoretical efforts
on nuclear structure
in Canada have produced cutting edge results using \textit{ab initio}
many-body techniques (No-Core Shell Model,  
In-Medium Similarity Renormalization Group, Quantum Monte Carlo) as well as more phenomenological
approaches (like Hartree-Fock-Bogoliubov theory).

Taking these in turn, the no-core shell model (NCSM),\footnote{B. R. Barrett, P. Navratil, and J. P. Vary, Prog. Part. Nucl. Phys. 69, 131 (2013)} employed by TRIUMF's P. Navratil, builds upon an \textit{ab initio} configuration-interaction expansion of the trial wave function in the square-integrable harmonic-oscillator basis. 
To describe unbound states and nuclear reactions, the NCSM has been extended to include continuum effects, in the no-core shell model with continuum (NCSMC)\footnote{S. Baroni, P. Navratil, and S. Quaglioni, Phys. Rev. Lett. 110, 022505 (2013); Phys. Rev. C 87, 034326 (2013), P. Navratil \textit{et al}., Physica Scripta 91, 053002 (2016)}. 
Medium-term goals of this program include investigations of reactions important for astrophysics and future fusion energy generation, including radiative capture reactions, the charge exchange and transfer reactions involving $\alpha$ particles. The plan is also to perform proton capture calculations with the electron-positron internal pair conversion relevant for the recent claims of the so-called X17 anomaly (which motivates the TRIUMF ARIEL DarkLight experiment).
Description of alpha clustering is key for understanding the above processes and structure of atomic nuclei in general.\footnote{K. Kravvaris \textit{et al}., Phys. Lett. B 856 (2024) 138930.}

TRIUMF's J. D. Holt employs the valence-space formulation of the in-medium similarity renormalization group (VS-IMSRG), which extends the scope of \textit{ab initio} theory to that of the nuclear shell model, and now beyond. Recent successes 
include the calculation of the proton and neutron drip lines,\footnote{S. R. Stroberg \textit{et al}., Phys. Rev. Lett. 126, 022501 (2021).}
connecting the neutron skin of $^{208}$Pb to nuclear forces,\footnote{B. Hu \textit{et al}., Nature Physics 18, 1196 (2022).}
and uncertainty quantification of neutrinoless double-beta decay.\footnote{A. Belley \textit{et al}., arXiv:2307.15156.}
In the near-term future, the goal is to explore the superheavy region and search for a potential island of stability. The structure of exotic nuclei will also be investigated, extending our predictions for the nuclear driplines and evolution of magic numbers to the heavy region of nuclei. 

\begin{SCfigure}
\centering
     \includegraphics[width=0.5\textwidth]{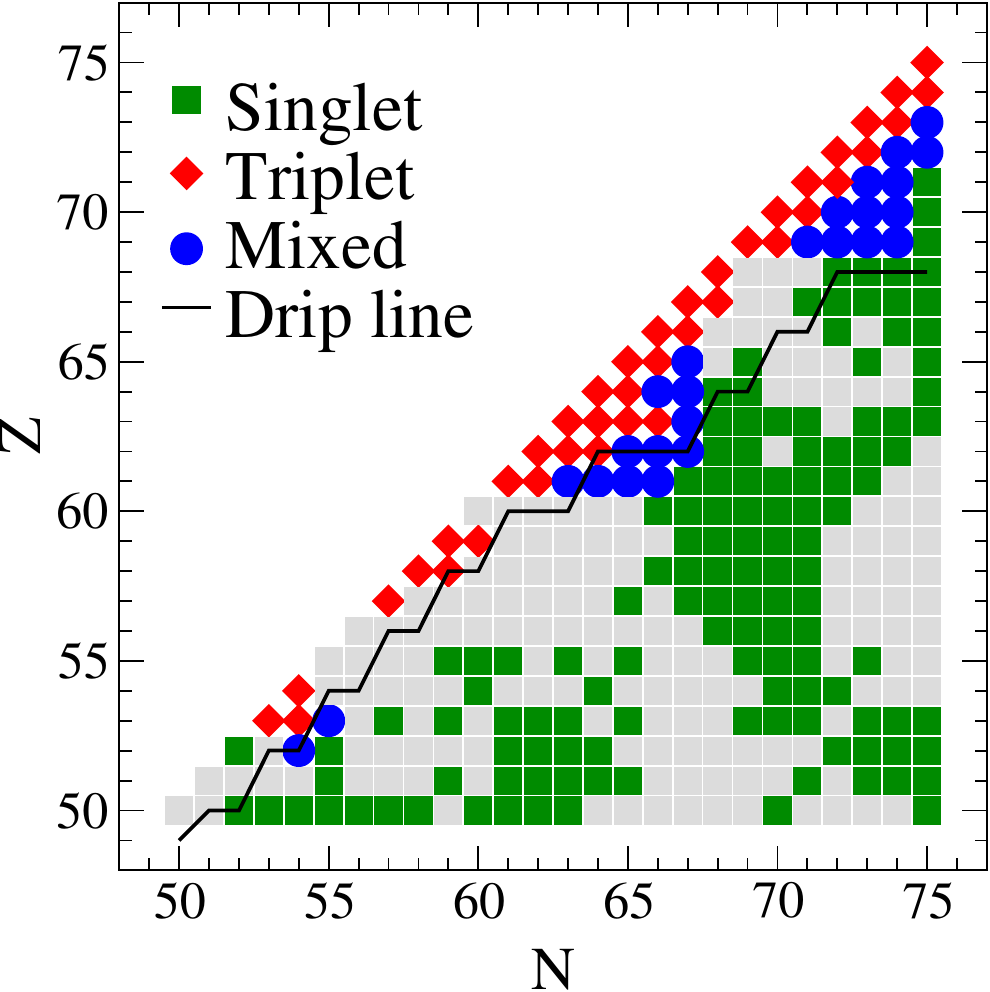}
\caption{Chart of nuclides with $Z \le N$ for neutron numbers from 50 to 75. Results from deformed mean-field studies of pairing. Shown is the spin character of the nuclear condensate: green squares, blue circles, and red diamonds, represent nuclei with singlet, mixed-spin, and triplet pairing, respectively, and gray squares denote nuclei with $E_{\rm corr}<0.5~\textrm{MeV}$.
The predicted proton-drip line is drawn in black. Note the presence
of (experimentally accessible) exotic pairing near and below the proton-drip line. Figure adapted from: G. Palkanoglou, M. Stuck, and A. Gezerlis, Phys. Rev. Lett. 134, 032501 (2025).
} \label{fig:nuctheo_struc}
\end{SCfigure}

UGuelph's A. Gezerlis employs stochastic quantum Monte Carlo
techniques, like auxiliary-field diffusion Monte Carlo, to study neutron
matter and light nuclei. 
Some highlights of recent work that will reach its fruition over the next
decade include i) studies of four-component fermions at unitarity; these results impact the prospects of developing nuclear physics as a perturbation around the physics of cold atoms,\footnote{W. Dawkins, J. Carlson, U. van Kolck, and A. Gezerlis, Phys. Rev. Lett. 124, 143402 (2020).} ii)  neural networks designed to incorporate complicated correlations for light nuclei, capturing, already
at the level of Variational Monte Carlo, the overwhelming majority of the ground-state energy estimated by Green's Function Monte Carlo (GFMC).\footnote{P. Wen, A. Gezerlis, and J. W. Holt, arXiv:2505.11442.} In addition to these
\textit{ab initio} studies, A. Gezerlis's group also uses 
more phenomenological techniques. A recent success involved
the study of mixed-spin pairing in heavy nuclei, using
the Hartree-Fock-Bogoliubov method and incorporating realistic nuclear deformation. Several experimental groups
(in Canada and France) are trying to check these predictions, 
as shown in Fig.~\ref{fig:nuctheo_struc}.\footnote{G. Palkanoglou, M. Stuck, and A. Gezerlis, Phys. Rev. Lett. 134, 032501 (2025); G. Palkanoglou and A. Gezerlis,  	arXiv:2505.08879.}

\subsubsection{Nuclear astrophysics}
\label{sec:nuctheo_astro}

The astrophysical origins of the elements observed to naturally occur in the galaxy is a longstanding
mystery. Nucleosynthesis studies leverage the role of nuclear physics in shaping astrophysical observables in order to
infer the properties of exotic nuclei that cannot presently be produced on Earth.
Crucially, multi-messenger astronomy has entered a new observational era, with the detection of GW170817 marking the first confirmed neutron-star merger observed via both gravitational waves and electromagnetic signals. Its interpretation, and those anticipated in the near future, depend on a deep understanding of nuclear matter, weak interactions, nuclear reactions, and gravitational waves from compact objects. A central goal 
in the field is to identify and characterize the imprints of subatomic physics - particularly nuclear and neutrino processes - on multi-messenger observations.

\begin{SCfigure}
\centering
     \includegraphics[width=0.5\textwidth]{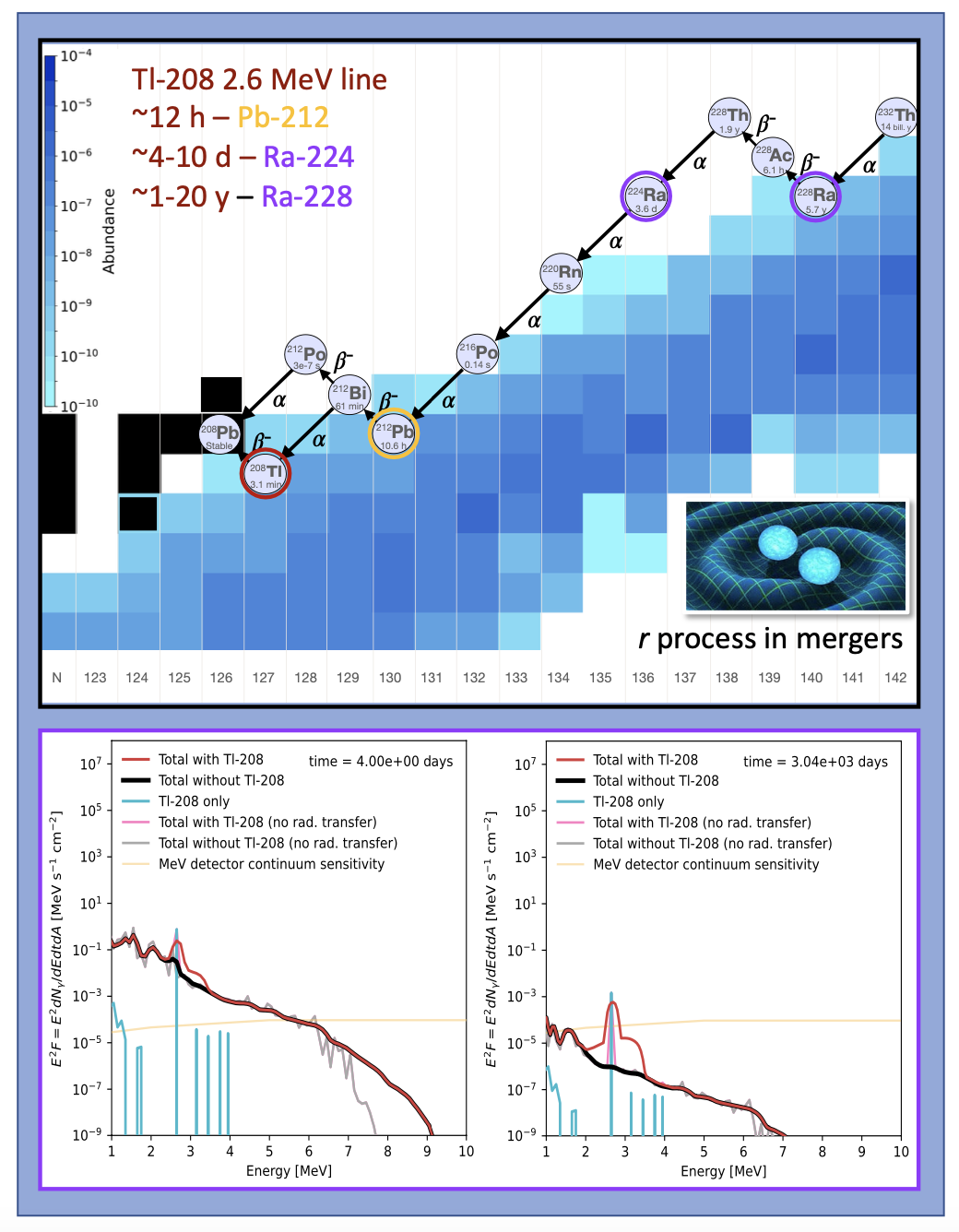}
\caption{(Top panel) A snapshot in time of a nucleosynthesis calculation showing the production of elements as neutron captures cease and decays convert these freshly synthesized species to nuclei along the well-known Th-232 decay chain, which ultimate produced Tl-208 at observable timescales. Tl signals on the order of hours and days (e.g., bottom left panel) point to the presence of Pb-212 and Ra-224 respectively. The decay of Ra-228 also presents the opportunity for future MeV gamma ray detectors such as COSI to make late-time observations on the order of years (e.g., bottom right panel).  Figure
adapted from: N. Vassh \textit{et al}., Phys. Rev. Lett. 132, 052701 (2024).
} \label{fig:nuctheo_astro}
\end{SCfigure}

TRIUMF's N. Vassh does research on the 
astrophysical origins of the elements, with
analysis pipelines which 
take inputs from experiment and theory for hundreds of nuclei combined with hydrodynamic simulations to predict observables from specific astrophysical events.
Recent accomplishments include the discovery of
correlations between silver (Ag, $Z=47$) and europium (Eu, $Z=63$), which suggest that the enrichment of some stars is connected to the fission of nuclei with mass number around A$\sim$260.\footnote{N. Vassh \textit{et al}., ApJ 896, 28 (2020); I. U. Roederer, \textit{et al}., Science 382, 1177 (2023).} 
This is the highest mass number so far inferred to be reachable in astrophysical environments from observables.
Another finding is
related to the possibility to observe the real-time synthesis of lead via gamma-ray emission from Tl-208 (due to a line at 2.6 MeV previously utilized in several other branches of science but not astrophysics), see Fig.~\ref{fig:nuctheo_astro}.\footnote{N. Vassh \textit{et al}., Phys. Rev. Lett. 132, 052701 (2024).}
 In the medium-term nuclear astrophysics is set to grow its inventory of astrophysical observables, nuclear data, and toolkits. For instance nucleosynthesis network approaches are diversifying (e.g., new uses of graph theory) and new analysis tools are emerging (e.g., machine learning and Bayesian methods).
 A recently published highlight is the first application of machine learning to classify stars as having been enriched by the r, s, or i process after training on nucleosynthesis calculations.\footnote{N. Vassh, et al., ApJ 992, 36 (2025).}
 A major goal is to overcome the traditional silos housing the knowledge of different processes. For instance, studies of $r$ vs $s$ vs $i$ process have often been performed by different nucleosynthesis groups which all apply their own networks and data. Future studies exploring the full range of possible neutron densities with consistent nuclear data are key to teasing out smoking gun signatures of distinct processes.

UGuelph's O. L. Caballero conducts cutting-edge theoretical studies of environments associated with neutron stars, focusing on how nuclear interactions and neutrino properties influence observable phenomena. By analyzing simulations of these systems - under conditions of extreme temperature, density, and gravity - predictive models of neutrino fluxes, gravitational wave signatures, and nuclear element synthesis are generated.\footnote{ B. Knight, O. L. Caballero, H. Schatz,  J. Phys. G: Nucl. Part. Phys. 51 (2014) 095201;  D. F. Rojas-Gamboa, N. G. Kelkar, O. L. Caballero, Phys. Rev. C 110, 035804 (2024).} A key focus will be the role of nuclear isomers in heavy element synthesis. Another important direction will explore the potential impact of dark matter interactions on nucleon behavior and neutrino emission in neutron star environments. Additionally, the aim is to uncover connections between relic neutrinos and stochastic gravitational wave backgrounds. Finally, the modeling of nuclear reaction rates under conditions of extreme density and low temperature — particularly involving radioactive and cluster decay — will 
be refined in order to enhance nucleosynthesis predictions.

Finally, the UGuelph's A. Gezerlis
has also carried out challenging studies of 
infinite/extended neutron matter. 
Most notably, this involved a fusion of (phenomenological)
energy-density functional calculations together
with quantum Monte
Carlo techniques, to tackle the problem of the 
compressibility sum rule in neutron matter.\footnote{M. Buraczynski, S. Martinello, A. Gezerlis,  Phys. Lett. B 818, 136347 (2021); M. Buraczynski, S. Martinello, A. Gezerlis, Phys. Rev. C 105, 025807 (2022).}
In the near term, such studies can be extended to
include protons, in the context of quantum Monte
Carlo for nucleonic matter. The pairing-gap calculations
can, similarly, be generalized to four components, 
leading to the problem of $p$-wave pairing in neutron
star crusts and cores. 
Another recent accomplishment was the calculation of
the $^1$S$_0$ pairing gap in neutron matter 
using auxiliary-field diffusion Monte Carlo,
finally putting to rest a long-standing controversy.\footnote{S. Gandolfi \textit{et al}., Condens. Matter 2022, 7(1), 19.}

\subsubsection{Fundamental symmetries}
\label{sec:nuctheo_funsym}

The research at the MemorialU (A. Aleksejevs and S. Barkanova) spans three interconnected directions - development of novel methods for multi-loop Feynman diagram calculations in electroweak and hadronic physics, support for ultra-precision experiments looking for BSM physics, and the study of hadronic structure using low-energy QCD. The group’s dispersive sub-loop insertion methodology addresses a key bottleneck in precision theory: reliable, cross-checkable two-loop electroweak computations. By reformulating messy nested loops into a simpler scalar integrals, it will be possible to streamline previously intractable calculations.\footnote{A. Aleksejevs, S. Barkanova, and A. I. Davydychev, arXiv:2510.23809.} Moreover, this brings modularity and automation to two-loop physics. The work is essential for MOLLER, P2 and Belle II experiments—sensitive to parity-violating asymmetries in electron scattering—which provide an indirect window to multi-TeV scale phenomena but require extensive theoretical input. An example of importance of loop calculations is the X17 anomaly, the MUN group has shown that next-to-leading order (NLO) corrections and the geometry of detectors can explain the peak around 17 MeV by means of SM contributions only.\footnote{A. Aleksejevs \textit{et al}., arXiv:2102.01127.} The two-loop input is also essential for ChPT models which describe hadronic structure in the non-perturbative regime of QCD. The group focuses on hadronic form factors and electromagnetic polarizabilities, which are key observables of static and dynamic hadronic response, and expanding their NLO Computational Hadronic Model (CHM) to study the SU(3) meson octet, where for neutral mesons the non-zero sum of electric and magnetic polarizabilities arises only at NNLO.

\begin{SCfigure}
\centering
     \includegraphics[width=0.6\textwidth]{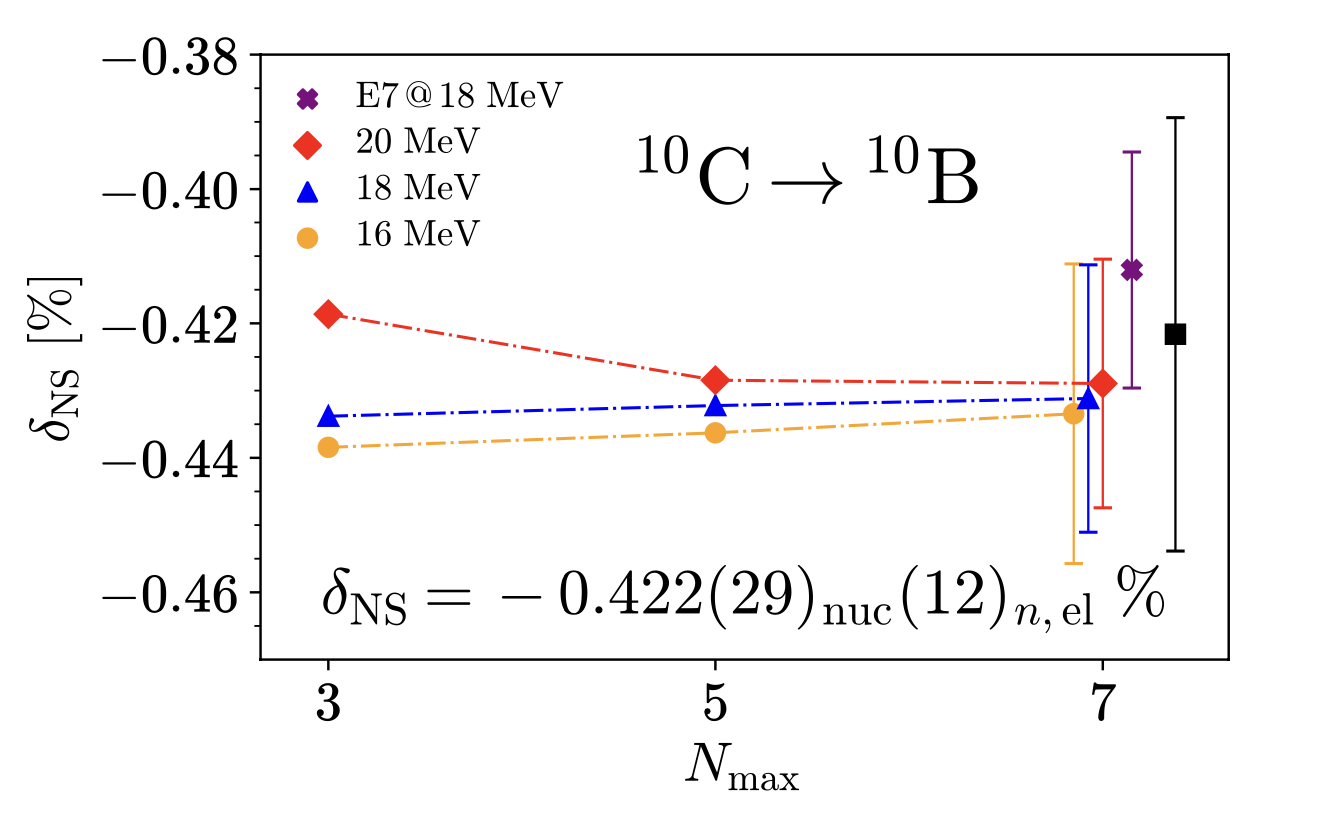}
\caption{Nuclear structure dependent radiative correction 
$\delta_{NS}$ relevant for the extraction of the CKM matrix element $V_{ud}$ from superallowed beta decays. Presented results are for the decay of $^{10}$C obtained within the \textit{ab initio} NCSM with $N_{max} = 3, 5, 7$ model space truncations and frequencies $\hbar \Omega = 16 - 20$ MeV for two different chiral NN+3N interactions.
Figure adapted from: M. Gennari \textit{et al}., Phys. Rev. Lett. 134, 012501 (2025).
} \label{fig:nuctheo_funsym}
\end{SCfigure}

TRIUMF's P. Navratil also works on projects
of relevance to the study of this big question in the field.
Medium term goals include applications of \textit{ab initio} nuclear theory for the interpretation of precision experiments testing fundamental symmetries and physics beyond the Standard Model. Examples include the determination of the $V_{ud}$ matrix element of the CKM matrix and its unitarity tests\footnote{M. Gennari \textit{et al}., Phys. Rev. Lett. 134, 012501 (2025).} (see Fig.~\ref{fig:nuctheo_funsym}), neutrinoless double beta decay experiments, as well as searches for nuclear parity-violating moments. Ongoing and planned atomic and molecular parity-violating experiments require support from nuclear theory in calculations of anapole moments and Schiff moments. Similarly, future calculations will involve nuclear-structure dependent radiative corrections needed for the inerpretation of experiments with muonic atoms and parity-violating electron scattering on nuclei.

Another TRIUMF researcher with a focus on
\textit{ab initio} nuclear many-body techniques,
J. D. Holt, will in the near-term future continue to study fundamental questions related to nuclear-weak physics. The most prominent of these is calculations of the nuclear matrix element for neutrinoless double beta decay and WIMP-nucleus structure functions for dark matter direct detection searches and neutrino-nucleus scattering for all experimentally relevant nuclei, with quantified uncertainties.\footnote{A. Belley \textit{et al}., arXiv:2408.02169.} Since both atomic and nuclear theory are vital to interpret searches for violations of fundamental symmetries in the universe, future work will tackle anapole moments, Schiff moments, and EDM, as well as calculating the atomic theory inputs for measurements of isotope shifts, which will give absolute nuclear charge radii and can be connected to other interesting searches for BSM physics.

\subsection{Beyond the next 7 years}

Owing to its very nature, theoretical research is less dependent on material infrastructure that
requires long-term advance planning than its experimental counterpart. However, some of the
projects outlined above are both long-term and labor-intensive, and thus require planning for
continued or additional personnel such as postdocs, especially if the theory input is required for
the timely progress of experimental projects. 
Canada has several faculty/staff members engaged
in such work, but the influx of new permanent positions
would help the country solidify its standing.
Given that we now have two new nuclear theorists
in the country (N. Vassh and G. Vujanovic) increased theory support is already a necessity.

Coast to coast, Canadian researchers are at the forefront of the many developing trends in
nuclear and hadronic theory with many examples of close involvement in the major breakthroughs
of the field, and it is clear that advances in experimental nuclear physics will be accompanied by comparable progress in theory. In addition to the many Canadian experimental endeavours receiving
theoretical support, theorists – during the coming years and beyond – will continue their association
with major off-shore laboratories such as EIC, GSI/FAIR, RIKEN, NSCL/FRIB, JLab, the LHC,
and RHIC. Our community will continue to make significant contributions in all the major areas
that define modern nuclear theory, in the time period relevant for this report.

\paragraph{Hadronic physics and QCD}

In addition to devising precise models of hydrodynamization and subsequent dynamical evolution of QCD flavors, all quark flavors carry electric charges and thus produce long-range electromagnetic (EM) fields. Thus, reliable magnetohydrodynamical (MHD) equations of motion need to be developed upon which novel simulations will rest, together with improved kinetic theories coupled to Maxwell equations. The ground work for a robust formulation of magnetohydrodynamics (MHD) equations is already underway, and the next major step lies in extending our semi-classical-based understanding to a more comprehensive quantum description. Taking into account all sources of angular momentum (spin and orbital) within an MHD simulation of a nuclear fluid, thus yielding a spin-MHD description, is an ambitious long-term goal.  There is also work underway to develop NLO chiral perturbation theory frameworks to the full SU(3) baryon octet to bridge theory and upcoming Compton scattering experiments. The wealth of high-precision data from the Electron-Ion Collider on Generalized Parton Distributions (GPDs) and gluon saturation will provide stringent benchmarks for lattice QCD and other non-perturbative approaches.

\paragraph{Nuclear structure}

At its broadest, the field's vision is to arrive at global calculations of all atoms and essentially all nuclei, using consistent nuclear forces and electroweak currents with quantified uncertainties. The region of
heavy nuclei has been reached only recently, so the full
ramifications will be investigated over the coming decade.
A specific future challenge for \textit{ab initio} theory is to describe nuclear deformation. 
Also on the horizon is the generalizing 
of existing nuclear wave function expansion 
schemes for
bound and unbound states by a component describing high-energy excitations such as giant
resonances. 
The study of pairing, similarly, is still in its infancy
at the \textit{ab initio} level, so synergy with
more phenomenological techniques will be helpful
here.
The nuclear quantum many-body problem will greatly benefit from the development of quantum-computing capabilities and
machine-learning technologies that are currently still
in their infancy.

\paragraph{Nuclear astrophysics}

A major goal is to develop a unified data analysis framework for major neutrino and gravitational wave observatories, such as Hyper-Kamiokande, DUNE, LIGO, LISA and Cosmic Explorer. This framework will enable consistent interpretation of multi-messenger signals.
Similarly, new telescopes such as the Compton Spectrometer and Imager (COSI) will reveal important information about the MeV gamma-ray sky.
Bayesian methods and ML applications in nuclear astrophysics have also begun to take root and will be a major focus of future studies synergizing observation and theoretical inputs.
The aforementioned consistent nuclear forces and electroweak currents with quantified uncertainties will lead to the first ab initio input, namely masses, $\beta$ decay, and neutron capture rates for use in $r$-process nucleosynthesis simulations.
Finally, a fascinating combination of the problems of
deformation and pairing (mentioned in the previous
paragraph) is the physical setting of \textit{nuclear pasta},
found inside neutron stars; a dependable
\textit{ab initio} study of nuclear pasta will certainly take more than 7 years.

\paragraph{Fundamental symmetries}

In the longer term, the work carried out by Canadian
researchers aims to develop the machinery for automated ingestion of two-loop topologies. The goal is to develop full gauge-invariant two-loop sets aimed at key experiments such as MOLLER, P2, and Belle II. Such work is foundational for next-generation precision electroweak theory, merging analytics, numerics, and dispersion. Another
promising avenue is to carry out the first \textit{ab initio} calculations of symmetry-violating moments (anapole,
Schiff), with quantifiable uncertainties. 
This work will be carried out while working closely with the experimentalist R. Garcia Ruiz from MIT, who is currently developing a setup to be situated at TRIUMF, leveraging the predicted unprecedented sensitivity of radioactive molecules to P- and CP-violating properties.

\subsection{Summary}

As outlined above, nuclear theory efforts in Canada
are distributed across the country, of wide interest,
and impactful. They range from traditional
studies of finite systems (light or heavy nuclei)
to the nucleonic matter of astrophysical significance,
as well as the structure of hadrons, 
and the importance of neutrinos. In addition to such astrophysical connections, interdisciplinary
aspects of this research are especially prominent when it comes
to dark-matter detection (particle physics) 
and clustering of unitary fermions (ultracold atomic physics).

It is essential to maintain a diverse program of research in theoretical nuclear and hadronic physics.
A large portion of this theoretical work demands the features associated with high-performance 
and quantum computing facilities; these
aspects are discussed in Section~\ref{sec:HPC}. In addition, much of the progress in theory
research is linked to the mentoring of the next generation of theorists. A great opportunity therefore
exists to further strengthen and grow this program with strategic investment into highly qualified
personnel who can accelerate the efforts of the recognized Canadian world leaders at the forefront
of an exciting and fast-moving discipline that is intimately linked to Canadian experimental efforts
in nuclear physics. With a wider regional representation, theorists are also an excellent resource
for increasing diversity in the field by attracting women, visible minorities and Indigenous students
to this fascinating field, in both theory and experiment.

\clearpage
\chapter{Facilities and Institutes}
\section{Canadian facilities}

\subsection{TRIUMF}
{TRIUMF}, located in Vancouver BC, is  Canada's particle accelerator center, and the only ISOL type facility in North America for rare isotope production.  At the heart of TRIUMF lies the 520-MeV cyclotron that delivers the primary proton beam that impinges onto various target material to produce RIBs though the  Isotope Separation Online (ISOL) method. The Isotope Separator and Accelerator (ISAC) I and II facilities deliver low/medium- and higher-energy beams to experiments, respectively. For example, in ISAC-I the TITAN and GRIFFIN facilities require low-energy beams of the order of 40-60 keV, while DRAGON and TUDA require beams accelerated to $0.15$--$1.5$ MeV/nucleon.  In ISAC-II, the TIGRESS, EMMA and IRIS experiments require higher-energy beams ranging from approximately $1$--$15$ MeV/nucleon. 
ISAC-I/II can also deliver stable beams from the offline ion source, which are used either as pilot beams for RIB measurements or directly for experiments.

With 
$\$$399.8M in funding from the Canadian federal government for the next 5 years, TRIUMF is on track to commission and exploit the new flagship multidisciplinary research facility called the Advanced Rare Isotope Laboratory (ARIEL), detailed in Section~\ref{sec:ARIEL}. Complementary to ISAC, ARIEL will triple TRIUMF’s rare isotope production capacity, reaching up to 9000 hours of beam delivery, and 
further enhance TRIUMF's position as a global leader in
isotope production through increased academic and international partner collaborations, industry partnerships, and technology transfer. 


\subsubsection{TRIUMF experimental facilities}

The beam-delivery capabilities of TRIUMF are exploited by a number of experimental facilities located at TRIUMF. These facilities perform a wide variety of measurements motivated by nuclear structure, astrophysics and fundamental symmetries.

\begin{itemize}
\item {\bf TITAN (TRIUMF's Ion Trap for Atomic and Nuclear science)} is a world-leading facility specializing in the precise mass measurement of short-lived, highly charged rare isotopes. TITAN consists of five unique individual ion traps coupled together: a radio-frequency quadrupole  (RFQ) cooler and buncher, a Multi-Reflection Time-Of-Flight (MR-TOF) isobar separator and spectrometer, an Electron Plasma Cooler trap, a precision Penning trap and an Electron Beam Ion Trap charge-state breeder to produce highly charged ions for their use to increase the precision of mass measurements. The MR-TOF can be used to provide further identification of mass-unknown species, to distinguish between a long- and shorter-lived species, or to determine half-lives for the first time. The MR-TOF-derived half-lives are complementary to more traditional methodologies like decay stations. TITAN started in 2003, originally funded for equipment and project through an NSERC RTI, and later augmented with additional equipment from CFI and NSERC SAP RTI. The operation of the system is supported in Canada through NSERC SAP project grants. International partners have contributed as well and the estimated total capital investment in TITAN to date is more than \$4.5M.

\begin{figure}
    \centering
    \includegraphics[width=0.85\linewidth]{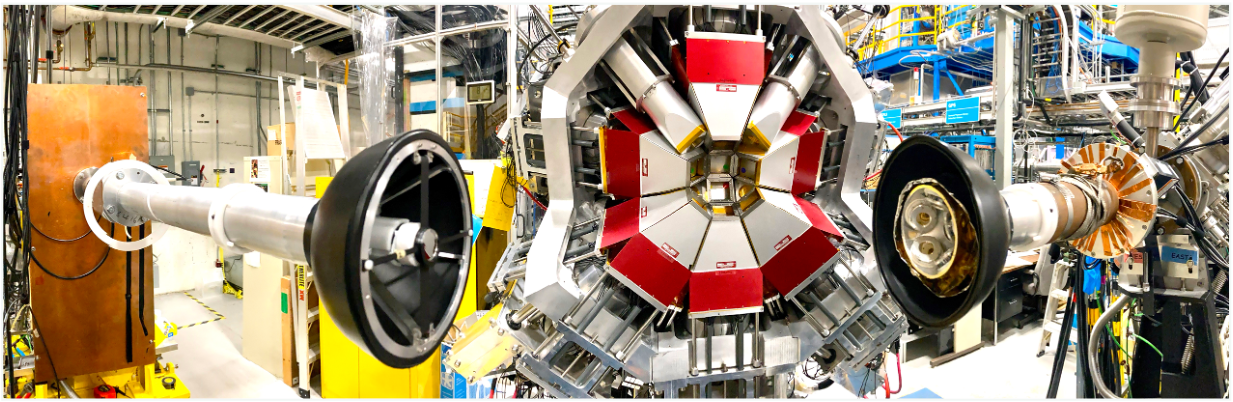}
    \caption{Half of the GRIFFIN array with its HPGe and LaBr$_3$(Ce) detectors equipped with Compton-suppression shields. The beam is traveling from right and is implanted in the mylar tape at the center of the target chamber (black ball) that hosts the Zero Degree Scintillator (ZDS) (on the left) and the Pentagonal Array for Conversion Electron Spectroscopy (PACES) array (on the right).}
\label{fig:GRIFFIN-all}
\end{figure}

\begin{figure}
    \centering
    \includegraphics[width=0.35\linewidth]{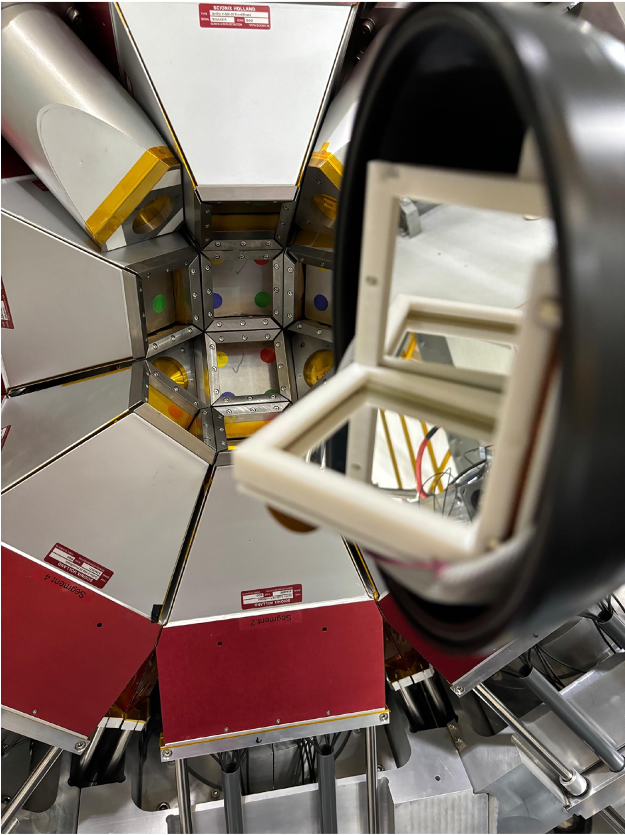}
    \includegraphics[width=0.35\linewidth]{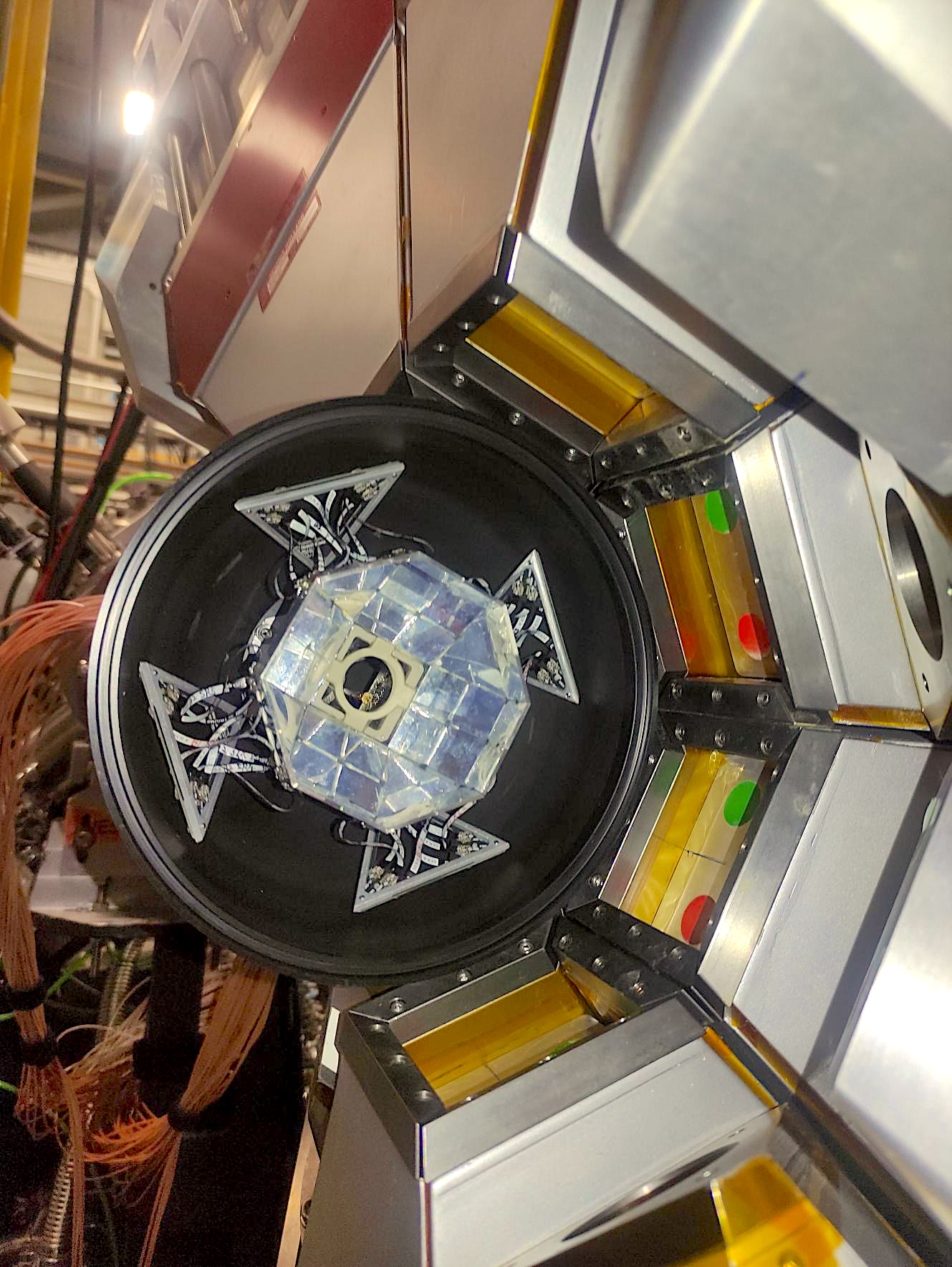}
    \caption{(Left) The new highly-segmented and high-efficiency double-sided silicon strip detector cube (RCMP) for highly-sensitive studies of $\beta$-delayed charged-particle (proton, alpha) emission used in conjunction with GRIFFIN. (Right) The new next-generation ARIES detector for $\beta$-tagging and fast-timing array of plastic-scintillator tiles mounted in the middle of GRIFFIN.}
    \label{fig:RCMP-ARIES}
\end{figure}

\item {\bf GRIFFIN  (Gamma-Ray Infrastructure For Fundamental Investigations of Nuclei)} is a high-efficiency $\gamma$-ray spectrometer comprised of up to 16 Compton-suppressed high-purity germanium (HPGe) clover detectors arranged in a rhombicuboctahedral geometry that is optimized
for $\gamma$-ray detection following the $\beta$ and $\beta$-delayed neutron decay of low-energy radioactive beams far from stability delivered by ISAC-I. It is augmented with a series of ancillary detectors such as .) five LN$_2$ - cooled lithium-drifted silicon detectors (PACES) for conversion electrons detection, b) two sets of 10 plastic scintillators used for $\beta$-tagging that can also be replaced with a ultra-fast plastic scintillator, c) a new highly-segmented and high-efficiency double-sided silicon strip detector cube (RCMP) for highly-sensitive studies of $\beta$-delayed charged-particle (proton, alpha) emission, d) LaBr$_3$(Ce) scintillators for fast lifetime measurements of nuclear levels populated in $\beta$ decay, and finally DESCANT, an array of up to 70 liquid scintillator neutron detectors that can be placed in a close-packed shell to enable high-efficiency $\beta$-delayed neutron emission studies. A picture of the opened GRIFFIN array showing the RCMP detector is shown in the left panel of Fig.~\ref{fig:RCMP-ARIES}.

The new RCMP detector provides both excellent particle energy resolution ($<$200 keV) and very high proton detection efficiency (77\%). Along with GRIFFIN, the RCMP can provide comprehensive decay spectroscopy measurement providing proton- and p-$\gamma$ coincidences, absolute branching ratios and half-life measurements in nuclei around the proton drip line.

The ARIES detector shown in the right panel of Fig.~\ref{fig:RCMP-ARIES} is a next-generation $\beta$-tagging and fast-timing array of plastic-scintillator tiles which replaces both SCEPTAR and ZDS with enhanced capabilities. In development are the Conversion
Electron Detection ARray (CEDAR) detector which will replace PACES with an array of up to 16 Si(Li) crystals, and the DAEMON detector, an array for energy measurements of neutrons, consisting of 1.5-cm thick fast-timing plastic scintillators that will be incorporated on the front faces of the Deuterated Scintillator Array for Neutron Tagging (DESCANT) detector for improved neutron energy resolution measurements of $\beta$-delayed neutron emitters.

The infrastructure was constructed and installed at ISAC-I during the 2011–2019 period in two funding phases totaling \$12.55M that was provided by CFI, the Ontario Ministry of Research and Innovation, the BC Knowledge Development Fund, TRIUMF, Simon Fraser University, and the University of Guelph. A comprehensive description of GRIFFIN and its ancillary detectors can be found in Nucl.\ Inst.\ Meths.\ A 918, 9 (2019), and the custom-built digital data acquisition system
with a 100 MHz sampling frequency can be found in Nucl. Inst. Meths. A 853, 85 (2017).

\begin{SCfigure}
      \centering
    \includegraphics[width=0.7\linewidth]{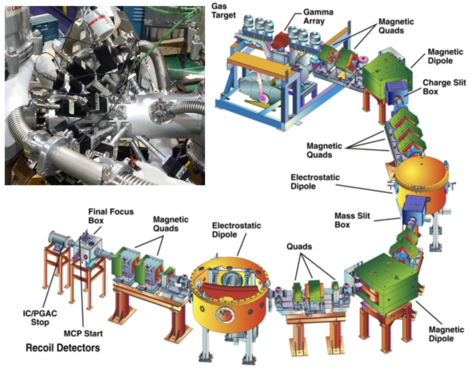}
    \caption{\emph{Main image:} Schematic of the DRAGON recoil separator. The total length of the device is ${\sim}21$ m from the gas target to the final focus. \emph{Inset:} Photograph of the SONIK precision scattering array installed at the DRAGON windowless gas target.}
    \label{fig:DRAGON_SONIK}
    \end{SCfigure}

\item {\bf DRAGON (Detector of Recoils and Gammas of Nuclear Reactions)} 
is recognized as the most successful recoil separator for nuclear astrophysics in the world, having made $80\%$ of the world’s measurements of radiative capture reactions involving radioactive nuclei. A schematic of DRAGON is shown in Fig.~\ref{fig:DRAGON_SONIK}. DRAGON uses two separate stages of magnetic and electrostatic dipoles to select ions based on their mass and charge, separating the heavy-ion recoils from inverse-kinematics $(p,\gamma)$ and $(\alpha,\gamma)$ reactions from the incident beam. The DRAGON reaction target is an extended recirculating volume of H or He gas and is surrounded by an array of Bismuth Germanium Oxide (BGO) detectors for detecting the $\gamma$~rays produced by the reactions of interest in coincidence with the recoils at the DRAGON focal plane. Recently, ancillary OGS neutron detectors have been installed at DRAGON to measure $(\alpha,n)$ reactions, and it is envisioned that in the future the separator can be switched from radiative capture mode to $(\alpha,n)$ mode and back again at the touch of a button.

DRAGON can be used in conjunction with several complementary detector systems for nuclear astrophysics, such as TUDA (TRIUMF-UK Detector Array), TACTIC (TRIUMF Annular Chamber for the Tracking of Charged Particles) and SONIK (Scattering of Nuclei in Inverse Kinematics) devices (Fig.~\ref{fig:DRAGON_SONIK} inset), all of which were designed primarily for nuclear reaction and scattering studies of astrophysical importance. Together these form the ``Astro-4'' suite of facilities. The TUDA, TACTIC, and SONIK facilities are each 
owned and operated by a consortium of collaborators expert in the field of nuclear astrophysics, including the local Canadian collaboration based at TRIUMF who maintain custodianship of the devices. The collaboration comprises mainly researchers from Canada, the United Kingdom and the United States, but also from Spain and France.

\item {\bf TIGRESS (TRIUMF-ISAC Gamma-Ray Escape Suppressed Spectrometer)} is an array of 16
Compton-suppressed 32-fold segmented clover-type HPGe detectors optimized for in-beam $\gamma$-ray spectroscopy with the accelerated radioactive ion beams provided by the ISAC-II at energies approaching or beyond the Coulomb barrier. The 32-fold
segmentation of the TIGRESS HPGe detectors enables precise translation of the $\gamma$-ray energies measured in the laboratory to those in rest frame of the nucleus. TIGRESS was funded by an \$ 8.06M NSERC RTI-3 grant over the 6-year period from 2003 to 2009, and has since been augmented by more than \$ 2.5M of associated detectors funded through NSERC RTI and CFI awards.

The sensitivity of TIGRESS 
is enhanced with a versatile suite of specialized charged-particle detector systems that are tailored, for each experiment, to the reaction of interest. The TIGRESS Integrated Plunger (TIP) has a Recoil Distance Method (RDM) ``plunger” integrated with a nearly hermetic array of radiation-hard CsI detectors, the latter for identification of weak reaction products or scattered particles amongst strong backgrounds. This device is designed for measuring lifetimes of long-lived states ($\tau > 1$ ps) populated through fusion-evaporation reactions and can be used both in the EMMA and standalone configurations. The Silicon Highly-segmented Array for Reactions and Coulex (SHARC) for standalone experiments and SHARC-II for EMMA experiments each are comprised of arrays of double-sided silicon strip detectors with approximately 1000 electronics channels. They are intended to detect
light charged particles ejected in single- or two-nucleon transfer with accelerated radioactive ion beams in inverse kinematics. The SPectrometer for Internal Conversion Electrons (SPICE) is a large-volume, highly-segmented lithium-drifted silicon (Si(Li)) detector located in vacuum that is shielded from direct sight of the target by a photon shield for in-beam internal-conversion-electron
spectroscopy. It allows detailed investigations of shape coexistence in exotic nuclei far from stability. The Bambino detector is comprised of 24 $\times$ 32-fold segmented annular Si detectors. The TRIumf Fast Ionization Chamber (TRIFIC) is a tilted-plane heavy-ion gas counter for counting and identifying
beam particles and beam-like reaction products, and is designed for rates approaching 10$^6$ pps. Together, Bambino and TRIFIC are ideal for Coulex measurements of rare and often contaminated radioactive ion beams. Lastly, DESCANT can be incorporated with TIGRESS for efficient tagging on emitted neutrons from reactions, including both fusion and direct reactions.

\begin{figure}
    \centering
    \includegraphics[width=0.75\linewidth]{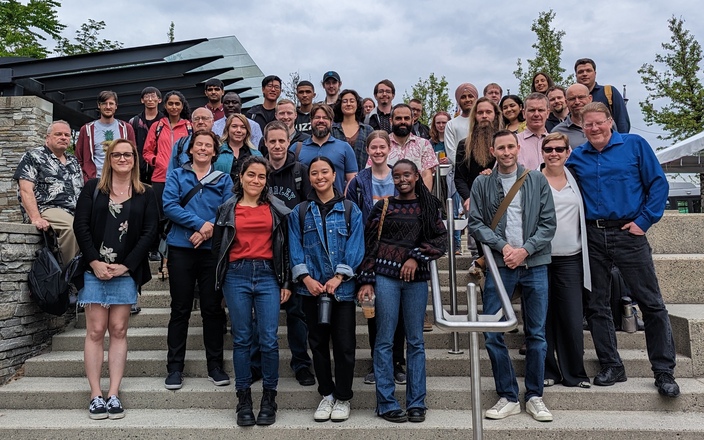}
    \caption{The Gamma-Ray Spectroscopy Meeting official photo at SFU in 2023. Front and center are several students and postdoctoral researchers.}
    \label{fig:GRS2023}
\end{figure}

\item {\bf EMMA (ElectroMagnetic Mass Analyzer) recoil separator} is a recoil separator at TRIUMF that is used to spatially separate the heavy products of nuclear reactions from the radioactive beams that produce them, and to disperse them in according to their mass/charge ratios. It focuses ions in both angle and energy for
high detection efficiency. This enables challenging measurements
involving extremely low cross-section reactions through recoil-$\gamma$ coincidences, such as those motivated by medium- to heavy-element production in stellar environments, where measuring either $\gamma$~rays or reaction products alone does not provide sufficient sensitivity. Since 2019, TIGRESS has been located at the target location of EMMA to facilitate experiments detecting $\gamma$~rays and heavy recoils in coincidence. Recent successful tests have also been performed incorporating OGS detectors for triple-coincidence $\mathrm{recoil}+n+\gamma$ measurements of $(\alpha,n)$ reactions.

\begin{figure}
    \centering
    \includegraphics[width=0.75\linewidth]{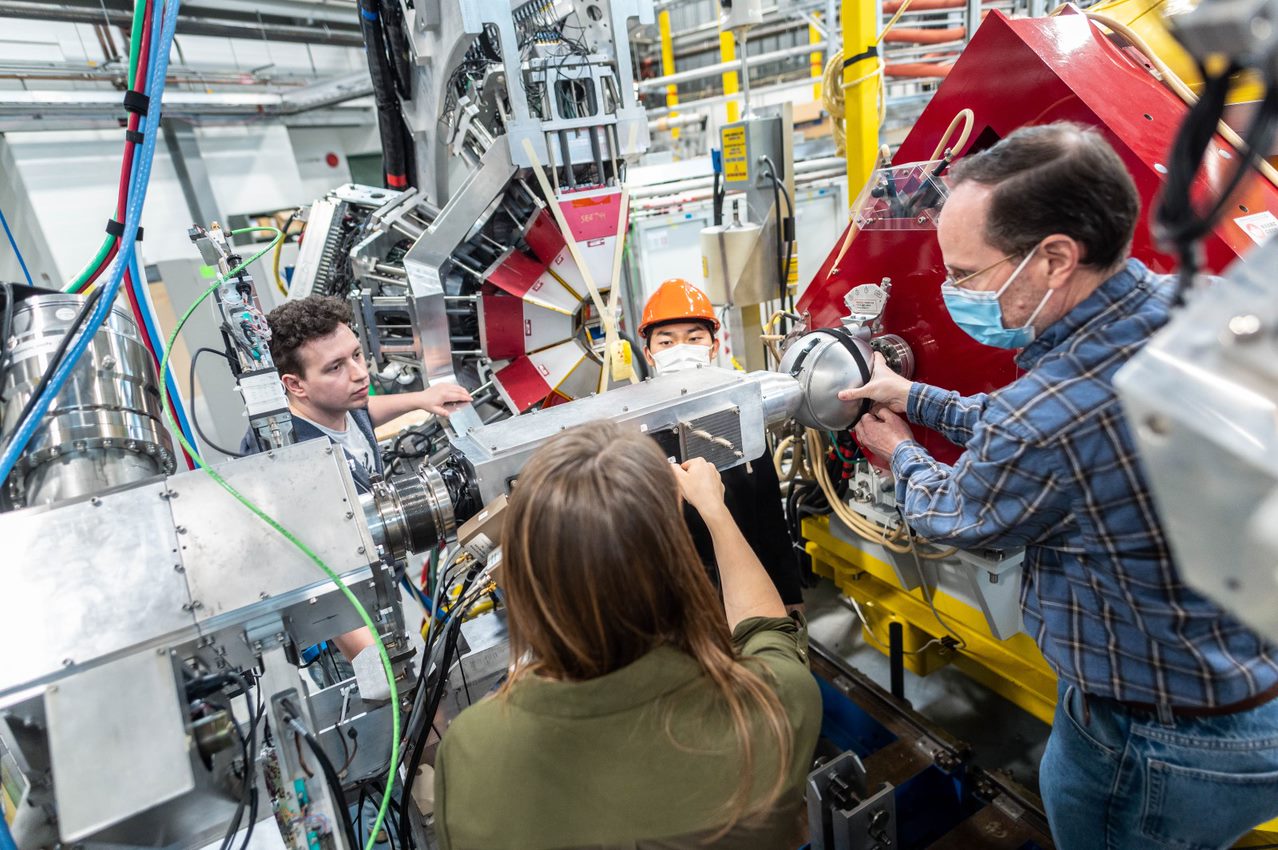}
    \caption{A team working on the EMMA+TIGRESS target chamber.}
    \label{fig:EMMA-TIGRESS}
\end{figure}

\item {\bf The GANIL Active Target and Time Projection Chamber (ACTAR TPC)} is a state-of-the-art
gas-filled detector system that is used to study nuclear reactions and decays of exotic nuclei. The
detector employs a large gas volume that acts as both a sensitive detection medium and a reaction
target, for the case of in-beam experiments, or as a gas stopper and delayed charged particle
detector for decay studies. The anticipated low beam intensities for the most exotic nuclei, combined
with the need to detect low-energy charged particles with excellent resolution and efficiency,
constitutes a demanding experimental challenge.

Funded via an European Research Council starting grant awarded to the PI (Dr. Gwen Grinyer now in Regina), the ACTAR TPC
detector was designed and constructed at GANIL in France during the period 2013 to 2019. The detector,
which was always intended to be portable, was involved in experiments at TRIUMF in 2019 and again in 2025 to perform 3 experiments. As such a detector does not exist in Canada, this campaign will be essential for building Canadian knowledge and expertise with active targets.
Future plans are in place in the medium to long term to develop a similar detector in Canada, optimized for ARIEL beams. 
The proposed detector is highly versatile and can be used both as a thick target for high efficiency in-beam
reaction experiments or as an active gas stopper for decay experiments.

\begin{figure}
    \centering
    \includegraphics[width=0.40\linewidth]{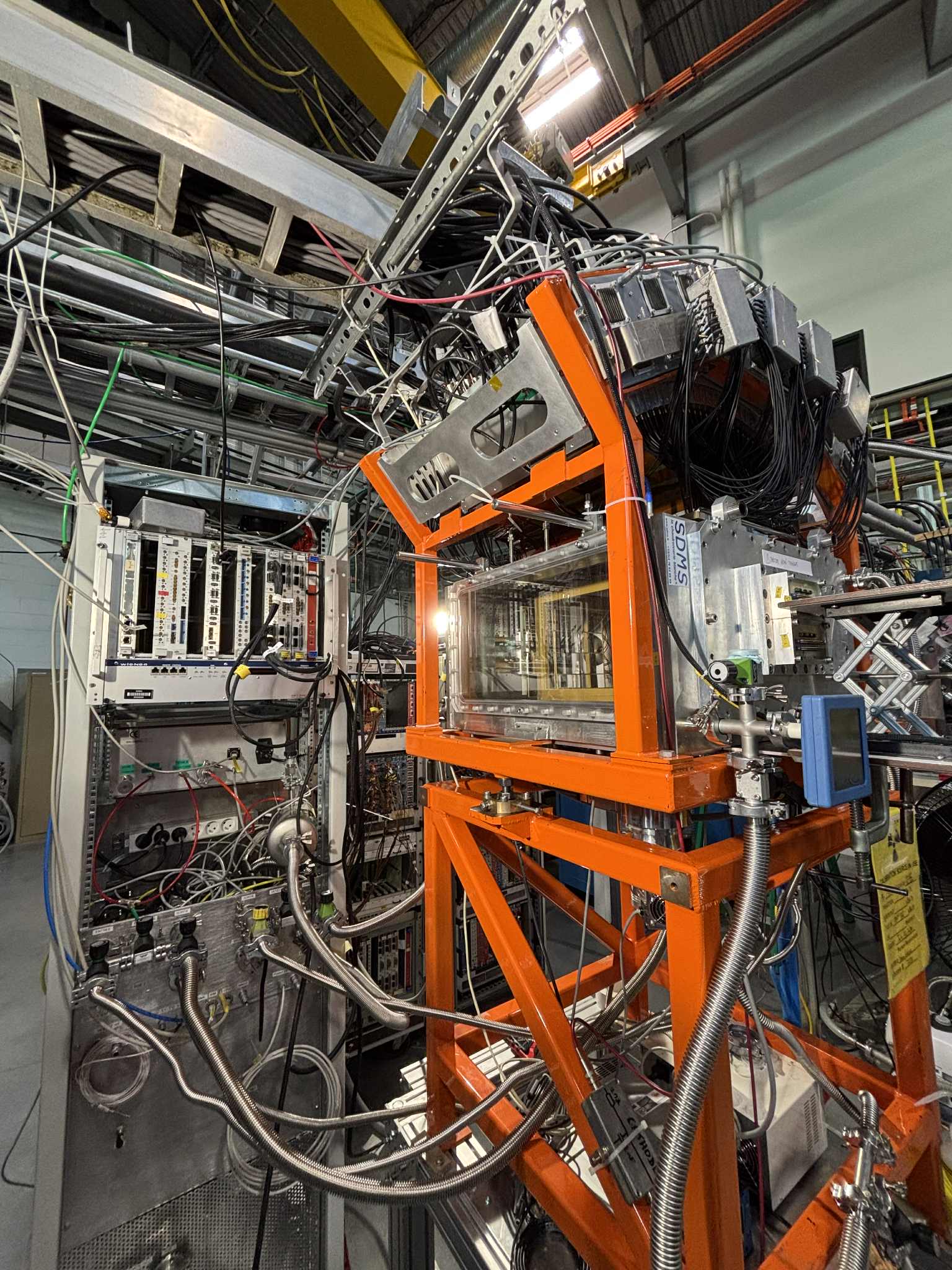}
    \includegraphics[width=0.40\linewidth]{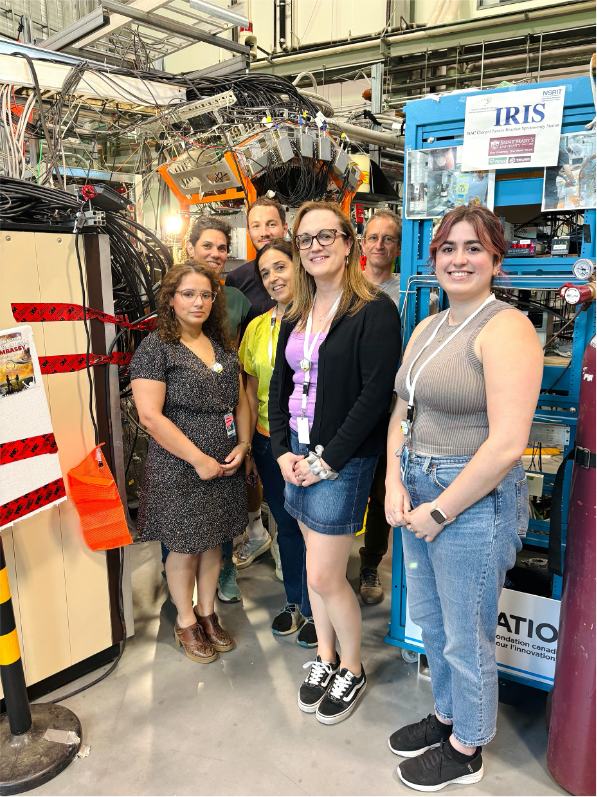}
    \caption{(Left) ACTAR TPC installed in the beam line at TRIUMF. (Right) A happy international team of researchers from France, Spain, and Canada after successfully installing the ACTAR TPC at TRIUMF in 2025.}
    \label{fig:ACTAR collaboration}
\end{figure}

\item {\bf IRIS (ISAC Charged Particle Reaction Spectroscopy Station)} is the charged particle reaction spectroscopy station that uses ISAC II at TRIUMF to studying transfer reactions and inelastic scattering of rare isotopes in inverse kinematics. 
The IRIS project is led by Saint Mary’s University in partnership with TRIUMF, McMaster University, Simon Fraser University, University of Regina and the University of Guelph in Canada, as well as other international
collaborators. This infrastructure is a dedicated standalone beamline facility for charged particle reaction spectroscopy of rare isotopes. The unique feature of IRIS is the use of solid H$^2$ and D$^2$ reaction targets that makes it possible to perform spectroscopy of highly neutron-rich or proton-rich nuclei that are available only as low intensity beams delivered by ISAC-II. Hence, the scientific program complements measurements that can be done with the TIGRESS and EMMA $\gamma$ spectroscopy project.

\begin{figure}
    \centering
    \includegraphics[width=0.6\linewidth]{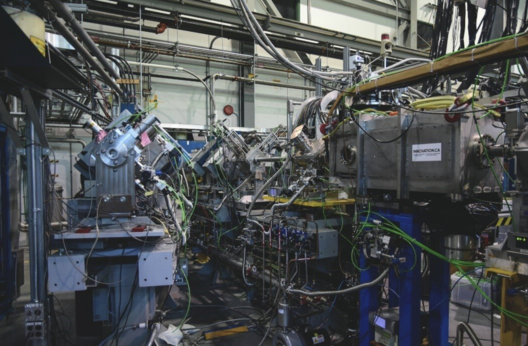}
    \caption{IRIS, the ISAC Charged Particle Reaction Spectroscopy Station.}
    \label{fig:IRIS-facility}
\end{figure}

\item {\bf Ultracold neutrons from the TUCAN source} 

The TUCAN source (see Fig.~\ref{fig:overhead}), located in the TRIUMF Meson Hall, is envisioned to become
a user facility with applications to use the source approved by TRIUMF
and supported by the TUCAN collaboration.  This will bring new users
to Canada, seeking to operate the UCN source for a variety of research
goals in fundamental neutron physics.  The PENeLOPE project will be
the first example of an experiment that will be conducted in this way.
In addition to experiments on the neutron lifetime (like PENeLOPE),
experiments measuring neutron gravitational levels and detailed
measurements of neutron $\beta$-decay have also been discussed.  All
such experiments have been chronically limited by the number of
ultracold neutrons (UCNs) that can be measured in the experiment.  The
world-leading UCN production of the TUCAN source would enable a new
generation of precision measurements for these experiments.

\begin{figure}[h!]
\centering
\includegraphics[width=0.6\textwidth]{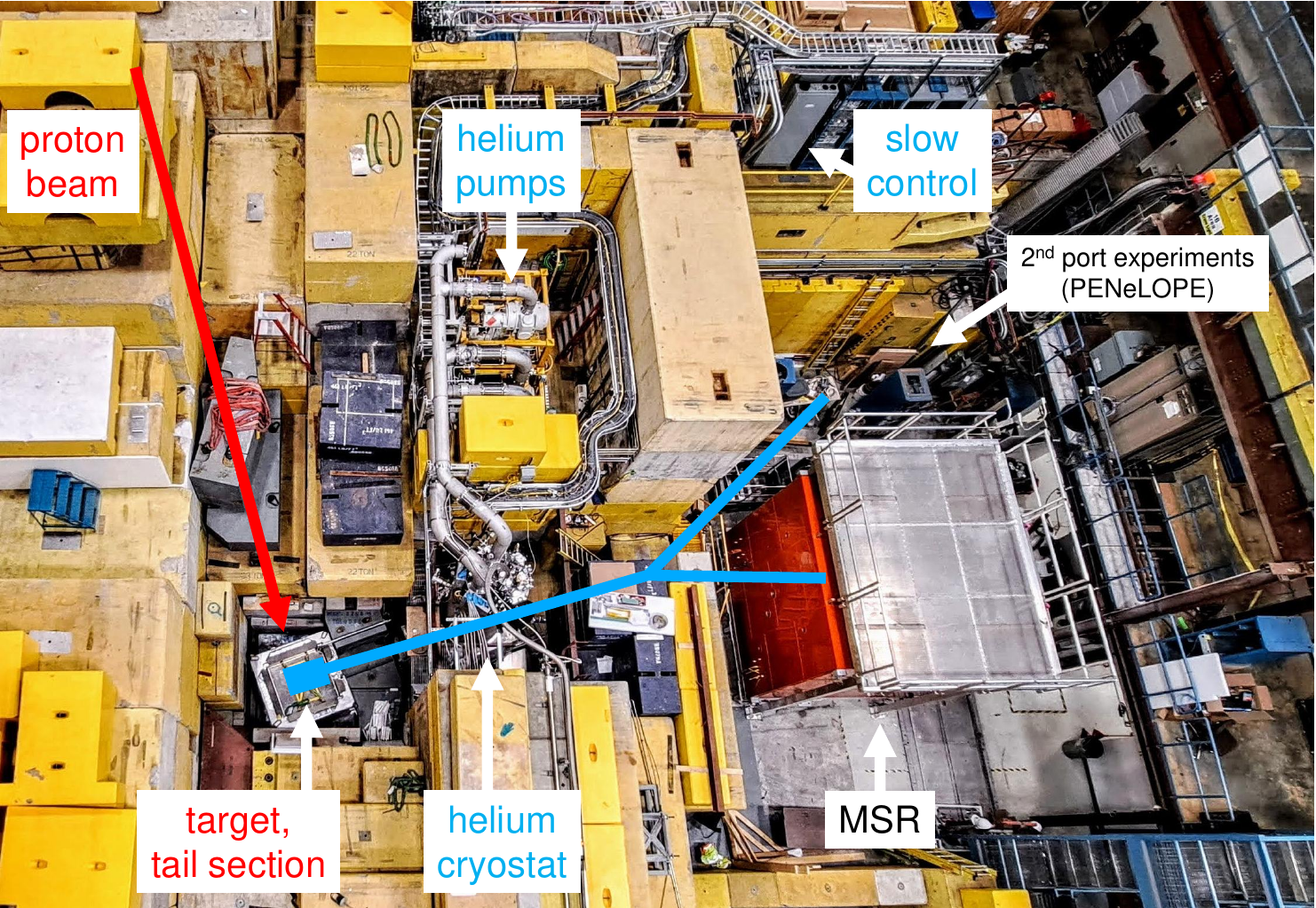}
\caption{Overhead view of the UCN source facility (April 2024) located in the
Meson Hall at TRIUMF.  Lines display the underlying proton beam path (red) and sketch the existing and planned UCN guide paths (blue).  While the nEDM experiment will be conducted inside the magnetically shielded room (MSR), the 2nd port will be available for other experiments, such as PENeLOPE.\label{fig:overhead}}
\end{figure}

\item {\bf HAICU: Hydrogen-Antihydrogen Infrastructure at Canadian
  Universities}

HAICU is a major research infrastructure project funded by the CFI and
currently under development at TRIUMF, in partnership with UBC,
UCalgary, SFU, and BCIT. The facility will provide a unique platform
for the development of advanced quantum sensing technologies
applicable to antimatter research. Using hydrogen atoms as proxies for
their antimatter counterparts, HAICU will enable the design, testing,
and improvement of techniques essential for next-generation
experiments. These include the realization of anti-atomic fountains,
anti-atom interferometers, optical trapping of antihydrogen, and the
formation of antimatter molecules. In addition, HAICU will serve as a
testbed for new technologies in laser and microwave systems, magnets,
and cryogenics—providing an environment for innovation in precision
control and measurement. Through these developments, HAICU will bridge
fundamental physics and quantum technology, enhancing Canadian
leadership in precision measurements with both matter and antimatter.

\end{itemize}

\subsubsection{The ARIEL facility at TRIUMF}
\label{sec:ARIEL}

\begin{table}[t]
\centering
\renewcommand{\arraystretch}{1.15}
\begin{tabular}{|p{4cm}|p{8cm}|p{2cm}|}
\hline
{Capability} & {Will deliver isotopes...} & {First experiments} \\
\hline
{CANREB} & ...to elucidate our fundamental understanding of atomic nuclei by enabling studies of the evolution of structure and dynamics of neutron-rich nuclei. & $\sim$2027 \\
\hline
{Two simultaneous RIBs} & ...as probes of magnetism at interfaces and surfaces of new functional materials using $\beta$-NMR. & $\sim$2028 \\
\hline
{Photo-fission} & ...to elucidate our fundamental understanding of atomic nuclei by enabling studies of the evolution of structure and dynamics of very neutron-rich nuclei approaching the $r$-process path. & $\sim$2028 \\
\hline
{Proton target station} & ...for molecular imaging of diseases and treatment of cancer in the ARIEL collection station and isotopes for developing a standard model for nuclear physics. & $\sim$2028 \\
\hline
{Three simultaneous RIBs} & ...to search for new forces in nature by searching for violations of Fundamental Symmetries. It will also mark the milestone of three simultaneous rare isotope beams delivered to users. & $\sim$2029 \\
\hline
{Routine 5000 RIB hours per year} & ...to deliver new capabilities of ARIEL for scientific discoveries. & $\sim$2029 \\
\hline
{Full driver beam intensities} & ...to determine how the heavy elements from iron to uranium were produced in the universe. & $\sim$2030 \\
\hline
{Routine 9000 RIB hours per year} & ...to fully exploit the new capabilities of ARIEL for all scientific programs. & $\sim$2032 \\
\hline
\end{tabular}
\caption{Timeline of ARIEL capabilities and their scientific goals.}
\label{tab:ARIEL-timeline}
\end{table}

Construction of ARIEL, the Advanced Rare Isotope Laboratory, is now underway at TRIUMF
with the goal to significantly expand TRIUMF’s Rare Isotope Beam (RIB) program for Nuclear
Physics and Astrophysics, Nuclear Medicine and Materials Science. At its heart, ARIEL contains a
350 kW, 35 MeV, 10 mA electron accelerator (eLINAC) for isotope production via photo-production
and photo-fission as well as a second proton beam line from TRIUMF’s 520-MeV cyclotron for
isotope production via proton-induced spallation and fission. The second stage of the project,
ARIEL-II, is a joint CFI funding initiative through all 19 TRIUMF member universities (at the time
of the application) and led by the University of Victoria. This will allow ARIEL to be completed
in the timeframe covered by the forthcoming SAP Long Range Plan.

The ARIEL scientific program will be implemented in phases beginning with advanced beam cleaning
and preparation capabilities for accelerated radioactive beam. These new capabilities will
drive forward the programs studying nuclear structure effects in exotic isotopes. This phase will be
followed by the implementation of a new production target station to receive first beams from the
eLINAC. Neutron-rich fission fragments produced from more than 10$^{13}$ fissions per second will be
possible in the final implementation. Photo-fission will enable the study of the very neutron-rich
nuclei involved in the astrophysical $r$ process responsible for the production of the heavy elements
from iron to uranium. The new proton beam-line (BL4N) will deliver up to 100~$\mu$A beam
onto an additional production target. In conjunction with the eLINAC production target TRIUMF
will therefore go from the current single ISAC RIB production target to the parallel production of
RIBs on three target stations. This new and worldwide unique multi-user capability will allow for
a much better exploitation of the available forefront experimental facilities at ISAC. Aside from
the tremendous gain in available time for the material science program also other experimental
programs that need large amounts of beam time will be enabled by this multi-user capability of
ARIEL. In addition, the capabilities for harvesting isotopes for investigation as potential medical
diagnostic and therapeutic isotopes will be implemented through a dedicated and symbiotic target
module integrated into the beam dump of the ARIEL proton beam target station.
Table~\ref{tab:ARIEL-timeline} identifies the key scientific deliverables for each new capability of the Ariel facility.

The ARIEL facility at TRIUMF, together with the existing ISAC production target and 18 current experimental facilities in the ISAC-I and ISAC-II halls, will deliver a rich and world-leading
user program in all three branches of rare isotope SAP science (nuclear structure/reactions, nuclear astrophysics, and fundamental symmetries). 
The experimental
facilities are predominantly operated by collaboration with strong Canadian involvement or under
Canadian leadership and thus the ARIEL facility with its dramatic increase in RIB availability and
further reach to the extremes of isospin, will elevate the Canadian nuclear physics community even
further.
The development and construction of the eLINAC, with Made-In-Canada SRF cavities and
in-house development of the electron gun, cryo-modules, beam diagnostics and machine protect
systems has been an enormous achievement in accelerator science. This has also benefited the
education of several graduate students in accelerator science, who are part of the only graduate
program in accelerator science in Canada at the University of Victoria.
The ARIEL program represents an investment of approximately \$200 million  by the federal and 6 provincial governments;
supported by 21 universities across Canada, and about 700 FTE years total effort. This
is a combination of four CFI projects; ARIEL-I (2012 - 2014), CANREB (2014 - 2019), ARIEL-II
(2017 - 2027), Therapeutic Isotopes (2020 - 2027) and other funding sources. The manpower to
design and construct the ARIEL-II infrastructure is provided by TRIUMF, funded through the
operating funds provided by the Canadian government via a contribution agreement by NRC. The operation of the ARIEL facility will be integrated into  TRIUMF operations.

\subsubsection{Proposed new TRIUMF facilities}

\paragraph{RadMol}

Beyond EDM searches, RadMol is envisioned to become a facility for
precision spectroscopy of (ultra-)cold radioactive molecules.  This
could lead to improved measurements of nuclear anapole moments,
applications to quantum chemistry and nuclear medicine, and
measurements of radioactive molecules useful for astrophysics.  The
facility will be available for these studies with a long-term future.

\paragraph{TRIUMF Storage Ring (TRISR)
\label{sec:TRISR}}

\begin{SCfigure}
    \centering
    \includegraphics[width=0.82\textwidth]
    {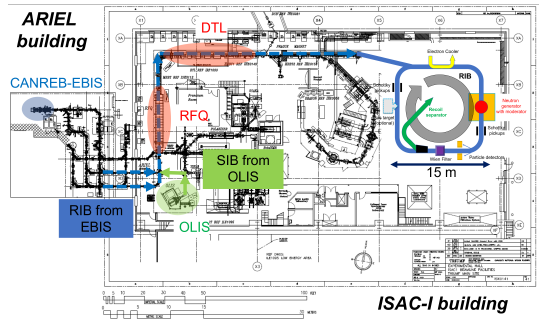}
    \caption{The TRIUMF ISAC-I experimental hall with the proposed location for the TRISR low-energy storage ring.}
    
    \label{fig:TRISR}
\end{SCfigure}

%
The Canadian nuclear astrophysics community has plans to develop a new storage ring, TRISR, coupled to a neutron-production device. The new ring is planned to be located in the ISAC-I hall at TRIUMF, as shown in Fig.~\ref{fig:TRISR}. The coupling of a storage ring with a neutron target would make TRISR a unique facility, one that would take advantage of existing infrastructure as well as the ARIEL upgrade, greatly expanding the scientific scope of the nuclear physics and astrophysics program at TRIUMF.

The concept behind TRISR is to couple a low-energy storage ring ($0.15-2$~MeV/nucleon) with a moderated neutron source in the ring matrix, creating a quasi-static neutron ``gas'' target that intersects with the (radioactive) ion beam.\footnote{A. Tarifeno-Saldiva \emph{et al.}, submitted to Phys.\ Rev.\ Acc.\ and Beams, \url{http://arxiv.org/abs/2508.15465}} The beam passes through the neutron target
area with a orbiting frequencies of hundreds of kHz and thus increases the luminosity by orders of magnitudes compared to ordinary one-pass experiments. Neutron capture
reaction products will be extracted into a highly-sensitive recoil separator and detected there. This
new technique would allow for the first time a direct measurement of neutron capture cross sections
of short-lived radionuclides down to half-lives of seconds. A dedicated storage ring at an ISOL facility like the TRISR could allow access to almost 1000 radioactive nuclei with half-lives down to 1 s. At the ISAC facility, more than 300 of these isotopes are already accessible now with high intensity. Many of these isotopes, with half-lives down to several hours, are located in the $s$- and $i$-process reaction path, and the direct measurements facilitated by TRISR would help to better constrain the reaction flow.  Plans for similar projects are in place at Los Alamos National Laboratory (USA), and others may be developed elsewhere worldwide. The construction of TRISR at TRIUMF would ensure Canada retains its spot as a world leader in direct measurements of astrophysical reactions in the next 10--15 years and beyond.

\paragraph{Two-Step RIB}

Long-term plans include development of a two-step RIB production facility in which neutron-rich ISOL beams such at ${}^{132}$Sn are re-accelerated to intermediate energies ($>100$ MeV/nucleon) and fragmented on a Be or Ta target. This technique is expected to produce beams of highly neutron-rich nuclides that cannot be produced at conventional fragmentation facilities using stable-beam drivers. These beams can then be used directly for intermediate-energy experiments, or they can be energy-degraded for experiments such as decay spectroscopy or mass measurements. R\&D supporting this new facility is expected to begin during the next seven years, with eventual construction planned beyond 2040.

\subsection{The Nuclear Science Laboratory (NSL) at Simon Fraser University (SFU)}
{\bf The Nuclear Science Laboratory (NSL)} at Simon Fraser University (SFU) supports research programs which include nuclear structure studies, development of techniques for radiation detection, production and separation of radioisotopes, neutron activation analysis, and studies of environmental radioactivity. The NSL provides access to a Thermo Scientific P-385 deuterium-tritium neutron generator producing nearly mono-energetic 14.2 MeV neutrons at a nominal rate of 3·10$^8$ neutrons/second. The NSL is one of very few university laboratories providing access to neutrons from deuterium/tritium fusion, and the only laboratory operating a neutron generator in Western Canada. The NSL also hosts the 8$\pi$ $\gamma$-ray spectrometer equipped with 20 HPGe Compton suppressed detectors. 

\subsection{SNOLAB}
\label{sec:snolab}

\begin{figure}[h!]
  \centering
  \includegraphics[width=\linewidth]{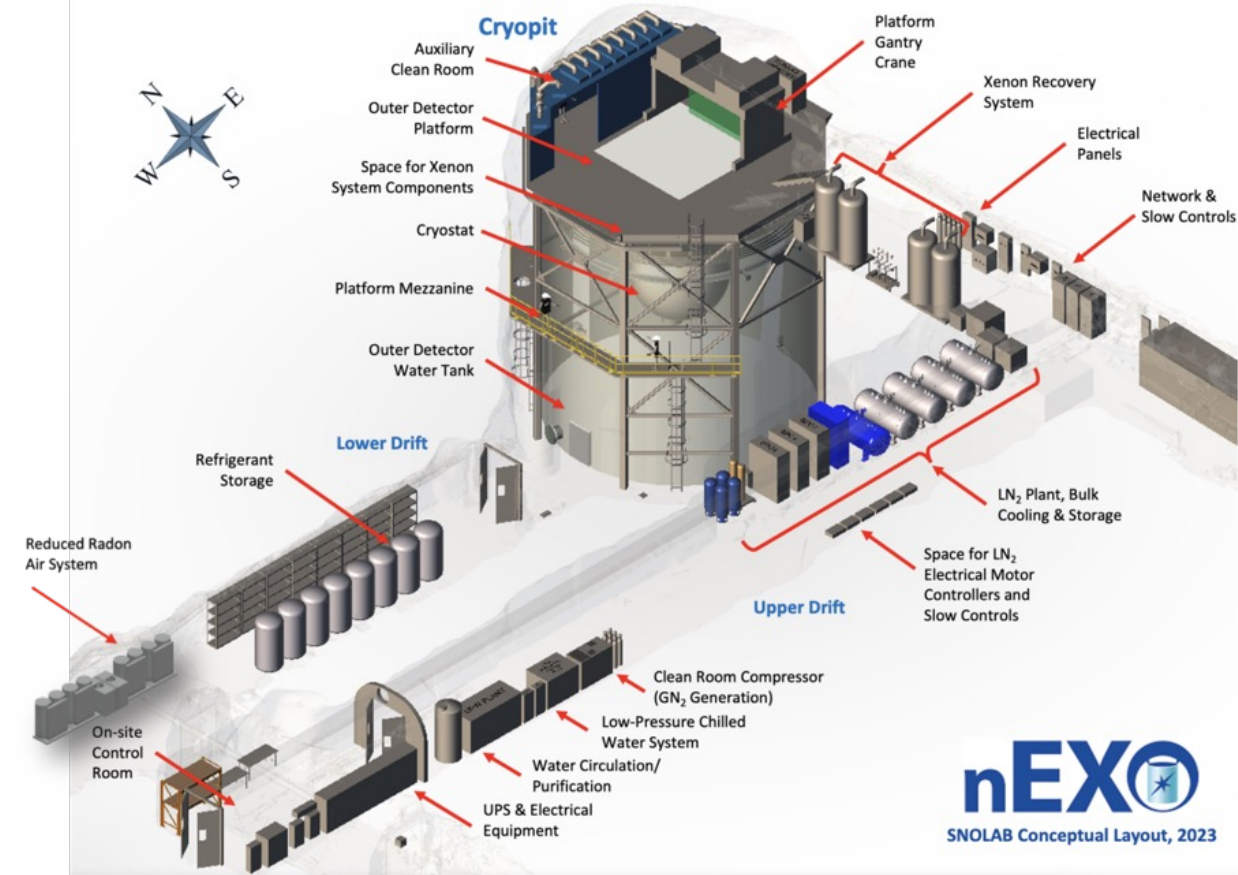}
  \caption{Engineering drawing of the proposed layout of nEXO in the
    SNOLAB Cryopit.}
  \label{fig:nexo}
\end{figure}

SNOLAB, as one of the world’s deepest and cleanest underground labs,
is an ideal host site for nEXO.  Its Cryopit has been 3-D modeled to
accommodate the development of the full detector and infrastructure
layout (see Fig.~\ref{fig:nexo}).  The Canadian team is actively
engaged in subsystem development and has secured CFI IF awards (2020
and 2023) for infrastructure critical to hosting and developing nEXO.

SNOLAB will serve as both the host site and a lead institution,
providing project management, safety certification, and logistics
support.  SNOLAB project management has enabled nEXO to quickly turn
around and develop a Canadian-led budget.  By hosting nEXO at SNOLAB,
Canada reinforces its position as a global leader in rare-event
physics and strengthens its international research partnerships, while
fostering economic development through high-tech manufacturing,
scientific infrastructure, and knowledge transfer.

\section{International facilities}

\subsection{CERN (Antiproton Decelerator and ELENA)}

\begin{SCfigure}
\centering
 \includegraphics[width=0.65\textwidth]{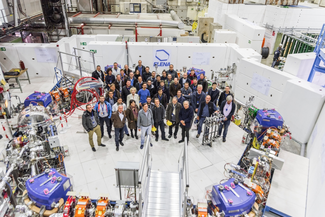}
 \caption{ELENA, a newly developed antiproton decelerator ring at CERN.}
 \label{fig:elena}
\end{SCfigure}

\begin{figure}[h!]
\centering
 \includegraphics[width=0.65\textwidth]{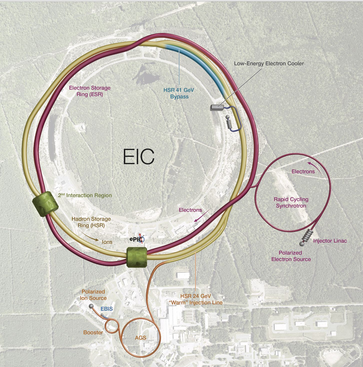}
 \caption{Layout of the EIC, with the ePIC detector and crab cavities in the 6~o'clock position.}
 \label{fig:eic}
\end{figure}

The Antiproton Decelerator (AD) at CERN is the world’s only facility
which provides low energy antiproton beams.  ELENA — Extra Low ENergy
Antiproton — is a major upgrade to the AD that was completed in 2021
(Fig.~\ref{fig:elena}).  It further slows antiprotons from the AD (5.3
MeV) to the 100 keV regime, while serving the beams simultaneously to
up to four experiments. The advent of ELENA will ensure antiproton
physics opportunities at CERN for the next 10+ years.

ALPHA–Canada is the single largest group in ALPHA, consisting of about
1/3 of the international collaboration.  The ALPHA-Canada program is
described in Section \ref{sec:alpha}.  In the medium-term (2026–2030),
taking advantage of CERN's Long Shutdown, a significant upgrade to the
ALPHA apparatus is planned -- ``ALPHA Next Generation'' is currently
under review by CFI. This will enable simultaneous confinement of and
measurements on antihydrogen and hydrogen, allowing direct comparison
experiments, and alleviating key systematic uncertainties. In the
2030–2034 period, the atomic fountain technique is proposed to be
deployed by ALPHA at CERN, initiating an antihydrogen quantum sensing
program—breaking new ground in antimatter physics. This program will
continue well into the period beyond 2034.

\subsection{Electron-Ion Collider (EIC)}

The Electron-Ion Collider (EIC) \href{https://www.bnl.gov/eic/} is a major new collider facility underway at Brookhaven National Laboratory in New York state, by the US Department of Energy.  Canadian physicists have participated intensively in the planning of this facility
and have chartered the EIC-Canada Collaboration to coordinate participation.
Canadian participation in the Electron-Ion Collider will focus
on detector design and physics program development (present-2028), detector construction (2027–
2032), and operations (2034 and beyond).  A recent EIC project timeline is given in Fig. \ref{fig:eic:schedule}.

The Electron-Ion Collider (EIC) will consist of a polarized electron ring with a variable beam energy from 10 to 18~GeV, and an ion ring with a variable beam energy from 41 to 250~GeV, allowing for beams of polarized protons, deuterons and $^3$He, as well as unpolarized nuclei up to lead and uranium. This range of energies will allow for center of mass energies $\sqrt{s}$ from 20 to 140~GeV with a collision luminosity $\mathcal{L}$ of 10$^{33-34}$~cm$^{-2}$ s$^{-1}$ (optimal luminosity of 10$^{34}$~cm$^{-2}$ s$^{-1}$ at $\sqrt{s} \approx 105$~GeV, about 1000 times larger than at HERA, the only previous electron--proton collider). Despite the high luminosity, the interaction rate and multiplicity/occupancy rate are manageable compared to the proton--proton collisions at the Large Hadron Collider (LHC). Event rates up to $10^5$~Hz per unit solid angle are expected.  The EIC physics program is briefly summarized in Sections \ref{sec:eic} and \ref{sec:eic2}.

The EIC-Canada Collaboration has members from Victoria, TRIUMF, Regina, Manitoba and Mount Allison universities.
EIC-Canada
anticipates that the next 7 years will be a period of growth. 
The start of the first North American collider of this century will be associated with significant scientific interest, and
opportunities exist for subatomic physics groups with detector technology and simulations expertise to join.
In the 2025 CFI-IF competition, 
EIC-Canada submitted a proposal supporting Canadian contributions to the ePIC Barrel Imaging Calorimeter and the superconducting RF cavities.
It is anticipated that by the start of EIC data taking in 2034, the Canadian participation in the EIC will number 21 HQP supervised by an integrated 5.6 FTE investigators and funded researchers.

There is significant synergy in the physics programs of the EIC and JLab. As the EIC program is ramping up, the JLab 12 GeV program continues to take advantage of the energy upgrade completed in 2017.  While EIC-Canada anticipates an increasing focus on the Electron Ion Collider program once it comes online, Canadian participants remain committed to the success of the Jefferson Lab parity program, a unique program world-wide.

\subsection{Thomas Jefferson National Accelerator Facility (Jefferson Lab, JLab)}

Jefferson Lab (JLab) \href{https://www.jlab.org/} is the world’s largest nuclear physics user facility, numbering over 1500 active
users.  Canadians are the third largest international
group at JLab, behind only France and Italy. The JLab 12-GeV Upgrade, enabling a doubling of the available electron beam
energy and the construction of a suite of new detectors,
was completed in
2017.  Canadians have leading roles in several high profile experiments that are either currently acquiring data, or scheduled to acquire data in the near future.  Over the course of this long range plan, Canadian efforts at JLab will primarily include:
JEF - JLab Eta Factory (Section \ref{sec:jef}), the Hall C meson structure program (Section \ref{sec:hallc}), 
the MOLLER experiment (Section \ref{sec:moller}),
and the Solenoidal Large Intensity Detector (SoLID) (Section \ref{sec:solid}).  The CFI-IF program has most recently provided funds in support of MOLLER (2021 competition) and SoLID (2023 competition).  The JLab scientific program has an approved experiment backlog exceeding 7 years, with continuing interest in new experimental proposals submitted.

Longer term, JLab intends two major upgrades of its capabilities.  
\begin{enumerate}
    \item A positron beam upgrade \url{https://arxiv.org/abs/2401.16223/}
    is motivated by the fact that interferences between many reaction amplitudes change sign between $e^-$ and $e^+$ beams, allowing reaction phase structures to be uncovered.  
    The positron beam upgrade would occur FY32--35, with the physics program planned for FY36--38.  
    \item The 22 GeV upgrade \url{https://doi.org/10.1140/epja/s10050-024-01282-x}
    would complement the EIC program, by filling in the gap between the JLab 12 GeV physics reach and the EIC.
    Advanced accelerator techniques, making use of fixed field alternating-gradient permanent magnets, would be utilized for the first time, enabling a near doubling of the JLab beam energy without increasing the linac RF gradient.
    While the positron program is running, advanced R\&D for the 22 GeV upgrade would be carried out, with the intent for construction to occur in FY35--40 and 22 GeV physics to begin in FY41.
    \end{enumerate}
    SoLID, Hall~C and GlueX would all make use of the positron and 22 GeV beams.

\subsection{Major Radioactive Beam Facilities}

\subsubsection{FRIB in the USA}
FRIB is a next-generation facility providing intense in-flight RIBs at intermediate energies, with additional options for experiments with stopped or re-accelerated RIBs. Canadian researchers currently perform experiments studying the structure of very neutron-rich isotopes at FRIB. In the future, experimental techniques developed at TRIUMF to study $(\alpha,n)$ and other astrophysical reactions may be ported to FRIB to take advantage of the complementary RIBs available at that facility.

\subsubsection{GSI Helmholtz Center and Facility for Antiproton and Ion Research (FAIR) in Darmstadt, Germany}
The GSI accelerator facility provides intense in-flight RIBs with energies $>$500 MeV/nucleon. The new FAIR facility with the SuperFRS fragment separator will start operation in 2028. Within the next 7 years, this facility will deliver some of the most intense radioactive ion beams in the world. The GSI facility will continue operation before many experimental setups will transfer into the FAIR experimental halls. One of the programs that will continue at GSI is the ILIMA@ESR program at the Experimental Storage Ring until the new Collector Ring at FAIR is built.
At GSI, Canadians lead a program using the ESR storage ring to perform decay measurements of highly-charged ions for nuclear astrophysics and nuclear structure. A proposal exists to install a neutron target at the planned new CRYRING storage ring facility, which would be a first development step towards the construction of the TRISR ring at TRIUMF (Section~\ref{sec:TRISR}).

\subsubsection{RIKEN in Japan}
\begin{figure}[h!]
\centering
 \includegraphics[width=0.70\textwidth]{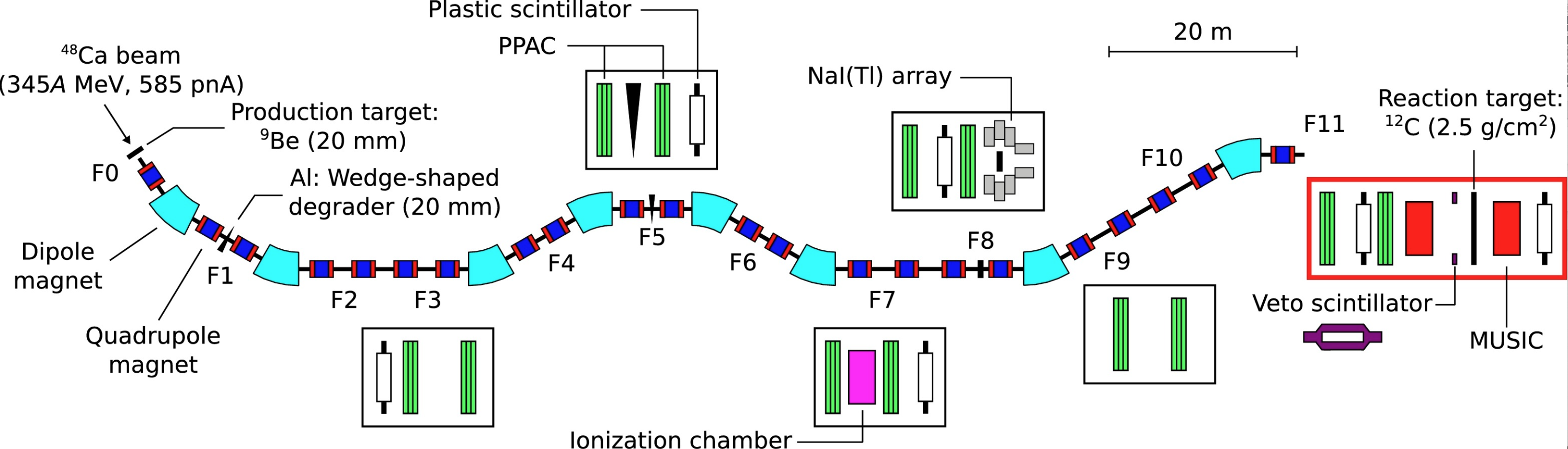}
 \caption{A sketch of a typical experiment at RIKEN showing the path of a $^{19}$B beam produced by the interaction of a 345A MeV $^{48}$Ca beam with a $^{9}$Be target and transported through BigRIPS (F0 to F7) and ZDS spectrometer (F8 to F11) to the $^{12}$C reaction target. Figure taken from Nucl. Phys. A 1053, 122977 (2025).}
 \label{fig:RIKEN}
\end{figure}

 RIKEN operates the RIBF facility (as seen in Fig.~\ref{fig:RIKEN}), providing intense in-flight RIBs at intermediate energies. Within the Canadian community, halo nuclei are studied at RIKEN/RIBF using the BigRIPS fragment separator and the Zero-Degree Spectrometer (ZDS) shown in Fig.~\ref{fig:RIKEN}. Another topic of study at RIKEN within the $\gamma$-ray spectroscopy is the evolution of shell structure in neutron-rich nuclei; for example shape coexistence in doubly-magic $^{78}$Ni nucleus. During the BRIKEN campaign from 2016-2021 the BELEN detector was merged with other neutron detectors to make up the world-wide most efficient neutron detector. Half-lives and neutron-branching ratios for more than 200 nuclei were measured, many of them for the first time.

 The CRIB facility at CNS/RIBF enables both direct and indirect measurements of astrophysically important reactions by producing low-energy radioactive beams in-flight. Typical examples include proton and $\alpha$-resonant scattering using the thick-target inverse kinematics method, as well as studies employing Coulomb dissociation and the Trojan Horse Method.

\subsection{Smaller-Scale Radioactive and Stable Beam Facilities}

\subsubsection{Argonne National Laboratory (ANL), close to Chicago, USA}
 The $\gamma$-ray spectroscopy program at ANL has been centered on the GAMMASPHERE spectrometer consisting of
 110 Compton-suppressed high-purity germanium detectors placed in a spherical arrangement. GAMMAsphere can be coupled with a variety of ancillary detectors and also with the Franment Mass Analyzer for recoil detection. More recently, ANL has also hosted GRETINA, a new type of $\gamma$-ray spectrometer built from large, electrically segmented hyper-pure germanium crystals, providing advanced $\gamma$-ray tracking capabilities.
 Coulomb excitation and fusion evaporation experiments are 
 enabled by combination GAMMASHERE or GRETINA with a suite of ancillary setups like the charged-particle arrays CHICO-II, short for Compact Heavy Ion COunter, ORRUBA short for Oak Ridge Rutgers University Barrel Array, or coupling GRETINA to the Fragment Mass Analyzer (FMA).

 Direct measurements of astrophysically important $(\alpha,n)$ and $(\alpha,p)$ reactions are carried out using the active-target system
MUSIC. These measurements complement those performed at TRIUMF using EMMA, TUDA,
and TACTIC. Studies of reactions near the $N = 50$ shell closure will further benefit from the commissioning of the nuCARIBU neutron-induced fission source.

\subsubsection{Heavy Ion Laboratory (HIL) in Warsaw, Poland}
Similarly to INFN, the Heavy Ion Laboratory in Warsaw  is a center for Coulex and fusion evaporation experiments. Its cyclotron offers a wide range of heavy ion beams and the flexibility of adjusting the beam energy according to the requirements. HIL hosts the EAGLE array (central European Array for Gamma Levels Evaluations), a multi-configuration detector setup which can accommodate up to 30 High Purity Germanium (HPGe) detectors in anti-Compton shields, and various ancillary devices.

\subsubsection{INFN Legnaro close to Padova in Italy}
The National Institute for Nuclear Physics – Legnaro National Laboratories (INFN-LNL) in Italy hosts two major $\gamma$-ray spectrometers, GALILEO and AGATA. These instruments are used primarily for fusion-evaporation and Coulomb-excitation experiments with stable beams, often in combination with advanced ancillary detectors for particle identification and neutron detection.

\subsubsection{Institut Laue-Langevin (ILL) Grenoble in France}
To study excited states in nuclei, one must explore various mechanisms for populating specific states
of interest since no single reaction can populate all excited states. The RIBs produced at
TRIUMF primarily $\beta$-decay, meaning the states which can be populated in the daughter nucleus are limited
based on the $\beta$-decay selection rules as well as the Q-value. An alternative technique is the use of thermal neutron capture at ILL, which only populates states at the neutron separation energy,
S$_n$, which then decay through many levels via $\gamma$-ray emission to the ground state.
At ILL, the most intense continuous flux of neutrons in the world (10$^{15}$ n/(cm$^2\mathrm{s})$ in the reactor moderator) 
is produced from the fission of highly-enriched $^{235}$U. The research reactor
operates at 58 MW and a collimated pencil-like beam (1.5 cm in diameter) of neutrons with a flux
of 10$^8$ n/(cm$^2$·s) is delivered to many experimental stations throughout the facility, including the FIssion Product Prompt $\gamma$-ray Spectrometer (FIPPS) array is the $\gamma$-ray spectrometer used in thermal neutron capture experiments.

\subsubsection{Institute of Nuclear Physics (IKP) at the University of Cologne in Germany}
The group in Cologne is specialized in the recoil distance Doppler-shift (RDDS) method with plungers for the measurement of lifetimes of excited states. 

The group also hosts the Cologne CATHEDRAL (Coincidence Array at the Tandem accelerator for high-efficiency Doppler recoil and LaBr fast-timing measurements) which is the $\gamma$-spectrometer for lifetime measurements of excited nuclear states.

\subsubsection{University of Jyv\"askyl\"a in Finland}
The $\gamma$-ray spectroscopy program in Jyv\"askyl\"a is centered around the JUROGAM3 spectrometer, an array of Compton-suppressed HPGe detectors which has been constructed for use at the target position of the RITU or MARA recoil separators. 

\subsubsection{Other laboratories}
In future, the Canadian teams will seek opportunities to diversify their experiments and pursue experiments at Australian National University in Canberra, Australia and Florida State University and Triangle Universities Nuclear Laboratory in the USA. Other opportunities to perform experiments at smaller University-based laboratories worldwide are also possible in the future.

\section{The Role of the Canadian Institute of Nuclear Physics (CINP)}

The Canadian Institute of Nuclear Physics (CINP) is a formal organization of
the national nuclear physics research community to promote excellence in
nuclear research and education, and to advocate the interests and goals of the
community both domestically and abroad.

The CINP has assumed a central role in the Canadian nuclear physics research community, bringing together nuclear
physics researchers from across the country and providing a common point of
contact for external agencies to communicate with.  There is no other national
organization that can serve in this role.  Prior to the existence of the CINP,
the nuclear physics community had no effective representation to SAPES and ACOT (Advisory Committee on TRIUMF).  Furthermore, the preparation of the LRP was largely carried out
under the CAP-DNP, which was hamstrung by the scant resources available
through the CAP.  The CINP now has both the mandate, and resources, to fulfill
these roles, along with many others.  The CINP is further developing
institutional memory that will greatly benefit the community so that data on
activities and participation can be gathered and analyzed for trends.

The CINP is incorporated under the Not-for-Profit Corporations
Act of Canada, and accommodates both institutional and individual membership.\ \ \ 
{\em Institutional Membership }
is open to any Canadian university, laboratory,
institute or establishment of the Government of Canada which is actively
involved in academic research in nuclear physics.  The CINP currently has 12 institutional members: Memorial University of Newfoundland, Saint Mary's University,
Mt. Allison University, McGill University, the University of Guelph, the
University of Manitoba, The University of Winnipeg, the University of Regina,
the University of Calgary, the University of Northern British Columbia, Simon Fraser University, and TRIUMF.  This is an increase of 3 institutional members since 2020, reflecting the growing importance and scope of the CINP.\ \ \ 
{\em Individual membership} is open to all Canadian nuclear physics
researchers, regardless of whether or not they are employees at a CINP member
institution.  Individuals who are eligible to apply for NSERC research grants
usually hold ``faculty-class'' individual membership (some exceptions are made
for faculty emeriti), while PDFs, grad students, and other non-NSERC eligible
researchers qualify for ``associate-level'' membership.  The CINP currently has 192 (+72) individual members, 
93 (+16) at the faculty level and 99 (+56) at the associate level, where the numbers in parentheses indicate the increases since 2020.

An important activity of the CINP is the organization of Scientific
Working Groups (SWGs) that facilitate collaboration among researchers with
common interests, plan future projects, and enhance the profile of specific
research areas within Canada. 
The variety of problems under study in
nuclear physics makes it a field where different innovative techniques are
continuously developed and employed, e.g., in the construction of detectors and
spectrometers.
This is facilitated by collaboration and coordination.  
The Subatomic Physics Long Range Plan (SAP-LRP) process
periodically focuses the attention of the community on its goals
in a wider sense. 
However, there is a need to provide such a perspective on a
continuing basis, especially in such a rapidly developing field as nuclear
physics.  The Canadian
Institute of Nuclear Physics fulfills that role, and it is unique in this
capacity.  The purpose of the CINP is not to
direct the scientific activities of nuclear
physics researchers in Canada in a top-down manner, but rather to facilitate 
new connections and integrate the Canadian nuclear physics efforts into a more
clearly cohesive whole.
The CINP has six SWGs: Nuclear Physics Education and Training, Hadronic Physics/QCD, Nuclear
Astrophysics, Nuclear Structure, 
Fundamental Symmetries, and Nuclear Theory.  The Nuclear Theory SWG was created in 2021, as a result of the CINP community input received for the 2022--2026 long range plan.
Members of the Institute may elect to belong to more
than one of these groups. 

The CINP has several initiatives to strengthen the level and quality of nuclear
physics research in Canada.  The CINP offers undergraduate research scholarships, 
graduate fellowships, as well as undergraduate and graduate
student travel awards to present research at conferences.  It also sponsors conferences and workshops consistent with the nuclear physics thematic area covered by CINP.

The CINP and the Institute of Particle Physics (IPP) have worked together very effectively in advancing the interests of the Canadian subatomic physics community as a whole.  As such, the CINP Executive Director communicates regularly with the IPP Director on issues of common concern to their respective communities.  
An example of such cooperation is this Canadian Subatomic Physics Long Range Plan, which we co-sponsor.  Another example is our coordination of communications to federal agencies (NSERC, CFI, ISED, and the Office of the Chief Science Advisor (OSCA)), including our periodic visits with officials in Ottawa.  The issues raised in Recommendation \ref{recommend:Support}
(Section \ref{sec:prio_reco}) reflect our joint discussions, in addition to the feedback we have received through our LRP consultation process (Appendix \ref{sec:CINP_proc}).  

To give some broader context, the IPP is roughly twice the size of CINP in terms of total number of PIs.  As such, the costs for this LRP are shared between CINP and IPP in a 1:2 ratio.
Canadian subatomic physicists are free to hold individual membership in both the CINP and IPP if they so choose.  Of the 43 briefs received for this strategic report, to the best of our knowledge 8 were submitted to both institutes.  These are: nEXO 2.0, MOLLER, EDM$^3$, PIONEER, quantum sensors, superconducting magnets for accelerators, and TRIUMF accelerator and computing groups.  Of the projects listed in this report, only MOLLER is an official IPP project.  This means that it is eligible to receive IPP Research Scientist support, although no such support has been provided, to date.

Both CINP and IPP receive funding through the NSERC-MRS program. However key differences in the strength and scope of support should be noted. Firstly, CINP funding is $<$10\% of that 
received by IPP. Additionally, the option to have a Research Scientist funded through the SAP 
envelope is not open to CINP, as is the case for IPP. 
%
For any faculty-level research scientist position, the ability to apply for funds and build up an independent research program is vital to attracting good applicants.  However, if NSERC funds are used to pay for any part of a researcher's remuneration, that person is ineligible to apply or co-apply for NSERC grants.  The IPP Research Scientist program received an exemption from this rule many years ago, when the circumstances were very different.  This causes an inequity between particle physics and nuclear physics experiments in Canada, places significant strain on the SAP envelope, and ``locks in'' the support of certain programs for a very long time frame.

The CINP is looking forward to working constructively with IPP towards a long-term solution of benefit to the entire subatomic physics community.
A good model might be offered by bridge faculty positions.  This would allow a strategic building of highly promising research areas within Canada in a more economical fashion, and would avoid the long term commitment of funds to a specific research program.  Either way, new funds would need to come from an agency other than NSERC, to reduce strain on the SAP envelope and avoid the applicant exemption problem. An ideal forward-thinking solution would be a national subatomic physics institute, such as is foreseen in Recommendation \ref{rec:SAP_institute}; 
this would be a particularly attractive prospect for the whole Canadian subatomic physics community to self-organize and direct its resources in the way that it deems appropriate.

\clearpage
\chapter{Project grids, budget scenarios, recommendations}

\section{Project grid and timelines}
\label{sec:prio_grid}

Based on the briefs submitted by project investigators, we have prepared several schematics illustrating the projected impact and scope, duration, timeline, and investigator commitment of all projects within the Canadian nuclear physics portfolio. 

\begin{figure}[t!]
\centering
  \includegraphics[width=0.825\textwidth]{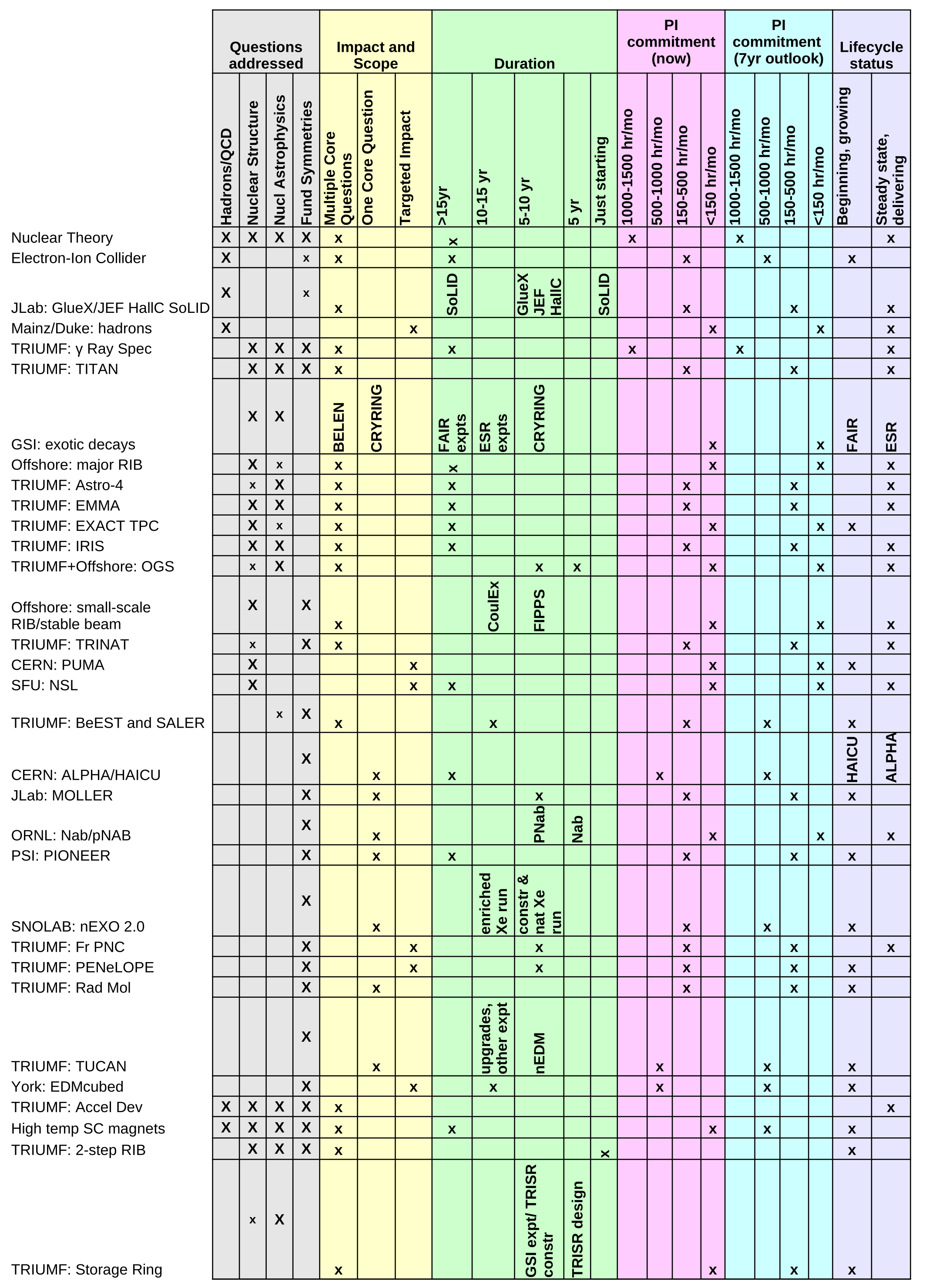}
 \caption{\label{fig:assessment_grid}
Schematic representation of the Canadian nuclear physics research portfolio according to big questions addressed, impact and scope, duration, current and projected PI commitment, and lifecycle status. An {\bf X} marks major thematic relevance and {\bf x} minor overlap.
Projects addressing similar sets of Big Questions are grouped together, in alphabetical order.}
\end{figure}

All projects were evaluated according to a consistent set of criteria, which is indicated in Fig. \ref{fig:assessment_grid}.  
As will be discussed in Sec. \ref{sec:prio_reco}, one of our highest recommendations is to maintain a diverse research program covering all of the Big Questions of Nuclear Physics (Chapter~\ref{sec:big_questions}) in all budget scenarios.
The first set of columns in grey indicates which of the Big Questions are addressed by the respective project, where larger program contributions are given by {\bf X} and smaller contributions by {\bf x}.  
The second set of columns in yellow addresses the projected scientific impact and scope of each project.  The categories are: projects with broad physics outcomes addressing multiple Big Questions, projects with strategic physics outcomes addressing primarily a single Big Question, and high-quality measurements of more limited scope.

The third set of columns indicates the projected duration of the project;
if different parts of the project have different timelines they are named separately.  The fourth and fifth columns indicate the primary investigator (Canadian grant eligible) commitment to the project, both  now and 7 years into the future; these provide a good indication on whether the investigator commitment is expected to grow substantially in the future.  The final column indicates the lifecycle status of the project.

\begin{figure}[t!]
\centering
  \includegraphics[width=0.825\textwidth]{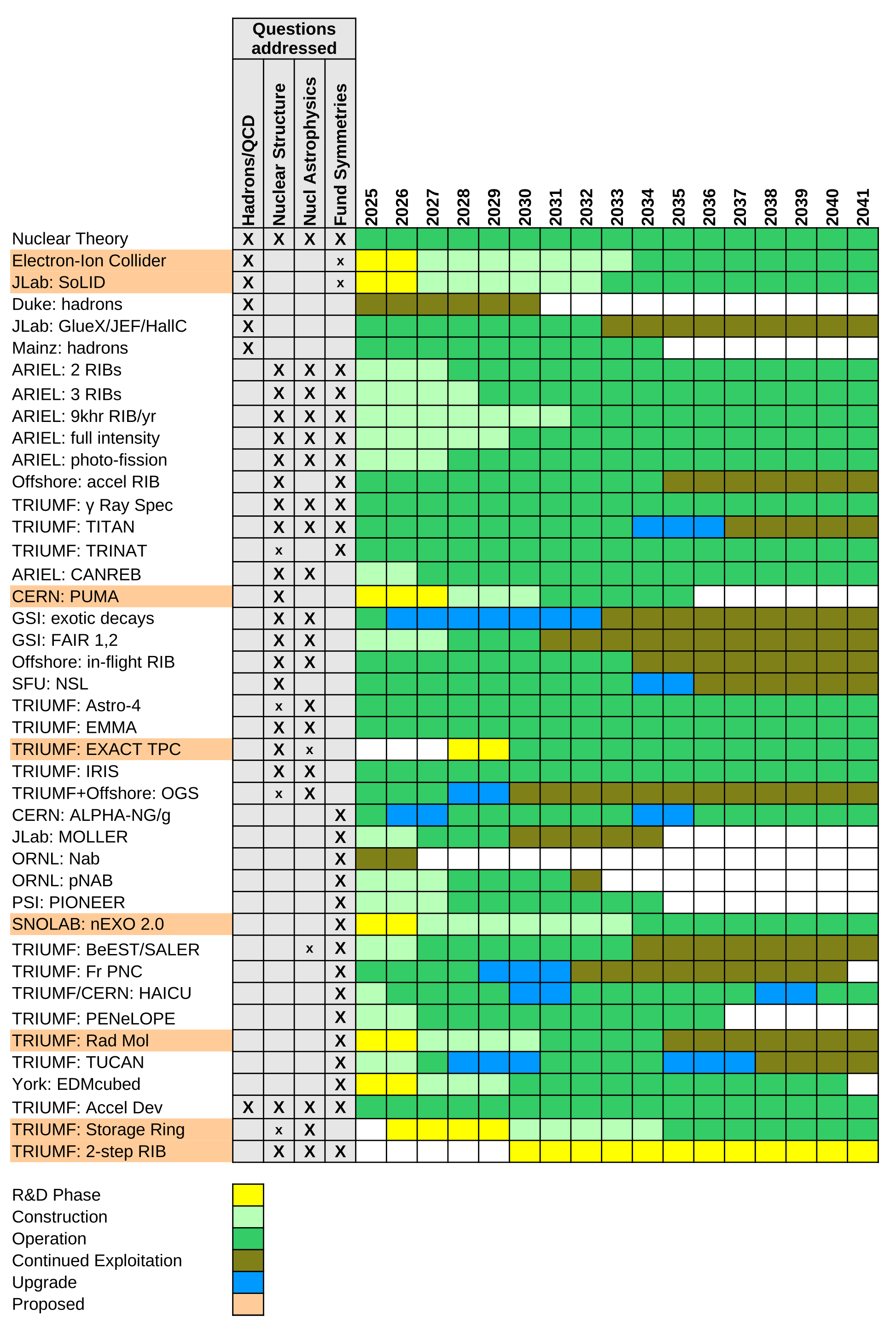}
 \caption{\label{fig:timeline}
Schematic representation of the Canadian nuclear physics research portfolio, for current and proposed projects, and projected timelines at the time of writing.
Projects addressing similar sets of Big Questions are grouped together, in alphabetical order.    The calendar year (Jan 1-Dec 31) is used.}

\end{figure}

Fig.~\ref{fig:timeline} gives the projected timeline for each project from now through the next LRP span, 2025-2041.  Planned projects that have not yet received their full project approvals are indicated in beige on the left, and anticipated future upgrades are given as blue in the timeline.  R\&D, construction, operation, and continued exploitation phases of each project are also indicated by various shades of yellow, light green, green and olive green, through to the end of the next LRP.
In certain cases, projects listed in a single row of Fig.~\ref{fig:assessment_grid} are divided into multiple rows of Fig.~\ref{fig:timeline}, if different parts of the larger project have substantially different completion or upgrade timelines.  This table provides a useful indication of when major capital requests for upcoming projects may occur. This also gives a clearer view of when some current projects are expected to be completed, providing capacity for newer projects as they come online in the future.

\clearpage

\section{Future budget needs}
\label{sec:prio_budget}

The terms of reference for the Canadian Subatomic Physics Long Range Plan require: {\em Budgetary estimates, both for new capital investments and for operations, must be provided,
including funding ranges for prioritized endeavours. These ranges should include funding levels
that would allow for a restrained, yet efficient, contribution to the ventures, as well as levels that
would enable a more extensive contribution. The scenarios also need to address the intellectual
capacity available in the community. The overarching goal needs to be to facilitate coalescence
around the most impactful projects.}  Thus, it is important for this strategic report to provide an assessment of the future budget needs of the Canadian nuclear physics research community.

\subsection{Operating funds
\label{sec:operating}}

\begin{figure}[b!]
\centering
  \includegraphics[width=0.7\textwidth]{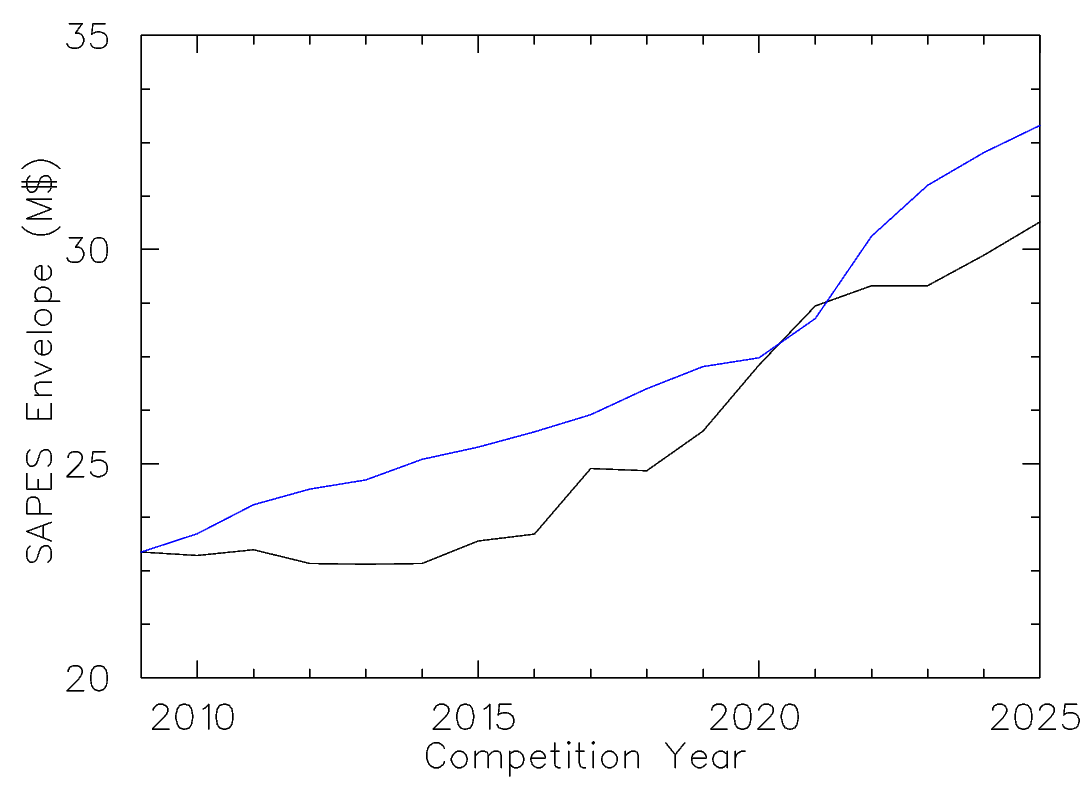}
   \caption{\label{fig:envelope}
Black curve: NSERC Subatomic Physics (SAP) base envelope, as taken from SAPES annual reports \url{https://cinp.ca/subatomic-physics-evaluation-section-chairs-reports}.  Blue curve: Canadian Consumer Price Index 
taken from \url{https://www150.statcan.gc.ca/t1/tbl1/en/tv.action?pid=1810000501}, normalized to the envelope value in 2009.
The difference in the integral of these two curves represents a large cumulative inflationary erosion of envelope resources, corresponding to missed scientific opportunity.  The Canadian fiscal year (Apr 1-Mar 30) is used.}
\end{figure}

As will be discussed in Sec. \ref{sec:prio_reco}, our highest recommendation is to increase the size of the subatomic physics funding envelope.  As part of making the case for increased funding,
Fig. \ref{fig:envelope} presents the current operating funds situation in comparison to the Canadian Consumer Price Index since 2009.  The increase in the SAPES envelope over this period averages significantly less than the accrued inflation, as indicated by the difference between the black and blue curves in the figure.  This has led to a significant cumulative operating funds deficit over this period.
Fortunately, there has been a recent funding increase, as
announced in the 2024 federal budget, of 2.4\% in the SAPES envelope from 2023 to 2024, and a further 2.6\% increase from 2024 to 2025.  While these new funds are deeply appreciated, they come nowhere close to addressing the cumulative deficit since 2009.  

\begin{figure}[htb!]
\centering
   \includegraphics[width=0.99\textwidth]{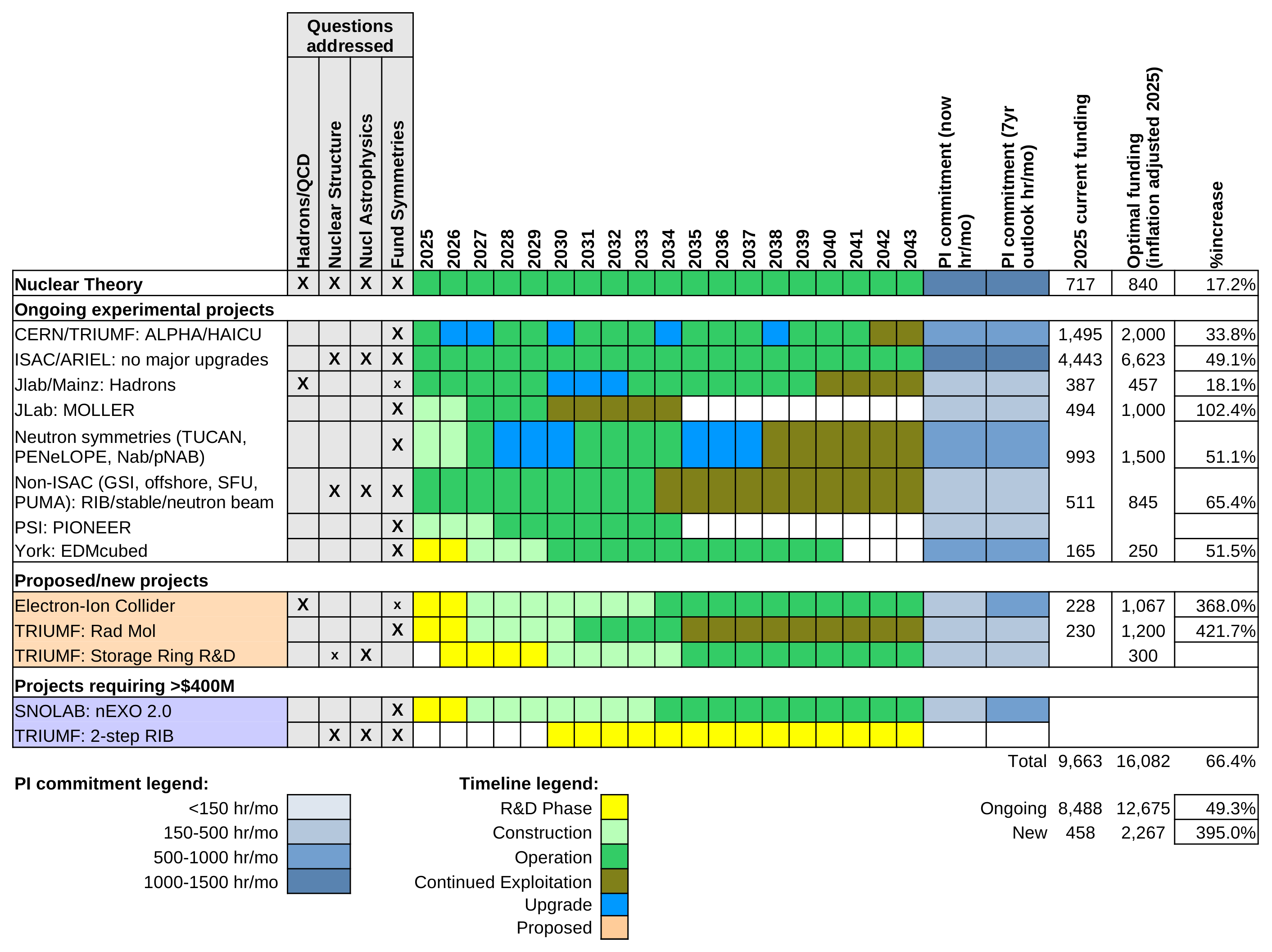}
    \caption{\label{fig:projectgrid}
Schematic representation of the Canadian nuclear physics operating fund needs, divided into: ongoing projects, new initiatives, and large scale investments.  Within each category, projects are listed alphabetically.  The project timeline is based on Fig. \ref{fig:timeline}.
At the far right are the estimated investigator time commitments now and 7 years from now (as provided to us in project briefs), and the current and optimal operational funding/year.  Please see the text for additional discussion.
}
\end{figure}

To gain some indication of the amount of increased funding that is needed by the Canadian nuclear physics research community, 
the project briefs to CINP solicited both the current funding for each project, and the investigator's estimate of what would constitute optimal funding for 2027-2034 (in inflation-adjusted 2025 dollars).  This is indicated in Fig. \ref{fig:projectgrid}, where we have grouped together those projects addressing similar Big Questions at comparable facilities.  Note that we have not independently vetted these numbers.  Overall, the Canadian nuclear physics community self-assesses an increase of funds of 66.4\% over the next 7 years to achieve an optimal situation (bottom right in the figure).  In inflation adjusted dollars, this corresponds to an increase of 6.5\%/year, which is similar to the 7.1\%/year projected in the 2020 CINP Brief.  Assuming 2\% annual inflation (the Bank of Canada target inflation), this would correspond to 8.5\%/year.  This is the ``optimal scenario'' indicated in green in the figure.  This demonstrates the strong need for additional operating funds, as we discuss further in Section \ref{sec:prio_reco}.

All of the projects indicated in the figure are producing important science, and in the optimal budget scenario they would each be funded at an optimal level, including necessary funds for expansion and upgrades.  In a more constrained scenario, projects will be funded at a less than optimal level, particularly with regards to project expansion or detector upgrades, leading to a loss of scientific output and non-optimal exploitation of prior investments.  The timelines for each major topic are indicated in the right half of Fig. \ref{fig:projectgrid}.  The projected timeline information is only approximate, but it gives a guide as to when these expansionary/upgrade pressures may be most pronounced.

The projects in Fig. \ref{fig:projectgrid} are categorized into four broad categories according to where they are in their respective lifecycles.
\begin{itemize}
    \item \emph{Nuclear Theory:} Canadian theory efforts cover all of the Big Questions of nuclear physics, and many theorists work closely with experimental collaborations. Theory is an integral part of the Canadian nuclear physics landscape. It is essential that nuclear theory receives enhanced funding from the subatomic physics envelope, as we discuss further in Section \ref{sec:prio_reco}.
   \item \emph{Ongoing Experimental Projects:} This category contains ongoing projects that are producing world-leading science
   and doing substantial HQP training. 
   ISAC/ARIEL experiments have the largest investigator commitment, and largest share of the current operational funding, and are projected to remain active during the full LRP term.  
   A few projects are expected to be completed in the coming 10-15 years, with the investigator effort moving to other projects in the table.
   
   The optimal operating funds in the ongoing category vary from increases of 18 to 102\%, with most in the 50\% range.  
   At a minimum, it is essential that each of these projects receive continued funding in order for Canada to maintain its status as a global leader in nuclear physics. At the same time, many of these projects require a funding expansion for upgrades and new developments.  The associated capital needs are discussed in more detail in Section \ref{sec:capital}.
    \item \emph{Proposed/New Projects:} This category contains major new initiatives requiring significant capital investment in order to become a reality.  They anticipate a need of substantially increased operating funds ($>$300\%) over the next 7 years, corresponding to growing projects near the beginning of their lifecycle.  Each of these projects promises significant and unique advances in the scientific output of the Canadian nuclear physics community. It is precisely these kinds of projects that make a case for modest-to-aggressive growth of the envelope.  That said, in a constrained scenario it may not be possible to fund each of these projects at their ideal level.  Their ramp-up will be heavily constrained in a greatly constrained financial scenario, corresponding to a loss of Canadian leadership and lost scientific opportunity in these projects.
\item \emph{Extraordinary Capital Projects:} This category contains the largest-scale projects, requiring $>$\$400 million in funds. These are bold, blue sky initiatives.  They may require either multiple CFI-IF awards to be implemented in full, special funds allocated directly by the Government of Canada, significant international contributions, or a combination of all of the above.
These are presumably the future flagships of Canadian nuclear science, bringing the best Canada has to offer to the international scientific stage. The continued support of major initiatives such as these is integral to maintain Canada's nuclear physics leadership well into the future.
\end{itemize}

\subsection{Major capital investments
\label{sec:capital}}

\begin{table}[ht!]
\begin{center}
\begin{tabular*}{5.99in}{|p{3.0in}|p{0.8in}|p{0.65in}|p{0.85in}|}
\hline
  &  CFI Capital (\$k) & Other Canadian (\$ k) & International Contribution (\$k) \\
\hline
\underline{Extraordinary capital}             &        &         &   \\    
nEXO 2.0 global project (anticipated 2025-26) & 25,507 & 434,100 & 14,020\\
TRIUMF 2-step RIB projected 2035)             &   TBD  &  TBD    &    \\
\hline
\underline{Large scale capital}               &        &         &   \\     
ALPHA Next Gen (CFI-IF submitted 2025)        & 6,598 & 6,598   & 3,108  \\ 
Anti-hydrogen fountain (CFI projected 2029)   &   TBD  &     TBD &   \\ 
EIC (CFI-IF submitted 2025)                   &  5,337 &   5,337 & 21,432\\
EXACT-TPC (CFI-IF projected 2029)             & 2,000 &     &    \\
RadMol-1 (CFI-IF submitted 2025)              &  6,915 &  10,372 &   400 \\
RadMol-2 (CFI-NFRF projected 2029)            & 24,000 &         &   \\
TUCAN (CFI-IF submitted 2025)                 &  9,500 &  12,204 &  3,400\\
TUCAN (CFI-IF projected 2035)                 &   TBD  &   TBD   &   TBD \\
TRISR (CFI-IF projected 2031)                 & 40,000 &         &   \\
\hline
\underline{Mid scale capital}                 &        &         &   \\
OGS array (CFI-IF projected 2029)             &    500 &         & 1,000 \\
LaBr3 array (CFI-IF projected 2029)           &    800 &     1,200 &   \\
Fr-PNC upgrade (CFI-IF anticipated 2030)      &    TBD &     TBD &   \\
TITAN (CFI-IF anticipated 2035)               &    TBD &     TBD &   \\     
\hline
\end{tabular*}
\end{center}
\caption{\label{table:capital} 
Projected capital needs for projects covered by this report. 
  Please note that we have not independently vetted any of these numbers, they are given 
  to us as a result of community consultations.  `Other Canadian' typically refers to provincial or NSERC-RTI funds.  `International Contributions' indicate the
  value of in-kind contributions or cash investments by foreign partners.}
\end{table}

Table~\ref{table:capital} lists the projected capital needs of the Canadian nuclear physics community, including CFI-IF and CFI-NFRF expected requests, other Canadian (typically provincial or NSERC-RTI) and international partner contributions.  
The success of these projects depends strongly on the acceptance of Recommendation 1, which is needed to position Canada well in terms of international scientific leadership in the coming decade.

As explained at the end of the previous subsection, two projects are listed as having ``extraordinary'' capital requirements, by which we mean their scope exceeds what can be supported through a single CFI-IF application.  In these cases, we anticipate an alternate funding route.  A recent example is afforded by Canadian contributions to the LHC-HL upgrade, which were funded via a special request to the federal government and not by application to CFI.  For the nEXO 2.0 global project, the current cost estimate for a 5 tonne enriched Xe experiment is \$434.1 M, in addition to the support already provided or applied for via the CFI-IF program.  It is not yet known what level of international commitments to this project will occur, but clearly significant interest is required as part of the case to the federal government.  A cost estimate for the TRIUMF 2-step RIB upgrade can only occur after a significant R\&D effort.  Special construction funds, not divided into separate work packages in multiple CFI-IF applications, would allow this upgrade to proceed more quickly than otherwise.

\section{Recommendations}
\label{sec:prio_reco}

\begin{enumerate}

\item{Top priority: Increasing the NSERC Subatomic Physics envelope:}

In our consultations, CINP received an extensive list of concerns and suggestions regarding the funding process, indicating that the nuclear physics community is encountering a significant number of problems in securing funding.  It is necessary to not only manage current funds in a more effective manner, but also to ``grow the pie''.

The NSERC Subatomic Physics envelope is essential to provide long-term support for national and international research projects in nuclear physics and should be enhanced.  Our analysis in Sec. \ref{sec:prio_budget} indicated a need of 66\% more funds over the next 7 years. There are several pressing issues caused by the current funding situation:

\begin{enumerate}

\item
{\em Competitive remuneration needed for students, PDFs and technicians.}  The SAPES envelope has not kept up with inflation (Fig. \ref{fig:envelope}), with the recent increases not sufficient to relieve the inflationary pressures that have built up. Highly qualified personnel supported by envelope funding continue to be compensated at insufficient levels.   The lack of international competitiveness makes it difficult to retain and attract talent to Canada.  {\em An increase in the subatomic physics envelope, associated with a concerted effort at increasing HQP stipends, would ensure that we can compensate them at the levels they deserve.}

\item 
{\em Operating funds for experimental programs.}
Substantive progress towards the resolution of the big questions in nuclear physics requires the significant efforts of undergraduate and graduate students, PDFs and technical staff.  It is common that the PI and students need to be on-site for an extensive period in support of setup, training, and data taking.  During these periods, these students typically cannot receive Teaching Assistantship support from their universities, and TA-relief funds are required to come from experimental grants.  Often, the support of a postdoc is essential, as it allows for a skilled individual to be stationed on-site to set up experiments in collaboration with lab scientists. In many cases, data analysis is intensive, with several years of effort after data taking needed for calibrations and systematic studies to be completed before the first papers can be published.  {\em It is essential that research funds permit experiments to operate with the critical number of HQP needed for program success.}  The financial needs for optimal operation are summarized in Fig. \ref{fig:projectgrid}, and projected future capital costs in Table \ref{table:capital}.

\item 
{\em Opportunities for training in transformative technologies:}
The field of subatomic physics has historically been at the forefront of developing and applying advanced technologies in the pursuit of fundamental scientific discovery. At present, there is substantial national and international interest in emerging technological domains such as artificial intelligence, quantum science, and superconductors. SAP researchers are playing key roles in both utilizing and advancing these technologies. These activities provide exceptional opportunities for the training of highly qualified personnel (HQP) in areas of strategic importance to Canada. However, such opportunities are presently constrained by the limited availability of NSERC funding. An increase in the SAPES envelope would strengthen Canada’s capacity to develop HQP expertise in these nationally prioritized and rapidly evolving fields.

\item 
{\em Increased RTI funds.}  
The \$150k ceiling for RTI-1 in place since 2003 should be increased, as there is currently a too significant gap between RTI-1 and CFI funding.  More avenues are also needed for funding medium-level equipment projects (e.g. \$100k-\$1M range), which are often passed over for funding but which potentially offer a significant return on investment.  While nominally RTIs of categories 2 and 3 exist, too few are awarded to effectively address this need.  {\em The RTI-1 ceiling needs to be increased, and more avenues are needed for the funding of medium-size projects (up to \$1M).}

\item 
{\em MRS support restored.}  
The MRS-funded infrastructure provides support to a wide variety of subatomic physics programs, and the recently implemented national oversight board ensures broad access and efficient management of these resources.  Despite these clear needs, MRS-funded infrastructure has received significant cuts.  {\em These cuts are alarming, and we require the envelope to expand to enable this support to be restored.}

\end{enumerate}

\item{How Canadian subatomic physics support should be directed.}\label{recommend:Support}

\begin{enumerate}

\item 
{\em Support a diverse program of excellence in experimental and theoretical nuclear physics research addressing the ``Big Questions'' in all funding scenarios:}
Nuclear physics addresses many of the most important scientific questions being pursued today.  We have listed in Chapter~\ref{sec:big_questions} what are considered internationally the big open questions in nuclear physics research.  By making best use of its established expertise and strengths, and seeking to contribute to the fields of greatest scientific opportunity, the Canadian nuclear physics research community has self-selected where to best concentrate its efforts.  Nuclear physics is a many-body problem, and history has shown that its surprises have  come from what might otherwise have been considered to be straightforward measurements, and advances towards the solution of one major question often lead from progress in a complementary area.   Despite this natural evolution and concentration of efforts, we caution it is important not to make the Canadian research contributions too narrow.  {\em It is essential that in all funding scenarios NSERC and CFI support a broad and diverse research program in nuclear physics that leads to high-profile research in the respective sub-fields.}

\item 
{\em Increased nuclear theory support:}
The advancement of nuclear physics is strongly dependent on the interplay between theory and experiment. For example, nuclear theory is indispensable for interpretation of experiments testing beyond the standard model physics that involve atomic nuclei such as measurements of neutrinoless double beta decay, tests of CKM unitary in beta decays, measurements of nuclear electric dipole and anapole moments. Theorists identify promising future directions for the experimental programs, participate in experimental proposals, develop new computational methodology, help to interpret the experimental data, and educate the future generation of researchers in both theory and experiment. The key to successful collaboration between theory and experiment in such areas is close coordination and rapid theory response to the needs of experimental programs. At the same time, excellence in theory depends on diversity of ideas and people, and it is essential to support a wide range of theoretical programs in all regions of Canada.  {\em Given the close linkage between progress in nuclear theory and experiment, it is essential that nuclear theory research be allocated increasing funds, to allow for support of postdocs and graduate students, and to maintain and further develop a leading edge in nuclear physics.}

\item 
{\em Support of small-scale impactful research:}
Support for individual researchers needs to be maintained and strengthened.   For example, in the study of fundamental symmetries, it is anticipated the Standard Model is an incomplete effective field theory; dark matter, dark energy and the matter-antimatter asymmetry all remain unexplained. Large numbers of creative high-risk high-reward small-scale experiments may yet be needed to find this unknown physics.  It is also important that research support be distributed equitably across Canada, without disadvantaging smaller campuses from coast to coast to coast.  {\em Resources should be allocated fairly to smaller projects with scientific promise, doing otherwise undermines Canada’s potential to foster innovation in fundamental science.}

\item
{\em Leverage the scientific opportunities enabled by the completion of ARIEL:}  
The completion of ARIEL will establish a new era for subatomic physics in Canada. Tremendous new opportunities for nuclear structure, nuclear astrophysics, isotope science, and fundamental physics including searches for beyond the Standard Model physics will open, such as the ground-breaking RadMol and BeEST experimental programs. The full exploitation of ARIEL is essential.  {\em Given the support recently announced by the Government of Canada, new positions associated with the opportunities afforded by 
ARIEL (both proton and electron driven beams, as well as theory support) are crucial.
ARIEL beam delivery will be tripled, and the community’s operating funds to acquire, analyze and interpret the data will need to keep pace.}

\item 
{\em Support for Canadians playing integral roles in international research endeavours:}

Subatomic physics has always been a highly international endeavor, a fact reflected in our community, where a significant portion of researchers have come to Canada specifically to participate in this field. Canada's excellent international reputation stems not only from programs based onshore, but also from our involvement in global projects. It should be emphasized that participation in international projects is a direct investment in Canadian universities, institutes, researchers, and highly qualified personnel, rather than merely a blank cheque sent abroad. This investment offers a unique return: it provides us with both informal and formalized representation and control over major international projects, and it fosters goodwill for reciprocal investment in Canadian initiatives.
For example, the major investments into the PENeLOPE neutron lifetime project at TRIUMF have been by the German Research Foundation  (DFG).  These funds supplement those requested from CFI or NSERC RTI.  Conversely, Canadians are making major contributions to projects abroad, such as MOLLER and ALPHA, and conduct experiments at offshore rare isotope facilities that have capabilities complementary to those in Canada.
{\em Given the strong budget pressure within the NSERC SAP envelope, it is essential that Canadian significant contributions to international nuclear physics research be supported.}

Regarding major international projects on the future horizon, we single out two for special mention:

\begin{enumerate}

\item 
{\em 
Electron-Ion Collider (EIC)} will uniquely address profound questions about nucleons (neutrons and protons) and how they are assembled to form the nuclei of atoms.  It will be the first collider to be built worldwide this century, and naturally builds on the decades of Canadian participation in the physics programs at Jefferson Lab and Mainz.  Canadians have been involved in the planning of the EIC program for some time. {\em A substantial involvement in the EIC project will confirm Canada’s leadership role in scientific research and development.}

\item 
{\em nEXO 2.0 global project in Canada} is part of a global strategy to deploy multiple isotope targets with complementary technologies to conclusively probe the full parameter space of $0\nu\beta\beta$ with an intermediate goal of 15 meV.  A phased approach for nEXO 2.0 is proposed, with phase 1 using a 5-tonne TPC filled with natural xenon, and later phases using enriched xenon.  The nEXO 2.0 global project at SNOLAB is an important opportunity for Canada with significant discovery potential.  {\em We encourage the realization of a global experiment utilizing $^{136}$Xe in a timely fashion at SNOLAB.}

\end{enumerate}

\item  
{\em Support for an inclusive Canadian nuclear physics community:}
The Canadian nuclear physics community has made significant progress in the last decade towards a more diverse and representative community.  It is important that this work continue.  Canadian physicists in international collaborations have a unique responsibility to ensure that Canadian values, such as integrity, equity, diversity, and inclusion are promoted.  {\em CINP strongly recommends equitable access to available resources and the promotion of equity, diversity and inclusion in the nuclear physics ecosystem in Canada, recognizing that this will contribute to more robust research and education outcomes. } 

\end{enumerate}

\item{Increased support to enable transformational technologies.}

\begin{enumerate}

\item  
{\em Quantum sensors and high temperature superconductors:}
In recent years, countries around the world have been investing substantial resources in quantum technologies -- in research aimed at commercial applications and in the use of quantum techniques for fundamental physics. For example, the United States, United Kingdom, Germany, and Japan have established dedicated funding programs for applying quantum technologies to fundamental research.  In addition, high temperature superconductor magnet technology is expected to play a major role in the upgrade of accelerator facilities worldwide.  Such superconducting magnets would be capital-intensive pieces of equipment, requiring much R\&D. 
To fully leverage quantum technologies in future subatomic physics research, there should be an increased funding via a combination of increased SAPES envelope, increased CFI funding, and a dedicated funding stream to support national scale development efforts aligning with federal government priorities, such as AMO sensors, superconducting technologies, and cryogenic engineering. Such an increase would enable direct and strategic support for the training of HQP in quantum technologies. The necessary funding mechanisms to support this work should be found to take advantage of this opportunity.

\item 
{\em Opportunities with high performance and quantum computing/AI:} 
Investment in high-performance computing, machine learning, and artificial intelligence is essential to progress in nuclear physics. These tools are rapidly becoming integral to accelerator operations, enabling more efficient facility management, improved data extraction, advanced automation, and optimized performance.   They are also being developed for applications in understanding the complicated signals from highly granular detectors. On the theory front, neural networks have been used to improve many-nucleon wave functions or to classify stars, while first-generation studies of the deuteron on quantum computers have also been carried out. While the current High-Performance Computing (HPC) facilities provided by the Digital Research Alliance of Canada (DRAC) are effectively being used to address current scientific problems (especially if 24/7 services to the research community are provided), future larger-scale computations will be significantly different, as scaling limitations of traditional computing approaches are reached.  Research into the applications of Artificial Intelligence is growing rapidly abroad, and Canada has the advantage of a longstanding strength in quantum information science.  {\em It is essential to have ready access to large-scale high-performance parallel computers, including national access to quantum computing infrastructure located in Canada. In addition, the JELF program should continue to support HPC infrastructure upgrades.}
\end{enumerate}

\item{The vital roles of TRIUMF.}

\begin{enumerate}

\item  
{\em TRIUMF’s infrastructure and international roles:}
TRIUMF is more than just the on-site program.  It also has an important role in supporting Canada’s international subatomic physics program.  It is vital that TRIUMF be provided adequate resources for to act in this capacity, such as detector and accelerator development in support of Canada’s off-shore program, as this benefits a large fraction of the subatomic physics community.

\item  
{\em Potential TRIUMF upgrades both medium-term and long-term:}

\begin{enumerate}

\item 
{\em TRIUMF Storage Ring (TRISR)} will allow long-standing problems in nuclear astrophysics to be addressed and will help to advance nuclear astrophysics simulations and nuclear reaction theory models.  TRISR would be a unique facility, and the connection to the existing ISAC facility would make it possible to take advantage of the existing infrastructure and facility upgrades of ARIEL. TRIUMF is in a unique position internationally to build such a facility, which will have significant impact upon the design of other facilities abroad.

\item
{\em 2-step RIB:} Once ARIEL is fully implemented, the production of neutron-rich nuclides, in particular in the $N=50$ and $82$ regions, through 2-step fragmentation would build upon the intense beams that ARIEL will enable.
This will involve the addition of a heavy ion accelerator with energy $E/A\geq 100$ MeV.  Access to these exotic isotopes would lead to pioneering results in nuclear structure and nuclear astrophysics.  Such a project would be very ambitious, and can only proceed after a significant R\&D effort.
\end{enumerate}

{\em Medium and long-term upgrades, such as the TRIUMF Storage Ring, and 2-step RIB beams at TRIUMF should be supported in a staged approach.}

\end{enumerate}

\item{Funding mechanism improvements.}

\begin{enumerate}

\item{Enhancements to SAPES operation.}

\begin{enumerate}

\item 
{\em In-person SAPES meetings:}  
It has been clearly established that face to face interactions offer richer discussions, allowing for more effective decision making. It is much harder to coherently deliberate on contentious issues via virtual meetings, which tend to lead participants back to their (pre-determined) entrenched  positions. This is particularly important for the international members of SAPES, who cannot develop an understanding of the Canadian context needed for their voting when meeting only online. We believe that
regional site visits by SAPES greatly benefited Canadian projects
(especially smaller ones), so we recommend that these be
reinstated, in addition to competition week taking place in Ottawa every year. Similarly, the expert review committees for large projects 
would benefit from returning to their pre-pandemic 
\textit{modus operandi}.
Other countries have moved back to the superior in-person model, so Canada should not maintain COVID-19 restrictions in perpetuity. {\em We strongly urge SAPES to resume in-person meetings, with the possibility of remote participation when special constraints are at play.  If NSERC is unable to fund these travel costs, then they should be paid from the Subatomic Physics envelope.}

\item 
{\em Enhanced SAPES voting:}
The subatomic physics ecosystem, due to its highly collaborative nature, is rife with conflicts of interest and necessitates foreign evaluators.  Having more than 5 people voting on grants would significantly help this issue.  {\em We recommend an increase to 7 reviewers per SAPES application.}

\item  
{\em Stronger role for academics within NSERC:}
The placement of academics seconded to NSERC, such as has been done at NSF, could allow such personnel to play a stronger role coordinating between NSERC evaluation committees, ensuring continuity of instructions and comments from one committee to the next, as well as provide clarity of instructions, and context on the Canadian research ecosystem.

\item 
{\em Longer Project Grants:}  
5 year project grants would permit a better harmonization between SAP-PJ and SAP-IN timelines and align with the 5 year funding of TRIUMF.  Such long-term, stable funding would provide the most efficient use of resources for both the researchers involved and the peer review process conducted by SAPES and its external expert review committees.  {\em NSERC should extend the maximum duration of Subatomic Physics Project Grants to 5 years for fully operational and ongoing research programs.  This reduction in reviews could then be used to increase the number of voting reviewers on SAP grants without unduly increasing committee workload.}

\item 
{\em SAP envelope management:}
SAPES should have full control of the management of the envelope, including the ability to carry forward funds for upcoming projects foreseen in the LRP.  These in fact are the prime rationales for the creation of both the envelope and the SAP LRP.  The ``need for funds'' criterion is also an important part of SAPES decision making and must be retained.  SAPES should also have the ability to provisionally award funds pending progress, particularly on projects that are ramping up.  It is also essential that SAPES have full control of the duration and amount of funds awarded for both SAP-PJ and SAP-IN grants.  Having an academic seconded to NSERC would help ensure consistent decision-making across different competition cycles.

\end{enumerate}

\item{CFI infrastructure support.}

There is a strong need for a clearer and more agile path to support sophisticated scientific infrastructure upgrades through mechanisms such as CFI.

\begin{enumerate}

\item 
{\em Sufficient CFI funding to sustain the international leadership of Canadian researchers:} Significant infrastructure investments from the CFI over the past two decades have enabled Canadian nuclear physics researchers to play a visible leadership role internationally, both at home and abroad. Examples include the ARIEL facility and the GRIFFIN detectors at TRIUMF, and the MOLLER and ALPHA-g experiments abroad. To maintain this Canadian leadership, current CFI funding levels should be maintained and enhanced.

\item 
{\em More frequent IF competitions:}
As RTI Category 2 and 3 funding is now extremely competitive and difficult to secure, the Innovation Fund is the only viable national source for major equipment funding.  CFI-IF competitions occur only once every 2-3 years.  As a result, it can take up to five years, even if successful, to secure the necessary support to begin impactful experimental work. This timeline is too long to remain internationally competitive. For comparison, European Research Council (ERC) grant competitions are held annually, allowing for significantly faster mobilization of new research initiatives.  {\em Predictable and more frequent (annual) CFI Innovation Fund timelines are essential to allow planning for submissions in the context of large international collaborations.}

\item 
{\em Flexibility to handle IF purchases that are not off-the-shelf:}  
Many experimental facilities increasingly require the integration of emerging technologies. Upgrades are no longer straightforward replacements but involve R\&D projects employing modern techniques (including machine learning, advanced controls, and energy-efficient systems) that require sustained investment to maximize impact and maintain technological sovereignty.  {\em When IF funds are awarded, it is essential that funding rules are sufficiently flexible to allow the necessary R\&D for program success.}

\item 
{\em Need for enhanced mid-scale CFI funding:}
CFI opportunities may become critical to obtain \$1M-scale upgrades not available through RTI grants.  The JELF program in its present form is not sufficient to fill this need, as it is primarily targeted to starting researchers and Canada Research Chairs.  {\em CFI should provide a national-wide competition for infrastructure costing below the IF threshold and open to researchers in all stages of their career.}

\end{enumerate}

\item  \label{rec:SAP_institute}
{\em A better mechanism is needed for large-scale, long-term projects:}
There is often ambiguity on ``who speaks for Canada'' in large international projects, with these roles being typically shared between NSERC, CFI and ISED.  Possibly the new ``umbrella organization'' announced in the 2024 federal budget will play this role, but that is yet to be seen.  Furthermore, there is still much uncertainty on how the new major research facilities (MRF) paradigm will be implemented, and in particular whether it will be appropriate for TRIUMF, SNOLAB and McDonald Institute.  Thus, there is a need for a national body that can provide life-cycle stewardship of major projects in a coordinated manner that also has the authority to speak for Canada with our international scientific partners.  This body could also permit investments relevant to quantum technologies in subatomic physics to be made.  Canada's peer countries use a variety of models to achieve these goals, with IN2P3 in France or INFN in Italy providing valuable experience on what works (and what doesn't).  {\em Canada should develop, within ISED, a mechanism to fund large-scale, long-term projects in subatomic physics, potentially in the form of a new national institute with its own dedicated funding and management.  It would also be appropriate to transfer other relevant infrastructures, such as the MRS supported technicians, research scientists, and institutes (including CINP and IPP) to this new entity.}

\end{enumerate}

\end{enumerate}

\clearpage
\appendix
\chapter{CINP's consultation process}
\label{sec:CINP_proc}

The consultation process we followed is outlined below.  
One of the fundamental aspects of the structure of the CINP are the Scientific 
Working Groups (SWG), which facilitate collaboration among researchers with common 
interests, and enhance the profile of specific research areas within Canada.
Each SWG is headed by a Chair, who is elected by its members and appointed by the CINP
 Board.
The SWG Chairs and the Executive Director form the CINP Strategic Report Committee, ensuring
representation from all of the sub-disciplines of nuclear physics.

In 2024, in preparation for the Long Range Plan, the CINP Board and Executive
Director undertook a review and renewal of the Institute's  Scientific Working Groups, including their leadership and terms of reference.  New Chairs were appointed for the Fundamental Symmetries, Nuclear Astrophysics, Nuclear Education and Training, Nuclear Structure working groups.  The Chair for the Hadronic Physics/QCD working group was re-elected to a second term.  The Nuclear Theory SWG was created off-cycle in 2021, following feedback received in the 2020 LRP consultations, so that group was not reviewed.

Due to their prominence in the community, two of the SWG Chairs (P. Garrett and N. Vassh) were also invited to be members of the Canadian Subatomic Physics Long Range Plan Committee (LRPC).  Following prior practice, it was strongly recommended that any LRPC member not in addition participate as a member of any committee that will prepare reports submitted to the LRPC. As a result, CINP carried out consultations within the Nuclear Structure and Nuclear Astrophysics SWGs, finding two replacement members on the CINP Strategic Report Writing Committee.  This committee thus consisted of: Corina Andreoiu (SFU, Nuclear Structure), Svetlana Barkanova (Memorial, Hadrons/QCD), Greg Christian (Saint Mary's, Nuclear Astrophysics), Alexandros Gezerlis (Guelph, Nuclear Theory), Jeffery W. Martin (Winnipeg, Fundamental Symmetries), Ruben Sandapen (Acadia, Nuclear Education), with Garth Huber (Regina, CINP Executive Director) serving as Chair and chief Editor.

A one-day hybrid Town Hall meeting was held immediately following the CAP Congress at the University of Saskatchewan, on June 13, 2025.  Participants were requested to submit a draft written
document on their activities, plans and HQP training before the Town
Hall meeting, and all were given an opportunity to revise their
written briefs afterward, reflecting the discussions at the meeting.  
The response to the call for input was excellent, with 43 briefs received by 
the final deadline.

The Town Hall meeting consisted of plenary sessions, where the SWG Chairs provided 
overviews of the submitted briefs, and proponents of new projects were invited to give 
presentations outlining their plans and opportunities for new research more thoroughly.
There were also breakout sessions, where each SWG Chair led discussions on the plans and 
priorities in each of their sub-fields of nuclear physics.  Finally, there was a plenary 
discussion going over the main points to be emphasized in the CINP Report.

Following this extensive community input, the committee members met several times over 
Zoom in June-July, 2025 to discuss issues and develop a cohesive plan.  All seven members of the
committee met in person at TRIUMF August 5-7, 2025, to prepare the main portions of the report. After
further writing, online discussion, and edits, the first draft of the report was released to the 
Canadian nuclear physics community on August 29.  A virtual town hall meeting to present the report and gather initial feedback was held on September 25, with written comments due October 3.
The committee continued to meet weekly to discuss changes to the report and a second draft incorporating these improvements was circulated on October 30.
A third virtual town hall meeting to gather further input and consensus
was held on November 6, with final comments due November 12.  The report was finalized
 and submitted on November 28.
Given this extensive process, the Strategic Report committee members are
confident that this final document reflects the consensus of the CINP community.

\section{List of submitted briefs}

\begin{itemize}

\item 
Ab-initio nuclear theory for applications in astrophysics and tests of fundamental symmetries, Petr Navratil (TRIUMF)

\item 
ALPHA/HAICU, Contact: Makoto Fujiwara (TRIUMF), Co-Investigators: 
Andrea Capra, Tim Friesen, Dave Gill, Mike Hayden, Alex Khramov, Scott Menary, Taka Momose, Art Olin, Chukman So, Rob Thompson

\item 
ARIEL, Adam Garnsworthy (TRIUMF)

\item 
Atlantic Canada nuclear theory, Contact: Ruben Sandapen (Acadia), Co-Investigator: Mohammad Ahmady

\item 
BeEST \& SALER: Nuclear Recoil Spectroscopy with Rare-Isotope-Doped Superconducting Sensors for High-Precision Tests of the Standard Model \& Novel Searches for keV-Scale Sterile Neutrinos, Contact: Annika Lennarz (TRIUMF), Co-Investigators: T. Brunner, P. Giampa, D. McKeen, V. Radchenko,  W. Rau, C. Ruiz

\item 
L. Caballero group nuclear theory, Liliana Caballero (Guelph)

\item 
Canadian participation in the Electron-Ion Collider, Contact: Wouter Deconinck (Manitoba), Co-Investigators: Michael Gericke, David Hornidge, Garth Huber, Tobias Junginger, Oliver Kester, Bob Laxdal, Savino Longo, Juliette Mammei, Zisis Papandreou, Aram Teymurazyan

\item 
Decay spectroscopy of exotic nuclei at GSI-FAIR, Iris Dillmann (TRIUMF)

\item 
Detailed nuclear structure studies of nuclei on or near the valley of stability, Contact: Paul Garrett (Guelph), Co-Investigator: Corina Andreoiu (SFU)

\item 
Development of Enabling Accelerator Technologies, Contact: Thomas Planche (TRIUMF), Co-Investigators: Oliver Kester, Bob Laxdal, Tobias Junginger, Alexander Gottberg, Jens Lassen

\item 
Direct and Indirect Measurements of Astrophysical Reactions at TRIUMF and International Facilities, Greg Christian (Saint Mary's)

\item
Direct measurements of astrophysical nuclear reactions, scattering and reaction spectroscopy for astrophysics and low-energy benchmarking of ab initio nuclear theory, using the DRAGON, TUDA and TACTIC facilities, Contact: Chris Ruiz (TRIUMF), Co-Investigators: B. Davids, A. Chen, G. Christian, A. Hussein, D.A. Hutcheon,  A. Lennarz,  P. Navratil, N. Vassh

\item 
EDMcubed: High-precision measurement of the electric dipole moment of the electron using polar molecules in a cryogenic argon solid, Contact: Eric Hessels (York), Co-Investigators: Amar Vutha, Cody Storry, Marko Horbatsch, Rene Fournier, Matthew George

\item 
EMMA, Contact: Barry Davids (TRIUMF), Co-Investigators: Corina Andreoiu, Greg Hackman; Adam Garnsworthy, Greg Christian, Alan Chen, Carl Svensson, Krzysztof Starosta, Chris Ruiz (Henderson), Iris Dillmann, Annika Lennarz

\item 
Experimental Probes of Hadron Structure at Jefferson Lab, Contact: David Hornidge (Mount Allison), Co-Investigators: Garth Huber, Zisis Papandreou

\item 
Exploring the exotic nuclear landscape with reactions using in- flight radioactive beams, Rituparna Kanungo (TRIUMF)

\item 
Fundamental symmetry tests with the francium laser trap facility at ISAC, Contact: Gerald Gwinner (Manitoba), Co-Investigators: 
John Behr, Stephan Malbrunot, Jens Lassen, Ruohong Li

\item 
Gamma-Ray Spectroscopy at TRIUMF-ISAC/ARIEL, Contact: Carl Svensson (Guelph), Co-Investigators: 
Corina Andreoiu, Gordon Ball, Barry Davids, Iris Dillmann, Thomas Drake, Adam Garnsworthy, Paul Garrett, Gwen Grinyer, Greg Hackman, Stephan Malbrunot, Krzystof Starosta

\item 
GANIL Active Target and Time Projection Chamber (ACTAR TPC): 2025 campaign at TRIUMF, Contact: Gwen Grinyer (Regina), Co-Investigators: Martin Alcorta, Corina Andreoiu, Adam Garnsworthy, Paul Garrett, Rituparna Kanungo, Carl Svensson

\item 
Hadron Polarizability Experiments at Mainz and Duke, David Hornidge (Mount Allison)

\item
High Performance Computing for Nuclear Theory, Gogko Vujanovic (Regina)

\item 
J. Holt group nuclear theory, Jason Holt (TRIUMF)

\item 
IRIS Reaction Spectroscopy Facility at TRIUMF, Contact: Rituparna Kanungo (TRIUMF), Co-Investigators: Greg Hackman, Barry Davids, Chris Ruiz, Annika Lennarz, Corina Andreoiu, Greg Christian, Gwen Grinyer, Alan Chen

\item
R. Lewis group nuclear theory, Randy Lewis (York)

\item 
MOLLER, Contact: Juliette Mammei (Manitoba), Co-Investigators: Michael Gericke, Savino Longo, Wouter Deconinck, Russell Mammei, Jeff Martin, Svetlana Barkanova, Alex Aleksejevs

\item 
Nab/pNAB, Russell Mammei (Winnipeg)

\item 
nEXO 2.0 Global Project, Thomas Brunner (McGill), Co-Investigators: Erica Caden, Serge Charlebois, Jacques Farine, Nasim Fatemighomi, Razvan Gornea, Jason Hold, Annika Lennarz, Caio Licciardi, Chloe Malbrunot, Fabrice Retiere, Stephen Sekula, Simon Viel, Ubi Wichoski

\item 
Nuclear many-body theory, Alex Gezerlis (Guelph)

\item 
Nuclear matter under extreme conditions, Contact: Gojko Vujanovic (Regina), Co-Investigators: Charles Gale, Sangyong Jeon

\item 
Nuclear Science Laboratory at Simon Fraser University, Contact: Krzysztof Starosta (SFU), Co-Investigator: Corina Andreoiu

\item 
PENELOPE – Precision Experiment on the Neutron Lifetime Operation with Proton Extraction, Contact: Ruediger Picker (TRIMF), Co-Investigators: Pietro Giampa, Noah Yazdandoost

\item 
PIONEER, Contact: Chloe Malbrunot (TRIUMF), Co-Investigators: Thomas Brunner, Douglas Bryman

\item 
PUMA - probing nuclear-surface effects with antiprotons, Stephan Malbrunot (TRIUMF)

\item 
Quantum Sensing for Fundamental Physics, Contact: Makoto Fujiwara (TRIUMF), Co-Investigators: Andrea Capra, Tim Friesen, Pietro Giampa, Annika Lennarz, Chloe Malbrunot, Ruediger Picker, Thomas Planche, Oliver Stelzer, Rob Thompson

\item 
Radioactive Molecules, Contact: Stephan Malbrunot (TRIUMF), Co-Investigators: Babcock Carla, Behr John, Borduas-Dedekind Nadine, Buchinger Fritz, Charles Chris, Gwinner Gerald, Holt Jason, Jamison Alan, Krems Roman, Kwiatkowski Ania, Madison Kirk, Momose Taka, Radchenko Valery, Stolow Albert, Vutha Amar 

\item 
Superconducting magnet technology for supporting research in subatomic physics, Frederic Sirois (Polytechnique Montreal)

\item 
TITAN: precision measurements for nuclear structure, nuclear astrophysics, and fundamental interactions, Contact: Ania Kwiatkowski (TRIUMF), Co-Investigators: Corina Andreoiu, Thomas Brunner, Gerald Gwinner, Stephan Malbrunot-Ettenauer, Rob Thompson, Nicole Vassh, Mike Wieser

\item 
TRINAT TRIUMF neutral atom trap for beta decay, Contact: John Behr (TRIUMF), Co-Investigator: Gerald Gwinner

\item 
TRISR storage ring for neutron captures on radioactive nuclei, Contact: Iris Dillmann (TRIUMF), Co-Investigators: 
Richard Baartman, Alan Chen, Barry Davids, Falk Herwig, Tobias Junginger, Annika Lennarz, Chris Ruiz, Nicole Vassh

\item 
TRIUMF Scientific Computing Activities \& Support, Contact: Reda Tafirout (TRIUMF), Co-Investigator: Wojciech Fedorko

\item 
TUCAN, Contact: Jeff Martin (Winnipeg), Co-Investigators: Ruediger Picker, Noah Yazdandoost, Thomas Lindner, Takamasa Momose, Kirk Madison, Elie Korkmaz, Michael Gericke, Juliette Mammei, Savino Longo, Russell Mammei, Blair Jamieson

\item
Two-step RIB facility with ISOL \& In-flight at TRIUMF, Contact: Rituparna Kanungo (TRIUMF), Co-Investigators: C. Andreoiu, J. Behr, A. A. Chen, G. Christian, B. Davids, I. Dillmann, P.E. Garrett, A.B. Garnsworthy, G. Grinyer, G. Gwinner, G. Hackman, O. Kester, A. Kwiatkowski, R.E. Laxdal, A. Lennarz, S. Malbrunot, C. Ruiz, K. Starosta, C. Svensson

\item 
N. Vassh group nuclear theory, Nicole Vassh (TRIUMF)

\end{itemize}

\chapter{List of Acronyms}
\parindent 0mm

{\bf ACTAR TPC}: Active Target and Time Projection Chamber

{\bf AD (Antiproton Decelerator)}: An antiproton faciity at CERN, currently the only in the world which provides high quality antiproton beams.

{\bf AGB star (Asymptotic Giant Branch)}: Region in the Hertzsprung-Russell diagram. Intermediate-mass stars with 0.6–10 solar masses leave the Main sequence after completion of their hydrogen burning phase and appear as Red Giant star in the Asymptotic Giant Branch region when they ignite helium burning and thus increase their luminosity.

{\bf ALPHA (Antihydrogen Laser PHysics Apparatus)}: An experimental
program at the CERN Antiproton Decelerator which performs antihydrogen
symmetry tests.

{\bf ALPHA-2}: The second generation apparatus of the ALPHA experiment focused primarily on precision laser spectroscopy.

{\bf ALPHA-g}: An experimental apparatus to measure the gravitational
interaction of antihydrogen, which is mostly Canadian funded.

{\bf ALPHA-NG (ALPHA Next Generation)}

{\bf ANL (Argonne National Laboratory)}: A DOE national laboratory in Argonne, Illinois, 
 is home to a number of facilities, including ATLAS the world’s first superconducting linear accelerator for heavy ions at energies in the vicinity of the Coulomb barrier.

{\bf APS (American Physical Society)}

{\bf ARC (Astronomy Research Center)}: Communication  platform  to increase awareness  and  opportunities  in  astronomical  research  at  the  University  of  Victoria. ARC hosts an NSERC-CREATE training program on New Technologies for Canadian observatories

{\bf ARIEL (Advanced Rare IsotopE Laboratory)}: A project to enhance
TRIUMF’s capabilities to produce rare isotope beams and to showcase new
Canadian accelerator technology.

{\bf ARIES (Ancillary detector for Rare-Isotope Event Selection)}: Ancillary detector subsystem of the GRIFFIN spectrometer for beta tagging and fast-timing.

{\bf Astro-4}: A suite of four detector systems for nuclear astrophysics experiments at TRIUMF: DRAGON, TUDA, SONIK, and TACTIC.


{\bf BeEST (Beryllium Electron capture in Superconducting Tunnel junctions Experiment}:  The BeEST (beast) experiment employs the decay–momentum reconstruction technique to precisely measure the 7Be  7Li recoil energy spectrum in superconducting tunnel junctions (STJs).

{\bf BELEN (BEta-deEayEd Neutron detector)} European $^{3}$He-long counter neutron detection array built for operation within the DESPEC collaboration at FAIR. It can be used at both, ISOL and in-flight fragmentation facilities.

{\bf BNL (Brookhaven National Laboratory)}: A DOE national laboratory in Upton, New
York, which is home to a number of facilities including RHIC and the future EIC.

{\bf BRIKEN (Beta-delayed neutron studies at RIKEN)} Large $^{3}$He-long counter neutron detection array with an implantation detector which will take data at RIKEN Nishina Center until 2021. 

{\bf CANREB (CANadian Rare-isotope facility with Electron-Beam ion source)}: A
CFI-funded initiative that will improve the purity of rare ion beams delivered
by ARIEL to ISAC.

{\bf CANS (Compact Accelerator-driven Neutron Source)}: A less expensive, alternative technology to nuclear reactors based on an accelerator-driven neutron source. 

{\bf CAP (Canadian Association of Physicists)}: The national organization representing physicists in Canada, promoting research, education, outreach, and the advancement of physics across the country.



{\bf CERN (Centre European pour la Recherche Nucleaire)}: The European
Organization for Nuclear Research, based in Geneva, Switzerland.

{\bf CFI (Canada Foundation for Innovation)}: Created by the Government of Canada in
1997, CFI makes investments in state-of-the-art research facilities and
equipment in a wide variety of scientific disciplines.

{\bf CINP (Canadian Institute of Nuclear Physics)}: The organization that gathered
input from the Canadian nuclear physics research community in order to put
together this document.

{\bf \textbf{CLAS-12 (CEBAF Large Acceptance Spectrometer for 12 GeV)}}: A large-acceptance, high-luminosity detector in Jefferson Lab’s Hall B designed to operate with the 12 GeV CEBAF electron beam.

{\bf CNO cycle}: Hydrogen burning cycle in the interior of stars involving isotopes of carbon (C), nitrogen (N), and oxygen (O). 

{\bf CNS}: Center for Nuclear Study at the University of Tokyo.


{\bf CP (Charge, Parity) symmetry}:  The combined symmetry of charge (particle-antiparticle) and parity (spatial inversion).  While most interactions seem to obey this symmetry, the weak force is known to violate it, and a similar violation is expected but not yet observed in the strong force (the strong CP problem). 

{\bf CPT (Charge, Parity, Time reversal) symmetry}: A fundamental symmetry of local relativistic quantum field theories.



{\bf CRYRING (Cryogenic Ring)}: a low-energy storage ring for heavy ions integrated into the GSI accelerator complex.

{\bf CRIB: CNS RI beam separator}: An in-flight radioactive beam facility providing low-energy, high intensity radioactive beams for nuclear astrophysics studies.

{\bf DCSB (Dynamical Chiral Symmetry Breaking}: The mechanism by which quark-gluon interactions are expected to dynamically generate most nucleon mass, ultimately accounting for $>98\%$ of the mass of the visible universe.

{\bf DEMAND (Direct Experimental Measurements of Astrophysical reactions using Neutron Detectors)}: An array of OGS neutron detectors for studying $(\alpha,n)$ and $(p,n)$ reactions together with the DRAGON or EMMA recoil separators at TRIUMF.

{\bf DESCANT (DEuterated SCintillator Array for Neutron Tagging)}: A neutron detector array to be used at ISAC.

{\bf DESPEC (DEcay SPECtroscopy)}: Decay spectroscopy collaboration at FAIR.  

{\bf DFT (Density Functional Theory)}

{\bf DOE (Department of Energy)}: The United States Department of Energy, which
operates a number of national laboratories across the USA.

{\bf DRAGON (Detector of Recoils And Gammas Of Nuclear reactions)}: A
detector designed to measure the rates of nuclear reactions important in
astrophysics, based at ISAC-I.


{\bf EBIT/S (Electron Beam Ion Trap/ Source)}

{\bf EDI (Equity, Diversity and Inclusion)}

{\bf EDM (Electric Dipole Moment)}: Permanent electric dipole moments
are a fundamental property of particles and are forbidden by
time-reversal symmetry.

{\bf EDM$^3$ (EDM-cubed)}: Electron EDM experiment performed with
species of interest embedded in noble-gas ice.

{\bf EFT:} Effective Field Theory

{\bf EIC (Electron-Ion Collider)}: A new DOE nuclear physics user facility to be 
constructed at Brookhaven National Laboratory.

{\bf ELENA (Extra Low ENergy Antiproton ring)}: Antiproton cooling and deceleration ring, under construction as an upgrade to the Antiproton Decelerator at CERN.

{\bf EMMA (ElectroMagnetic Mass Analyzer)}: A device being constructed to
study the products of nuclear reactions involving rare isotopes at ISAC-II.

{\bf EOS (Equation of State)}: Thermodynamic equation relating variables which describe the state of matter under a given set of physical conditions, in astrophysics for example for the relation between the pressure and the density, and the resulting radius of neutron stars.

{\bf \textbf{ePIC} \textbf{(electron–Proton/Ion Collider)}} detector: A detector designed for the era of artificial intelligence being built for the EIC.

{\bf ESR (Experimental Storage Ring}: Heavy-ion storage ring at the GSI Helmholtz Center for Heavy Ion Research in Germany.

{\bf EXACT-TPC}: Exotic Nuclei Active Target Time Projection Chamber

{\bf EXO (Enriched Xenon Observatory)}: An experiment seeking to measure
neutrinoless double beta-decay in $^{136}$Xe.  The experiment is currently
located at the WIPP facility in New Mexico, USA. A substantially larger
next-generation detector nEXO is proposed for SNOLAB.

{\bf FAIR (Facility for Antiproton and Ion Research)}: An accelerator facility for
studying nuclear structure and nuclear matter, presently under construction as upgrade of the GSI facility in Darmstadt/ Germany.


{\bf FIPPS (FIssion Product Prompt $\gamma$-ray Spectrometer) is a $\gamma$-ray spectrometer used at ILL Grenoble.}

{\bf FRIB (Facility for Rare Isotope Beams)}: a next-generation facility providing intense in-flight RIBs at intermediate energies, with additional options for experiments with stopped or re-accelerated RIBs.

{\bf FrPNC (Francium Parity Non-Conservation)}: An experiment to study
atomic parity non-conservation in francium, based at ISAC-I.

{\bf GANIL (Grand Accelerateur National d'Ions Lourds)}:
France's national heavy-ion facility specializing in the production of a wide variety of intense stable and rare-isotope beams for nuclear physics research and applied nuclear science. 

{\bf GlueX (Gluonic Excitations Experiment)}: An experiment seeking to identify
hybrid mesons with explicit gluonic degrees of freedom at Jefferson Lab Hall D.

{\bf GDP (Gross Domestic Product)}

{\bf GPD (Generalized Parton Distribution)}:  A framework to better understand hadron structure by representing the parton distributions as functions of more variables, such as transverse momentum and parton spin.  They can be used to study the spin structure of the proton, and will enable a tomographic 3D picture of the proton to be built up.

{\bf GRIFFIN (Gamma-Ray Infrastructure For Fundamental Investigations of
Nuclei)}: Compton-suppressed High-purity germanium clover array for gamma-ray spectroscopy with stopped radioactive ion beams.


{\bf GSI}: Formerly "Gesellschaft fuer Schwerionenforschung", now GSI Helmholtz Center for Heavy Ion Research in Darmstadt, Germany.

{\bf HAICU (Hydrogen-Antihydrogen Infrastructure at Canadian
  Universities} for quantum innovations in antimatter science):
CFI-funded initiative to establish infrastructure in Canada for the
development of quantum sensing techniques for antimatter research,
such as anti-atomic fountains and antimatter wave interferometers.

{\bf HERA}: A former electron proton collider at the DESY laboratory in Hamburg, Germany.

{\bf \textbf{HERMES} (\textbf{HERa MEasurement of Spin)}}: A polarized fixed-target experiment at DESY that studied the spin structure of the nucleon.



{\bf HIGS (High Intensity Gamma-Ray Source)}: Compton gamma-ray source by operated by the Triangle Universities Nuclear Laboratory at Duke.

{\bf HPC (High-Performance Computing)}:  Leveraging large scale computing facilities
or networks towards problems on scales that are prohibitive on individual systems.

{\bf HQP (Highly Qualified Personnel)}: Personnel obtaining advanced skills as a
result of NSERC-funded research, including students, postdocs and technicians.

{\bf HyperK or HK (Hyper-Kamiokande)}:  Future neutrino observatory in Japan involving a considerably larger detector than the previous Super-Kamiokande experiment.

{\bf IAEA (International Atomic Energy Agency)}:  Set up within the United Nations family in 1957 as the world's centre for cooperation in the nuclear field, the Agency works with its Member States and multiple partners worldwide to promote the safe, secure and peaceful use of nuclear technologies.


{\bf IF (Innovation Fund)}: The primary research mechanism for large scale scientific infrastructure funding through the Canada Foundation for Innovation (CFI).



{\bf ILIMA (Isomers, LIfetimes, and MAsses)}: Collaboration for the measurement of lifetimes and masses of radioactive isotopes in storage rings like the ESR at GSI and the new CR at FAIR in Germany. 

{\bf ILL}: The Institut Laue–Langevin is an internationally scientific facility, situated on the Polygone Scientifique in Grenoble, France. It is one of the world centres for research using neutrons.


{\bf IN2P3}: French National Institute of Nuclear and Particle Physics.  \url{https://www.in2p3.cnrs.fr/en}

{\bf INFN}: Italian National Institute of Nuclear and Particle Physics.  \url{https://www.infn.it/en/}

{\bf IPP (Institute of Particle Physics)}: the sister institute to CINP, overseeing particle physics research in Canada.  \url{https://particlephysics.ca/}


{\bf ISAC (Isotope Separator and ACcelerator)}: A rare isotope accelerator facility, based at TRIUMF. There are two experimental halls, ISAC-I and ISAC-II.

{\bf ISED (Innovation, Science and Economic Development Canada}: The ministry of the Government of Canada that is responsible for all federal science funding within Canada, such as NSERC, CFI and TRIUMF.

{\bf ISOL (Isotope Separation On-Line)}: A technique of radioactive ion production in which spallation and fission of thick targets is used to produce a wide range of radioactive nuclei.

{\bf ISOLDE (Isotope Separator On-Line DEtector)}: An On-Line Isotope Mass Separator facility at CERN for the study of low-energy beams of radioactive isotopes .

{\bf JLab (Jefferson Lab)}: The Thomas Jefferson National Accelerator Facility,
located in Newport News, Virginia.

{

{\bf JELF (John R. Evans Leaders Fund)}: A CFI program that enables a select number of an institution’s researchers to undertake research by providing them with the infrastructure required to be or become leaders in their field. 




{\bf LANL (Los Alamos National Laboratory)}

{\bf LHC (Large Hadron Collider)}: The world's largest and most powerful particle accelerator and collider, located at CERN.

{\bf LIGO (Laser Interferometer Gravitational-wave Observatory)}: NSF-funded gravitational wave observatory to detect cosmic gravitational waves and to develop gravitational-wave observations as an astronomical tool. The two LIGO interferometers are located in the USA in Hanford, Washington and Livingston, Louisiana.

{\bf LLNL}: Lawrence Livermore National Laboratory, located in Livermore, California.

{\bf LQCD}: Lattice QCD.

{\bf \textbf{MAGIX (Mainz Academic Gas Internal target eXperiment)}}: A high-precision, internal-target experiment at the Mainz Energy-recovering Superconducting Accelerator (MESA)

{\bf MAMI (Mainz Microtron)}: An electron accelerator facility, located on the
campus of the Johannes Gutenberg University of Mainz, Germany.



{\bf MESA (Mainz Energy-Recovering Superconducting Accelerator)}: A new electron synchrotron under construction on the campus of the Johannes Gutenberg University of Mainz, Germany.

{\bf MINIBALL} - a gamma-ray detector array based at ISOLDE-CERN.

{\bf MOLLER (Measurement Of a Lepton-Lepton Electroweak Reaction)}: An
experiment to measure the parity-violating asymmetry in
electron-electron (Moller) scattering at Jefferson Lab.

{\bf MR-TOF (Multi-Reflection Time-of-Flight)}: A MR-TOF traps ions between two electrostatic mirrors, folding the flight path into an extremely compact device, while achieving resolving powers on the order of 10$^5$.

{\bf MUSIC (MUlti Sampling Ionization Chamber)}: An active target device consisting of a segmented ionization chamber, sensitive to the mass difference between the incident beam and reaction products. 

{\bf Nab/pNAB:}  Experiment measuring the $a$ and $b$ coefficients in neutron decay, conducted at the SNS.  The term pNAB refers to a future upgrade involving polarized neutrons, which will enable the experiment to access the $A$ and $B$ coefficients.

{\bf NASA}: National Aeronautics and Space Administration, founded in 1958, is an independent agency of the U.S. federal government responsible for the civilian space program, as well as aeronautics and space research.



{\bf nEDM}: neutron electric dipole moment (see: EDM).

{\bf NERSC (The National Energy Research Scientific Computing Center)}: The primary scientific computing facility for the Office of Science, US Department of Energy.  It is located at Lawrence Berkeley National Laboratory, in California.

{\bf NLO/NNLO (Next to Leading Order/Next-to-Next to Leading Order)}: increasing
level of complexity of various types of loop diagrams.

{\bf NSAC (Nuclear Science Advisory Committee)}: An advisory committee that provides official advice on basic nuclear science research to the U.S.  Department of Energy (DOE) and the U.S. National Science Foundation (NSF).

{\bf NSCL (National SuperConducting Laboratory)}: Radioactive beam facility at Michigan State University. Now hosting FRIB. 


{\bf NSERC (Natural Sciences and Engineering Research Council of Canada)}: An agency of the Government of Canada that supports university students in their advanced studies, promotes and supports discovery research, and fosters innovation by encouraging Canadian companies to participate and invest in postsecondary research projects.


{\bf NSF}: U.S. National Science Foundation.

{\bf NuPECC (Nuclear Physics European Collaboration Committee)}: Co-ordinates nuclear
physics research planning as an Expert Committee of the European Science
Foundation.

$\boldsymbol{\nu p}$ {\bf Process}: A nucleosynthesis process posited to occur during a core-collapse supernova explosions.

{\bf OGS}: Organic glass scintillator, a new scintillator material with $n$/$\gamma$ pulse shape discrimination capabilities and the ability to be cast into arbitrary shapes.

{\bf ORNL}: Oak Ridge National Laboratory, located in Oak Ridge, Tennessee.



{\bf PENeLOPE (Precision Experiment on the Neutron Lifetime
Operating with Proton Extraction):} A precision experiment to measure the neutron lifetime.

{\bf PIENU:} precision experiment conducted at TRIUMF that measured the ratio of decay rate of pions to electrons as compared to muons.

{\bf PIONEER:} a next-generation rare pion decay experiment, conducted at PSI.

{\bf POLARIS} Polarized Radioactive Isotope Science; a future facility that will produce polarized radioactive beams at TRIUMF.

{\bf PSI (Paul Scherrer Institute)}: the largest research institute for natural and engineering sciences in Switzerland.

{\bf pQCD (Perturbative QCD)}: QCD in the hard-scattering regime, where perturbative
methods can be reliably employed, as opposed to the non-perturbative regime
where they cannot.


{\bf PVES}: Parity Violating Electron Scattering.

{\bf PWA}: Partial Wave Analysis. PWA is a technique for solving scattering problems by decomposing each wave into its constituent angular momentum components and carrying out a spin-amplitude analysis, that includes production and decay elements.

{\bf QCD (Quantum ChromoDynamics)}: The theory describing the fundamental interactions
between quarks and gluons.

{\bf RadMol (Radioactive Molecules):} A facility to search for
symmetry-violating effects (EDMs) in radioactive molecules planned at
ISAC/ARIEL.

{\bf RCMP (Regina Cube for Multiple Particles)}: Ancillary detector subsystem of the GRIFFIN spectrometer for charged-particle spectroscopy.
 
{\bf RCNP (Research Centre for Nuclear Physics)}: A national centre for nuclear
physics, based in Osaka, Japan.

 {\bf RDM} - Recoil Distance Method (RDM) for measuring pico-second lifetimes of nuclear levels.

{\bf RHIC (Relativistic Heavy-Ion Collider)}: A high-energy heavy-ion collider
facility based at Brookhaven National Laboratory.

{\bf RIB}: Rare/ radioactive ion beam.

{\bf RIBF (Rare Isotope Beam Factory)}: A user facility for nuclear science,
located at RIKEN Nishina Center in Japan.

{\bf RIKEN (The Institute of Physical and Chemical Research)}: Japan's largest
comprehensive research institution that performs research in a diverse range of
scientific disciplines, including physics, chemistry, medical science, biology
and engineering.  Founded in 1917 as a private research foundation in Tokyo,
RIKEN has grown in size and scope, and now encompasses a network of research
centers and institutes across Japan.

$\boldsymbol{rp}$ {\bf Process}: A series of rapid proton capture reactions occurring in astrophysical sites such as type-I X-ray bursts.


{\bf SALER (Superconducting Array for Low Energy Radiation)}:
Experiment performing on-line measurements of sub-keV radiation by
embedding the most exotic rare isotopes in superconducting quantum
sensors.

{\bf SAP (SubAtomic Physics)}: The broader field of nuclear and
particle physics, comprising all knowledge taking place at scales
smaller than that of the atom.

{\bf SAPES (SubAtomic Physics Evaluation Section)}: The committee that is responsible for making all funding decisions within NSERC portfolio, including individual grants (SAP-IN), project grants (SAP-PJ), major resource support grants (SAP-MRS), and research tools and instruments grants (SAP-RTI).


{\bf SCEPTAR (SCintillating Electron-Positron Tagging ARray)}: Ancillary detector subsystem of the GRIFFIN spectrometer for beta tagging.

{\bf SECAR (SEparator for CApture Reactions)}: A recoil separator in the re-accelerated beam hall at FRIB.

{\bf SHARC (Silicon Highly-segmented Array for Reactions and Coulex)}: Ancillary detector subsystem of the TIGRESS spectrometer for charged-particle detection.


{\bf SiPM}: Silicon Photo Multiplier.

{\bf SLAC}: The Stanford Linear Accelerator Center, in Stanford, California.

{\bf SM (Standard Model)}: The standard model of elementary particle interactions.

{\bf SoLID (Solenoidal Large Intensity Device)}: A high luminosity, large acceptance
detector for Jefferson Lab Hall A that makes use of the former CLEO
solenoid magnet.

{\bf SNO+}: An experiment under construction at SNOLAB, whose objective is
to use the infrastructure from SNO to study double beta-decay and lower-
energy solar neutrinos using a liquid scintillator instead of heavy water.

{\bf SNOLAB}: An underground science laboratory specializing in neutrino and
dark matter physics, based in Sudbury, Canada.

{\bf SNS (Spallation Neutron Source)}: Accelerator-based neutron source facility at ORNL that provides the most intense pulsed neutron beams in the world.

{\bf SONIK (Scattering Of Nuclei in Inverse Kinematics): A specialized scattering chamber for measuring elastic scattering on gaseous H or He targets.}

{\bf SPICE (SPectrometer for Internal Conversion Electrons)}: Ancillary detector subsystem of the TIGRESS spectrometer for internal conversion electron spectroscopy. 

{\bf SPIRAL 2}: A heavy-ion accelerator facility in Caen, France.

{\bf SRF (Superconducting Radio Frequency)}: Acceleration of charged particles
via the use of superconducting cavities operating in the radio frequency range.
Several examples include the ISAC-II and ARIEL accelerators at TRIUMF, and the
Continuous Electron Beam Accelerator at Jefferson Lab.

{\bf STJ (Superconducting Tunnel Junction)}:  A type of Josephson junction used in a variety of superconducting devices; can be used in nuclear physics as a quantum-enhanced particle detector.

{\bf SuperFRS (Superconducting FRagment Separator)}: A large-acceptance superconducting fragment separator at GSI, with three branches serving different experimental areas including a new storage-ring complex.

{\bf TACTIC (TRIUMF Annular Chamber for Tracking and Identification of Charged particles)}: A device used in conjunction with TUDA.


{\bf tauSPECT:} Experiment on the neutron lifetime conducted initially at Mainz and later at PSI.

{\bf TIGRESS (TRIUMF-ISAC Gamma-Ray Escape-Suppressed Spectrometer)}:
Compton-suppressed and segmented high-purity germanium clover array for gamma-ray spectroscopy with accelerated stable and radioactive ion beams.

{\bf TIP (TIGRESS Integrated Plunger)}: Ancillary detector subsystem of the TIGRESS spectrometer for heavy-ion recoil detection.

{\bf TITAN (TRIUMF’s Ion Trap for Atomic and Nuclear science)}: An ion trap
facility at ISAC for high-precision mass measurements of rare isotopes.


{\bf TPC (Time Projection Chamber)}: Particle detector that uses a  combination of electric fields and magnetic fields together with a sensitive volume of gas or liquid to perform a three-dimensional reconstruction of a particle trajectory or interaction.


{\bf TRINAT (TRIUMF Neutral Atom Trap)}: A device to trap and study the
radioactive decays of neutral atoms, based at ISAC-I.

{\bf TRISR (TRIUMF Storage Ring)}: Proposed low-energy, heavy ion storage ring coupled to the ISAC facility at TRIUMF. 

{\bf TRIUMF}: Canada's national laboratory for particle and nuclear physics, based in Vancouver, BC. TRIUMF is owned and operated by a consortium of (presently 21) Canadian universities.

{\bf TUCAN (TRIUMF UltraCold Advanced Neutron)}: Japan-Canada
collaboration performing a neutron electric dipole moment measurement
at TRIUMF.  Can alternatively refer to the ultracold neutron source
(TUCAN source), the nEDM experiment (TUCAN EDM experiment), or the
scientific collaboration (TUCAN collaboration).

{\bf TUDA (TRIUMF U.K. Detector Array)}: A detector designed to measure
the rates of nuclear reactions important in astrophysics, based at ISAC-I.

{\bf TUNL (Triangle Universities Nuclear Laboratory)}:  U.S. Dept. of Energy Center of Excellence, consisting of a consortium of Duke University, North Carolina Central University, North Carolina State University, and the University of North Carolina at Chapel Hill.

{\bf UCNs (UltraCold Neutrons)}: Extremely low-energy neutrons ($<$ 300
neV) which can be trapped in material, magnetic, and gravitational
fields.

{\bf UCNtau (UltraCold Neutron tau)}: An experiment conducted at LANL to measure the neutron lifetime.

{\bf UNESCO (United Nations Educational, Scientific and Cultural Organization)}



{\bf WIMP (Weakly Interacting Massive Particle)}: one of the hypothesized possibilities for dark matter.


\end{document}